\documentclass[a4paper,titlepage,twoside,openright,12pt]{book}
\usepackage[british]{babel}
\usepackage[tbtags]{amsmath}
\usepackage{mathtools}
\usepackage{amstext}
\usepackage{booktabs}  
\usepackage[font = small, format=hang, label font={sf, bf}]{caption}
\usepackage{amssymb}
\usepackage{amsthm}
\usepackage{amsmath}
\allowdisplaybreaks 
\usepackage[utf8]{inputenc}
\usepackage{xcolor}
\usepackage{cite}
\usepackage{listings}
\usepackage{color}
\usepackage{tocbibind}  
\usepackage{verbatim}
\usepackage{float}
\usepackage[T1]{fontenc}
\usepackage{ color}
\usepackage{lmodern}
\usepackage{braket}
\usepackage{mathrsfs}
\usepackage{lipsum}
\usepackage[toc,page]{appendix}
\usepackage[outercaption]{sidecap}
\usepackage{graphicx}
\usepackage{color}
\usepackage{slashed}
\usepackage{url}
\usepackage{hyperref}
\numberwithin{equation}{section}
\hypersetup{
colorlinks=true,
linkcolor=blue,
linktocpage=true,
citecolor=red,
}
\usepackage[a4paper,
bindingoffset=0.6cm
]{geometry}
\usepackage{microtype}
\usepackage{emptypage}
\usepackage{transparent}
\usepackage{eso-pic}
\usepackage{titlesec, blindtext, color}
\definecolor{gray75}{gray}{0.75}
\definecolor{verdescuro}{RGB}{0,128,0}
 
\usepackage[ PetersLenny ]{ fncychap }
\ChNameVar{\Huge \bfseries \rm \color{\grey} }
\ChTitleVar{\Huge \bfseries \centering \color{\grey} }
\titleformat{\chapter}[hang]{\Huge\bfseries}{\textcolor{gray}{\thechapter}\hspace{5pt}\textcolor{gray75}{|}\hspace{9pt}}{0pt}{\Huge\bfseries}

\makeatletter
\renewcommand{\@chapapp}{}

\makeatother

 \newpagestyle{main}{
    \sethead[\thepage][\textmd{ \chaptertitle}][\thesection]  
            {\thesection}{\sectiontitle}{\thepage}
	\setheadrule{.15pt}
            } 

 \newpagestyle{intro}{
    \sethead[\thepage][\textmd{ Introduction and motivation}][] 
            {\thesection}{Introduction and motivation}{\thepage}
	\setheadrule{.15pt}
            } 

 \newpagestyle{app}{
    \sethead[\thepage][\textmd{ Appendices}][]  
            {}{Appendices}{\thepage}
	\setheadrule{.15pt}
            }

 \newpagestyle{biblio}{
    \sethead[\thepage][\textmd{ Bibliography}][]  
            {}{Bibliography}{\thepage}
	\setheadrule{.15pt}
            } 

\newpagestyle{conclusion}{
    \sethead[\thepage][\textmd{ Conclusions}][]  
            {}{Conclusions}{\thepage}
	\setheadrule{.15pt}
            } 
\newpagestyle{acknowledgement}{
    \sethead[\thepage][\textmd{ Acknowledgements}][]  
            {}{Acknowledgements}{\thepage}
	\setheadrule{.15pt}
            } 

\newpagestyle{ringraziamenti}{
    \sethead[\thepage][\textmd{ Ringraziamenti}][]  
            {}{Ringraziamenti}{\thepage}
	\setheadrule{.55pt}
            } 

\makeatletter
\renewcommand*\env@matrix[1][*\c@MaxMatrixCols c]{%
  \hskip -\arraycolsep
  \let\@ifnextchar\new@ifnextchar
  \array{#1}}
\makeatother

\newsavebox\MBox

\newlength{\myleftlen}

\begin{document}

\frontmatter

\newgeometry{margin=2cm}
\begin{titlepage}

\center 
\textsc{\large Sede amministrativa}\\[.5cm]
\textsc{\Large Universit\`a degli studi di Padova}\ \
\rule{\linewidth}{0.8pt} 
\\[0.3cm]

\vspace{0.3cm} 
\vspace{0.3cm} 
\textsc{\large Dipartimento di Fisica e Astronomia "G. Galilei"}\\

\vspace{1.5cm}
\centering{\transparent{0.5}\includegraphics[width=0.25\textwidth]{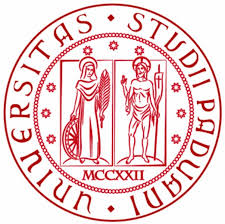}}
\vspace{1cm}

\textsc{\LARGE Corso di Dottorato di Ricerca in Fisica}\\[.5cm]
\textsc{\Large XXXI Ciclo}

\vspace{1cm}
\centering{\huge \bfseries \boldmath Non-linear realisations in global and\\[.5cm] local supersymmetry}\\ 
\vspace{2.5cm}

\begin{flushleft}
{\large
\textbf{Vice Coordinatore}\\
\textsf{Ch.ma Prof. Cinzia Sada}\\[.5cm]
\textbf{Supervisore}\\
\textsf{Ch.mo Prof. Gianguido Dall'Agata}\\
}
\end{flushleft}

\begin{flushright}
{\large
\textbf{Dottorando}\\
\textsf{Niccol\`o Cribiori}
}
\end{flushright}

\vfill

{\large \textsc{September 2018}}

\end{titlepage}

\restoregeometry

\pagestyle{plain}
\newpage $ $ 
\newpage $ $ 
\vspace{4cm}
\begin{flushright}
\emph{To Ica.\\
All she ever wanted,\\
was the ball back.}
\end{flushright}
\newpage $ $

\newpage

\begin{center}
{\textsc{\Large \bf Abstract} }\\[1cm]
\end{center}

In modern physics the role of symmetries is central and, even when they are broken, their remnants can pose constraints on the interactions. In considering effective descriptions of certain physical models, the symmetry group can be indeed spontaneously broken to some subgroup, but it is still possible to write symmetry transformations which leave a given action invariant. These transformations are usually realised non-linearly on the fields, while in the unbroken regime they act linearly.

The Standard Model of particle physics can be though of as an effective theory. In this respect, beyond the Standard Model scenarios have been proposed, of which supersymmetry is one of the best motivated at present. 
Having been yet no direct detection in the experiments, supersymmetry is postulated to be spontaneously broken at some energy scale above the TeV. For this reason, it is important to study supersymmetry breaking mechanisms and to consider low energy effective theories with spontaneously broken and non-linearly realised supersymmetry.
An essential ingredient of this class of models is the presence of a massless goldstone fermion in the spectrum, namely a goldstino, which is associated to each broken generator.

The subject of this thesis is the construction and the study of four-dimensional effective theories with spontaneously broken and non-linearly realised global and local supersymmetry. 
In the first part, the global supersymmetric case is analysed. The discussion starts from the supersymmetry breaking sector, describing the goldstino and its interactions, in the case of minimal supersymmetry and it is then generalised to a generic number of spontaneously broken supersymmetry generators.
A systematic procedure is given in order to construct effective theories with non-linearly realised supersymmetry and with any desired spectrum content.
In the second part of the thesis, non-linear realisations are analysed in the case of local supersymmetry, namely supergravity. The coupling of the goldstino sector to gravity is presented first and the superhiggs mechanism on a generic background is discussed. Matter couplings are then constructed in some simple examples. A new class of models is introduced in which supersymmetry is spontaneously broken and non-linearly realised already on the gravity sector. As a consequence, it is possible to construct actions in which the cosmological constant is bounded to be positive, which can be motivated for studying inflation. 
Two applications of non-linear realisations in local supersymmetry are discussed in detail. In the first one it is shown how to implement K\"ahler invariance in a way that mimics the global supersymmetric case. In particular, the K\"ahler--Hodge restriction on the scalar manifold, which is typical of supergravity, is avoided. In the second application the role of non-linear realisations in the construction of de Sitter vacua is discussed. Attention is devoted to the relationship with known de Sitter uplift constructions and with a new, recently proposed D-term in supergravity.

\newpage $ $
\newpage

\pagestyle{main}
\pagenumbering{Roman}
\ChNameVar{\Huge \bfseries \rm }
\ChTitleVar{\Huge \bfseries \rm }
\titleformat{\chapter}[hang]{\Huge\bfseries}{\textcolor{gray}{\thechapter}\hspace{5pt}\textcolor{gray75}{|}\hspace{9pt}}{0pt}{\Huge\bfseries}
\phantomsection
\tableofcontents

\mainmatter

\pagestyle{intro}

\chapter{Introduction and motivation}

Symmetry principles are central in modern physics. They represent an essential tool in the construction of theoretical models and they have revealed to be a fundamental mean in the understanding of Nature. Particles, for example, can be classified according to their symmetry properties while, in the Standard Model of particle physics, interactions are described by gauge symmetries and the causal structure of spacetime is captured by Lorentz symmetry. This is the most accurate theory of Nature we have been able to develop so far.

Symmetries are important also when they are broken. Even in those cases in which symmetry breaking occurs, in fact, symmetries or their remnants can still pose constraints on the model under investigation and they can be most conveniently used as an organising principle.
To understand better the point, suppose that a given theoretical model possesses a certain amount of symmetry in some high energy regime. In a realistic scenario however, the experimental access to such regime could be very challenging. 
In this situation, physical aspects of the theory can nevertheless be tested by constructing an effective description for it, which captures information about the low energy region, where experiments can be performed. In practice this is obtained by reducing the number of degrees of freedom of the original theory and by keeping only those whose affection by high energy phenomena is negligible. To quantitatively govern the description, a cut-off parameter can be introduced, which bounds the regime of validity of the effective theory and suppresses its interactions. 
It is therefore not surprising that, together with the number of degrees of freedom, also the original amount of symmetry can be reduced, along the procedure. The symmetry group is said to be spontaneously broken to some subgroup and the transformations become non-linearly realised, in contrast to the unbroken case in which they are realised linearly on the fields of the theory.   
In the presence of such a type of breaking, the information on the original symmetry might not be completely lost, as it should be possible in principle to follow the reverse path and understand how to reconstruct the original theory from its effective description. Such bottom-up approach can be tough to pursue, but the analysis of the symmetry in its non-linearly realised form can give precious hints along this direction. 
The concepts of spontaneous symmetry breaking and that of effective theory are therefore intimately related and the interest in them is motivated both from a theoretical and a phenomenological perspective: by studying the low energy description, where symmetries are non-linearly realised and where experimental data could be available, information can be gained about the nature of the complete theory in the high energy regime, which might not be accessible experimentally.

The Standard Model itself can be viewed as an effective theory. Even though it is renormalisable and thus it could be in principle properly applied at any energy scale, phenomena are known which are not described within it. The most vivid reason for this interpretation is perhaps that gravity is excluded a priori in any of the interactions. The ultimate problem in theoretical physics at present is indeed to understand which is the form of the theory, in the high energy regime, from which the Standard Model might be originated. To this purpose, several proposals have been formulated and are currently under investigation: they are generically named Beyond the Standard Model scenarios. In the so called Grand Unification Theories, for example, the gauge group of the Standard Model is embedded into a larger symmetry group, which is supposed to be spontaneously broken at low energies. In this class of models, different gauge coupling constants become unified in the unbroken phase, as the gauge group at high energies is unique by construction. The simplest versions of such models however, seem to be ruled out, as they predict a lifetime for the proton which is in tension with the current experimental bounds. Beside gravity, the Standard Model cannot describe other phenomena for which there is experimental evidence, as dark matter and neutrino oscillations. A further aspect, which is not satisfactory from a more theoretical perspective, is the presence of a large gap between the electroweak breaking scale and the natural cut-off of the theory, namely the Planck scale. The difference indeed is roughly of seventeen orders of magnitude. The electroweak breaking scale, moreover, is related to the mass of the Higgs boson, which is the only particle of spin zero in the Standard Model. Being a scalar, it receives corrections to the mass growing quadratically with the energy and in principle running up to the cut-off of the theory. Dramatic cancellations among these corrections have therefore to occur, in order to reproduce the measured value of the Higgs mass, which is extremely small if compared to the Planck scale.

Supersymmetry is one of the best motivated beyond the Standard Model scenarios, as it provides a mechanism to explain the occurrence of such cancellations, but it can also suggest possible candidates for dark matter particles and it allows for the unification of the gauge couplings. It is a spacetime symmetry, extending the Poincar\'e group and relating bosons to fermions. In addition, theories with local supersymmetry automatically include gravity among their interactions and, for this reason, they are called theories of supergravity. From a more theoretical point of view, supersymmetry is an essential ingredient of string theory, which is at present one of the best proposals for a complete theory of Nature.

At the present stage, there is no direct evidence for supersymmetry at the energy regime probed by the experiments. Since its theoretical motivations are solid, however, supersymmetry is still not discarded, but it is supposed to be spontaneously broken at some scale above the TeV. In the spirit of the previous discussion, it is therefore important to study the supersymmetry breaking mechanism and, more in general, to consider low energy effective theories with broken and non-linearly realised supersymmetry, in order to extract information concerning the fully supersymmetric theory in the high energy regime.

This thesis is about the construction and the studying of four-dimensional effective theories with spontaneously broken and non-linearly realised global and local supersymmetry. Both technical and physical aspects of the subject are presented. In particular, a systematic recipe to build models with any desired particle content in the low energy is given. This prescription is the starting point for the studying of systems of phenomenological interest and some applications are also discussed within this work. 

\vfill

This thesis is based on the following publications by the author.

\begin{itemize}
\item[\bf I] \emph{Interactions of $\mathcal{N}$ Goldstini in Superspace},\\ with G. Dall'Agata and F. Farakos,\\ Phys.\ Rev.\ D {\bf 94} (2016) no.6,  065019, \href{https://arxiv.org/abs/1607.01277}{arXiv:1607.01277 [hep-th]};
\item[\bf II] \emph{Minimal Constrained Supergravity},\\ with G. Dall'Agata, F. Farakos and M. Porrati,\\ Phys.\ Lett.\ B {\bf 764} (2017) 228, \href{https://arxiv.org/abs/1611.01490}{arXiv:1611.01490 [hep-th]};
\item[\bf III] \emph{From Linear to Non-linear SUSY and Back Again},\\ with G. Dall'Agata and F. Farakos,\\ JHEP {\bf 1708} (2017) 117, \href{https://arxiv.org/abs/1704.07387}{arXiv:1704.07387 [hep-th]};
\item[\bf IV] \emph{Fayet--Iliopoulos terms in supergravity without gauged R-symmetry},\\ with F. Farakos, M. Tournoy and A. van Proeyen,\\
JHEP {\bf 1804} (2018) 032, \href{https://arxiv.org/abs/1712.08601}{arXiv:1712.08601 [hep-th]}.
\end{itemize}

The author also contributed to
\begin{itemize}
\item[\bf V] \emph{On the off-shell formulation of N=2 supergravity with tensor multiplets},\\
with G. Dall'Agata,\\
JHEP {\bf 1808} (2018) 132, \href{https://arxiv.org/abs/1803.08059}{arXiv:1803.08059 [hep-th]}. 
\end{itemize}

Part of the material in the thesis has never been published before.

\section{Supersymmetry and its breaking}
Supersymmetry is a spacetime symmetry which relates bosons and fermions, viewing them as a single object: a supersymmetric multiplet. It is generated by a spin-1/2 parameter, which can become a local function of spacetime. In this case supersymmetry becomes local and it is called supergravity. According to a result of Haag, Lopuszanski and Sohnius, within the framework of quantum field theory, supersymmetry is the only extension of the Poincar\'e group which can be a consistent symmetry of the S-matrix.

For what concerns interactions, supersymmetric theories are usually more constrained with respect to non-supersymmetric ones and in general they display certain common features. For example, a given spectrum is said to be supersymmetric if it contains the same number of bosons and fermions and if particles which are related by supersymmetry transformations, namely superpartners, have the same mass. The mass degeneracy within supersymmetric multiplets can have important consequences, when calculating quantum corrections. Since in supersymmetric theories pairs of particles exist with the same mass, but with opposite statistic, cancellations occur in the calculation of loop diagrams. When applied to a supersymmetric extension of the Standard Model, these cancellations can therefore give a solid explanation for the smallness of the predicted masses in the theory, like for example the one of the Higgs boson, with respect to the Planck scale. In other words, supersymmetry can give a motivated solution to the so called hierarchy problem of the Standard Model. 
Supersymmetric extensions of the Standard Model require moreover the existences of particles yet to be discovered: at least all the superpartners of the observed ones, in a minimal setup. Some of these additional particles, for example, can be sensible candidates for dark matter.
These are among the main reasons why supersymmetry is at present one of the best motivated beyond the Standard Model scenarios. 

Supersymmetry is not observed in the experiments at present. For this reason it is justified the assumption that, if realised in Nature, supersymmetry has to be spontaneously broken above the TeV scale and therefore the mechanism of spontaneous supersymmetry breaking becomes of phenomenological interest.
The class of models with spontaneously broken supersymmetry can be large but, since there should exist a regime of validity in which supersymmetry is restored, even in the broken phase certain common features can be present.
The essential ingredient of any spontaneous breaking of supersymmetry is the presence in the spectrum of a massless goldstone fermion, which is called goldstino, whose supersymmetry transformation is non-linear on the fields and non-homogeneous. More in general, spectra can contain a different number of bosons and fermions and the mass degeneracy among superpartners can be removed by the presence of some mass gap. In any case, supersymmetry still constraints the form of the interactions, since it is possible to define supersymmetry transformations which leave a given action invariant.

\section{Effective theories with spontaneously broken supersymmetry}
Given the general properties of models with spontaneously broken supersymmetry, one can wonder which is the simplest example which realises them.
In 1973 Volkov and Akulov \cite{Volkov:1973ix} proposed the following Lagrangian for a single spin-1/2 massless field $\lambda_\alpha$:
\begin{equation}
\label{LVA}
\mathcal{L}_{VA}=-f^2 - i\left(\lambda \sigma^m \partial_m \bar\lambda-\partial_m\lambda\sigma^m\bar\lambda\right)+\mathcal{O}(f^{-2}),
\end{equation}
where $f$ is a parameter of mass dimension 2, which is suppressing the interactions. This Lagrangian is left invariant by the transformation
\begin{equation}
\label{susyLVA}
\delta_\epsilon \lambda_\alpha = f \epsilon_\alpha - \frac if\left(\lambda \sigma^m \bar\epsilon-\epsilon \sigma^m\bar\lambda\right)\partial_m\lambda_\alpha,
\end{equation} 
which is generated by a spin-1/2 parameter $\epsilon_\alpha$ and it consists of a non-homogeneous part, namely the first term, and a non-linear part in $\lambda_\alpha$, namely the second term.
It can be checked that the transformation \eqref{susyLVA} closes the algebra of supersymmetry
\begin{equation}
[\delta_\epsilon,\delta_\eta]\lambda_\alpha = 2i
(\epsilon\sigma^m\bar\eta-\eta \sigma^m \bar \epsilon)\partial_m
\lambda_\alpha.
\end{equation}

The motivation behind this model was the interpretation of neutrinos, which at that time were believed to be massless, as Goldstone modes of a certain spontaneously broken symmetry. Since neutrinos are fermions, however, such symmetry has to be generated by a fermionic parameter. One of the very first appearences of supersymmetry was therefore at the spontaneously broken and non-linearly realised level.
Within this interpretation, the fermion $\lambda_\alpha$ is the goldstino and the supersymmetry breaking scale is given by $\sqrt f$. 

The model \eqref{LVA} presents a non-supersymmetric spectrum, since it does not contain bosonic degrees of freedom, and it is organized as a power series in the parameter $f$. It is therefore tantalising to think of \eqref{LVA} as an effective theory, where the cut-off is some scale $\Lambda_\text{cut-off}\lesssim \sqrt f$. This intuition is correct, even though at this stage it is not clear what could be the form of the theory in the high energy regime. It turns out eventually that the model proposed by Volkov and Akulov is universal in the sense that, as it is going to be discussed further in the thesis, it captures (part of) the low energy information of any generic model with spontaneously broken supersymmetry. In other words, the class of Lagrangians whose (deep) infrared description is given by \eqref{LVA} is infinite.

Before proceeding with the discussion, a technical comment is in order. In the original paper, the derivation of \eqref{LVA} employed a geometric method. Given a goldstino with supersymmetry transformation \eqref{susyLVA},  the following differential form can be defined
\begin{equation}
\label{VA1form}
dx^m {A_m}^a = dx^m \left[\delta_m^a - \frac{i}{f^2}\partial_m\lambda\sigma^a\bar\lambda + \frac{i}{f^2}\lambda\sigma^a\partial_m\bar\lambda\right],
\end{equation}
in terms of which the Volkov--Akulov Lagrangian \eqref{LVA} is
\begin{equation}
\label{LVA2}
\mathcal{L}_{VA} = -f^2 \det {A_m}^a
\end{equation}
and the invariance under supersymmetry follows from the fact that the determinant transforms as a density, namely
\begin{equation}
\label{deltadetA}
\delta_\epsilon \det {A_m}^a = -\frac{i}{f}\partial_n\left[\left(\lambda \sigma^n \bar\epsilon-\epsilon \sigma^n\bar\lambda\right)\det {A_m}^a\right]. 
\end{equation}

The construction can be extended to include other fields which are usually called matter, in order to distinguish them from the goldstino. The result is a systematic strategy to build models with spontaneously broken and non-linearly realised supersymmetry, which generalises the so called CCZW procedure adopted in the case of bosonic symmetries \cite{Callan:1969sn}. The underlying logic is the following: given a non-supersymmetric Lagrangian, supersymmetry can be realised non-linearly on it by introducing a goldstino, together with its sector \eqref{LVA}, and by replacing the other fields with some redefined ones, the field redefinitions including appropriate goldstino interactions.
The precise recipe is discussed in several publications and it is not going to be reviewed within this work. This procedure is systematic and general enough to reproduce any desired model, nevertheless it could be not completely satisfactory, as the supersymmetric structure of the theory is not transparent. The reason is rooted in the fact that, within this approach, operations are performed onto components of supersymmetric multiplets, but the structure of the multiplets themselves is lost from the very beginning. 

The situation can be different when using other techniques. Superspace, for example, is a very convenient tool for the investigation of supersymmetric theories. In this formulation, fields are embedded into superfields and operations are performed onto superfields as a whole. Supersymmetry is therefore manifest at every step of the computations. It is quite remarkable then that superspace methods can be used also in the case in which supersymmetry is broken and non-linearly realised. In particular, after the work of \cite{Rocek:1978nb,Lindstrom:1979kq}, it has been understood that, by imposing certain constraints on superfields, it is possible to eliminate some of their components and to construct representations of supersymmetry with a different number of bosons and fermions.
For these reasons, superspace is the formalism adopted in this thesis, in the conventions of \cite{WessBagger}.

\section{Supergravity}

The very existence of gravity is perhaps the most urgent reason why beyond the Standard Model scenarios have to be pursued on. In this respect, supersymmetry seems to be a very promising candidate, as it can lead automatically to an extension of General Relativity which is left invariant by supersymmetry transformations.
The algebra of a generic number $\mathcal{N}$ of supersymmetry generators $Q_\alpha^I$ and $\bar Q_{\dot\beta \, J}$, where $I,J=1,\ldots,\mathcal{N}$, closes in fact on the generator $P_m$ of spacetime translations
\begin{equation}
\{ Q_\alpha^I, \bar Q_{\dot\beta \, J}\} = \delta^I_J \sigma^m_{\alpha \dot\beta} P_m.
\end{equation}
As a consequence, a theory invariant under local supersymmetry is automatically invariant under local spacetime translations and it is therefore a theory of gravity. It is called supergravity and it reduces to General Relativity when all the amount of supersymmetry is broken. It requires the existence of a spin-3/2 particle, which is the superpartner of the graviton and it is called gravitino.
Theories of supergravity appear also in ten and eleven dimensions, as the classical limit of string and M-theory respectively.

For these reasons, supergravity is a natural framework to study the effects of gravity on the four-dimensional world, in particular in the low energy region. One of the main open problems in this context is the explanation, from a theoretical perspective, of the observed value of the cosmological constant \cite{Aghanim:2018eyx}: $\Lambda \sim 10^{-122}$ in reduced Planck Units. Besides the problem of explaining its smallness, even more compelling is perhaps the urge for motivating its positive sign. It is known that any de Sitter background, namely a solution of Einstein equations with positive cosmological constant, breaks supersymmetry. 
The mechanism of supersymmetry breaking within supergravity becomes then of phenomenological interest while, from a theoretical level, what occurs is a supersymmetric version of the Brout-Englert-Higgs mechanism which predicts a massive gravitino.

The starting point for constructing models of supergravity with spontaneously broken supersymmetry is to generalise the goldstino sector \eqref{LVA} to curved space, coupling it to the graviton and to the gravitino \cite{Farakos:2013ih,Bergshoeff:2015tra,Hasegawa:2015bza}. Matter fields can then be added and a possible strategy is to operate at the component level. For the same reasons expressed before, however, working in a manifestly supersymmetric setup could be preferable and therefore superspace techniques can be adopted, which are the generalisation of those used in the rigid case.  

A characteristic of supergravity theories is the presence of scalars in the spectrum, which can be natural candidates for the inflaton in applications to inflation and cosmology, for example. Dealing with the complete setup of supergravity can be technically demanding, as the number of fields in the spectrum is usually large, in particular in the case of extended supersymmetry. Once again, however, supersymmetry breaking comes to the rescue. If the spontaneous breaking of supersymmetry is assumed, in fact, an effective theory of supergravity can be produced by integrating out the massive fields and leaving only the light degrees of freedom. In this way the model becomes approachable, in principle, and predictions can be made. This is the main lesson of \cite{Antoniadis:2014oya,Ferrara:2014kva}, where non-linear realisations of supersymmetry are implemented in models of inflation, in order to eliminate undesired fields from the spectrum. These works inaugurated indeed the studying of inflation with non-linear supersymmetry, which is an active field of research at present.

\section{An example: Pions as goldstone bosons}
One of the purposes of this introduction is to show that the concepts of effective theory, spontaneous symmetry breaking and non-linear realisation of a certain symmetry are related one another. An explicit example is given in the present section for a purely bosonic system, namely a model describing pions, where a $\rm SO(4)$ global symmetry is spontaneously broken to $\rm SO(3)$. The rest of the thesis is going be devoted to the study of effective theories with spontaneously broken and non-linearly realised supersymmetry.

Consider the following $\rm SO(4)$-invariant Lagrangian
\begin{equation}
\label{eq:c1:Lpions}
\mathcal{L}=-\frac12 \partial_m \Phi_I\partial^m \Phi^I - V(\sigma),\qquad V(\sigma) = \frac{\mu^2}{2}\sigma^2+\frac{\lambda}{4}\sigma^4,
\end{equation}
where $\Phi_I$, with $I=1,\ldots,4$, are four fields transforming in the fundamental of $\rm SO(4)$, $\mu$ and $\lambda$ are parameters and the invariant quantity $\sigma^2=\Phi_I \Phi^I$ has been defined.  To study the theory perturbatively, a specific vacuum has to be chosen among the infinitely many admitted by the scalar potential $V(\sigma)$. Consider therefore the $\rm SO(3)$-invariant vacuum
\begin{equation}
\langle \Phi_I\rangle = v
\left(
\begin{array}{c} 
0\\ 0\\ 0\\ 1\\
\end{array}
\right),
\end{equation}
where $v=\sqrt{-\mu^2 / \lambda}$ and $\mu^2<0$ has been assumed. This vacuum is breaking spontaneously the original $\rm SO(4)$ global symmetry to its subgroup $\rm SO(3)$. By applying the Goldstone theorem and by counting the number of broken generators, three goldstone modes are expected in the theory. This is in agreement with the fact that three pions are going to be identified precisely with these goldstone bosons.

To proceed with the analysis and to read the spectrum, take the field $\Phi_I$ which is breaking the original symmetry and parametrise it as
\begin{equation}
\label{eq:c1:phipar}
\Phi_I = \Pi_I\left(v+\rho(x)\right).
\end{equation}
The logic underlying this parameterisation is the following: the degrees of freedom associated to the will-be goldstone modes, which are embedded into the four fields $\Pi_I$, are being separated from all the remaining degrees of freedom, represented in this simple example by the sole field $\rho(x)$. To preserve the number of degrees of freedom on both sides of \eqref{eq:c1:phipar}, the field $\Pi_I$ has to be constrained in order that not all of its components are independent, but one of them is function of the others. The parameterisation \eqref{eq:c1:phipar} therefore has to be supplemented with the condition $\Pi_I \Pi^I=1$, which can be solved in terms of $\Pi_4 = \sqrt{1-\Pi_i^2}$.   
Because of this constraint, the symmetry transformations on the unconstrained $\Pi_i$, which represent the pions, act non-linearly
\begin{equation}
\Pi_i \quad \longrightarrow \quad \Pi_i^\prime = 
{\Lambda_i}^j\Pi_j+ {\Lambda_i}^4\sqrt{1-\Pi_i^2}, \qquad {\Lambda_I}^J \in {\rm SO(4)}.
\end{equation}             
In other words, as a consequence of the spontaneous breaking the symmetry transformations on the goldstone modes become non-linearly realised.  On the contrary, the constrained fields $\Pi_I$ allow to describe the spontaneously broken symmetry in a linear way. Inserting the parameterisation \eqref{eq:c1:phipar} into the original Lagrangian \eqref{eq:c1:Lpions} gives
\begin{equation}
\begin{aligned}
\mathcal{L} =&-\frac12 (v+\rho)^2\left(\delta_{ij}+\frac{\Pi_i\Pi_j}{1-\Pi_k^2}\right) \partial_m \Pi^i \partial^m \Pi^j\\
&-\frac12 \partial_m \rho \partial^m \rho-\lambda v^2 \rho^2-\lambda v \rho^3-\frac{\lambda}{4}\rho^4,
\end{aligned}
\end{equation}
It is now possible to read the spectrum and realise that the field $\rho$ is massive, with mass $m_{\rho}=\sqrt{2 \lambda}v$, while the fields $\Pi_i$ are massless. The result is consistent with the interpretations of the pions $\Pi_i$ as goldstone bosons. After the insertion of the parameterisation \eqref{eq:c1:phipar} into the Lagrangian therefore, the mass gap in the spectrum has become evident.

At this point an effective theory can be produced from this model with spontaneously broken symmetry. Since $\rho$ is massive, it can be integrated out by restricting the analysis to an energy regime $E\ll m_\rho$. An equivalent procedure would be to take the formal limit in which $\lambda\to \infty$, with $v$ fixed, and the mass of $\rho $ becomes infinite. The equations of motion for $\rho$ in the zero-momentum limit are solved by $\rho=0$ and, after rescaling $\Pi_i \rightarrow \Pi_i/v$ the following effective Lagrangian is obtained
\begin{equation}
\label{eq:Leffpions}
\mathcal{L}_{eff}=-\frac12 g_{ij} \partial_m \Pi^i \partial^m \Pi^j, 
\end{equation}
where 
\begin{equation}
g_{ij}=\delta_{ij}+\frac{1}{v^2}\frac{\Pi_i\Pi_j}{1-\frac{\Pi_k^2}{v^2}}
\end{equation}
is the metric of the non-linear sigma model which captures the low energy information of the original system. The first corrections to the Lagrangian \eqref{eq:Leffpions}  are terms $\mathcal{O}\left(1/\lambda\right)$, which can be safely neglected for all the purposes of the effective description.

To conclude the analysis, notice that the parameter $v$ is still present in the effective theory and it gives information on the scale at which the symmetry breaking occurs. The Lagrangian can be expanded as a power series in $1/v^2$, for $v\to\infty$, to obtain
\begin{equation}
\label{eq:Leffpions2}
\mathcal{L}_{eff} =- \frac12 \partial^m \Pi_i\partial_m \Pi^i-\frac{1}{2v^2}(\Pi_i \partial_m \Pi^i)^2+\mathcal{O}\left(\frac{1}{v^4}\right),
\end{equation}
which has a strong similarity with the Volkov--Akulov Lagrangian \eqref{LVA}.
At this level, the particular model \eqref{eq:c1:Lpions} from which the analysis started is not relevant anymore and, in order to write the most general effective theory, in the expression $\mathcal{O}\left(1/v^4\right)$ can be inserted any interaction which is invariant under $\rm SO(3)$.

\clearpage

\pagestyle{main}

\chapter{The supersymmetry breaking sector}
\label{c1:susybreakingsec}

The analysis of theories with spontaneously broken and non-linearly realised supersymmetry is initiated. The attention is devoted first to global and then to local supersymmetry.
In this chapter the supersymmetry breaking sector in rigid supersymmetry is described. It is known as goldstino sector and it is a common ingredient of any theory with spontaneously broken supersymmetry.
Using superspace techniques, models with a goldstino are constructed and it is shown then how the corresponding Lagrangians can be related one another. A superspace formulation of the Volkov--Akulov model \eqref{LVA} is going to be given as well. 
The minimal supersymmetric case is analysed first and then the discussion is generalised to extended supersymmetry.

\section{The $\mathcal{N}=1$ goldstino in superspace}
In this section the supersymmetry breaking sector is described in the case of minimal supersymmetry. This is going to be the essential building block for all the rest of the discussion.

\subsection{Decoupling the sgoldstino} 
\label{cap1:sec:Adecoupling}
For the sake of simplicity, start from considering a chiral superfield $X$, which is an irreducible representation of supersymmetry. It can be expressed as an expansion in the coordinates $\theta^\alpha$, $\bar \theta_{\dot \alpha}$ of $\mathcal{N}=1$ superspace as
\begin{equation}
X = A + \sqrt 2 \theta^\alpha G_\alpha + \theta^2 F
\end{equation} 
and it contains a complex scalar $A$, a Weyl fermion $G_\alpha$ and a complex auxiliary field $F$. With only this ingredient, the simplest Lagrangian that breaks supersymmetry has a canonical K\"ahler potential and a linear superpotential:
\begin{equation}
\begin{aligned}
\mathcal{L} &= \int d^4\theta X \bar X +\left(f \int d^2\theta X + c.c.\right)\\
&= -\partial_m A\partial^m\bar A - i\bar G\bar \sigma^m\partial_m G +  F \bar F +f F+ f \bar F,
\end{aligned}
\end{equation}
where the parameter $f$ is assumed to be real, without loss of generality.
The field $F$ is called auxiliary because it is not propagating, namely it does not have a kinetic term, nevertheless it can be used to break supersymmetry. Indeed by taking its equations of motion, $F=-f$, and by integrating it out, a (constant) positive definite scalar potential is produced
\begin{equation}
\mathcal{L} =  -\partial_m A\partial^m\bar A - i\bar G\bar \sigma^m\partial_m G-f^2
\end{equation}
and supersymmetry is spontaneously broken at the scale $\sqrt f$. This is confirmed by the fact the supersymmetry transformation of the fermion $G_\alpha$ becomes non-homogenous
\begin{equation}
\delta_\epsilon G_\alpha = -\sqrt 2 f \epsilon_\alpha + \sqrt 2i \,(\sigma^m \bar\epsilon)_\alpha\,\partial_m A
\end{equation}
and therefore the field is a goldstone mode of the broken supersymmetry, or a goldstino. 
A model with spontaneous breaking of supersymmetry and with a massless goldstino in the spectrum has therefore been described.

Following the logic of the example of the pions, it should be possible to construct an effective theory from this model, by integrating out the massive degrees of freedom. At this stage, however, both the scalar $A$ and the fermion $G_\alpha$ are massless. The mass of the fermion, moreover, is protected by the goldstone theorem. The only field which can acquire mass is then the sgoldstino $A$. By modifying the original Lagrangian and by adding a curvature term in the K\"ahler potential directly in superspace, a mass for $A$ can be generated. Consider therefore the model
\begin{equation}
\begin{aligned}
\label{eq:c1:Lmod}
\mathcal{L} &= \int d^4\theta \left(X \bar X-\frac{1}{\Lambda^2} X^2 {\bar X}^2\right) +\left(f \int d^2\theta X + c.c.\right)\\
&=-\partial_m A\partial^m\bar A - i \left(G\sigma^m\partial_m \bar G\right) + F \bar F +f (F+\bar F)\\
&\phantom{=.}-\frac{1}{\Lambda^2}\bigg[G^2 \bar G^2 - 2 G^2 \bar A \bar F - 2 \bar G^2 A F+4 A\bar A F\bar F\\
&\phantom{\frac{1}{\Lambda^2}aaa} - 4 A\bar A \partial_m A\partial^m \bar A+4i A\partial_m \bar A (\bar G\bar \sigma^m G)+4i A\bar A (\partial_m \bar G\bar\sigma^m G) \bigg]
\end{aligned}
\end{equation}
where $\Lambda$ is a parameter of mass dimension one. 
To shed light on the relation between the off-shell and the on-shell structure, it could be useful to rewrite the component expansion as 
\begin{equation}
\begin{aligned}
\label{eq:c1:Lmod2}
\mathcal{L} &= -\left(1-\frac{4 A\bar A}{\Lambda^2}\right)\partial_m A\partial^m\bar A - i \left(1-\frac{4 A\bar A}{\Lambda^2}\right)\left(G\sigma^m\partial_m \bar G\right)\\
&\phantom{=.}+\frac{4i}{\Lambda^2}A\partial_m\bar A(G\sigma^m \bar G)-\frac{G^2\bar G^2}{\Lambda^2}-\frac{2f}{\Lambda^2}\frac{G^2\bar A + \bar G^2 A}{1-\frac{4 A\bar A}{\Lambda^2}}-\frac{4}{\Lambda^4}\frac{G^2\bar G^2 A\bar A}{1-\frac{4 A\bar A}{\Lambda^2}}\\
&\phantom{=.}-\mathcal{V}(A,\bar A)+\left|F\sqrt{1-\frac{4 A\bar A}{\Lambda^2}}+\frac{f}{\sqrt{1-\frac{4 A\bar A}{\Lambda^2}}}+\frac{2}{\Lambda^2}\frac{G^2\bar A}{\sqrt{1-\frac{4 A\bar A}{\Lambda^2}}}\right|^2,
\end{aligned}
\end{equation}
where 
\begin{equation}
\mathcal{V}(A,\bar A) = \frac{f^2}{1-\frac{4 A\bar A}{\Lambda^2}}
\end{equation}
is the scalar potential. Using this form of the Lagrangian in fact, the integration of the auxiliary fields is immediate: it is sufficient to drop the term with the squared modulus. 
The modification of the K\"ahler potential produces a mass for $A$, which is $m_A^2=4f^2/\Lambda^2$, and creates a mass gap in the spectrum.
By restricting the analysis to an energy regime $E\ll {f / \Lambda}$, or equivalently by taking the formal limit $1/\Lambda \to \infty$, the scalar $A$ can be integrated out. Its equation of motion in the zero momentum limit gives
\begin{equation}
\label{eq:c1:solA}
A = \frac{G^2}{2F}.
\end{equation}
Notice the two important facts that this solution does not depend on the details of the theory in the high energy regime, namely on the parameter $\Lambda$, and that it requires $\langle F \rangle\neq 0$ for its consistency. This last requirement, in particular,  is a signal of spontaneous supersymmetry breaking in the vacuum. The meaning of the expression \eqref{eq:c1:solA} is that, in the low energy regime, the massive scalar $A$ can be integrated out and eliminated from the theory by replacing it with a composite expression built out of the remaining degrees of freedom, namely the goldstino and the auxiliary field which is breaking supersymmetry.
Inserting \eqref{eq:c1:solA} into \eqref{eq:c1:Lmod} gives 
\begin{equation}
\label{eq:c1:Leffoffshell}
\mathcal{L}_{eff} = -i \left(G\sigma^m\partial_m \bar G\right) + F \bar F +f (F+\bar F)+ \frac{\bar G^2}{2\bar F}\partial^2  \frac{G^2}{2F}
\end{equation}
This is an effective theory with spontaneously broken supersymmetry, which describes the interactions of the goldstino. The spectrum is manifestly not supersymmetric, as it contains only one fermion. Notice that the information on the theory in the high energy regime, which is encoded into the parameter $\Lambda$, disappeared after the insertion of \eqref{eq:c1:solA}.
The Lagrangian \eqref{eq:c1:Leffoffshell} is left invariant by the supersymmetry transformations
\begin{align}
\delta_\epsilon G_\alpha &= \sqrt 2 F\epsilon_\alpha +  \sqrt 2i\,  (\sigma^m \bar\epsilon)_\alpha \,  \partial_m \left(\frac{G^2}{2F}\right),\\
\delta_\epsilon F &= \sqrt 2 i \, \bar\epsilon\bar\sigma^m\partial_m G,
\end{align}
which are non-linearly realised on the fields. By integrating out the auxiliary field $F$, whose equation of motion is
\begin{equation}
\label{eq:c1:eomF}
F = -f -\frac{1}{4 f^3}\bar G^2\partial^2 G^2+\frac{3}{16 f^7}\bar G^2 G^2\partial^2\bar G^2\partial^2 G^2,
\end{equation}
the following on-shell Lagrangian can be obtained
\begin{equation}
\label{eq:c1:Leffonshell}
\mathcal{L}_{eff} = -f^2 + i \partial_m\bar G\bar\sigma^m G + \frac{1}{4f^2}\bar G^2\partial^2 G^2-\frac{1}{16f^6}G^2\bar G^2\partial^2 G^2\partial^2\bar G^2.
\end{equation}
This is invariant under the non-homogeneous and non-linear supersymmetry transformation
\begin{equation}
\delta_\epsilon G_\alpha = \sqrt 2 F\epsilon_\alpha + i \sqrt 2 (\sigma^m \bar\epsilon)_\alpha\, \partial_m \left(\frac{G^2}{2F}\right),
\end{equation}
where $F$ is given by \eqref{eq:c1:eomF}. These calculations can be performed iteratively and by using the anticommuting properties of the field $G_\alpha$. For example, the variation of the Lagrangian \eqref{eq:c1:Leffoffshell} with respect to $F$ gives
\begin{equation}
\label{c1:eq:eomF2}
F = - f +\frac{\bar G^2}{2 \bar F^2}\partial^2 \frac{G^2}{2F}.
\end{equation}
From this one can calculate iteratively
\begin{equation}
\begin{aligned}
\frac{G^2}{2F} &= -\frac{G^2}{2f}\left(1-\frac{\bar G^2}{2f \bar F^2}\partial^2 \frac{G^2}{2F}\right)^{-1} = -\frac{G^2}{2f}+\frac{1}{4 f^5}G^2 \bar G^2 \partial^2 G^2
\end{aligned}
\end{equation}
and
\begin{equation}
\begin{aligned}
\frac{G^2}{2F^2} &= \frac{G^2}{2f^2}\left(1-\frac{\bar G^2}{2f \bar F^2}\partial^2 \frac{G^2}{2F}\right)^{-2} = \frac{G^2}{2f^2}-\frac{1}{4 f^6}G^2 \bar G^2 \partial^2 G^2
\end{aligned}
\end{equation}
which, inserted into \eqref{c1:eq:eomF2}, give \eqref{eq:c1:eomF}.

The model \eqref{eq:c1:Leffonshell} has been derived from \eqref{eq:c1:Lmod} by decoupling the sgoldstino $A$. It describes the interactions of the goldstino in the low energy regime and, in this respect, it is similar to the Volkov--Akulov model \eqref{LVA}. The Lagrangian \eqref{eq:c1:Leffonshell} in fact can be recast into the form of \eqref{LVA} with a field redefinition, as it is showed in \cite{Kuzenko:2010ef, Kuzenko:2011tj} at the component level and as it is going to be proved in superspace in the following. This is a manifestation of the universality of the Volkov--Akulov model, namely the fact that any system with spontaneous breaking of supersymmetry contains a goldstino sector which can be described by \eqref{LVA}.

\subsection{The constraint $X^2=0$}
\label{cap1:sec:X2=0}
In the example of the pions and in the previous supersymmetric model, the standard procedure for constructing an effective theory has been employed, which requires the integration of the massive modes through the solution of their equations of motion. Since the interest in this thesis is in supersymmetric systems, superspace methods can be used in order to outline a more efficient strategy.

The main result of the previous example is that, in the low energy limit, the scalar $A$ is integrated out and replaced by the specific combination
\begin{equation}
\label{eq:c1:solA2}
A = \frac{G^2}{2F}.
\end{equation}
The crucial step, at this point, is to observe that it is possible to construct a superfield which contains the fermion $G_\alpha$ and the auxiliary field $F$ and whose lowest component is given by $G^2/(2F)$. Such a superfield turns out to be a chiral superfield $X$ satisfying an additional constraint. In analogy with the example of the pions in fact, in which the constraint $\Pi_I\Pi^I=1$ has been used to remove one of the four components of $\Pi_I$, this additional constraint is going to eliminate the scalar in $X$ and replace it with the particular combination given by \eqref{eq:c1:solA2}.
The constraint to be imposed on $X$ to remove the scalar $A$ is \cite{Rocek:1978nb, Casalbuoni:1988xh}
\begin{equation}
\label{eq:c1:X2=0}
X^2=0.
\end{equation}
This is indeed a supersymmetric constraint and its solution is a superfield. It is possible to solve \eqref{eq:c1:X2=0} directly in superspace, by acting on the left with superspace derivatives. The action of $D^2$ gives
\begin{equation}
D^\alpha X D_\alpha X +X D^2 X=0,
\end{equation}
which, assuming that $\langle D^2X|\rangle \neq 0$, can be solved for
\begin{equation}
\label{eq:c1:solX}
X = -\frac{D^\alpha XD_\alpha X}{D^2X}.
\end{equation}
By taking the projection of this solution onto the surface $\theta=\bar\theta=0$, the expression \eqref{eq:c1:solA2} is recovered. This means  that \eqref{eq:c1:solX} is the superfield whose lowest component is \eqref{eq:c1:solA2}. Its higher components can be found by acting with superspace derivatives and projecting then to spacetime
\begin{align}
D_\alpha\left(-\frac{D^\beta XD_\beta X}{D^2X}\right)\bigg|=D_\alpha X|&=\sqrt 2 \, G_\alpha,\\
D^2\left(-\frac{D^\alpha XD_\alpha X}{D^2X}\right)\bigg|=D^2 X|&=-4F.
\end{align}
Viceversa, given a generic chiral superfield $X$, in the case in which the condition $\langle D^2X|\rangle \neq 0$ holds, it is always possible to construct a superfield 
\begin{equation}
-\frac{D^\alpha XD_\alpha X}{D^2X} \equiv X_{NL},
\end{equation}
such that
\begin{equation}
\begin{aligned}
D_\alpha X_{NL} &= D_\alpha X,\\
D^2 X_{NL} &= D^2 X.
\end{aligned}
\end{equation}
It can be checked then that this superfield is chiral and it satisfies the nilpotent constraint $X_{NL}^2=0$. Notice finally that, by acting on the constraint with $D_\alpha$, the relation $X D_\alpha X=0$ is obtained. This is not a new constraint, but it is a consistency condition for the solution \eqref{eq:c1:solX}.
To summarise, it has been found that the chiral superfield $X$ such that
\begin{equation}
X^2 = 0 \quad \Longleftrightarrow \quad X = \frac{G^2}{2F}+\sqrt 2 \theta^\alpha G_\alpha +\theta^2 F,
\end{equation}
contains only one fermion and one auxiliary field. Since the auxiliary field must be non-vanishing everywhere on the vacuum, $\langle D^2X|\rangle =\langle -4F\rangle\neq 0$ , in order for the superfield to be well defined, supersymmetry is spontaneously broken and the fermion $G_\alpha$ is a goldstino. It has also been shown explicitly that, by imposing constraints on superfields, it is possible to remove some of their components and express them as functions of the remaining ones. This is the general logic behind the so called constrained superfield approach.

Given a nilpotent chiral superfield $X$, the simplest Lagrangian to consider is
\begin{equation}
\label{eq:c1:LXNL}
\mathcal{L} = \int d^4 \theta X \bar X + \left(f \int d^2\theta X + c.c.\right), \qquad X^2=0.
\end{equation}
As stressed above, for the constraint $X^2=0$ to be imposed consistently, it is essential that supersymmetry is spontaneously broken by the auxiliary field $F$ in $X$, as it is the case in \eqref{eq:c1:LXNL}. 
To calculate the component expansion of this model, it is possible to proceed in the usual manner and to substitute then $A=\frac{G^2}{2F}$ in the final expression. The result is 
 \begin{equation}
 \mathcal{L} =-i \left(G\sigma^m\partial_m \bar G\right) + F \bar F +f (F+\bar F)+ \frac{\bar G^2}{2\bar F}\partial^2  \frac{G^2}{2F}
 \end{equation}
and it corresponds precisely to the Lagrangian \eqref{eq:c1:Leffoffshell}, which is invariant under non-linearly realised supersymmetry transformations and which has been obtained as an effective description of a system where the sgoldstino has been decoupled. Notice that, with the use of constrained superfields, such an effective theory has been constructed without the need for solving the equation of motion of the massive scalar in the zero momentum limit. This is one of the advantages of the constrained superfields approach, namely the possibility of obtaining effective theories avoiding all the potential complications of the standard procedure. In addition, as in the example of the pions, the constraint $X^2=0$ effectively allows the use of the language and of the tools developed for linear supersymmetry for studying instead models in which supersymmetry is spontaneously broken and non-linearly realised. The constraint can also be imposed at the Lagrangian level using a Lagrange multiplier, as it is discussed in \cite{Ferrara:2016een}.
In \cite{Komargodski:2009rz} the superfield $X$ is conjectured to be the infrared limit 
of the superfield violating the supercurrent conservation equation in a supersymmetry-breaking setup, but this conjecture might not be valid in more general systems, as it going to be discussed in the subsection \ref{cap2:subsec:supercurrent}.

\subsection{On the relation with other goldstino superfields}
\label{subsec:equivalenceVA}
At this stage, two different descriptions of the same goldstino model have been introduced. The one that appeared first, namely the Volkov--Akulov model, has been constructed at the component level, while the one with the nilpotent chiral superfield $X$ has been embedded into superspace from the very beginning. Other descriptions for the goldstino sector are known, as for example those studied in \cite{Kuzenko:2011ti, Farakos:2014iwa, Farakos:2015vba}, and the relations among some of them are going to be explored in this subsection. The purpose is to take advantage of superspace in order to show that these descriptions are all equivalent, up to field redefinitions. 
As a consequence of this result, the formulation of the supersymmetry breaking sector in terms of a nilpotent chiral superfield $X$ is going to be adopted for the rest of the thesis and without loss of generality.\footnote{There is actually a subtle point in this statement, which is the following. In this section the equivalence is going to be proved only for models which are known. In principle then, one cannot exclude the existence of a yet undiscovered description of the goldstino sector which might not be related to the Volkov--Akulov, or one could not be convinced that such equivalence holds also in the presence of other superfields, besides the goldstino. The question about the generality of the Volkov--Akulov or, equivalently, of the nilpotent chiral superfield construction, is going to be addressed again in section \ref{cap2:sec:alwaysX2=0}, where it is going to be given evidence for the fact that, in any generic model in which supersymmetry is spontaneously broken, it is always possible to recombine the degrees of freedom in order to let a chiral superfield $X$, such that $X^2=0$, appear in the low energy regime.\label{footgenX2}} The reason behind this choice is that, in a modern language, chiral superfields are an essential ingredient in the construction of supersymmetric models, since both the K\"ahler potential and the superpotential in general are functions of them.

The first step is to promote the Volkov--Akulov description to superspace. This requires the introduction of a superfield with a spinor index. At first instance it might seem quite exotic, but such a superfield is going to be an essential ingredient for passing from one description of the goldstino sector to another and it is going to provide also a new superspace goldstino Lagrangian on its own.
Consider therefore a spinor superfield $\Lambda_\alpha$ satisfying the constraints 
\begin{eqnarray}
\begin{split}
\label{eq:c1:VASWrepr}
D_\alpha \Lambda_\beta & = f\,\epsilon_{\beta\alpha}+\frac{i}{f}\,\sigma^m_{\alpha\dot\beta}\bar\Lambda^{\dot\beta}\partial_m\Lambda_\beta,\\
\bar D_{\dot \alpha} \Lambda_\beta & = -\frac{i}{f}\,\Lambda^\rho \sigma^m_{\rho\dot\alpha}\partial_m\Lambda_\beta,
\end{split}
\end{eqnarray}
where $f$ is a parameter of mass dimension one, which eventually is associated to the supersymmetry breaking scale. This superfield $\Lambda_\alpha$ is closely related to the one introduced by Ivanov--Kaputsnikov \cite{Ivanov:1978mx} and by Samuel--Wess in \cite{Samuel:1982uh}, as it is going to be discussed in a while. More details on the Samuel--Wess formalism are contained in the appendix \ref{appASW}.
Due to the constraints \eqref{eq:c1:VASWrepr}, which are effectively eliminating component fields, the only independent degree of freedom in $\Lambda_\alpha$ is the Volkov--Akulov goldstino in the lowest component
\begin{eqnarray}
\Lambda_\alpha | = {\lambda_\alpha},
\end{eqnarray}
and its supersymmetry transformation is given precisely by \eqref{susyLVA}. The other components can be found by acting with superspace derivatives and projecting onto the surface $\theta=\bar\theta=0$. For example, some of them are
\begin{equation}
\begin{aligned}
D_\alpha\Lambda_\beta | &= f \epsilon_{\beta \alpha}+\frac{i}{f}\,\sigma^m_{\alpha\dot\rho}\bar\lambda^{\dot\rho}\partial_m\lambda_\beta,\\ 
\bar D_{\dot\alpha}\Lambda_\beta | &= -\frac{i}{f}\,\lambda^\rho \sigma^m_{\rho\dot\alpha}\partial_m\lambda_\beta,\\
D^2\Lambda_\alpha| &= -\frac{1}{f^2}\partial^m(\bar\lambda^2\partial_m\lambda_\alpha)+\frac{2}{f^2} {({\bar\sigma}^{mn})^{\dot\gamma}}_{\dot\beta}(\bar\lambda_{\dot\gamma}\partial_m\bar\lambda^{\dot\beta}-\partial_m{\bar\lambda}_{\dot\gamma}\bar\lambda^{\dot\beta})\partial_n\lambda_\alpha
\end{aligned}
\end{equation}
and notice that they are all functions of $\lambda_\alpha$, as expected.
With this new ingredient it is possible to define the superspace analogous of the matrix $A_m^a$ appearing in \eqref{VA1form} to be
\begin{equation}
\mathbb{A}_m^a = \delta_m^a-\frac{i}{f^2}\partial_m\Lambda\sigma^a\bar\Lambda+\frac{i}{f^2}\Lambda\sigma^a\partial_m\bar\Lambda 
\end{equation}
and the superspace Lagrangian density associated to \eqref{LVA2} has then the form 
\begin{equation}
\label{LVAsuperspace}
\mathcal{L}=-f^2\det {\mathbb{A}_m}^a |.
\end{equation} 
The superfield $\Lambda_\alpha$ can be used also to write another supersymmetric Lagrangian in superspace, namely 
\begin{eqnarray}
\label{LSW}
{\cal L} = -\frac{1}{f^2} \int d^4 \theta \Lambda^2 \bar \Lambda^2 .
\end{eqnarray} 
It can be checked, at the component level, that this reduces to the Volkov--Akulov action \eqref{LVA}, however the equivalence between \eqref{LVAsuperspace} and \eqref{LSW} can be proved directly in superspace.  
Notice in fact that, due to the particular form of $\mathbb{A}_m^a$, it holds that
\begin{equation}
\int d^4\theta \Lambda^2\bar\Lambda^2 = \int d^4\theta \Lambda^2\bar\Lambda^2\det\mathbb{A}_m^a = \frac{1}{16} D^2 \bar D^2\left(\Lambda^2\bar\Lambda^2\det\mathbb{A}_m^a\right)|, 
\end{equation}
up to boundary terms. The first equality in particular follows because of the fact that terms in $\det\mathbb{A}_m^a$ containing either $\Lambda_\alpha$ or $\bar\Lambda_{\dot\alpha}$ are annihilated by the factor $\Lambda^2\bar\Lambda^2$ and only the constant term has an effective role in the computation. Concentrating then on the right hand side and acting with the covariant derivatives inside the parenthesis, the superfields $\Lambda_\alpha$ are removed and, from the properties
\begin{equation}
\begin{aligned}
D_\rho \det\mathbb{A}_m^a &= \frac{i}{f}\partial_m\left(\sigma^m_{\rho\dot\rho}\bar\Lambda^{\dot\rho}\det\mathbb{A}_m^a\right),\\
\bar D^{\dot{\rho}}\det\mathbb{A}_m^a &=\frac{i}{f}\partial_m\left(\bar\sigma^{m\,\dot{\rho}\rho}\Lambda_\rho\det\mathbb{A}_m^a\right),
\end{aligned}
\end{equation}
which are the superspace generalisation of \eqref{deltadetA},
the equivalence between the two Lagrangians follows up to total derivatives
\begin{equation}
\int d^4x d^4\theta\, \Lambda^2\bar\Lambda^2 = f^4 \int d^4x\det\mathbb{A}_m^a | \, .
\end{equation} 
To perform the calculation the following results have been used
\begin{align}
\bar D^{\dot\rho}(\Lambda^2\bar\Lambda^2 \det \mathbb{A}) &= 2\, \Lambda^2 \bar \Lambda^{\dot \rho}\det \mathbb{A},\\
\bar D^2(\Lambda^2\bar\Lambda^2 \det \mathbb{A}) &= -4\, \Lambda^2 \det \mathbb{A},\\
D_\alpha (\Lambda^2 \det \mathbb{A})&=2\, \Lambda_\alpha \det \mathbb{A} + \text{total derivative},\\
D^2(\Lambda^2 \det \mathbb{A})&=-4 \,\det \mathbb{A} + \text{total derivative}.
\end{align}

The relation between the Volkov--Akulov and the formulation in terms of a nilpotent chiral superfield is now investigated in superspace, with the explicit use of the superfield $\Lambda_\alpha$ as a tool to connect the two descriptions. 
Notice first of all that, by using $\Lambda_\alpha$, it is possible to define a chiral superfield
\begin{equation}
\begin{aligned}
\Phi &= \frac{1}{4f^3} \bar D^2 \left( \Lambda^2 \bar \Lambda^2 \right)\\ 
&=- f^{-3}\,\Lambda^2 \left( f^2 - i \partial_m \Lambda \sigma^m \bar \Lambda  
-  f^{-2}\bar \Lambda^2 \partial_m \Lambda \sigma^{mn} \partial_n \Lambda  \right). 
\end{aligned}
\end{equation}
The superfield $\Phi$ contains all the supersymmetry breaking information and it satisfies the constraints 
\begin{eqnarray}
\begin{aligned}
\label{Rocekconstr}
\Phi^2 &= 0, 
\\[0.1cm] 
\Phi \bar D^2 \bar \Phi &= 4 f\Phi,  
\end{aligned}
\end{eqnarray} 
which have been introduced in \cite{Rocek:1978nb}. 
Their role can be understood by assuming that $\Phi$ is an unconstrained chiral superfield. 
By imposing then the first constraint in \eqref{Rocekconstr}, as discussed before, the scalar component is removed from the spectrum, while by imposing the second the supersymmetry breaking scale is fixed. As usual, the constraints reduce the number of independent component fields and provide them as functions of the remaining ones.

The goldstino inside the superfield $\Phi$ is defined as the component $D_\alpha \Phi |$ and, since
\begin{eqnarray}
\label{DPhilambda}
D_\alpha \Phi | \equiv G^\Phi_\alpha= -2 \lambda_\alpha + \dots\,, 
\end{eqnarray}
where dots stand for higher order terms in $\lambda_\alpha$,
it is related to the Volkov--Akulov goldstino $\lambda_\alpha$ via a field redefinition. For this reason the supersymmetric Lagrangian for the constrained chiral superfield $\Phi$,
\begin{eqnarray}
{\cal L} =  f\int d^2 \theta \, \Phi ,
\end{eqnarray}
at the component level does not reduce directly to the Volkov--Akulov Lagrangian \eqref{LVA}, as the Lagrangian \eqref{LSW} does, 
because the goldstini have to be mapped into each other. 
The proper field redefinition can be found by inverting the relation \eqref{DPhilambda} and by finding $\lambda_\alpha$ in terms of $G^\Phi_\alpha$. Notice that such an inversion can always be performed, due to the nilpotent property of the fermionic field $\lambda_\alpha$.

The superfield $\Phi$ is similar to the nilpotent $X$, however it satisfies one constraint more with respect to the latter, since also its auxiliary field is eliminated. To relate the Volkov--Akulov description to the one with $X$, which contains also $F$ as independent component, a further step is needed.
It has been shown that the description in terms of $\Lambda_\alpha$ reproduces exactly the Volkov--Akulov Lagrangian, either in components or directly in superspace. In the original paper \cite{Samuel:1982uh}, Samuel and Wess introduce a spinor goldstino superfield $\Gamma_\alpha$ which is not satisfying \eqref{eq:c1:VASWrepr}, rather it satisfies the quite similar constraints 
\begin{eqnarray}
\label{SWchiralrepr}
\begin{split}
D_\alpha \Gamma_\beta & = 
\,f\,\epsilon_{\beta\alpha},
\\[0.1cm]
\bar D_{\dot \alpha} \Gamma_\beta & = -  
\,\frac{2i}{f}\,\Gamma^\rho \sigma^m_{\rho\dot\alpha}\partial_m\Gamma_\beta.
\end{split}
\end{eqnarray}
The relation between $\Lambda_\alpha$ and $\Gamma_\alpha$ is analogous to the one between the Volkov--Akulov goldstino $\lambda_\alpha(x)$ and the same goldstino expressed in the chiral coordinates $y^m = x^m + i \theta \sigma^m \bar \theta$, as discussed in the appendix \ref{appASW} or in \cite{Samuel:1982uh, Liu:2010sk}. 
The Lagrangian for $\Gamma_\alpha$ in superspace has the same structure as the one in \eqref{LSW}: 
\begin{equation}
{\cal L} = -\frac{1}{f^2} \int d^4\theta \, \Gamma^2 \, \bar\Gamma^2 . 
\end{equation}
To move from the description in terms of $\Lambda_\alpha$ to the one in terms of $\Gamma_\alpha$, one can set 
\begin{equation}
\label{LambdaGamma}
\Gamma_\alpha = - 2 f \, \frac{D_\alpha \bar D^2 ( \Lambda^2 \bar \Lambda^2 ) }{D^2 \bar D^2 ( \Lambda^2 \bar \Lambda^2 ) } 
\end{equation} 
and, using this relation, the equivalence between the two Lagrangians follows directly, 
because 
\begin{equation}
\Gamma^2\bar\Gamma^2 = \Lambda^2 \bar \Lambda^2 . 
\end{equation}

With this new ingredient, the relation between the Volkov--Akulov and the nilpotent chiral superfield $X$ can be proved. It can be checked in fact that, given $X$, one can construct \cite{Luo:2009ib}
\begin{equation}
\label{gamma}
\Gamma_\alpha = -2\,f\,\frac{D_\alpha X}{D^2 X},
\end{equation}
which satisfies the constraints \eqref{SWchiralrepr}, when $X^2=0$. The superspace Lagrangian \eqref{eq:c1:LXNL} for $X$ can now be written as 
\begin{equation}
{\cal L}  = \frac{1}{16 f^4} \int d^4 \theta \, \Gamma^2 \bar \Gamma^2 
\left( D^2 X \bar D^2 X  + 4 f D^2 X + 4 f \bar D^2 \bar X \right),
\end{equation} 
where $\Gamma_\alpha$ is given by  \eqref{gamma}.
Once the superspace integration is performed and up to boundary terms,  
this Lagrangian is equivalent to 
\begin{equation}
\label{XNL33}
{\cal L} = 
\left(  \mathbb{F} \, \overline{ \mathbb{F}} +  f  \, \mathbb{F} +  f \, \overline{ \mathbb{F}} \right) 
\det\mathbb{A}_m^a | ,
\end{equation} 
where the superfield 
\begin{equation}
\mathbb{F} = -\frac{1}{16 f^2}  (D - i \sigma^n \bar \Lambda \partial_n )^2 (\bar D + i \Lambda \sigma^n  \partial_n )^2 \left( X \, \bar \Gamma^2 \right)
\end{equation}
has been defined. As a consequence of these last steps, the auxiliary degrees of freedom, which are encoded into $\mathbb{F}$, have been separated from the rest of the fields and the Lagrangian has been written in the factorised form \eqref{XNL33}. 
The integration of the complex auxiliary scalar $\mathbb{F}|$ is straightforward and it gives
\begin{equation}
\mathbb{F}|=-f. 
\end{equation}
After substituting it back into the Lagrangian, \eqref{XNL33} reduces exactly to \eqref{LVAsuperspace}. 
This concludes the equivalence between the known goldstino descriptions and in particular between the models \eqref{eq:c1:LXNL} and \eqref{LVAsuperspace} . 
Notice that, to relate these two formulations, the integration of a complex scalar is expected, because the model with the nilpotent chiral superfield $X$ contains also the auxiliary field $F$ in the Lagrangian. More details can be found in the appendix B of \cite{Cribiori:2016hdz}, where the origin of the scalar superfield $\mathbb{F}$ is explained. 

To summarise, the chain of equivalences presented in this subsection proceeds as follows. The discussion has started from the original Volkov--Akulov description, which has been embedded into superspace after the introduction of the goldstino superfield $\Lambda_\alpha$. Using this ingredient it is possible to construct other two goldstino superfields, $\Phi$ and $\Gamma_\alpha$, in terms of which two equivalent formulations of the supersymmetry breaking sector are given. The description in terms of $\Gamma_\alpha$ is finally related to the one with the nilpotent chiral superfield $X$, but this last step requires the use of the equations of motion of the auxiliary degrees of freedom.

\section{Extended supersymmetry}

In this section the goldstino sector of theories with extended supersymmetry is investigated. The analysis starts from the case of $\mathcal{N}=2$ supersymmetry and then it is generalised to an arbitrary number $\mathcal{N}$ of spontaneously broken supersymmetries. The results are going to be organised eventually in the more familiar language of $\mathcal{N}=1$ superspace.
Notice that, for ${\cal N}>4$ supersymmetries, some superpartners of the goldstini are going to have spin higher than one. Due to the non-linear realisation, however, these fields can be removed in terms of the ${\cal N}$ goldstini, in the same way as the sgoldstino is removed from the spectrum in the ${\cal N}=1$ theory. In other words, in contrast with the typical situation, when supersymmetry is spontaneously broken and non-linearly realised, there is no upper bound on the number of fermionic generators admitted in the theory and $\mathcal{N}$ can be arbitrary. 
As the number of supersymmetries increases, the number of constraints to be imposed on the goldstini superfields increases as well, since more and more goldstini superpartners have to be eliminated. It turns out however that all the necessary information can be obtained by solving one single constraint in superspace, all the other constraints being consistency conditions. This is one of the results contained in \cite{Cribiori:2016hdz}.

The interest in models with a generic number of goldstini can be motivated by the fact that they have been used to describe composite quarks and leptons, under the assumption that the observed fermionic particles are goldstone modes of spontaneously broken supersymmetry \cite{Bardeen:1981df}. $\mathcal{N}$ of the Standard Model fermions, indeed, can be interpreted as pseudo-goldstini \cite{Liu:2016idz} and a Lagrangian invariant under $\mathcal{N}$ non-linearly realised supersymmetries can be employed in phenomenological investigations \cite{Bellazzini:2017bkb}.

\subsection{$\mathcal{N}=2$ supersymmetry}

The strategy is to promote the Samuel--Wess formalism, namely the description in terms of the superfield $\Lambda_\alpha$, or equivalently $\Gamma_\alpha$, to ${\cal N}=2$ superspace and to use this to find the appropriate constraints to be imposed on more familiar objects, as for example chiral superfields. 
The first part of the analysis is similar to \cite{Kandelakis:1986bj, Kuzenko:2011ya}, but it is worked out in a different superspace representation.
Once the minimal set of constraints is determined, the theory is reformulated in the language of ${\cal N}=1$ supersymmetry and the complete expression for the Lagrangian in terms of the nilpotent chiral superfield $X$ and other constrained superfield is given.
Total breaking of $\mathcal{N}=2$ supersymmetry with constrained superfields has been discussed also in \cite{Dudas:2017sbi}.

The algebra satisfied by the $\mathcal{N}=2$  superspace derivatives, without central charges, is
\begin{eqnarray}
	\begin{split}
		\{ D_\alpha , \bar D_{\dot \alpha} \} &=\{ \tilde D_\alpha , \bar{\tilde D}_{\dot \alpha} \} = -2i\, \sigma^m_{\alpha \dot \alpha} \partial_m , 
		\\[0.2cm]
		\{ D_\alpha , D_\beta \} &= \{ \tilde D_\alpha , \tilde D_\beta \}= \{ D_\alpha , \tilde D_\beta \} = \{ D_\alpha , \bar{\tilde D}_{\dot \beta} \}=  0 ,
	\end{split}
\end{eqnarray}
where $\tilde D_\alpha$ generates the second supersymmetry.
The first step consists in determining the minimal set of constraints needed to remove from the spectrum all the undesired component fields. 
When $\mathcal{N}=2$ supersymmetry is spon\-ta\-ne\-ous\-ly broken, two goldstini are present in the spectrum. To describe them in superspace, two spinor superfield $\Lambda_\alpha$ and $\tilde \Lambda_\alpha$ can be used, which satisfy the constraints 
\begin{eqnarray}
	\begin{split} 
		D_\alpha \Lambda_\beta & = f\,\epsilon_{\beta\alpha} +\frac{i}{f}\sigma^m_{\alpha\dot{\rho}}\bar{\Lambda}^{\dot\rho}\partial_m\Lambda_\beta \, ,  
		\qquad 
		\bar D_{\dot \alpha} \Lambda_\beta  = -\frac{i}{f}\, \sigma^m_{\rho \dot \alpha} \Lambda^{\rho} \partial_m \Lambda_\beta ,
		\\[0.1cm] 
		\tilde D_\alpha \tilde \Lambda_\beta & =f\, \epsilon_{\beta\alpha} +\frac{i}{f}\sigma^m_{\alpha\dot{\rho}}\bar{\tilde\Lambda}^{\dot\rho}\partial_m\tilde{\Lambda}_\beta \, ,
		\qquad 
		\bar{\tilde D}_{\dot \alpha} \tilde \Lambda_\beta  = -\frac{i}{f}\, \sigma^m_{\rho \dot \alpha} \tilde \Lambda^{\rho} \partial_m \tilde \Lambda_\beta 
	\end{split}
\end{eqnarray}
and 
\begin{eqnarray}
	\begin{split} 
		\tilde D_\alpha \Lambda_\beta & = \frac{i}{f}\sigma^m_{\alpha\dot{\rho}}\bar{\tilde{\Lambda}}^{\dot\rho}\partial_m\Lambda_\beta \, ,\phantom{-}
		\qquad 
		D_\alpha \tilde \Lambda_\beta  = \frac{i}{f}\sigma^m_{\alpha\dot{\rho}}\bar{\Lambda}^{\dot\rho}\partial_m\tilde{\Lambda}_\beta ,
		\\[0.1cm] 
		\bar{\tilde D}_{\dot \alpha} \Lambda_\beta & = -\frac{i}{f}\, \sigma^m_{\rho \dot \alpha} \tilde \Lambda^{\rho} \partial_m \Lambda_\beta \, , 
		\qquad 
		\bar D_{\dot \alpha} \tilde \Lambda_\beta  = -\frac{i}{f}\, \sigma^m_{\rho \dot \alpha} \Lambda^{\rho} \partial_m \tilde \Lambda_\beta .
	\end{split}
\end{eqnarray}
In analogy with the $\mathcal{N}=1$ case, $\Lambda_\alpha$ and $\tilde \Lambda_\alpha$ contain all the information on the breaking of supersymmetry.
The only independent component fields they have are the goldstini, defined as 
\begin{equation} 
	\Lambda_\alpha |_{\theta=\tilde \theta=0} =\lambda_\alpha \, , \qquad 
	\tilde \Lambda_\alpha |_{\theta=\tilde \theta=0} =\tilde{\lambda}_\alpha 
\end{equation}
and whose supersymmetry transformations are non-linearly realised 
\begin{eqnarray}
	\begin{split}
		\delta \lambda_\alpha  =& f\, \epsilon_\alpha-\frac{i}{f}\left(\lambda \sigma^m\bar\epsilon-\epsilon\sigma^m\bar\lambda\right)\partial_m\lambda_\alpha-\frac{i}{f}\left(\tilde\lambda \sigma^m\bar{\tilde\epsilon}-\tilde\epsilon\sigma^m\bar{\tilde\lambda}\right)\partial_m{\lambda}_\alpha,
		\\[0.1cm]
		\delta \tilde \lambda_\alpha  =&  f\,\tilde \epsilon_\alpha 
		-\frac{i}{f}\left(\tilde\lambda \sigma^m\bar{\tilde\epsilon}-\tilde\epsilon\sigma^m\bar{\tilde\lambda}\right)\partial_m \tilde \lambda_\alpha 
		-\frac{i}{f}\left(\lambda \sigma^m\bar\epsilon-\epsilon\sigma^m\bar\lambda\right)\partial_m \tilde \lambda_\alpha, 
	\end{split}
\end{eqnarray} 
where $\epsilon_\alpha$, $\tilde \epsilon_\alpha$ are the two supersymmetry parameters.
A supersymmetric and $\textup{U}(2)_R$-invariant Lagrangian for $\Lambda_\alpha$ and $\tilde \Lambda_\alpha$ can be written as a direct generalisation of \eqref{LSW} and it is
\begin{equation}
\begin{aligned}
	{\cal L} &=  -\frac{1}{f^6}\int d^4 \theta d^4\tilde\theta\, \Lambda^2\tilde{\Lambda}^2\bar\Lambda^2\bar{\tilde{\Lambda}}^2\\
 &= -f^2-i(\lambda\sigma^m\partial_m\bar\lambda-\partial_m\lambda \sigma^m\bar\lambda)-i(\tilde{\lambda}\sigma^m\partial_m\bar{\tilde\lambda}-\partial_m\tilde{\lambda} \sigma^m\bar{\tilde\lambda})+ O(f^{-2}) .
\end{aligned}
\end{equation}

In the previous section the superfield $\Lambda_\alpha$ has been used to construct a chiral goldstino superfield satisfying the constraints \eqref{Rocekconstr}. With the same logic, it is possible to define a $\mathcal{N}=2$ superfield $\Phi$\footnote{The same notation of the $\mathcal{N}=1$ case is used for this superfield. It is going to be clear from the context whether $\Phi$ is an object in $\mathcal{N}=1$, $\mathcal{N}=2$ or generic $\mathcal{N}$ superspace.}
\begin{equation}
\begin{aligned}
\Phi &= -\frac{1}{16f^7} 
	\bar D^{2} \bar{ \tilde D}^{2}  
	\left( \Lambda^{2}  \bar \Lambda^{2} \tilde \Lambda^{2}  \bar{ \tilde \Lambda}^{2} \right)\\
&=-f^{-5}\Lambda^2 \tilde \Lambda^2 \left( f^2 - i \partial_a \Lambda \sigma^a \bar \Lambda 
	- i \partial_a \tilde \Lambda \sigma^a \bar{ \tilde \Lambda}   + O(f^{-2}) \right),
\end{aligned}
\end{equation}
which is chiral,
\begin{eqnarray}
	\bar D_{\dot \alpha} \Phi = \bar{\tilde D}_{\dot \alpha} \Phi =0,
\end{eqnarray}
and, as it can be checked directly by inspection, it satisfies the following set of constraints
\begin{equation}
	\label{phi2}
	\begin{aligned}
		\Phi^2 &= 0,
		\\[0.09cm]
		\Phi D_\alpha  \Phi &= \Phi \tilde D_\alpha  \Phi   = 0 ,
		\\[0.1cm]
		\Phi \tilde D_\alpha D_\beta \Phi & 
		= \Phi \tilde D_\alpha \tilde D_\beta \Phi 
		= \Phi D_\alpha D_\beta \Phi = 0 ,
		\\[0.11cm]
		\Phi \tilde D_\alpha D_\beta \tilde D_\gamma \Phi & 
		= \Phi \tilde D_\alpha  D_\beta D_\gamma \Phi=0,
	\end{aligned}
\end{equation}
together with 
\begin{equation}
\label{eq:c1:constrfPhi2}
	\bar \Phi D^2 \tilde D^2 \Phi = -16\,f \, \bar \Phi .
\end{equation}
As a consequence of these constraints, the only independent fields in $\Phi$ are the two goldstini, which can be found in the components
\begin{equation}
	\begin{aligned}
		-\frac14 D^2 \tilde D_\alpha \Phi |_{\theta=\tilde \theta=0} \equiv G^\Phi_\alpha=& -2 \tilde \lambda_\alpha + \dots ,
		\\
		-\frac14 \tilde  D^2  D_\alpha \Phi |_{\theta=\tilde \theta=0} \equiv \tilde G^\Phi_\alpha=& -2 \lambda_\alpha  + \dots ,
	\end{aligned}
\end{equation}
where dots stand for terms with more fermions. Notice that, as anticipated, the number of constraints has increased with respect to the description in the $\mathcal{N}=1$ case.
The $\mathcal{N}=2$ Lagrangian for $\Phi$ can be written as 
\begin{equation}
	\label{lagrangian2phi} 
	{\cal L} = f \int d^2 \theta d^2 \tilde \theta \, \Phi .
\end{equation}

At this stage the crucial observation is that, as for the analogous discussion in the previous section, the constraint \eqref{eq:c1:constrfPhi2} implies that the highest component of the superfield $\Phi$, namely the auxiliary field breaking supersymmetry in the vacuum, is removed and expressed in terms of the supersymmetry breaking scale and of the remaining fields. By imposing therefore only the set of constraints \eqref{phi2}, without \eqref{eq:c1:constrfPhi2},  to a generic ${\cal N}=2$ chiral superfield $\mathcal{X}$, all of its components are removed, except the goldstini and, this time, even the auxiliary field which acquires the non-vanishing vacuum-expectation-value. This is indeed the generalisation to $\mathcal{N}=2$ supersymmetry of the field content of the $\mathcal{N}=1$ nilpotent chiral superfield $X$. The strategy that is going to be adopted is the following: to get the desired constrained superfield description, the set of constraints \eqref{phi2}, without \eqref{eq:c1:constrfPhi2}, is going to be imposed to a generic $\mathcal{N}=2$ chiral superfield.
It is important to notice however that, as proved in \cite{Cribiori:2016hdz}, only the last constraint in \eqref{phi2} is really essential, in the sense the its solution solves also the remaining constraints, which indeed are just consistency conditions.
Consider therefore a ${\cal N}=2$ chiral superfield ${\cal X}$ 
\begin{eqnarray}
	\bar D_{\dot \alpha} {\cal X}  = \bar{\tilde D}_{\dot \alpha} {\cal X}=0 .
\end{eqnarray} 
Imposing the constraint
\begin{equation}
	{\cal X} \tilde D_\alpha D_\beta \tilde D_\gamma {\cal X}
	= {\cal X} \tilde D_\alpha  D_\beta D_\gamma {\cal X}=0 
\end{equation}
and solving it in superspace, a unique solution can be obtained
\begin{equation}
	\label{XN2sol}
	{\cal X}=\frac{(D^2\tilde D_\alpha \mathcal{X})^2(\tilde D^2 D_\alpha \mathcal{X})^2}{(D^2\tilde D^2 \mathcal{X})^3}\equiv\frac{1}{4}\frac{G^2\,\tilde{G}^2}{{\cal F}^3}.
\end{equation}
In this formula, which is the analogous of \eqref{eq:c1:solX}, the ${\cal N}=2$ superfield ${\cal F}$ is defined such that
\begin{eqnarray}
	\label{FN2}
	F={\cal F}|_{\theta=\tilde \theta=0} = \frac{1}{(-4)^2} D^2 \tilde D^2 {\cal X} |_{\theta=\tilde \theta=0}
\end{eqnarray}
is the complex scalar auxiliary field breaking supersymmetry, while the superfields $G_{\alpha}$ and $\tilde{G}_\alpha$ are defined such that
\begin{eqnarray}
	\label{GN2}
	\begin{split}
		g_\alpha=\tilde G_\alpha|_{\theta=\tilde \theta=0} =&  -\frac{1}{4 \sqrt 2} D^2 \tilde D_\alpha {\cal X} |_{\theta=\tilde \theta=0},
		\\
		{\tilde{g}}_\alpha=G_\alpha |_{\theta=\tilde \theta=0} =&  - \frac{1}{4 \sqrt 2} \tilde D^2 D_\alpha {\cal X} |_{\theta=\tilde \theta=0} ,
	\end{split}
\end{eqnarray}
are two the goldstini.
From the explicit form of the solution \eqref{XN2sol} and from the properties \eqref{FN2}, \eqref{GN2} one can realise that all the component fields in ${\cal X}$
are effectively solved in terms of the auxiliary scalar ${F}$ and of the goldstini $g_\alpha$, $\tilde g_\alpha$. In the chiral coordinates $y^m = x^m  + i \theta \sigma^m \bar \theta + i \tilde \theta \sigma^m \bar{ \tilde \theta}$, in fact, ${\cal X}$ can be expressed as 
\begin{equation}
	\begin{aligned}
		{\cal X} & = \frac{g^2\,\tilde{g}^2}{4F^3}+\frac{\tilde{g}^2\, g^\alpha}{\sqrt{2}F^2}\theta_\alpha+\frac{{g}^2\, \tilde{g}^\alpha}{\sqrt{2}F^2}\tilde{\theta}_\alpha +\frac{\tilde{g}^2}{2F}\theta^2 +  2\,\frac{\tilde{g}^\alpha\tilde{\theta}_\alpha\,g^\beta\theta_\beta }{F}+\frac{g^2}{2F}\tilde{\theta}^2\\[0.1cm]
		        & +\sqrt{2}\,\tilde g^\alpha\tilde{\theta}_\alpha \theta^2+ \sqrt{2}\,g^\alpha\theta_\alpha \tilde{\theta}^2+F\,\theta^2\tilde{\theta}^2.
	\end{aligned}
\end{equation}
The Lagrangian for the supersymmetry breaking sector is
\begin{eqnarray}
	\label{lagrangian2X}
	{\cal L} = \int d^4 \theta d^4 \tilde \theta \, {\cal X} \bar {\cal X} 
	+ \left( f \int d^2 \theta d^2 \tilde \theta \, {\cal X}  + c.c. \right). 
\end{eqnarray}
To get the explicit goldstini action, after projecting to components one has to solve the equation of motion for the auxiliary field $F$ via an iterative procedure, giving 
\begin{eqnarray}
	\label{X2F}
	{ F} = - f + \text{fermions}, 
\end{eqnarray}
and then replace the solution back into \eqref{lagrangian2X}. Additional details on this Lagrangian are going to be given in the following subsection, where the results are expressed in the language of $\mathcal{N}=1$ superspace.

\subsection{$\mathcal{N}=2$ goldstini in $\mathcal{N}=1$ superspace}
\label{cap1:sec:N2inN1}

The previous analysis has been performed entirely at the full $\mathcal{N}=2$ superspace level, which might appear not completely transparent at first glance. It is therefore instructive to rephrase the results in terms of the more familiar language of $\mathcal{N}=1$ superspace, which is the one adopted in most of the applications. In this respect, it turns out that there are different possibilities for describing the two goldstini, depending on the choice of the superfields in which they are embedded. As a consequence of the discussion in subsection \ref{subsec:equivalenceVA} (see also the footnote \ref{footgenX2}), it is assumed without loss of generality that one goldstino resides into a nilpotent chiral superfield. Three descriptions for the other goldstino are given then using different constrained superfields. The analysis can be relevant in the case one is interested in studying the possible ultraviolet completions of the model in terms of $\mathcal{N}=2$ multiplets.
For a better understanding of the constraints imposed on the superfields in this subsection, the result that are going to be exposed in section \ref{sec:generalconstr} are needed. Even though some of the formulas are not completely proved within this section, the logic underlying them is explained in order for the discussion to be self-contained as much as possible.

The first step is to write the $\mathcal{N}=2$ chiral superfield $\mathcal{X}$ in terms of a set of $\mathcal{N}=1$ superfields, among which a nilpotent chiral $X$ is inserted. One possibility is to expand ${\cal X}$ as a series in the coordinates $\tilde \theta$, corresponding to the second supersymmetry:
\begin{equation}
	\label{XinNis1}
	{\cal X} = S(y,\theta) + \sqrt 2 \,  \tilde \theta^\beta W_\beta (y,\theta) + \tilde \theta^2 X(y,\theta)  ,
\end{equation} 
where $S$, $W_\alpha$ and $X$ are $\mathcal{N}=1$ chiral superfields
\begin{equation}
	\bar D_{\dot \alpha} S = 0 , \qquad \bar D_{\dot \alpha} W_\alpha = 0 , \qquad \bar D_{\dot \alpha} X = 0.
\end{equation}
Notice that $W_\alpha$ in \eqref{XinNis1} is not the superfield strength of a vector superfield $V$, namely it cannot be expressed as $W_\alpha = -\frac14 \bar D^2 D_\alpha V$, since it does not satisfy the superspace Bianchi identities $D^\alpha W_\alpha= \bar D_{\dot \alpha} \bar W^{\dot \alpha}$.
The expansion above is reminiscent of the analogous for a $\mathcal{N}=1$ chiral superfield: the lowest component is a complex scalar, then there is a fermion and finally another complex scalar. In this case, however, these objects are all $\mathcal{N}=1$ superfields.
The first supersymmetry acts on them as usual,
\begin{eqnarray}
	\delta_1 {\cal O} = \epsilon^\alpha D_\alpha {\cal O}  + \bar \epsilon_{\dot \alpha} \bar D^{\dot \alpha} {\cal O} ,    
\end{eqnarray}
and one can derive from here the supersymmetry transformations of the component fields.
The second supersymmetry acts by transforming the $\mathcal{N}=1$ superfields into each other 
\begin{equation}
	\label{secondsusy}
	\begin{aligned}
		\delta_2 S &=\sqrt 2 \tilde \epsilon^\alpha W_\alpha,
		\\[0.1cm]
		\delta_2 W_\alpha&= \sqrt 2 i \sigma_{\alpha \dot \alpha}^m \bar{\tilde \epsilon}^{\dot \alpha} \partial_m S + \sqrt 2 \tilde \epsilon_\alpha X ,
		\\[0.1cm]
		\delta_2 X &= \sqrt 2 i \bar{\tilde \epsilon}_{\dot \alpha} \bar \sigma^{m \dot \alpha \alpha} \partial_m W_\alpha .
	\end{aligned}
\end{equation} 
The auxiliary field acquiring a non-vanishing vacuum-expectation-value is now expressed as 
\begin{equation}
	{ F}  = -\frac14  D^2 X |   
\end{equation}
and, from the supersymmetry transformations
\begin{equation}
	\begin{aligned}
		\delta_1 g _\alpha  &= \frac{1}{\sqrt 2} \delta_1 D_\alpha X | = \sqrt 2 \epsilon_\alpha { F} + \dots  ,  
		\\
		\delta_2 \tilde g _\alpha  &= -\frac{1}{4} \delta_2 D^2 W_\alpha | = \sqrt 2 \tilde \epsilon_\alpha { F} + \dots ,  
	\end{aligned}
\end{equation}
it can be understood that the goldstini are accommodated inside the superfields $X$ and $W_\alpha$. The Lagrangian \eqref{lagrangian2X} can be written as
\begin{equation}
\label{lagrangian2XN=1}
{\cal L} = 
	\int d^4 \theta \left( X \bar X 
	- \partial_m S\partial^m\bar S
	+ i \partial_m W^\alpha \sigma^m_{\alpha \dot \alpha} \bar W^{\dot \alpha}  \right)+ f \left( \int d^2 \theta X + c.c. \right) .
\end{equation}

Inserting the explicit expression \eqref{XinNis1} in \eqref{phi2}, a large number of constraints emerges for the $\mathcal{N}=1$ superfields
\begin{equation}
	\label{constr1}
	\begin{aligned}
		S^2=W^2=X^2 &= 0 ,
		\\
		S D_\beta X&= 0 ,
		\\
		W_\alpha D_\beta X  &=0, 
		\\
		W^\alpha D^2 W_\alpha &= 2S D^2 X  .
	\end{aligned}
\end{equation} 
As proved in \cite{Cribiori:2016hdz}, however, some of these constraints are not essential, rather they are just consistency conditions, the set of essential constraints being considerably smaller:
\begin{align}
\label{fondconstrN2inN1_1}
X^2&=0,\\ 
\label{fondconstrN2inN1_2}
S &=\frac{X}{2} \frac{ (D^2 W)^2}{(D^2 X)^2},\\
\label{fondconstrN2inN1_3}
W_\alpha &= X \frac{D^2 W_\alpha }{D^2 X}.
\end{align}
Their role is going to be more clear after the results of chapter \ref{cap:mattsec}, but for the time being it is possible to interpret them in the following way.  The first constraint has already been discussed in the previous section. The second constraint implies that the full superfield $S$ is removed and expressed in terms of $W_\alpha$ and $X$, containing in turn the goldstini and the auxiliary field ${F}$. This is to some extent similar to the fate of the lowest component $A$ which, in the $\mathcal{N}=1$ case, has been entirely removed from the spectrum by the constraint $X^2=0$. The interpretation of the third constraint is less transparent at this level. A better insight can be gained by acting on it with superspace derivatives and projecting the result to spacetime. Being a generic chiral superfield, $W_\alpha$ has some $\theta-$expansion. Acting with $D^2$ on 
\eqref{fondconstrN2inN1_3} gives an identity, which means that the component field $D^2W_\alpha|\equiv \zeta_\alpha$ is not removed from the superfield. This is going to be in fact the goldstino of the second supersymmetry. Acting with only $D_\alpha$ gives
\begin{equation}
D_\alpha W_\beta|=-\frac{\sqrt 2G_\alpha}{4F}\zeta_\beta,
\end{equation}
which means that the component $D_\alpha W_\beta|$ is removed from the superfield and expressed as a function of the two goldstini $G_\alpha, \zeta_\alpha$ and of the auxiliary field $F$ which breaks supersymmetry. The same reasoning applies to the lowest component
\begin{equation}
W_\alpha| = -\frac{G^2}{8F^2}\zeta_\alpha,
\end{equation}
which is removed as well.
To sum up, as a consequence of \eqref{fondconstrN2inN1_1}, \eqref{fondconstrN2inN1_2} and \eqref{fondconstrN2inN1_3}  the superfield $S$ is entirely removed, $W_\alpha$ contains only a goldstino as independent component, while $X$ contains the other goldstino and the auxiliary field which breaks supersymmetry.

The Lagrangian \eqref{lagrangian2XN=1} can now be written in the ${\cal N}=1$ constrained superfield language, replacing $S$ with its expression in terms of $W_\alpha$ and $X$.
The result is the following low energy theory with $\mathcal{N}=2$ spontaneously broken supersymmetry 
\begin{equation}
\begin{aligned}
	\label{N2LL}
	{\cal L} = 
	&\int d^4 \theta \left( X \bar X 
	- \Big{|}\partial_m \left( \! \frac{X}{2} \frac{ (D^2 W)^2}{(D^2 X)^2} \right) \Big{|}^2
	+ i \partial_m W^\alpha \sigma^m_{\alpha \dot \alpha} \bar W^{\dot \alpha}  \right)\\ 
	&+ f \left( \int d^2 \theta X + c.c. \right) .
\end{aligned}
\end{equation}
Notice in fact that, on top of the manifest $\mathcal{N}=1$ supersymmetry, \eqref{N2LL} has a second supersymmetry given by 
\begin{eqnarray}
	\label{secondsusynilpotent}
	\begin{split}
		\delta_2 W_\alpha&= \sqrt 2 i \sigma_{\alpha \dot \alpha}^m \bar{\tilde \epsilon}^{\dot \alpha} \partial_m \left( \! \frac{X}{2} \frac{ (D^2 W)^2}{(D^2 X)^2} \right) 
		+ \sqrt 2 \tilde \epsilon_\alpha X ,
		\\
		\delta_2 X &= \sqrt  2  i \bar{\tilde \epsilon}_{\dot \alpha} \bar \sigma^{m \dot \alpha \alpha} \partial_m W_\alpha ,
	\end{split}
\end{eqnarray} 
and since the superfields $W_\alpha$ and  $X$ are constrained, this second supersymmetry is non-linearly realised.
In component form the Lagrangian \eqref{N2LL} reduces to 
\begin{equation}
\begin{aligned}
	\label{N2LLL}
	{\cal L} = &{F} \bar {F} + f {F} + f \bar { F}  
	+ i \partial_m g^\alpha \sigma^m_{\alpha \dot \alpha} \bar g^{\dot \alpha}
	+ i \partial_m \tilde g^\alpha \sigma^m_{\alpha \dot \alpha} \bar{ \tilde g}^{\dot \alpha}\\ 
	&+ \text{higher order fermion terms}.
\end{aligned}
\end{equation}
As a final step one can integrate out $F$ and replace its expression into \eqref{N2LLL}, obtaining a theory with only goldstini.

A possible description of the two goldstini in $\mathcal{N}=1$ superspace has been given in terms of the chiral superfields $X$ and $W_\alpha$. As the Volkov--Akulov model can be described in different ways, which are equivalent up to field redefinitions, the same logic is expected to apply also in this case. There should exist therefore other descriptions of the same system which use different superfields and which should be related by a superfield redefinition. In the following two of them are presented.

Given $W_\alpha$ and $X$, one can define the spinor superfield 
\begin{equation}
\label{eq:c1:HdefW}
	H_{\dot \alpha} = \frac{\bar D^2 \bar W_{\dot \alpha}}{\bar D^2 \bar X} .
\end{equation} 
This is a chiral superfield,
\begin{equation}
	\bar D_{\dot \beta} H_{\dot \alpha}  = 0 ,
\end{equation}
satisfying the property
\begin{equation}
	\label{XH}
	\bar D_{\dot \beta} \left( X \bar H_{\dot \alpha}  \right) = 0  .
\end{equation}
Since $H_{\dot \alpha}$ is chiral, it is known \cite{Komargodski:2009rz} that the constraint \eqref{XH} removes all of its higher components leaving just the lowest one, namely the goldstino of the second supersymmetry, unconstrained. This statement is going to be justified more concretely in chapter \ref{cap:mattsec}. The low energy theory of $\mathcal{N}=2$ spontaneously broken supersymmetry can be expressed in this description in terms of $X$ and $H_{\dot{\alpha}}$ as
\begin{equation}
\begin{aligned}
	\label{N2LL33}
	{\cal L} = 
	&\int d^4 \theta \left( X \bar X 
	- \Big{|}\partial_m \left( \! \frac{X \bar H^2}{2}  \right) \Big{|}^2
	+ i \partial_m (X \bar H^\alpha) \sigma^m_{\alpha \dot \alpha} (\bar X H^{\dot \alpha} ) \right)\\ 
	&+ f \left( \int d^2 \theta X + c.c. \right) .
\end{aligned}
\end{equation}

Another description of the $\mathcal{N}=2$ supersymmetry breaking sector that it is presented is in terms of the nilpotent chiral superfield $X$ and of another chiral superfield $Y$ which is constrained in the following way
\begin{align}
\label{eq:c1:constrY1}
XY &= 0,\\
\label{eq:c1:constrY2}
\bar X D^2 Y  &= 0 .
\end{align}
The origin of these constraints is going again to be clarified in chapter \ref{cap:mattsec}, but at this level it is nevertheless possible to understand their role. Acting on them with $D^2$, in fact, their solutions can be found, as it has been done in section \ref{cap1:sec:X2=0} for the constraint $X^2=0$. They are
\begin{equation}
	\label{ConstrY}
	\begin{aligned}
		Y & = -2 \frac{D^\alpha X D_\alpha Y}{D^2 X} - X \frac{D^2 Y}{D^2 X},
		\\[0.1cm]
		D^2 Y&= \frac{- 16 \bar X \partial^2 Y + 8 i \bar D_{\dot \alpha} \bar X \partial^{\dot\alpha \alpha} D_\alpha Y}{\bar D^2 \bar X}   , 
	\end{aligned}
\end{equation}
where the notation $\partial_{\alpha \dot \alpha} \equiv \sigma^m_{\alpha \dot \alpha} \partial_m$ has been used. By projecting the first of these expressions to spacetime, it can be understood that the lowest component $Y|$ is removed from the spectrum, while by projecting the second expression one can realise that also the auxiliary field in the highest component, $D^2 Y|=-4 F^Y$, is removed. The constrained superfield $Y$ contains therefore only one fermion as independent component. This fermion is indeed providing the second goldstino. More information concerning the nilpotent and orthogonal superfields $X$ and $Y$ are going to be given in section \ref{cap2:sec:XY}, but at this stage it is necessary only to keep in mind that $X$ contains one goldstino and the auxiliary field which is breaking supersymmetry, while $Y$ contains only the other goldstino.
Since also the superfield $W_\alpha$ introduced in the previous description contains only one goldstino, it should be possible to use it to build a superfield having exactly the properties of $Y$. The expression 
\begin{equation}
	\label{WY}
	Y = - \frac{1}{\sqrt 2} D^\alpha W_\alpha + \sqrt 2 \frac{\bar D^{\dot \rho} \bar X \bar D_{\dot \rho} D^\rho W_\rho}{\bar D^2 \bar X} ,
\end{equation}
satisfying \eqref{eq:c1:constrY1} and \eqref{eq:c1:constrY2} when $W_\alpha$ satisfies the constraint \eqref{fondconstrN2inN1_3}, is the desired one.

To rewrite the Lagrangian \eqref{N2LL} in the $X,Y$ system, the expression \eqref{WY} has to be inverted, in order to express $W_\alpha$ in terms of the chiral constrained superfield $Y$:
\begin{eqnarray}
	\label{WofY}
	\begin{split}
		W_\beta = & 2\sqrt 2 \frac{X D_\beta Y}{D^2 X} 
		+ 16 \sqrt 2 i X \frac{D^\rho Y}{D^2 X} \frac{\bar D^{\dot \rho} \bar X}{\bar D^2 \bar X} \partial_{\rho \dot \rho} \left( \frac{D_\beta X}{D^2 X} \right) -
		\\
		& - 128 \sqrt 2 X \frac{D^\sigma Y}{D^2 X} \frac{\bar D^{\dot \sigma} \bar X}{\bar D^2 \bar X} \partial_{\sigma \dot \sigma} \left( \frac{D^\rho X}{D^2 X} \right) 
		\frac{\bar D^{\dot \rho} \bar X}{\bar D^2 \bar X} \partial_{\rho \dot \rho} \left( \frac{D_\beta X}{D^2 X} \right) .
	\end{split}
\end{eqnarray} 
By replacing \eqref{WofY} into \eqref{N2LL} the following Lagrangian is obtained
\begin{equation}
	{\cal L}  = \int d^4 \theta \left( X \bar X + Y \bar Y [ 1 + {\cal A} ] + S \partial^2 \bar S \right) 
	+ f \left( \int d^2 \theta X + c.c. \right) ,
\end{equation}
where 
\begin{equation}
	\label{Aterm}
	{\cal A } = - 64 \frac{ \bar D_{\dot \gamma} \bar X \partial_{\rho}^{\dot \gamma} \bar D^{\dot \rho}  \bar X
		D_\gamma X \partial_{\dot \rho}^{\gamma}  D^\rho X }{|D^2 X|^4} 
\end{equation}
and 
\begin{equation}
	S = Y^2  \frac{ D^2 X}{ 
		\left( \delta_{\epsilon}^{\rho} 
		+ 8 i \frac{\bar D_{\dot \rho} \bar X \partial_{\epsilon}^{\dot \rho} D^{\rho} X}{|D^2 X|^2} \right) 
		\left( \delta^{\epsilon}_{\rho} 
		- 8 i \frac{\bar D_{\dot \gamma} \bar X \partial^{\dot \gamma \epsilon} D_\rho X}{|D^2 X|^2} \right) 
	} .
\end{equation}
This is the complete expression of the ${\cal N}=2$ supersymmetry breaking Lagrangian in the system of nilpotent and orthogonal superfields.

To summarise, three different descriptions of the goldstino sector for $\mathcal{N}=2$ spontaneously broken supersymmetry have been given, using the language of $\mathcal{N}=1$ superspace. Both supersymmetries are non-linearly realised and in the low energy regime only the two goldstini are present in the spectrum. These descriptions can be related one to the other by using field redefinitions, much in the same way as the Volkov--Akulov model has been related to alternative descriptions in subsection \ref{subsec:equivalenceVA}. The superspace form of these redefinitions has been calculated and given explicitly in formula \eqref{eq:c1:HdefW}, to pass from the first to the second, in formula \eqref{WY}, to pass from the second to the third, and vice versa in formula \eqref{WofY}.

\subsection{$\mathcal{N}$ supersymmetry}

In this subsection the goldstino sector of a generic number $\mathcal{N}$ of spontaneously broken supersymmetries is described. The discussion is the direct generalisation of the previous analysis of the $\mathcal{N}=2$ case. The Samuel-Wess formalism, in particular, is again going to be used in order to obtain the desired constraints which remove all the fields from the spectrum except the $\mathcal{N}$ goldstini. The results are going then to be expressed in the language of $\mathcal{N}=1$ superspace.

The  algebra satisfied by ${\cal N}$ superspace derivatives, without central charges, is
\begin{equation}
	\begin{aligned}
		\{ D^I_\alpha , \bar D_{J\,\dot \alpha} \} &= -2i \, \delta^I_J \, \sigma^m_{\alpha \dot \alpha} \partial_m , 
		\\[0.1cm]
		\{ D^I_\alpha , D^J_\beta \} &= 0 , 
	\end{aligned}
\end{equation}
where the indices $I,J,K,\dots$ run from 1 to $\mathcal{N}$ and label the supersymmetries.
Lower indices refer to the fundamental representation of the R-symmetry group $\textup{U}({\cal N})_R$, while upper indices refer to the antifundamental one.
Being ${\cal N}$ spontaneously broken supersymmetries, the theory contains ${\cal N}$ goldstini. To describe them in superspace, define ${\cal N}$ spinor superfields $\Lambda_{I\,\alpha}$, which satisfy the constraints
\begin{equation}
	\label{NSWrepr}
	\begin{aligned} 
		D_\alpha^I \Lambda_{J\,\beta} & = f\,\epsilon_{\beta\alpha} \, \delta^I_J+\frac{i}{f}\sigma^m_{\alpha\dot{\rho}}\bar\Lambda^{I\,\dot{\rho}}\partial_m \Lambda_{J\,\beta} ,
		\\
		\bar D_{I\,\dot \alpha} \Lambda_{J\,\beta} & = -\frac{i}{f} \sigma^m_{\rho \dot \alpha} \Lambda^{\rho}_I
		\partial_m \Lambda_{J\,\beta} .
	\end{aligned}
\end{equation}
The superfields $\Lambda_{I\,\alpha}$ contain all the information on the supersymmetry breaking.
Their only independent component fields are the goldstini, defined as 
\begin{equation}
	\Lambda_{I\,\alpha} |_{\theta^I=0} = \lambda_{I\, \alpha}, 
\end{equation}
and whose supersymmetry transformations are non-linearly realised
\begin{equation}
	\delta \lambda_{I\,\alpha} = f\epsilon_{I\,\alpha} - \frac{i}{f} \sum_J\left( \lambda_J \sigma^m \, \bar \epsilon^J-\epsilon_J\sigma^m\bar\lambda^J \right) \partial_m \lambda_{I\,\alpha}  ,
\end{equation}  
where $\epsilon_{I\,\alpha}$ are the ${\cal N}$ supersymmetry parameters. A supersymmetric and $\textup{U}({\cal N})_R$-invariant Lagrangian for $\Lambda_{I\,\alpha}$ is
\begin{equation}
\begin{aligned}
\label{eq:c1:LSWN}
	{\cal L} &= -\frac{1}{f^{4\mathcal{N}-2}}\int d^{4\mathcal{N}}\! \theta \,  \Lambda^{2\mathcal{N}} \bar \Lambda^{2 \mathcal{N}}\\ 
	&=-f^2 - \sum_I i (\lambda_I \sigma^m \partial_m\bar\lambda^I-\partial_m\lambda_I\sigma^m\bar\lambda^I) + O(f^{-2}).
\end{aligned}
\end{equation}

The constrained superfield approach in ${\cal N}$ superspace is now developed. The superfields $\Lambda_{I\,\alpha}$ can be used to define the superfield $\Phi$
\begin{equation} 
	\label{Finn}
\begin{aligned}
	\Phi &= -\frac{1}{(-4)^{\mathcal{N}}f^{4\mathcal{N}-1}} \bar D^{2\mathcal{N}} \left( \Lambda^{2\mathcal{N}} \bar \Lambda^{2 \mathcal{N}} \right)\\ 
&= -f^{-(2\mathcal{N}+1)}\Lambda^{2\mathcal{N}} \left( f^2 + \sum_I\bar\Lambda^I(\dots)+\sum_{I,J}\bar\Lambda^{I}\bar\Lambda^{J}(\dots)+\dots \right) ,
\end{aligned}
\end{equation}
where $(\dots)$ contains terms with derivatives of the superfields $\Lambda^I_\alpha$.
This superfield $\Phi$ is chiral 
\begin{equation}
	\bar D_{I\,\dot{\alpha}}\Phi =0
\end{equation}
and, as it can be checked by direct inspection, it satisfies the following set of constraints\footnote{The notation $(\Psi_{I})^{2{\cal N}}\equiv\Psi^{2{\cal N}}=\Psi_1^2\,\Psi_2^2\dots \Psi_{{\cal N}}^2$ is used to indicate the product of ${\cal N}$ squared spinor superfields. The dummy index $I$ is not summed and, to avoid confusion, whether sums occur they are going to be explicitly written.}
\begin{equation}
	\label{phi2N}
	\begin{aligned}
		\Phi^2 &= 0,
		\\
		\Phi D^I_\alpha  \Phi &= 0, 
		\\
		\Phi D^I_\alpha D^J_\beta \Phi & =0 ,
		\\
		\Phi D^I_\alpha D^J_\beta D^K_\gamma \Phi & =0 ,
		\\
		\cdots & 
		\\
		\Phi (D^{I\neq J})^{2\mathcal{N}-2}D^J_\alpha  \Phi & =0 ,
	\end{aligned}
\end{equation}
together with 
\begin{equation}
\label{eq:cap:1NRocekconstr}
	\bar \Phi  (D^I)^{2\mathcal{N}} \Phi =- (-4)^\mathcal{N}\,f\, \bar \Phi.
\end{equation}
As a consequence of these constraints, the only independent fields in $\Phi$ are the $\mathcal{N}$ goldstini, which reside in the components
\begin{equation}
	\frac{1}{(-4)^{{\cal N}-1}}(D^{I\neq J})^{2{\cal N}-2}D^J_{\alpha}\Phi|_{\theta^I=0}\equiv G^\Phi_{J\,\alpha}=-2\lambda_{J\, \alpha}+\dots,
\end{equation}
where dots stand for terms with more fermions.
The Lagrangian for $\Phi$ can be written as 
\begin{equation}
	\label{NRocek}
	{\cal L} =  f \int d^{2\mathcal{N}} \theta \, \Phi .
\end{equation}

As in the $\mathcal{N}=2$ case, consider now a generic chiral superfield $\mathcal{X}$ in $\mathcal{N}$ superspace and impose on it the constraints \eqref{phi2N}, without \eqref{eq:cap:1NRocekconstr}. As a result, all the component fields of $\mathcal{X}$ are going to be removed, except for the $\mathcal{N}$ goldstini and the auxiliary field which is breaking supersymmetry. Even in this general case, however, only the last constraint in \eqref{phi2N} is essential, the remaining ones being consistency conditions, as proved in \cite{Cribiori:2016hdz}. Imposing therefore the constraint
\begin{equation}
	\label{constrN}
	{\cal X} (D^{I\neq J})^{2\mathcal{N}-2}D^J_\alpha  {\cal X} =0
\end{equation}
and solving it in superspace, one can obtain the unique solution
\begin{equation}
	\label{solXN}
	\mathcal{X} = \left(\frac{1}{2}\right)^\mathcal{N} \frac{(G_1^{\alpha} G_{1\alpha})(G_2^{\alpha} G_{2\alpha})\dots(G_{\mathcal{N}}^\alpha G_{\mathcal{N} \alpha})}{\mathcal{F}^{2\mathcal{N}-1}}\equiv \left(\frac{1}{2}\right)^\mathcal{N} \frac{(G_I)^{2\mathcal{N}}}{\mathcal{F}^{2\mathcal{N}-1}}.
\end{equation}
In this formula the superfield ${\cal F}$, which lives in $\mathcal{N}$ superspace, is defined such that
\begin{equation}
	\label{auxF_N}
	F={\cal F}|_{\theta^I=0} = \frac{1}{(-4)^\mathcal{N}} (D^I)^{2\mathcal{N}} {\cal X} |_{\theta^I=0}   
\end{equation}
is the complex scalar auxiliary field breaking supersymmetry, while the superfields $G_{I \alpha}$ are defined such that
\begin{equation}
	\label{GoldN}
	g_{I\alpha}=G_{I\alpha} |_{\theta^I=0} = \frac{1}{\sqrt 2 (-4)^{\mathcal{N} -1}} (D^{J \ne I})^{2\mathcal{N}-2} D^I_\alpha {\cal X} |_{\theta^I=0}  
\end{equation}
are the $\mathcal{N}$ goldstini.
The Lagrangian describing the supersymmetry breaking sector is now
\begin{equation}
	\label{lagrangianNX} 
	{\cal L} = \int d^{4 \mathcal{N}} \theta  \, {\cal X} \bar {\cal X} 
	+ \left( f \int d^{2\mathcal{N}} \theta \, {\cal X}  + c.c. \right) 
\end{equation}
and ${ F}$ can be integrated out to obtain a theory including only the $\mathcal{N}$ goldstini.

\subsection{$\mathcal{N}$ goldstini in ${\cal N}=1$ superspace}
The previous analysis of the model with $\mathcal{N}$ goldstini is going now to be reformulated in the language of $\mathcal{N}=1$ superspace, which is one of the most employed for practical applications. As discussed in the analogous subsection on the $\mathcal{N}=2$ case, different ways of describing $\mathcal{N}$ goldstini in $\mathcal{N}=1$ superspace exist, depending on the particular choice of the superfields in which they are embedded. The generalisation of the three descriptions presented in \ref{cap1:sec:N2inN1} is constructed in the following. All of them are going to contain a nilpotent $\mathcal{N}=1$ chiral superfield $X$ as an essential ingredient.

As a first step, break the $\textup{U}({\cal N})_R$-covariance explicitly by splitting the set of superspace derivatives as
\begin{equation}
	\label{Dsplit}
	D^I_\alpha \rightarrow \{D_\alpha \, , \, D^i_\alpha \},
\end{equation}
where the indices $i,j,\dots$ take values from $1$ to $\mathcal{N}-1$. Introduce then the nilpotent chiral superfield $X$ defined as\footnote{Since projections to the surface $\theta^i_\alpha=0$ are omitted in the definition of $X$, this object lives in the full ${\cal N}$ superspace. The same observation applies also to $W_{i \, \alpha}$ in \eqref{Wi} and in other formulas thereafter. If this is source of confusion, it is possible to continue thinking of them as ${\cal N}=1$ superfields, because this is the role they have been introduced for. 
The $X$, $Y_i$ appearing in \eqref{NNLL} and thereafter are indeed properly projected ${\cal N}=1$ superfields.\label{foot:notSF}}
\begin{equation}
	X = \frac{1}{(-4)^{\mathcal{N}-1}}(D^i)^{2\mathcal{N}-2} {\cal X}. 
\end{equation}
It contains the auxiliary field $F$ which is breaking supersymmetry,
\begin{equation}
F=-\frac{1}{4}D_\alpha X|_{\theta^I=0},
\end{equation}
and the goldstino of the first supersymmetry, namely the one related to the first superspace derivative in \eqref{Dsplit}:
\begin{equation}
	g_\alpha=G_\alpha|_{\theta^I=0} =\frac{1}{\sqrt{2}}D_\alpha X|_{\theta^I=0}.
\end{equation}

In the first description discussed in this subsection, the remaining $\mathcal{N}-1$ goldstini are embedded into the $\mathcal{N}=1$ chiral superfields 
\begin{equation}
	\label{Wi}
	W_{i\, \alpha} = \frac{1}{\sqrt 2 (-4)^{\mathcal{N}-2}} (D^{j \ne i})^{2\mathcal{N}-4} D^i_\alpha  {\cal X} , 
\end{equation} 
satisfying the constraints 
\begin{eqnarray}
	W_{i\,\alpha} = X \frac{D^2 W_{i\,\alpha} }{D^2 X} .
\end{eqnarray}
In particular they reside in the highest component:
\begin{equation}
	\qquad G_{i\,\alpha} = -\frac{1}{4}D^2 W_{i\,\alpha}.
\end{equation}
In the full $\mathcal{N}$ superspace, the solution \eqref{solXN} can be expressed in terms of the goldstini and of the auxiliary field as 
\begin{equation}
	\label{solXN,1}
	\mathcal{X}=2^{5-5{\cal N}}X\frac{\prod_{i=1}^{\mathcal{N}-1}\left(D^2 W_{i\alpha}\right)^2}{\mathcal{F}^{2\mathcal{N}-2}}.
\end{equation}
The lower components of ${\cal X}$ are constrained $\mathcal{N}=1$ superfields and, similarly to $S$ of the ${\cal N}=2$ case, they are removed from the spectrum in terms of $X$ and  $W_{i \, \alpha}$.
They organise in representations of the group $\textup{U}(\mathcal{N}-1)$, which acts as a flavour symmetry after the breaking
\begin{equation}
	\textup{U}(\mathcal{N})_R \longrightarrow U(1)_R \times \textup{U}(\mathcal{N}-1).
\end{equation}
More details on them are going to be given in a while but, already at this stage, it is possible to realise that the structure of the Lagrangian in the $X$, $W_{i\,\alpha}$ description is of the type
\begin{equation}
	\label{NNLL}
	{\cal L} = 
	\int d^4 \theta \left( X \bar X 
	+ i \sum_{i} \partial_m W_i^{\alpha} \sigma^m_{\alpha \dot \alpha} \bar W^{i \dot \alpha}  \right) 
	+ f \left( \int d^2 \theta X + c.c. \right) 
	+ \dots  ,
\end{equation}
where dots stand for higher order terms, which are essential for the non-linear realisation of the $\mathcal{N}$ supersymmetries.

Mimicking the discussion of the $\mathcal{N}=2$ theory, a set of chiral superfields $Y_i$ can be defined such that
\begin{equation}
	\label{newYi}
	\begin{aligned}
		XY_i &= 0 , 
		\\
		\bar X D^2 Y_i  &= 0,  
	\end{aligned}
\end{equation}
where $X^2=0$ is always understood.
As a consequence of the constraints, these superfields $Y_i$ contain only fermions, which are going to be the goldstini of the spontaneously broken supersymmetries.
The explicit superfield redefinition which connects $W_{i\,\alpha}$ and $Y_i$ is the generalisation of \eqref{WofY}:
\begin{equation}
	\label{WiofYi}
	\begin{aligned}
		W_{i\,\beta} = & 2\sqrt 2 \frac{X D_\beta Y_i}{D^2 X} 
		+ 16 \sqrt 2 i X \frac{D^\rho Y_i}{D^2 X} \frac{\bar D^{\dot \rho} \bar X}{\bar D^2 \bar X} \partial_{\rho \dot \rho} \left( \frac{D_\beta X}{D^2 X} \right) -
		\\[0.15cm]
		& - 128 \sqrt 2 X \frac{D^\sigma Y_i}{D^2 X} \frac{\bar D^{\dot \sigma} \bar X}{\bar D^2 \bar X} \partial_{\sigma \dot \sigma} \left( \frac{D^\rho X}{D^2 X} \right) 
		\frac{\bar D^{\dot \rho} \bar X}{\bar D^2 \bar X} \partial_{\rho \dot \rho} \left( \frac{D_\beta X}{D^2 X} \right) 
	\end{aligned}
\end{equation}
and it can be used in order to rewrite the Lagrangian \eqref{NNLL} in terms of the superfields $X$ and $Y_i$
\begin{equation}
	{\cal L}  = \int d^4 \theta \left( X \bar X + \sum_{i} Y_i \bar Y^i [1 + {\cal A} ] \right) 
	+ f \left( \int d^2 \theta X + c.c. \right) 
	+ \dots  , 
\end{equation}
where ${\cal A}$ is defined as in \eqref{Aterm} and dots stand for higher order terms, making the theory invariant under the non-linearly realised additional supersymmetries.

A third alternative way to express the model is in terms of the chiral superfields $X$ and $H^i_{\dot \alpha}$ satisfying
\begin{equation}
	\begin{aligned}
		X^2                                       & =0 ,  
		\\ 
		\bar D_{\dot \alpha} (X \bar H_{i\, \beta} ) & = 0 .
	\end{aligned}
\end{equation}
The Lagrangian is of the type
\begin{equation}
\begin{aligned}
	\label{N2LL33}
	{\cal L} = 
	&\int d^4 \theta \left( X \bar X 
	- \Big{|}\partial_m \left( \! \frac{X \bar H^2}{2}  \right) \Big{|}^2
	+ i \sum_{i}\partial_m (X \bar H^\alpha_i) \sigma^m_{\alpha \dot \alpha} (\bar X H^{i\,\dot \alpha} ) \right)\\ 
	&+ f \left( \int d^2 \theta X + c.c. \right) +\dots,
\end{aligned}
\end{equation} 
where dots stand for higher order interactions which are again necessary for the non-linear realisation.

With some more effort, one can calculate the complete form of all these Lagrangians and realise explicitly how all the lowest components are removed from the spectrum.
To deal with all the fields in ${\cal X}$ in the general ${\cal N}$ case it is possible to proceed as follows.
The generic component can have a number $p$ of fermionic indices, with $0\leq p \leq {2{\cal N}}$, and some of them can be contracted in pairs. In particular, being fermionic, the same index cannot appear more than twice.
An efficient way to handle with this situation consists in distinguishing between the indices in the set $M_1=\{i_1,\dots,i_k\}$ which are all different, and the ones in $M_2=\{j_1,\dots,j_l\}$ which are equal two by two. One can therefore split the number of indices $p$ as $p=k+l$ and construct
\begin{equation}
	\label{Sdef}
	\left(S^{\,\,i_1\dots i_k,\, j_1\dots j_l}_{(k+l)}\right)_{\alpha_1\dots\alpha_k,\,\beta_1\dots \beta_l}=\left(\frac{1}{\sqrt{2}}\right)^{k+l} D_{\alpha_1}^{i_1} \dots D_{\alpha_k}^{i_k}\,D_{\beta_1}^{j_1} \dots D_{\beta_l}^{j_l}\,\mathcal{X}|_{\theta^I=0},
\end{equation}
which describes collectively all the components of $\mathcal{X}$. The explicit expression of this object \eqref{Sdef} is now given for the three different sets of $\mathcal{N}=1$ constrained superfields introduced before.

In the description in terms of $X$ and $W_{i\,\alpha}$, the form of the lowest component is obtained directly using \eqref{Wi} and no superfield redefinition is involved.
In the formulation in terms of $X$ and $Y_i$, the form of the lower superfields $S_{(p)}$ is more involved, since superfield redefinitions are needed. It turns out that the correct expression is
\begin{equation}
\label{SNN}
\begin{aligned}
(S^{\,\,i_1\dots i_k,\, j_1\dots j_l}_{(p)} & )_{\alpha_1\dots\alpha_k,\,\beta_1\dots \beta_l}=2^{2-2{\cal N}+\frac{3k}{2}-l}\epsilon_{\beta_1\beta_2}\dots\epsilon_{\beta_{l-1}\beta_l} X\\[0.15cm]
& \times \frac{(D_{\rho_1}Y_{i_1}Z^{\rho_1}_{\alpha_1})\dots (D_{\rho_k}Y_{i_k}Z^{\rho_k}_{\alpha_k})}{\mathcal{F}^{2{\cal N}-2-k-l}}(D_{\sigma}Y_{i\notin M}Z^{\sigma})^{2{\cal N}-2-2k-l}, 
\end{aligned}
\end{equation}
where $p=k+l$, $M=M_1+M_2$ and $Z^\rho_\alpha$ is a quantity containing only $X$ 
\begin{equation}
\begin{aligned}
Z^\rho_\alpha &= \delta^\rho_\alpha + 8i \epsilon^{\beta \rho}\frac{\bar D^{\dot\beta} \bar X}{\bar D^2 \bar X}\partial_{\beta \dot\beta}\left(\frac{D_\alpha X}{D^2 X}\right)\\
&-64 \epsilon^{\sigma \rho}\frac{\bar D^{\dot\sigma}\bar X}{\bar D^2 \bar X}\partial_{\sigma \dot\sigma}\left(\frac{D^\beta X}{D^2 X}\right) \frac{\bar D^{\dot \beta}\bar X}{\bar D^2 \bar X}\partial_{\beta \dot \beta}\left(\frac{D_\alpha X}{D^2 X}\right) .
\end{aligned}
\end{equation}
This proves in fact that the only independent superfields in $\mathcal{X}$ are the nilpotent $X$ and the orthogonal $Y_i$, while all the others are given in terms of these as a consequence of \eqref{SNN}.
In the description in terms of the constrained superfields $X$ and  $H^i_{ \dot\alpha}$, which can be obtained using the superfield redefinition
\begin{equation}
H^i_{\dot \alpha} = \frac{\bar D^2 \bar W^i_{\dot \alpha}}{\bar D^2 \bar X},
\end{equation}
the form of the lowest components is
\begin{equation}
	\begin{aligned}
		S^{\,\,i_1\dots i_k,\, j_1\dots j_l}_{(k+l)\alpha_1\dots\alpha_k,\,\beta_1\dots \beta_l} & =      
		(-4)^{2\mathcal{N}-2-k-l} 2^{5-5{\cal N}+3k+\frac{5}{2}l} \epsilon_{\beta_1\beta_2}\dots\epsilon_{\beta_{l-1}\beta_l}X
		\\[0.1cm]
& \quad \times 
		\bar H_{i_1 \alpha_1}   
		\dots 
		\bar H_{i_k \alpha_k} 
		(\bar H_{i\notin M})^{2{\cal N}-2-2k-l}.
	\end{aligned}
\end{equation}

The last step consists in giving an expression for the Lagrangian in terms of \eqref{Sdef}, which can eventually be specialised to the case of interest. After some reordering the theory can be recast into the compact form
\begin{equation}
	\label{Lcompact}
	{\cal L}=\int d^4\theta \,\sum_{l=0}^{{\cal N}-1}\,\sum_{k=0}^{{\cal N}-1}L^{({\cal N})}_{k,2l}+ f \left( \int d^2 \theta X + c.c. \right) , 
\end{equation}
where the quantity
\begin{equation}
	\begin{aligned}
		L^{({\cal N})}_{k,2l} & =\sum_{i_1,\dots,i_k=1}^{{\cal N}-1}\,\,\,\sum_{j_l>j_{l-2}>\dots>j_2=1}^{{\cal N}-1}(-i)^k\,2^{-l}\,  (S^{i_1\dots i_k,j_1\dots j_l})^{\alpha_1\dots \alpha_k,\,\beta_2\beta_4\dots\beta_l}_{\phantom{\alpha_1\dots \alpha_k,\beta_2\beta}\beta_2\beta_4\dots\beta_l}\\
		                    &\quad \times\partial_{\alpha_1\dot{\alpha}_1}\dots\partial_{\alpha_k\dot{\alpha}_k}(\partial^2)^{{\cal N}-1-k-\frac{l}{2}}\,(\bar{S}_{i_1\dots i_k,j_1\dots j_l})^{\dot{\alpha}_1\dots \dot{\alpha}_k,\,\phantom{\beta_2}\dot{\beta}_2\dot{\beta}_4\dots\dot{\beta}_l}_{\phantom{\alpha_1\dots \alpha_k}\,\dot{\beta}_2\dot{\beta}_4\dots\dot{\beta}_l} 
	\end{aligned}
\end{equation} 
has been introduced and the complex conjugate $\bar{S}$ has been defined with the fermionic indices in the opposite order, namely 
\begin{equation}
	(\bar{S}_{\,\,i_1\dots i_k,\, j_1\dots j_l}^{(k+l)})_{\dot{\alpha}_1\dots\dot{\alpha}_k,\,\dot{\beta}_1\dots \dot{\beta}_l}=\left(\frac{1}{\sqrt{2}}\right)^{k+l} \bar{D}_{\dot{\alpha}_k}^{i_k} \bar{D}_{\dot{\alpha}_{k-1}}^{i_{k-1}}\dots \bar{D}_{\dot{\alpha}_1}^{i_1}\,\bar{D}_{\dot{\beta}_l}^{j_l} \bar{D}_{\dot{\beta}_{l-1}}^{j_{l-1}}\dots \bar{D}_{\dot{\beta}_1}^{j_1}\,\bar{\mathcal{X}}| .
\end{equation}
This is the complete form in $\mathcal{N}=1$ superspace of the Lagrangian describing the interactions of $\mathcal{N}$ goldstini. Notice that, using this formula, one can effectively extract the contribution of the desired derivative order without necessarily compute the whole action.
In the following table \ref{tabb3}, the $\mathcal{N}=1$ superfield content of the goldstino sector for the different descriptions presented above is summarised. 
\begin{table}[h]
\begin{center}
$$
\begin{array}{ccccc}
\hline 
\text{\bf SUSY }              & {\cal N}=1 & {\cal N}=2    & {\cal N}=4  & \text{generic ${\cal N}$}                \\
\hline 
\text{Goldstini Superfields}  & X          & X ,  W        & X , W_1 ,  W_2 W_3       & X ,  W_1 , \dots, W_{{\cal N}-1} \\
\hline
\text{Goldstini Superfields}  & X          & X ,  H        & X , H^1 ,  H^2 , H^3       & X ,  H^1 , \dots, H^{{\cal N}-1} \\ \hline
\text{Goldstini Superfields}  & X          & X ,  Y        & X , Y_1 ,  Y_2 , Y_3       & X ,  Y_1  , \dots, Y_{{\cal N}-1} \\ \hline
\text{Eliminated Superfields} & -          & S^{(0)}       & S^{(0)},  S^{(1)}, S^{(2)} &          
S^{(0)},S^{(1)},\,\dots\,,S^{(2{\cal N}-4)}\\
			\hline
			\text{Residual Flavor Group}  & -          & \textup{U}(1) & \textup{U}(3)              & \textup{U}({\cal N}-1)                  \\
			\hline
		\end{array}
		$$
	\end{center}
	\caption{\small The ${\cal N}=1$ chiral superfields content of a minimal ${\cal N}$ goldstini theory.
The shorthand notation $S^{(p)}$ indicates all the possible components \eqref{Sdef} with $p$ fermionic indices contracted in all the allowed ways.} 
\label{tabb3} 
\end{table}

As a check, in the ${\cal N}=2$ case the term $L^{(2)}_{1,2}$ vanishes and the Lagrangian is given by 
\begin{equation}
	\begin{aligned}
		{\cal L} & = \int d^4\theta \left(L^{(2)}_{0,0}+L^{(2)}_{1,0}+L^{(2)}_{0,2}\right)+f\left(\int d^2\theta X + c.c.\right) =                                                \\
		        & =\int d^4\theta \left(S \partial^2 \bar{S}-iW^\alpha\partial_{\alpha\dot{\alpha}}\bar{W}^{\dot{\alpha}}+ X\bar{X}\right)+f\left(\int d^2\theta X + c.c.\right) ,
	\end{aligned}
\end{equation}
which matches exactly with \eqref{N2LL}.
One can calculate, for example, the Lagrangians for the goldstini in the $\mathcal{N}=3$ and $\mathcal{N}=4$ cases, which are
\begin{equation}
\begin{aligned}
\mathcal{L}_{\mathcal{N}=3} &=\int d^4\theta \left[L_{0,2}^{(3)} +L_{1,0}^{(3)}+L_{0,0}^{(3)}+L_{1,2}^{(3)}+L_{0,4}^{(3)}+L_{2,0}^{(3)}\right]\\
&+\left(f\int d^2\theta {X}+c.c.\right)\\
&= \int d^4 \theta \bigg[ \frac14 S^{0,33}\partial^2 \bar S^{0,33} - i (S^{3,0})^\alpha\partial^2\partial_{\alpha\dot\alpha}(\bar S^{3,0})^{\dot\alpha}\\
& + S^{0,0}\partial^4\bar S^{0,0}-\frac{i}{4}(S^{2,33})^\alpha\partial_{\alpha \dot\alpha}(\bar S^{2,33})^{\dot\alpha}\\
&-i(S^{2,0})^\alpha\partial_{\alpha\dot\alpha}\partial^2(\bar S^{2,0})^{\dot\alpha}+\frac{1}{16}S^{0,22,33}\bar S^{0,2233} \\
&-\frac{i}{4}(S^{3,22})^\alpha\partial_{\alpha \dot\alpha}(\bar S^{3,22})^{\dot\alpha}+\frac14 S^{0,22}\partial^2 \bar S^{0,22}\\
&-(S^{32,0})^{\alpha\beta}\partial_{\alpha\dot\alpha}\partial_{\beta\dot\beta}(\bar S^{32,0})^{\dot\alpha\dot\beta}
\bigg] \\
&+\left(f\int d^2\theta {X}+c.c.\right)
\end{aligned}
\end{equation}
and
\begin{align}
\nonumber
\mathcal{L}_{\mathcal{N}=4} &=\int d^4\theta \,\sum_{l=0}^{3}\,\sum_{k=0}^{3}L^{({ 4})}_{k,2l}+ f \left( \int d^2 \theta X + c.c. \right)\\
\nonumber
&=\int d^4 \theta \bigg[\frac{1}{16}S^{0,3344}\partial^2 \bar S^{0,3344}-\frac{i}{4}(S^{4,33})^\alpha \partial_{\alpha \dot\alpha}\partial^2 (\bar S^{4,33})^{\dot\alpha}\\
\nonumber
&+\frac14 S^{0,33}\partial^4 \bar S^{0,33}-\frac{i}{4}(S^{3,44})^\gamma \partial^2\partial_{\gamma \dot\gamma}(\bar S^{3,44})^{\dot\gamma}\\
\nonumber
&-(S^{43,0})^{\alpha\beta}\partial^2\partial_{\beta\dot\beta}\partial_{\alpha\dot\alpha}(\bar S^{43,0})^{\dot\alpha\dot\beta}-i (S^{3,0})^\alpha \partial^4\partial_{\alpha\dot\alpha}(\bar S^{3,0})^{\dot\alpha}\\
\nonumber
&+\frac14 S^{0,44}\partial^4\bar S^{0,44}-i(S^{4,0})^\alpha \partial^4 \partial_{\alpha\dot\alpha}(\bar S^{4,0})^{\dot\alpha}\\
\nonumber
&+S^{0,0}\partial^6 \bar S^{0,0}-\frac{i}{16}(S^{2,3344})^\alpha\partial_{\alpha\dot\alpha}(\bar S^{2,3344})^{\dot\alpha}\\
\nonumber
&-\frac14 (S^{42,33})^{\beta\alpha}\partial_{\alpha\dot\alpha}\partial_{\beta\dot\beta}(\bar S^{42,33})^{\dot\beta\dot\alpha}-\frac{i}{4}(S^{2,33})^\alpha \partial_{\alpha\dot\alpha}\partial^2 (\bar S^{2,33})^{\dot\alpha}\\
&-\frac{i}{4}(S^{2,44})^\alpha \partial_{\alpha\dot\alpha}\partial^2(\bar S^{2,44})^{\dot\alpha}-(S^{42,0})^{\beta\alpha}\partial_{\alpha\dot\alpha}\partial^2\partial_{\beta\dot\beta}(\bar S^{42,0})^{\dot\beta\dot\alpha}\\
\nonumber
&-i(S^{2,0})^\alpha\partial_{\alpha\dot\alpha}\partial^4(\bar S^{2,0})^{\dot\alpha}+\frac{1}{64} S^{0,223344}\bar S^{0,223344}\\
\nonumber
&-\frac{i}{16}(S^{4,2233})^\alpha \partial_{\alpha\dot\alpha}(\bar S^{4,2233})^{\dot\alpha}+\frac{1}{16}S^{0,2233}\partial^2 \bar S^{0,2233}\\
\nonumber
&-\frac{i}{16}(S^{3,2244})^\alpha \partial_{\alpha\dot\alpha}(\bar S^{3,2244})^{\dot\alpha}-\frac14 (S^{43,22})^{\alpha\beta}\partial_{\beta\dot\beta}\partial_{\alpha\dot\alpha}(\bar S^{43,22})^{\dot\alpha\dot\beta}\\
\nonumber
&-\frac{i}{4}(S^{3,22})^\alpha S_{\alpha\dot\alpha}(\bar S^{3,22})^{\dot\alpha}+\frac{1}{16}S^{0,2244}\partial^2\bar S^{0,2244}\\
\nonumber
&-\frac{i}{4}(S^{4,22})^\alpha\partial^2\partial_{\alpha\dot\alpha}(\bar S^{4,22})^{\dot\alpha}+\frac14 S^{0,22}\partial^4\bar S^{0,22}\\
\nonumber
&-\frac14 (S^{32,44})^\beta\partial_{\alpha\dot\alpha}\partial_{\beta\dot\beta}(\bar S^{32,44})^{\dot\beta\dot\alpha}+i(S^{432,0})^{\gamma\beta\alpha}\partial_{\alpha\dot\alpha}\partial_{\beta\dot\beta}\partial_{\gamma\dot\gamma}(\bar S^{432,0})^{\dot\gamma\dot\beta\dot\alpha}\\
\nonumber
&-(S^{32,0})^{\beta\alpha}\partial_{\alpha\dot\alpha}\partial_{\beta\dot\beta}\partial^2 (\bar S^{32,0})^{\dot\beta\dot\alpha} 
\bigg]\\
\nonumber
&+\left(f\int d^2\theta {X}+c.c.\right)
\end{align}

The discussion on the supersymmetry breaking sector is terminated. In the next chapter the $\mathcal{N}=1$ goldstino is going to be coupled to other fields which have the role of matter and which are in general present in any realistic scenario.

\chapter{Matter sector}
\label{cap:mattsec}

In this chapter the goldstino sector is coupled to matter. For simplicity, the analysis is restricted only to the minimal case of $\mathcal{N}=1$ supersymmetry.  Results in extended supersymmetry can be found for example in \cite{Bagger:1994vj, Bagger:1996wp,Antoniadis:2008uk, Antoniadis:2017jsk,Farakos:2018aml}

When using superspace methods, matter fields have to be embedded into superfields, as it has been the case for the goldstino. If supersymmetry is spontaneously broken and non-linearly realised, some matter component fields can be removed from the the spectrum, in much the same way as the sgoldstino has been eliminated in the previous chapter. 
This can be performed in an efficient manner by imposing constraints on matter superfields, avoiding thus the procedure of calculating the equations of motion in the zero-momentum limit. 

Depending on which is the particular matter component field that has to be removed, the constraints to be imposed are different. In this respect, indeed, several constraints have been proposed and discussed in literature \cite{Komargodski:2009rz}. 
Their origin, however, has remained not completely clear until the work of \cite{DallAgata:2016syy}, in which the authors proposed a general constraint to be imposed on matter superfields in order to eliminate any desired component.
Being a central ingredient in the constrained superfield approach to non-linear supersymmetry, this general constraint is going to be introduced and analysed in some detail.  

In particular, it relies only on the presence of a nilpotent goldstino chiral superfield $X$. For this reason, in the second part of the chapter the existence of such a superfield $X$ is discussed and it is shown, by giving several examples, that a nilpotent $X$ is present in a large class of models with spontaneously broken supersymmetry, either in the case of F-term or D-term breaking. 

It is possible to conclude then that these two constraints, namely the nilpotent one for the goldstino and the general constraint for matter, are the only two ingredients which are needed for writing any low energy effective model with spontaneously broken and non-linearly realised supersymmetry and with any desired spectrum content. The results presented in this chapter constitute therefore the theoretical basis for phenomenological applications.

\section{Two chiral superfields}
Given the superymmetry breaking sector of $\mathcal{N}=1$ supersymmetry, described in terms of a nilpotent chiral superfield $X$, the more immediate generalisation consists in coupling it to another chiral superfield. Two particular examples are going to be analysed: the case in which the second chiral superfield is a Lorentz scalar, $Y$, and the one in which it has a spinorial index, $W_\alpha$. This last example is going to motivate the introduction of a procedure for removing matter component fields, which is different from their integration in the zero-momentum limit via the equations of motion.

\subsection{$X$ and $Y$ system}
\label{cap2:sec:XY}
Consider two chiral superfields $X$ and $Y$ with superspace expansions
\begin{align}
X &= A + \sqrt 2 \,\theta^\alpha G_\alpha + \theta^2 F,\\
Y &=B+\sqrt 2 \,\theta^\alpha\chi_\alpha +\theta^2 F^Y.
\end{align}
The simplest Lagrangian which can be written with only these two ingredients and in which supersymmetry is spontaneously broken is
\begin{equation}
\begin{aligned}
\mathcal{L} &= \int d^4\theta \left(X \bar X+Y\bar Y\right) + f\left(\int d^2 \theta X + c.c.\right)\\ 
&=-\partial_m A\partial^m \bar A-\partial_m B\partial^m \bar B - i\bar G\bar\sigma^m\partial_m G-i\bar \chi\bar\sigma^m\partial_m \chi\\
&\phantom{=}+F\bar F+fF+f\bar F+F^Y\bar F^Y.
\end{aligned}
\end{equation}
In this model there are two massless scalar, $A$ and $B$, two massless fermions $G_\alpha$ and $\chi_\alpha$ and two auxiliary fields, $F$ and $F^Y$. Since the auxiliary field which acquires a non vanishing vacuum-expectation-value is $\langle F \rangle = -f \neq 0$, while $\langle F^Y \rangle =0$, supersymmetry is broken by $F$ and the fermion $G_\alpha$ is the goldstino. The component fields of $Y$ are then called matter fields, to distinguish them from the goldstone mode and from the auxiliary field describing the supersymmetry breaking sector.

To integrate out some of the degrees of freedom and to produce an effective theory, a mass splitting in the spectrum has to be created. The only component of $X$ which can be removed is the sgoldstino $A$ and its decoupling can be performed exactly in the same way as it has been discussed in subsection \ref{cap1:sec:Adecoupling}. For what concerns the components of $Y$, instead, alternatives are allowed. One possible choice, which is going to be pursued in the following, is to eliminate the scalar $B$ in the lowest component, in order to obtain a theory with only two fermions: $G_\alpha$ and $\chi_\alpha$. This theory is going to be similar to the one presented in \ref{cap1:sec:N2inN1}, with the difference that, a priori, there is no reason why it has to be invariant also under a second non-linearly realised supersymmetry.

As done in the goldstino sector, to create a mass gap the original model is modified with curvature terms in the K\"ahler potential. Consider therefore the following Lagrangian 
\begin{equation}
\begin{aligned}
\label{c2:eq:modLXY}
\mathcal{L} &= \int d^4\theta \left(X \bar X+Y\bar Y\right) + f\left(\int d^2 \theta X + c.c.\right)\\ 
&\phantom{=}-\frac{1}{\Lambda^2}\int d^4\theta X^2\bar X^2 -\frac{1}{\Lambda^2}\int d^4\theta X\bar X Y\bar Y\\
&=-f^2 +|F+f|^2 + F^Y\bar F^Y\\
&\phantom{=}-\frac{1}{\Lambda^2}|2AF-G^2|^2-\frac{1}{\Lambda^2}|B F+A F^Y-G\chi|^2 \\
&\phantom{=}+\text{terms with derivatives},
\end{aligned}
\end{equation}
where $\Lambda$ is a parameter of mass dimension one. Due to the modification, the masses of $A$ and $B$ become respectively $m_A^2 = 4f^2/\Lambda^2$ and $m_B^2 = f^2/\Lambda^2$. By restricting the analysis to an energy regime $E\ll f/\Lambda$, or equivalently by taking the formal limit $1/\Lambda\to \infty$, the scalars $A$ and $B$ can be integrated out. Their equations of motion in the zero-momentum limit give
\begin{align}
\label{eq:c2:solX}
A &= \frac{G^2}{2F},\\
\label{eq:c2:solY}
B &= \frac{G\chi}{F}-\frac{G^2 F^Y}{2F^2}.
\end{align}
Notice that both these solutions do not depend on the details of the theory in the high energy regime, namely on the parameter $\Lambda$. The effective theory can be obtained then by substituting \eqref{eq:c2:solX} and \eqref{eq:c2:solY} back into the Lagrangian \eqref{c2:eq:modLXY}.

As it has been done in the previous chapter, superspace techniques can be employed to define an alternative procedure to eliminate degrees of freedom and to produce a consistent effective theory. The starting point is again to understand which are the superfields whose lowest components are given by \eqref{eq:c2:solX} and \eqref{eq:c2:solY}. For the particular case under consideration, in which two scalars are removed from two chiral superfields, it turns out that also the sought superfields are chiral and they are solutions to the constraints
\begin{equation}
\begin{aligned}
\label{eq:c2:XY=0}
X^2&=0,\\ XY&=0.
\end{aligned}
\end{equation}
The nilpotent constraint on $X$ has been already discussed, while the second constraint, namely the orthogonality condition between the chiral superfields, has been first introduced in \cite{Brignole:1997pe}. The strategy for solving these constraints in superspace is the usual one: act on them with superspace derivatives and find
\begin{align}
X &= - \frac{D^\alpha X D_\alpha X}{D^2 X},\\
Y &= -2 \frac{D^\alpha X D_\alpha Y}{D^2 X}- X\frac{D^2 Y}{D^2 X},
\end{align}
besides some consistency conditions. By projecting these solutions to spacetime, the expressions \eqref{eq:c2:solX} and \eqref{eq:c2:solY} are recovered. The projections $D_\alpha Y| $ and $D^2Y|$ do not change with respect to the unconstrained case, since the fermion and the auxiliary field in $Y$ are not eliminated from the spectrum. The $\theta$-expansions of the constrained superfields $X$ and $Y$ therefore are
\begin{align}
X &= \frac{G^2}{2F}+\sqrt 2 \,\theta^\alpha G_\alpha + \theta^2 F,\\
\label{eq:c2:solY2}
Y &=\frac{G\chi}{F}-\frac{G^2 F^Y}{2F^2}+\sqrt 2 \,\theta^\alpha\chi_\alpha +\theta^2 F^Y,
\end{align}
and the Lagrangian governing their interactions in superspace has the form
\begin{equation}
\mathcal{L} = \int d^4\theta(X\bar X+Y\bar Y)+f\left(\int d^2\theta X+c.c.\right), 
\end{equation}
where $X^2=0$ and $XY=0$ are assumed.
Notice that the constraints \eqref{eq:c2:XY=0} imply also $Y^3=0$, as it can be seen from inserting the explicit expansion \eqref{eq:c2:solY2}. As a consequence, the most general model involving only $X$ and $Y$ has a K\"ahler potential and a superpotential of the type:
\begin{equation}
\begin{aligned}
K &= X\bar X+Y\bar Y+a\,(X\bar Y^2+\bar X Y^2)\\
&+b\,(Y\bar Y^2+\bar Y Y^2)+ c\, (Y\bar Y)^2,\\
W &= f X + g Y + h Y^2,
\end{aligned}
\end{equation}
where $a,b,c,f,g$ and $h$ are parameters.
The generalisation of this model to supergravity with spontaneously broken supersymmetry has been studied in \cite{DallAgata:2015pdd,DallAgata:2015zxp} and it is reviewed in section \ref{sec:XYsugra}.

\subsection{$X$ and $W_\alpha$ system}
The next system which is analysed is the one in which the gaugino, namely the fermionic superpartner of a gauge vector field, is decoupled from the spectrum. The discussion can be adapted to supersymmetric extensions of the Standard Model, where some of the gaugino fields are much heavier than the gauge vector bosons. In this context, in fact, it could be desirable to have an equivalent description in which the gauginos are removed, in order to facilitate the study of the low energy effective action.

The first step is the embedding of component fields into superfields. The goldstino is going to reside again in a chiral superfield $X$, while the gaugino $\lambda_\alpha$ and the vector $v_m$ are components of another chiral superfield $W_\alpha$, which can be expressed as the superfield strength of an abelian vector superfield $V$, namely
\begin{equation}
W_\alpha = -\frac14 \bar D^2 D_\alpha V.
\end{equation}
In chiral coordinates, its expansion in superspace is 
\begin{equation}
\label{eq:c2:Wexpansion}
W_\alpha = -i \lambda_\alpha + \left[\delta_\alpha^\beta D - \frac{i}{2}{(\sigma^m\bar\sigma^n)_\alpha}^\beta v_{mn}\right]\theta_\beta + \theta^2 \sigma^m_{\alpha\dot\alpha}\partial_m \bar\lambda^{\dot\alpha},
\end{equation}
where $v_{mn}=\partial_m v_n-\partial_n v_m$ and $\lambda_\alpha=i W_\alpha|$ is the gaugino. Notice that this superfield satisfies the constraint
\begin{equation}
D^\alpha W_\alpha = \bar D_{\dot \alpha} \bar W^{\dot \alpha},
\end{equation}
which implies that the auxiliary field $D=-\frac12 D^\alpha W_\alpha|$ is real.

A minimal Lagrangian for the superfields $X$ and $W_\alpha$ in which supersymmetry is spontaneously broken by $X$ is 
\begin{equation}
\begin{aligned}
\mathcal{L} &= \int d^4 \theta X \bar X + f\left(\int d^2 \theta X +c.c.\right)+\frac14 \left(\int d^2 \theta W^\alpha W_\alpha +c.c.\right)\\
&= -\partial_m A\partial^m \bar A -i \bar G \bar \sigma^m \partial_m G+ F\bar F+ fF+f\bar F\\
&\phantom{=}-\frac14 v_{mn}v^{mn}-i\lambda \sigma^m\partial_m\bar \lambda+\frac12 D^2.
\end{aligned}
\end{equation}
In this model all the fields in the spectrum are massless. In particular, the masses of the goldstino $G_\alpha$ and of the $\rm U(1)$ vector $v_m$ are protected by supersymmetry and gauge symmetry respectively. A mass gap can nevertheless be created between these modes and their superpartners, namely the sgoldstino $A$ and the gaugino $\lambda_\alpha$, by considering the modified Lagrangian
\begin{equation}
\label{eq:c2:Llambdadec}
\begin{aligned}
\mathcal{L} &= \int d^4 \theta \left(X \bar X-\frac{1}{\Lambda^2}X^2\bar X^2\right) + f\left(\int d^2 \theta X +c.c.\right)\\
&\phantom{.=}+\frac14 \left(\int d^2 \theta \left(1-\frac{X}{\Lambda}\right)W^\alpha W_\alpha +c.c.\right)\\
&=-f^2+|F+f|^2-\frac{1}{\Lambda^2}|2A F-G^2|^2+\frac12 D^2\\
&\phantom{.=}+\frac{1}{4\Lambda}\left[F\left(\lambda_\alpha -i\frac{\sqrt 2}{2}G_\alpha \frac{D}{F}\right)^2+c.c.\right]\\
&\phantom{.=}+\text{terms with derivatives}.
\end{aligned}
\end{equation}
The equations of motion in the zero-momentum limit give
\begin{align}
\label{eq:c2:solAAA}
A &= \frac{G^2}{2F},\\
\label{eq:c2:solgaugino}
\lambda_\alpha &= i\frac{\sqrt 2}{2}G_\alpha \frac{D}{F}.
\end{align}
Notice that the last steps were not completely rigorous. The correct way to proceed would have been to introduce two different cut-off parameters: $\Lambda$, governing the mass of $A$ and $\tilde \Lambda$, governing the mass of $\lambda$. The decoupling should proceed then in the following order: first the sgoldstino, by taking $\Lambda\to \infty$ while $\tilde \Lambda$ fixed, and after the gaugino, by taking $\tilde \Lambda\to \infty$ in the resulting Lagrangian. Formulae \eqref{eq:c2:solAAA} and \eqref{eq:c2:solgaugino} are indeed consistent with this specific procedure.

It is crucial to observe that the solution \eqref{eq:c2:solgaugino} of the equations of motion of the gaugino in the zero-momentum limit differs from the one reported in literature, see for example \cite{Komargodski:2009rz}. The difference amounts precisely to terms containing derivatives. Since the solution given in \cite{Komargodski:2009rz} has been obtained with a different procedure, the strategy of taking the zero-momentum equations of motion in order to decouple degrees of freedom in the low energy regime, has to be applied with care, as it can be misleading.

Even if not completely correct, however, the result \eqref{eq:c2:solgaugino} can still give some intuition on the form of the superfield $W_\alpha$ in the infrared regime, after the decoupling of its component $\lambda_\alpha$. To have the correct solution, a more systematic procedure is required and it is going to be described in the section \ref{sec:generalconstr}. In particular, the correct form of the superfield, together with the constraints it satisfies, are going to be found and discussed directly in superspace. It can be anticipated that the constraint which is eliminating the gaugino $\lambda_\alpha$ from $W_\alpha$ is $XW_\alpha=0$, as given in \cite{Komargodski:2009rz}.

\subsection{Comment on the zero-momentum limit}
The reason why the procedure of decoupling a massive mode by taking its equations of motion in the zero-momentum limit has failed for the gaugino is that, in general, it has to be applied with care, since by neglecting derivatives interactions supersymmetry could be broken explicitly. This is precisely what occurred in the previous example, even though it did not happen for the decoupling of the sgoldstino or for the scalar in the chiral superfield $Y$. The solution \eqref{eq:c2:solgaugino} cannot be the lowest component of a constrained superfield $W_\alpha$, as opposite to \eqref{eq:c1:solX} and \eqref{eq:c2:solY}, which have the correct structure to be inserted into superfields. In other words, the solution to supersymmetric constraints can in general contain derivatives, indeed in order for \eqref{eq:c2:solgaugino} to be the lowest component of a superfield, also derivatives interactions are needed. This means that, more generally, it is not guaranteed that the integration at zero-momentum is going to preserve supersymmetry, not even at the non-linear level. In the following, an alternative and manifestly supersymmetric approach is presented.

\section{The generic constraint for matter}
\label{sec:generalconstr}
In this section a general constraint to be imposed on matter superfields in order to remove any desired component is given. This is one of the central results of the constrained superfield approach and it is going to be used extensively in the rest of the discussion.

In the previous example it has been shown that the zero-momentum limit has to be performed with care, in order to get the desired result. This observation motivates the need for a more straightforward procedure for constructing low energy effective theories with spontaneously broken supersymmetry. Notice moreover that, at this stage, several constraints have already been introduced and discussed. An organising principle allowing to understand their origin would therefore be welcome. The answer to both these issues has been given in \cite{DallAgata:2016syy}, where the authors proposed one single constraint to be applied on any generic superfield and from which all the known constraints can be derived. This is going to be reviewed in the following.

In chapter \ref{c1:susybreakingsec} it has been shown that the constraint 
\begin{equation}
X^2=0
\end{equation}
eliminates the lowest component from the chiral superfield $X$ and expresses it as a function of the other degrees of freedom. In the previous section it has been discussed analogously that, assuming $X^2=0$, the constraint 
\begin{equation}
\label{eq:c2:conXY}
XY = 0
\end{equation}
eliminates the lowest component of the chiral superfield $Y$ and replaces it with a function of the remaining fields of $X$ and $Y$. It has also been anticipated that, assuming again $X^2=0$, the constraint
\begin{equation}
\label{eq:c2:conXW}
XW_\alpha=0
\end{equation}
is removing the lowest component $\lambda_\alpha=iW_\alpha|$ in a similar way. An educated guess would therefore be that, given a generic matter superfield $Q$ and under the sole assumption that $X$ is a nilpotent chiral superfield, the constraint 
\begin{equation}
XQ =0
\end{equation}
is removing the lowest component $q=Q|$ in a similar manner. The guess is meaningful but it is not completely correct: it works only for the case in which $Q$ is a chiral superfield. The constraint needed to remove the lowest component from any desired superfield $Q_L$, where $L$ is representing some set of Lorentz indices, has been proposed in \cite{DallAgata:2016syy} and it is 
\begin{equation}
\label{eq:c2:genconstr}
X\bar X Q_L =0.
\end{equation}
This is one of the central results in the framework of constrained superfields and the rest of the section is going to be devoted to comment and discuss it further.
Notice, first of all, that the constraints \eqref{eq:c2:conXY} and \eqref{eq:c2:conXW}  can be obtained directly from the general one. Specialising \eqref{eq:c2:genconstr} to the case in which $Q_L = Y$, it results indeed in $X\bar X Y=0$, which can be reduced to \eqref{eq:c2:XY=0} by acting with $\bar D^2$ and dividing by the expression $\bar D^2 \bar X\neq 0$. The same steps can be performed for the constraint on $W_\alpha$. In the work \cite{DallAgata:2016syy} it is shown how to derive also all the other known constraints in a similar way.

The constraint \eqref{eq:c2:genconstr} can be solved directly in superspace and, being supersymmetric, its solution is going to be a superfield. The strategy is again to act on it with the maximum number of superspace derivatives, obtaining
\begin{equation}
D^2 \bar D^2 (X\bar X Q_L)=0,
\end{equation}
and solving then the resulting expression to find
\begin{equation}
\begin{aligned}
Q_{L} &= -2 \frac{\bar{D}_{\dot \beta} \bar{X} \, \bar{D}^{\dot \beta} Q_{L}}{\bar{D}^2 \bar{X}} 
- \frac{\bar{X} \, \bar{D}^2 Q_{L}}{\bar{D}^2 \bar{X}}\\
&-2 \frac{D^\alpha X D_\alpha \bar{D}^2 \left( \bar X Q_{L} \right) }{D^2 X \bar{D}^2 \bar{X}} 
- X \frac{D^2 \bar{D}^2 \left( \bar X Q_{L} \right) }{D^2 X \bar{D}^2 \bar{X}} .
\end{aligned} 
\end{equation}
Acting on the constraint with less derivatives gives consistency conditions for the solution, which guarantee that the supersymmetric properties of the superfield $Q_L$ are not altered. This is precisely the point at which the zero-momentum procedure failed in the previous example, when decoupling the sgoldstino: in that case the solution of the equations of motion was not a superfield.

The constraint \eqref{eq:c2:genconstr} is removing the lowest component of $Q_L$:
\begin{equation}
Q_L = q_L + \theta^\alpha \chi^Q_{L\,\alpha}+\dots,
\end{equation}
but it can be adapted to eliminate also the higher components in a straightforward way.
In order to remove, for example, the field $\chi^Q_L$ in the $\theta$-component, it is sufficient to impose the slightly modified constraint
\begin{equation}
X\bar X D_\alpha Q_L=0.
\end{equation}
The reason why this constraint is the correct one is that $D_\alpha Q_L$ can be though of as a superfield with lowest component $\chi^Q_{L\, \alpha}=D_\alpha Q_L| $. Along with the same reasoning, the needed constraint to remove the $\theta^2$-component of $Q_L$ is
\begin{equation}
X\bar X D^2Q_L=0
\end{equation}
and the logic applies also to the higher components.

The role and the origin of the constraints introduced in the previous chapter can now be understood more properly. In section \ref{subsec:equivalenceVA}, for example, the constraint \eqref{Rocekconstr}
\begin{equation}
\Phi \bar D^2 \bar\Phi = 4f \Phi
\end{equation}
has been used to set the auxiliary field to $F^\Phi=-f+\text{fermions}$, thus fixing the supersymmetry breaking scale. By multiplying this constraint with $\bar\Phi$, it can be recast into the form of \eqref{eq:c2:genconstr}, namely
\begin{equation}
\Phi \bar \Phi \left(\bar D^2\bar\Phi-4f\right)=0.
\end{equation}
This proves then that the interpretation of \eqref{Rocekconstr} given above is correct. Notice moreover that, being chiral and nilpotent, $\Phi$ can be used instead of $X$ in the expressions of all the previous constraints. It has also been argued that the constraint \eqref{XH}
\begin{equation}
\bar D_{\dot \beta}(X\bar H_{\dot \alpha})=0
\end{equation}
removes both the $\theta$- and the $\theta^2$-component from the chiral superfield $H_\alpha$. Even such statement can be justified at this point. By multiplying the constraint with $\bar X$, in fact, it reduces to
\begin{equation}
X\bar X\bar D_{\dot \beta}\bar H_{\dot \alpha}=0,
\end{equation}
which confirms that the $\theta$-component is removed. By multiplying it with $\bar X \bar D^{\dot \beta}$, instead, the constraint becomes
\begin{equation}
X\bar X\bar D^2 \bar H_{\dot\alpha }=0,
\end{equation}
which has the correct form to eliminate the $\theta^2$-component. Since the general constraint \eqref{eq:c2:genconstr} can be applied to superfields with any generic set of Lorentz indices, in complete analogy with this last example it is possible to realise that, a scalar chiral superfield $H$, such that $\bar D_{\dot\alpha}(X\bar H)=0$, contains just the scalar in the lowest component as independent degree of freedom. This superfield is going to be employed in section \ref{c2:sec:Ftermbreak}.

The analysis of the general constraint \eqref{eq:c2:genconstr} proposed in \cite{DallAgata:2016syy} is concluded. Additional details can be found in the original paper. Evidence has been given that such an ingredient is central in the studying of non-linear realisations of supersymmetry with the constrained superfields approach. This constraint, in particular, can be used directly in supergravity without any further modification.

\section{Emergence of $X^2=0$ in the low energy}
\label{cap2:sec:alwaysX2=0}
The only assumption on which the general constraint $X\bar X Q_L=0$ is relying is the existence, in the low energy regime, of a nilpotent goldstino chiral superfield $X$, which contains all the information on the spontaneous breaking of supersymmetry. Since one of the purposes of this work is to outline a systematic procedure to build models with spontaneously broken supersymmetry and with any desired spectrum content, the generality of the existence of such a superfield $X$ has to be investigated.

It has been discussed previously how the nilpotent constraint on $X$ arises as a consequence of the decoupling of the sgoldstino, the scalar superpartner of the goldstino, from the low energy theory. Despite the fact that in various known examples the decoupling can be performed in a smooth way, even the generality of such a procedure can be questioned \cite{Dudas:2011kt, Ghilencea:2015aph}. The aim of this section is therefore to address the issues on how general is the description of the supersymmetry breaking sector in terms of the nilpotent $X$ and also if the appearance of the constraint $X^2=0$ has to be necessarily related to the decoupling of the sgoldstino. The answer is going to be that the description in terms of $X$ is completely general, namely in any desired model the goldstino can be embedded inside such a superfield, while, for what concerns the decoupling, it is going to be shown that  the nilpotent constraint can arise also in those systems in which the sgoldstino is not decoupled. 

The strategy adopted is to consider both F-term and D-term supersymmetry breaking and to proceed by examples. To analyse them, a particular approach is proposed, which could be of interest in its own right: it is going to be explained how each component field of a given superfield can be promoted to a full constrained superfield or, vice versa, how any unconstrained superfield arising from linear representations of supersymmetry can be parametrised as a combination of constrained ones. Using this strategy, which is not going to be altered by the decoupling of the heavy modes in the infrared regime, it is shown that the low energy description of any spontaneously broken supersymmetric theory contains the nilpotent chiral superfield $X$. 

It has to be stressed that, even in the case of pure D-term breaking, namely when only a vector superfield is present, the degrees of freedom can be reorganised in order to parametrise the model with the nilpotent chiral superfield $X$. In other words, D-term breaking itself can be interpreted as F-term breaking. This applies also to the new type of D-term introduced in \cite{Cribiori:2017laj} and reviewed in section \ref{c5:sec:newDterm}.

\subsection{The goldstino and the Ferrara--Zumino supercurrent}
\label{cap2:subsec:supercurrent}
To motivate further the forthcoming discussion, some known examples are given of models in which a goldstino is present in the spectrum, but its embedding into a nilpotent chiral superfield seems not to be viable. One possibility for this to happen is when the sgoldstino cannot acquire a mass, for example due to the presence of a shift symmetry protecting it in the low energy regime. Another scenario in which there seems to be some obstruction is the case in which the breaking of supersymmetry is not sourced solely by a nilpotent chiral superfield, but it receives also other contributions. The goldstino is then a combination of the fermions in the model and its relation with the nilpotent $X$ might not be completely transparent.
Even if $X$ is nilpotent, therefore, it is not describing totally the goldstino sector.  
A third possibility is when no chiral superfields are present from the very beginning, for example in the case of a model with only a vector superfield and pure D-term breaking. In the present subsection these situations are analysed in more detail, while in the next subsection it is going to be shown how, even in all these models, it is nevertheless possible to let a nilpotent $X$ emerge.

It is known that, for any globally supersymmetric theory with K\"ahler potential $K$ and superpotential $W$, there exists a supercurrent superfield ${\cal J}_{\alpha \dot \alpha}$ that satisfies the 
conservation equation  
\begin{equation}
\bar D^{\dot \alpha}  {\cal J}_{\alpha \dot \alpha} = D_\alpha \mathcal{Z} \,  , 
\end{equation} 
where $\mathcal{Z}$ is a chiral superfield given by the combination
\begin{equation}
\mathcal{Z} =4 W - \frac13 \bar D^2 K \, . 
\end{equation}  
Following the reasoning of \cite{Komargodski:2009rz}, it is possible to realise that, when supersymmetry is spontaneously broken, the low energy supercurrent is expressed in terms of the goldstino. In particular the $\theta$-component of ${\cal J}_{\alpha \dot \alpha}$, and in turn the one of $\mathcal{Z}$, should be identified with it.
When the sgoldstino ${\mathcal{Z}}|_{\theta=0}$ is not present in the low energy theory, then the corresponding operator creates composite states of the goldstino. Since on these composite states supersymmetry is acting linearly, the transformations become non-linear when specialised to the single objects constituting them. As a consequence, one can derive that the superspace expansion of $\mathcal{Z}$ satisfies
\begin{equation}
\mathcal{Z}^2 =0 \,  
\end{equation} 
and this would suggest the possibility of identifying it with the nilpotent goldstino chiral superfield even in more general systems.

It is possible however to construct some situations in which this argument does not seem to be straightforward and thus the existence of a nilpotent chiral superfield $X$ in the infrared regime can be questioned.
For instance, in the simple model
\begin{equation}
K = \Phi \bar \Phi \, , \qquad W = f  \, \Phi \, ,
\end{equation}
where $c$ is a complex constant, the shift symmetry $\phi\to\phi + c$ protects the mass of the scalar $\phi=\Phi|$. 
This implies that  $\mathcal{Z} = \frac{8}{3} f \Phi $ contains a massless scalar and therefore 
\begin{equation}
\mathcal{Z}^2 = \frac{64}{9} f^2 \Phi^2 \ne 0.
\end{equation}
In addition, this is suggesting a possible relationship between the decoupling of the sgoldstino and the presence of the nilpotent constraint. 

Even in those cases in which a nilpotent $X$ is present, it is not guaranteed that $\mathcal{Z}$ is going to be nilpotent as well.
For example, in a model with two orthogonal constrained superfields $X$ and $Y$, satisfying
\begin{equation}
\label{XY}
X^2 = 0 = XY  , 
\end{equation}
such a situation occurs whenever a linear term in $Y$ in the superpotential is considered. 
For the simplest choice
\begin{equation}
K = X \bar X + Y \bar Y  , \qquad W = f  \, X + g \, Y ,
\end{equation}
it is possible to obtain $\mathcal{Z} = \frac83 (f X +  g Y)$, which satisfies
\begin{equation}
\mathcal{Z}^2 = \frac{64}{9} g^2 Y^2 \ne 0  , 
\end{equation}
since \eqref{XY} implies $Y^3=0$, but $Y^2$ is still non-vanishing. In this model the goldstino is not completely identified with $G_\alpha = \frac{1}{\sqrt 2} D_\alpha X|$, but it is a combination of the two fermions. It is therefore evident that, even though a nilpotent $X$ is present in the model, $\mathcal{Z}$ cannot be completely identified with it in general situations. 

A third example is when there are no chiral superfields and supersymmetry is broken by a pure Fayet--Iliopoulos D-term $\xi$. In this case it can be calculated that
\begin{equation} 
\mathcal{Z}= -\frac{\xi}{3} \bar D^2 V ,
\end{equation}
which is not nilpotent. The standard strategy for identifying a nilpotent $X$ seems to be failing again.

In the following it is going to be shown that, even in these models in which $\mathcal{Z}^2\neq 0$ in the low energy regime, it is always possible to parametrise the breaking of supersymmetry by means of a constrained chiral superfield $X$, whose square vanishes. This superfield may not contain the full goldstino, but it is going to contribute in any case to the goldstino interactions and its auxiliary field will be selected by the dominant source of supersymmetry breaking.

\subsection{Chiral multiplets and F-term breaking} 
\label{c2:sec:Ftermbreak}

The first examples of supersymmetry breaking scenarios which are analysed, are models with interacting chiral multiplets, with F-term breaking. 
They have been already discussed previously \cite{Casalbuoni:1988xh,Komargodski:2009rz}, but the following presentation is going to emphasise the salient features of the approach used to prove the general existence of the nilpotent superfield $X$ in the low energy effective theory. The analysis is initially restricted to the supersymmetry breaking sector and then generalised with the inclusion of matter couplings.

Start by considering the simplest model with a single chiral superfield $\Phi$\footnote{This $\Phi$ is a generic chiral superfield and it is not related to the constrained superfield field $\Phi$ introduced and used in the previous chapter.} and with a supersymmetry breaking linear superpotential 
\begin{equation}
\label{LPhilinear}
\mathcal{L}= \int d^4 \theta \, \Phi \bar \Phi + f\left(\int d^2 \theta\, \Phi + c.c. \right) .
\end{equation}
As usual, supersymmetry is broken because $\langle F^\Phi \rangle = -f \neq 0$ and the fermionic component of $\Phi$ is the goldstino.  
The sgoldstino $\phi = \Phi|_{\theta=0}$ is massless but, as it has already been discussed extensively, one can amend this by introducing an additional operator in the Lagrangian suppressed by a scale $\Lambda > \sqrt f$, whose only net effect is to generate a mass for $\phi$. For completeness the Lagrangian modified by the addition of this new term is reported once more
\begin{equation}
\label{mu1}
\mathcal{L}= \int d^4 \theta \, \left(\Phi \bar \Phi -\frac{1}{\Lambda^2}\, \Phi^2 \bar \Phi^2\right) + f\left(\int d^2 \theta\, \Phi + c.c. \right) .
\end{equation}
The sgoldstino acquires a mass of the order $f/\Lambda$ and the infrared effective theory valid at energy below this scale is expected to be described by the goldstino alone.

The different approach proposed now to study this system consists in splitting the degrees of freedom of $\Phi$ between two constrained superfields $X$ and $S$, satisfying
\begin{equation}
\label{XH1}
X^2=0 \ ,\qquad X {\bar D}_{\dot \alpha} \bar S =0.
\end{equation} 
The first constraint removes the scalar from $X$ and expresses it in terms of its fermion $G_\alpha$ and the auxiliary field $F$, delivering   
\begin{equation}
\label{Xexpansion}
X = \frac{G^2}{2F} + \sqrt{2}\, \theta^\alpha G_\alpha + \theta^2 F.
\end{equation}
The second constraint, as discussed at the end of section \ref{sec:generalconstr}, removes the fermionic component $\chi^S_\alpha$ and the auxiliary field $F^S$ of $S$, giving 
\begin{equation}
\label{Sexpansion}
S = s + \sqrt 2 i  \, \theta \sigma^m  \left( \frac{\bar G }{\bar F}  \right) \partial_m s + \theta^2 \left( \frac{\bar G^2}{2 \bar F^2} \partial^2 s 
- \partial_n \left( \frac{\bar G}{\bar F}  \right) \bar \sigma^m \sigma^n \frac{\bar G}{\bar F}  \partial_m s   \right) \, . 
\end{equation}
Notice that the only independent field is the complex scalar $s$ in the lowest component.
Using these constrained superfields it is possible to construct the combination
\begin{equation}
\label{FXH}
\Phi= X + S \,,
\end{equation}
\emph{i.e.}
\begin{equation}
\begin{aligned}
\Phi&= \frac{G^2}{2F}+s\\
&+\sqrt 2\theta^\alpha\left(G_\alpha+i {\sigma^m}_{\alpha\dot\beta} \left(\frac{\bar G^{\dot\beta}}{\bar F}\right)\partial_ms\right)\\
&+ \theta^2 \left(F+ \frac{\bar G^2}{2 \bar F^2} \partial^2 s 
- \partial_n \left( \frac{\bar G}{\bar F}  \right) \bar \sigma^m \sigma^n \frac{\bar G}{\bar F}  \partial_m s   \right),
\end{aligned}
\end{equation}
which indeed contains the same degrees of freedom as the original unconstrained superfield $\Phi$. 
In other words, the unconstrained chiral superfield $\Phi$ has been parametrised in terms of two constrained chiral superfields: the nilpotent superfield $X$, given by \eqref{Xexpansion}, and the sgoldstino superfield $S$, given by \eqref{Sexpansion}.
The important fact is that this decomposition does not depend on the details of the model under consideration: it can always be performed and it is valid at any energy regime.  
Notice also that the parameterisation \eqref{FXH} has been constructed precisely with the same spirit of \eqref{eq:c1:phipar}: the goldstone modes degrees of freedom, contained in $X$, are separated from the remaining degrees of freedom, contained in $S$.

The Lagrangian of the model \eqref{mu1}, can be expressed in terms of the parameterisation \eqref{FXH} as
\begin{equation}
\label{LLXH}
\begin{aligned}
\mathcal{L} &= \int d^4 \theta \left(X\bar X + S \bar S - \frac{1}{\Lambda^2}(4 X \bar X S \bar S + S^2 \bar S^2)\right)\\ 
&+ f \left(\int d^2 \theta (X + S)+c.c.\right). 
\end{aligned}
\end{equation}
Following the reverse reasoning, this procedure shows explicitly that a non-linearly realised theory of supersymmetry, with very specific couplings as those in \eqref{LLXH}, can behave like a linearly realised one. The very expression \eqref{LLXH} can also be used to derive an effective description for the model under investigation. 
If energies well below $f/\Lambda$ are probed, the massive scalar $s$ in the sgoldstino superfield $S$ is decoupled and, since this is the only independent component of $S$, the effective theory should be obtained by setting $S=0$ directly in superspace. 
In the zero-momentum limit indeed, which can be performed safely in this simple example, the Lagrangian \eqref{LLXH} in component form becomes
\begin{equation}
\label{zp1}
{\cal L} = - f^2 + |F + f|^2 - 4 \frac{|F|^2}{\Lambda^2} | s |^2  + \text{terms with derivatives}\,
\end{equation} 
and this is minimised for configurations where
\begin{equation}
\label{s==0}
 s = 0 \,,
\end{equation}
which implies $S=0$, because of \eqref{Sexpansion}. The effective description of the original model is given then by the expected Lagrangian for the goldstino:
\begin{equation}
	\label{c2:eq:VA}
	{\cal L} = \int d^4 \theta\, X \bar X + f \left(\int d^2 \theta\, X + c.c.\right)\,.
\end{equation}

In the case in which the starting Lagrangian for $\Phi$ is more complicated, the parameterisation \eqref{FXH} can still be used in the same form, but the expression for the decoupling of $S$ can be less transparent than \eqref{s==0}. It is going to be possible to express again the low energy Lagrangian in terms of the nilpotent $X$, albeit in a more involved way than \eqref{c2:eq:VA}.

The same effective model can be derived by working entirely at the superspace level. Assuming in fact that the operator governed by $\Lambda$ in \eqref{mu1} dominates the equations of motion, their superspace formulation becomes
\begin{equation}\label{eomphi}
\Phi \bar D^2 \bar \Phi^2 =0.
\end{equation}
Decomposing $\Phi$ as in \eqref{FXH} and acting with $X\bar X D^2$ gives
\begin{equation}
X\bar X \left( \bar S |D^2 X|^2 + 8 S \partial^2 \bar S^2 + 16 |S|^2 \partial^2 \bar X \right) = 0 \,.
\end{equation}
This equation can then be used to find an expression for $X \bar X S$ in terms of operators including derivatives on $S$ and on $X$:
\begin{equation}
	X \bar X \bar S = - 8 \,\frac{ X \bar X S}{|D^2 X|^2} \left( \partial^2 \bar S^2 + 2 \bar S \partial^2 \bar X\right). 
\end{equation}
By substituting iteratively the resulting expression into itself it is possible to arrive at
\begin{equation}
X\bar X S=0
\end{equation} 
and therefore the desired result
\begin{equation}
 S=0
\end{equation} 
is obtained.

Another example involving solely the chiral superfield $\Phi$ can be constructed. It is known in fact that supersymmetry breaking in rigid supersymmetric models is often related to $R$-symmetry breaking, which implies the existence of an $R$-axion in the effective theory. For this reason, it is meaningful to be able to construct a low energy Lagrangian describing the interactions between the goldstino and one real massless scalar. Both supersymmetry and $R$-symmetry are going to be spontaneously broken.
Consider therefore the model
\begin{equation}
\label{mulambda}
\begin{aligned}
\mathcal{L}&= \int d^4 \theta \, \left(\Phi \bar \Phi + \frac{\mu}{4\Lambda^2} \, \Phi^2 \bar \Phi^2 
- \frac{\lambda}{9\Lambda^4}\, \Phi^3 \bar \Phi^3 \right) + f\left(\int d^2 \theta\, \Phi + c.c. \right) , 
\end{aligned}
\end{equation}
where $\mu$ and $\lambda$ are positive real constants.
This model has an $R$-symmetry and the superfield $\Phi$ has $R$-charge 2. 
The scalar potential 
\begin{equation}
\mathcal{V} = \frac{f^2}{1 + \frac{\mu}{\Lambda^2} \phi \bar \phi - \frac{\lambda}{\Lambda^4} \phi^2 \bar \phi^2 } \, ,
\end{equation}
where $\phi = \Phi|$, admits a stable vacuum at 
\begin{equation}
\langle \phi \bar \phi \rangle  = \frac{\mu}{2 \lambda} \Lambda^2 \equiv v^2 \, , 
\end{equation}
with 
\begin{equation}
	\langle \mathcal{V}\rangle = \frac{f^2}{1+ \lambda \frac{v^4}{\Lambda^4}}.
\end{equation}
The spectrum at this vacuum consists of a massless real scalar, which is the $R$-axion, another real scalar, with mass $m^2 = 128 \, \frac{f^2}{\Lambda^2} \frac{\lambda^3 \mu}{\left(4 \lambda + \mu^2 \right)^3}$ and the goldstino.

To construct the effective theory, the superfield $\Phi$ is parametrised using constrained superfields and separating the goldstone mode degrees of freedom from all the rest of the fields. The decoupling is different with respect to the previous example, since the effective theory should contain one real scalar, while before $s$ was complex. The way $\Phi$ is going to be parametrised is different as well.
One could still use the parameterisation \eqref{FXH}, but then the decoupling procedure would not work transparently. The parameterisation in fact reflects the mass gap in the spectrum and consequently the decoupling one wants to achieve. For this example the following parameterisation is proposed
\begin{equation}
\label{eq:c2:parPhiaxion}
\Phi = X + \left(v + {\cal B} \right) {\cal R}^2 \, , 
\end{equation}
where $X$ is the nilpotent goldstino superfield, while ${\cal B}$ and ${\cal R}$ are chiral constrained superfields satisfying
\begin{align}
\label{eq:c2:constrXB}
X {\cal B} = X \bar{\cal B}&,  \\ X \left( {\cal R} \bar{\cal R} -1 \right) = 0& \, . 
\end{align} 
In particular, the chiral superfield ${\cal B}$ has vanishing $R$-charge, while the chiral superfield ${\cal R}$ has $R$-charge one. 
The constraint on $\mathcal{B}$ implies $X\bar D_{\dot \alpha}\bar{\mathcal{B}}=0$, therefore its solution is a superfield of the form \eqref{Sexpansion}, where however the scalar in the lowest component is split into real and imaginary part 
\begin{equation}
B| = b + i h.
\end{equation}
Since \eqref{eq:c2:constrXB} is more stringent than \eqref{XH1}, the field $h$ is replaced by the composite expression 
\begin{equation}
\begin{aligned}
h &= \frac12 \left(\frac{G}{F}\sigma^m \frac{\bar G}{\bar F}\right) \partial_m b-\left(\frac{i}{8}\frac{G^2}{F^2}\partial_m \left(\frac{\bar G}{\bar F}\right)\bar \sigma^n \sigma^m \frac{\bar G}{\bar F}\partial_n b + c.c.\right)\\
&-\frac{G^2 \bar G^2}{32 F^2 \bar F^2}\partial_m \left(\frac{\bar G}{\bar F}\right)\left(\bar\sigma^n\sigma^m\bar\sigma^p+\bar\sigma^m\sigma^p\bar\sigma^n\right)\partial_p\left(\frac{G}{F}\right)\partial_n b.
\end{aligned}
\end{equation}
while $b$ is the real massive scalar of the model which eventually is going to decouple.
Assuming that $\langle {\cal R} \rangle \ne 0$, also the constraint on $\mathcal{R}$ implies $X \bar D_{\dot \alpha} \bar {\cal R} = 0$.
It is then possible to express $\mathcal{R}$ as
\begin{equation}
{\cal R} = e^{i \mathcal{A}},
\end{equation} 
where $\mathcal{A}$ is a chiral superfield satisfying $X \mathcal{A} = X \bar{\mathcal{A}}$. The lowest component of ${\cal R}$ is therefore
\begin{equation}
{\cal R}| = \text{e}^{ia} + {\cal O}(G,\bar G)  \, , 
\end{equation}
where the real scalar $a$ is the $R$-axion, which remains massless in the model.
To sum up, as a consequence of \eqref{eq:c2:parPhiaxion}, the goldstino and the auxiliary field which breaks supersymmetry are described by $X$, while the complex scalar $\phi$ is splitted into two real degrees of freedom. One of them is embedded into $\mathcal{B}$ and, being massive, it is going to decouple in the low energy regime, while the other one, namely the $R$-axion, is encoded into $\mathcal{R}$, or equivalently into $\mathcal{A}$.

In terms of this parameterisation, the zero-momentum limit of the Lagrangian becomes 
\begin{equation}
\begin{aligned}
\label{Lha}
\mathcal{L} &= - \frac{f^2}{1+ \lambda \frac{v^4}{\Lambda^4}} +  \left(1+ \lambda \frac{v^4}{\Lambda^4}\right)\left|\bar F + \frac{f}{1+ \lambda \frac{v^4}{\Lambda^4}}\right|^2 
- \lambda \frac{|F|^2}{\Lambda^4} 
\left[ b^2 \left( 2 v + b \right)^2\right] \,\\
&+\text{terms with derivatives} . 
\end{aligned}
\end{equation}
This is minimised when $b=0$ which, because of supersymmetry, implies that the complete superfield ${\cal B}$ vanishes, namely ${\cal B}=0$. The low energy effective theory is then described by the Lagrangian
\begin{equation}
\begin{aligned}
{\cal L} = & \int d^4 \theta \left( |X|^2 + f_a^2 |{\cal R}|^2 \right) 
 + \left( \int d^2 \theta \left(\hat{f}  X + \tilde{f}  {\cal R}^2 \right) + c.c. \right) ,
\end{aligned}
\end{equation}
where it has been used the fact that $ X \bar{\cal R}^2$ is chiral and 
that $({\cal R} \bar{\cal R})^n - n^2\, {\cal R} \bar{\cal R}$ is the real part of a chiral function.
This model coincides with the one presented in \cite{Komargodski:2009rz,Dine:2009sw}, upon identification of the parameters as follows:
\begin{equation}
	\hat{f} = \frac{f}{\sqrt{\alpha}}, \qquad f_a^2 = 4 v^2 \alpha, \qquad \tilde{f} = v f\,, \qquad \alpha = 1+ \lambda \left(\frac{v}{\Lambda}\right)^4.
\end{equation}
As expected from the general analysis in \cite{Dine:2009sw}, the vacuum-expectation-value of the superpotential is at the threshold of the bound
\begin{equation}
	|\langle W \rangle| \leq \frac12 f_a F,
\end{equation}
since, in the model presented here $|\langle W \rangle| = \tilde f$ and $F = \hat f$.

It has been shown that, starting from a single chiral superfield which breaks supersymmetry with an F-term, it is possible to parametrise the degrees of freedom in such a way that a nilpotent chiral superfield $X$ appears. This $X$ is usually partnered by other constrained superfields, which can decouple in the low energy regime.

Up to this point however the discussion was devoted only to the supersymmetry breaking sector. It is now shown how, even in the presence of matter couplings, the nilpotent chiral superfield $X$ can be introduced with the same logic. 
In a general situation, in the low energy regime there are going to be both complete and incomplete supermultiplets. Even if with the procedure introduced before both complete and not complete supermultiplets can be parametrised in terms of constrained superfields, usually a complete supermultiplet does not need to be described using non-linear realisations of supersymmetry, since none of its component is going to decouple in the effective theory. 
In particular if only complete matter supermultiplets are present, solely the goldstino superfield is going to realise supersymmetry in a non-linear way \cite{Antoniadis:2010hs}. 
On the other hand, if a given multiplet is incomplete, one can use the generic constraint 
\begin{equation}
\label{XXQ}
X\bar X Q = 0 \, 
\end{equation}
discussed in the previous section to remove any undesired component field from the spectrum. In the remaining part of this subsection a simple example in this direction is discussed. 

Consider a model with two chiral superfields $\Phi$ and $\Sigma$. 
In general they can both contribute to the spontaneous breaking of supersymmetry, but assume for simplicity that only $\Phi$ breaks supersymmetry. 
Since the scalar components of both the superfields $\Phi$ and $\Sigma$ are going to get a mass in the model that it is considered, in terms of constrained superfields it is possible to parametrise $\Phi$ and $\Sigma$ as 
\begin{equation}
\label{XHYH}
\begin{aligned}
\Phi & = X + S \, , 
\\
\Sigma & = Y + H \, , 
\end{aligned}
\end{equation}
where $X$, $S$, $Y$ and $H$ are chiral superfields satisfying
\begin{equation}
\label{XY=0}
X^2=0\, , \qquad X \bar D_{\dot \alpha}\bar S=0\, , \qquad X \, Y = 0 \, ,  \qquad  X {\bar D}_{\dot \alpha} \bar H =0 \,  
\end{equation}
and their superspace expansion has already been given and discussed previously.
The meaning of the parameterisation \eqref{XHYH} is the following. 
The superfield $\Phi$ is decomposed into the goldstino superfield $X$ and into the sgoldstino superfield $S$. The superfield $\Sigma$ instead is decomposed into the superfield $Y$, which contains the physical fermion $\chi_\alpha$ and the scalar auxiliary component field $F^Y$, and into the superfield $H$, which contains the scalar component field $h$.

A simple model with spontaneously broken supersymmetry is 
\begin{equation}
\label{Model2}
{\cal L} =  \int d^4 \theta \left( |\Phi|^2 + |\Sigma|^2 
- \frac{|\Phi|^4}{\Lambda^2}
- \frac{|\Phi|^2 |\Sigma|^2}{\Lambda^2} \right) 
+ f \left(\int d^2 \theta \Phi +c.c.\right) \, ,
\end{equation}
where the couplings proportional to the parameter $\Lambda$ are inserted in order to give a mass to the scalars in the chiral superfields.
As in the previous example, after the replacement \eqref{XHYH} this Lagrangian can be written only in terms of constrained superfields
\begin{equation}
\label{LXSYH}
\begin{aligned}
{\cal L} = &  \int d^4 \theta \left( |X|^2 + |S|^2 + |Y + H|^2 
- 4 \frac{|X|^2 |S|^2}{\Lambda^2} -  \frac{|S|^4 }{\Lambda^2}
-\frac{|X+S|^2}{\Lambda^2} |Y + H|^2 \right)  \\[3mm] 
& + f \left(\int d^2 \theta (X +S) +c.c.\right) \, .  
\end{aligned}
\end{equation} 
In particular the pure $X$ sector has again the form \eqref{c2:eq:VA} and this fact is not going to change in the case in which also $Y$ contributes to the spontaneous breaking of supersymmetry. 

It is possible to analyse once more the low energy effective limit by first looking at the zero-momentum equations. In component form, this gives the Lagrangian
\begin{equation}
\label{zp22}
\begin{aligned}
{\cal L} &= - f^2 + |F + f|^2 + |F^Y|^2 - 4 \frac{|F|^2}{\Lambda^2} | s |^2  - \frac{|F h +  F^Y s |^2}{\Lambda^2}\\
&+\text{terms with derivatives}.
\end{aligned} 
\end{equation}  
It is again clear that in the low energy limit this Lagrangian is extremized by 
\begin{equation}
\label{hs=0}
h = 0 \qquad {\rm and} \qquad  s  = 0 \,,
\end{equation} 
which eventually imply $S=H=0$. 

As in the previous example, one can also obtain the same result by considering the formal limit in which the terms suppressed by $\Lambda$ are large.
The superspace equation of motion of $\Phi$ is going to be the same as \eqref{eomphi}, which is solved by $S = 0$ once the decomposition $\Phi = X+S$ is used. 
The superspace equation of motion for $\Sigma$ in this limit is 
\begin{equation}
\Phi \, \bar D^2 \left( \bar \Sigma \, \bar \Phi \right) = 0 \, 
\end{equation} 
which, once the parameterisation \eqref{XHYH} is inserted, reduces to
\begin{equation}
\label{XHX} 
X \, \bar D^2 \left( \bar H \, \bar X \right) = 0 \, , 
\end{equation}
where $S=0$ and $XY=0$ have been used. 
Due to the properties of $H$, equation \eqref{XHX} gives directly $X \bar H=0$, which implies 
\begin{equation}
H=0 \, . 
\end{equation}  

It has been shown that, with the superfield parameterisation proposed above, the decoupling of the massive states in the low energy regime can be obtained in an efficient way. 
The equations of motion at zero-momentum, in particular, have a straightforward solution.
When changing the parameterisation, it is not guaranteed that the decoupling can still be performed and in general, even in the cases in which it can, the calculations are going to be more involved.
The importance of the parameterisation is also related to the fact that it allows the introduction of a nilpotent chiral superfield in the model and, in turns, it enables the use of all the tools of the constrained superfield approach to non-linear supersymmetry, like the general constraint \eqref{eq:c2:genconstr}, in order to study low energy effective models with spontaneously broken supersymmetry.

\subsection{A parameterisation for the nilpotent superfield $X$}
\label{c2:sec:XZA}

It has been shown that, in generic examples with F-term breaking, even though the nilpotent chiral superfield $X$ is not present from the beginning, it is possible to introduce it by parametrising appropriately the superfields in the model. The other counterexample mentioned in subsection \ref{cap2:subsec:supercurrent} to the appearance of a nilpotent $X$ is related to models in which there are no chiral superfields from the very beginning at all and supersymmetry is broken by the auxiliary field of a vector superfield.
Before proceeding with the analysis of models with D-term breaking or with mixed sources of supersymmetry breaking, however, it is necessary to discuss a particular parameterisation for the nilpotent superfield $X$ that is going to be used to simplify the derivation of some of the forthcoming results. 
Since this is a rather technical intermezzo, the reader interested in the physics of the supersymmetry breaking mechanisms and in the discussion of the resulting low energy effective theories can skip this subsection at first.

The nilpotent superfield $X$ contains, as degrees of freedom, a fermion field and a complex scalar auxiliary field. The idea is to separate further this three modes and to construct three constrained superfields, each one containing only one of them as independent component. This is going to be needed in order to be able to keep track of the auxiliary degrees of freedom when passing from a vector superfield, which has a real auxiliary field, to a chiral superfield, with a complex auxiliary field.

Define first the chiral superfield
\begin{equation}\label{definitionZ}
Z = \bar D^2 \left( \frac{X\bar X}{D^2 X\,\bar D^2
\bar X} \right) ,
\end{equation}
which satisfies 
\begin{equation}
\label{constrZ}
Z^2 =0 , \qquad  Z \, \bar D^2 \bar Z = Z. 
\end{equation}
This superfield has been proposed in \cite{Rocek:1978nb} and it has been introduced and discussed in \eqref{Rocekconstr}, where it has been called $\Phi$.
The constraints \eqref{constrZ} imply that $Z$ contains only a fermion field, which is related to the fermion in $X$ by a field redefinition such that the original goldstino always appears in the fixed combination $G/F$. To understand which is the desired parameterisation for $X$ according to the logic described above, observe that it is possible to invert \eqref{definitionZ} and express $X$ in terms of $Z$. The needed expression is indeed \cite{Liu:2010sk,Cribiori:2016hdz} 
\begin{equation}
X = Z \frac{D^2 X}{D^2 Z} \,,
\end{equation}
which means that $X$ is proportional to $Z$ times an antichiral superfield $D^2 X/D^2 Z$. Since $Z$ contains only the goldstino degree of freedom, the information on the auxiliary field should be encoded into the quantity $D^2 X/D^2 Z$. It is indeed possible to construct constrained superfields built from the chiral projections of the real and imaginary parts of $D^2 X/D^2Z$, namely
\begin{equation}
{\cal A}_1 =  \bar D^2 \left( \frac{\bar Z}{\bar D^2
\bar Z}
\left[  \frac{D^2 X}{D^2 Z} +  \frac{\bar D^2 \bar
X}{\bar D^2 \bar Z} \right]
\right)  
\end{equation}
and
\begin{equation}
{\cal A}_2 = - i \,  \bar D^2 \left( \frac{\bar Z}{\bar
D^2 \bar Z}
\left[  \frac{D^2 X}{D^2 Z} -  \frac{\bar D^2 \bar
X}{\bar D^2 \bar Z} \right]
\right)  \, .
\end{equation}
They are manifestly chiral and satisfy the constraints
\begin{equation}
\label{constrXAZA}
X ({\cal A}_i - \bar{\cal A}_i )=0\,,  \qquad Z ({\cal A}_i -
\bar{\cal A}_i)=0\,.
\end{equation}
In particular, as explained before, the only independent field in ${\cal A}_i$ is a real scalar, which resides in its lowest component, namely
\begin{equation}
{\cal A}_i |_{\theta = \bar \theta = 0} = a_i + {\cal O}(G,\bar G) \,  .
\end{equation}
Since these two superfields contain therefore two real scalars, they can be used to describe the complex auxiliary field of $X$. The parameterisation of $X$ in terms of the three constrained superfields $Z$, $\mathcal{A}_1$ and $\mathcal{A}_2$ is then
\begin{equation}
\label{XZA}
X = \frac{Z}{2} \left({\cal A}_1  + i {\cal A}_2  \right) .
\end{equation}
The independent degrees of freedom in $X$, namely one fermion and two real scalars, have therefore been isolated and promoted to constrained superfields $Z$, $\mathcal{A}_i$, which can be treated as independent. 
In other words, the nilpotent goldstino superfield $X$ can be decomposed into three constrained chiral superfields: one pure goldstino superfield $Z$, containing only one fermion, and two auxiliary superfields $\mathcal{A}_{i}$, each one containing one real (auxiliary) scalar.

Inserting this parametrization \eqref{XZA} into the goldstino Lagrangian \eqref{c2:eq:VA} gives
\begin{equation}
\label{LZA}
\mathcal{L} = \frac14 \int d^4 \theta \, Z \bar Z \left( |{\cal
A}_1|^2 + |{\cal A}_2|^2 \right)
+ \frac{f}{2} \left(\int d^2 \theta Z ({\cal A}_1 + i {\cal A}_2)
+c.c.\right) \, ,
\end{equation}
where for simplicity $f$ is assumed to be real. The properties of the superfields $Z$ and ${\cal A}_2$ imply the interesting fact that their combination drops from the superpotential:
\begin{equation}
\int d^2 \theta \left( i f Z  {\cal A}_2 \right)  + c.c.= -4 i f \,
\int d^4 \theta \, Z \bar Z ({\cal A}_2  - \bar{\cal A}_2  )
=  0 .
\end{equation}
The Lagrangian \eqref{LZA} simplifies then to
\begin{equation}
\label{LAA12}
\mathcal{L} = \frac14 \int d^4 \theta \, Z \bar Z \left( |{\cal
A}_1|^2 + |{\cal A}_2|^2 \right)
+ \frac{f}{2} \left(\int d^2 \theta Z {\cal A}_1 +c.c.\right) \,
\end{equation}
and the variation with respect to ${\cal A}_2$ gives\footnote{The reader is referred to the appendix of \cite{Cribiori:2017ngp} for a detailed proof.}
\begin{equation}
\label{A2=0}
{\cal A}_2 = 0 \, .
\end{equation}
This shows that ${\cal A}_2$ is a trivial auxiliary superfield, whose net effect on the calculation of the final Lagrangian is null.
It is then clear that the model \eqref{c2:eq:VA} is equivalent to
\begin{equation}
\label{onlyA1A1}
\mathcal{L} = \frac14 \int d^4 \theta \, |Z|^2 |{\cal A}_1|^2
+ \frac{f}{2} \left(\int d^2 \theta Z {\cal A}_1 +c.c.\right) \,
\end{equation}
and, using the inverse relation for ${\cal A}_1$ in terms of $X$
\begin{equation}
Z {\cal A}_1 = X \left( 1 + \bar D^2 \left( \frac{\bar
X}{D^2 X } \right) \right) \, ,
\end{equation}
this can be finally expressed as:
\begin{equation}
\label{onlyA1}
\mathcal{L}=\frac14 \int d^4 \theta \, X \bar X \left(2 + \frac{D^2 X}{\bar D^2 \bar X} + \frac{\bar D^2 \bar X}{D^2 X} \right) + \left(f \int d^2 \theta\, X+c.c.\right) .
\end{equation}
With respect to \eqref{c2:eq:VA}, this Lagrangian contains two new terms, which have the form of higher derivatives. As it has been shown, however, the two models are effectively equivalent on-shell and the new terms are present in order to cancel the degree of freedom encoded into the imaginary part of the auxiliary field $F$ of the superfield $X$.
It is important to stress that this result does not imply that the imaginary part of $F$ is set to zero by the equations of motion: it is instead replaced by a composite expression built out of goldstini. 
Following the reverse reasoning, when the parameter $f$ is real the higher derivative terms in \eqref{onlyA1} can be eliminated by restoring the part containing ${\cal A}_2$ in the Lagrangian. 
This step can always be performed, at least at the classical level, because the models \eqref{LAA12} and \eqref{onlyA1A1} are equivalent due to \eqref{A2=0}.
It is in fact straightforward to show that 
\begin{equation}
Z {\cal A}_2 = -i\,X \left( 1 -\bar D^2 \left( \frac{\bar
X}{D^2 X } \right) \right) \, ,
\end{equation}
which implies
\begin{equation}
\label{ZZA2}
\frac14 \int d^4 \theta\, Z \bar Z |\mathcal{A}_2|^2 = \frac14 \int d^4 \theta \, X \bar X \left(2 -\frac{D^2 X}{\bar D^2 \bar X} - \frac{\bar D^2 \bar X}{D^2 X}
\right) \, ,
\end{equation}
which, added to \eqref{onlyA1}, reproduces precisely the Lagrangian \eqref{c2:eq:VA}.
To recapitulate, it has been shown that, if the parameter $f$ is real, the model \eqref{c2:eq:VA}  and \eqref{onlyA1} are equivalent on-shell.

Before ending this subsection, a comment is in order. The previous discussion concerning the imaginary part of the auxiliary field $F$ involved only the pure goldstino sector, without taking into account matter. In particular it can be questioned if the solution \eqref{A2=0} holds also in more general cases, where matter superfields are present and some of their components might be removed.
To answer this question notice first that the generic way to eliminate matter component fields is described by constraints of the form given in \eqref{XXQ}.
Since the constraint \eqref{XXQ} is equivalent to
\begin{equation}
Z\bar Z Q =0 \, ,
\end{equation}
the eliminated components are not going to depend neither on the fields of ${\cal A}_1$ nor on the ones of ${\cal A}_2$, but only on the components of $Z$. The constraints on the matter superfields are therefore independent of $a_2$ and the result presented above holds also in the presence of matter couplings.

\subsection{Vector multiplets and D-term breaking}
\label{c2:eq:DtermisFterm}

All the ingredients have now been introduced, which are needed to discuss the case of D-term breaking and to show that, even in the situation in which no chiral multiplets are present from the beginning, it is possible to parametrise the degrees of freedom in order to let a nilpotent chiral $X$ emerge. In other words, it is going to be proved that any model with D-term breaking can be recast into an equivalent one with F-term breaking.
The case of a pure D-term breaking is considered first and then more general situations are analysed in which a mixing between D-term and F-term breaking occurs. 

Given a vector superfield $V$, a simple model realising pure D-term supersymmetry breaking is
\begin{equation}
\label{LWV}
\begin{aligned}
\mathcal{L} &= \frac14 \left(\int d^2 \theta \, W^\alpha W_\alpha + c.c. \right) + \xi \int d^4\theta \,V\\
&= -\frac14 v^{mn}v_{mn}-i \lambda \sigma^m \partial_m \bar \lambda + \frac12 {\rm D}^2+\xi {\rm D}
\end{aligned}
\end{equation}
where $W_\alpha = -\frac14 \bar D^2 D_\alpha V$ is the chiral super-field strength of the vector $V$ and $\xi$ is a Fayet--Iliopoulos parameter. 
Supersymmetry is spontaneously broken by the auxiliary field $\langle{\rm D}\rangle=-\langle\frac12 D^\alpha W_\alpha|\rangle=-\xi$, whenever $\xi\neq 0$. 

In the spirit of the previous discussion, the first step consists in parametrising this theory with spontaneously broken but linearly realised supersymmetry in terms of constrained superfields. 
To this purpose, consider one chiral superfield $X$ and one real superfield $\tilde{V}$ such that
\begin{equation}
\label{constrXW}
X^2=0,\qquad X \tilde{W}_\alpha =0, \qquad X\bar X D^\alpha \tilde{W}_\alpha=0,
\end{equation}
where $\tilde{W}_\alpha=-\frac14 \bar D^2 D_\alpha \tilde V$.
The first constraint removes the scalar component from $X$, while the second and the third constraints remove the fermion $\tilde \lambda_\alpha$ and the auxiliary field $\tilde {\rm D}$ from $\tilde{V}$. 
The only independent component field in $\tilde V$ is therefore a real vector field. 
The superspace expansion of $\tilde W_\alpha$, in particular, is
\begin{equation}
\label{eq:c2:Wexpansion}
\tilde W_\alpha = -i \tilde \lambda_\alpha + \left[\delta_\alpha^\beta \tilde {\rm D} - \frac{i}{2}{(\sigma^m\bar\sigma^n)_\alpha}^\beta v_{mn}\right]\theta_\beta + \theta^2 \sigma^m_{\alpha\dot\alpha}\partial_m \bar{\tilde \lambda}^{\dot\alpha}
\end{equation}
and, due to the constraints, the gaugino $\tilde\lambda_\alpha$ and the auxiliary field $\tilde {\rm D}$ are expressed as composite combinations of the other degrees of freedom:
\begin{align}
\nonumber
\tilde \lambda_\alpha &= i L_\alpha^\beta \left(\frac{G_\beta}{\sqrt 2 F}\right)-\frac{G^2}{2F^2}\partial_m \left(\frac{\bar G}{\sqrt 2\bar F}\bar L\bar\sigma^m\epsilon\right)_\alpha \\
&-i \frac{G^2}{2 F^2}\left(\sigma^m\bar\sigma^n\partial_m\left(\frac{\bar G^2}{2 \bar F^2}\partial_n\left(L\frac{G}{\sqrt 2 F}\right)\right)\right)_\alpha\\
\nonumber
&-2\left(\frac{G^2}{2 F^2}\right) \left(\frac{\bar G^2}{2 \bar F^2}\right) \left(\partial_m \frac{G^\beta}{\sqrt 2 F}\right) \left(\partial^m \frac{G_\beta}{\sqrt 2 F}\right)\left(\partial_n\frac{\bar G}{\sqrt 2 \bar F}\bar L \bar \sigma^n\epsilon\right)_\alpha,\\
\nonumber
\tilde{\text{D}} &=  \frac12 \left[ \partial_p \left( \frac{G}{\sqrt 2 F} \right) \sigma^m \bar \sigma^n \sigma^p 
\left( \frac{\bar G}{\sqrt 2 \bar F} \right) \right] v_{mn} 
\\
&-  \frac12 \left[  \left( \frac{G}{\sqrt 2 F} \right) \sigma^p \bar \sigma^m \sigma^n 
\partial_p \left( \frac{\bar G}{\sqrt 2 \bar F} \right) \right] v_{mn} 
\\
\nonumber
& -  \frac12 \left[ \left( \frac{G}{\sqrt 2 F} \right) \sigma^p \bar \sigma^m \sigma^n 
\left( \frac{\bar G}{\sqrt 2 \bar F} \right) \right] \partial_p v_{mn}  
+ \cdots  ,
\end{align}
where dots stand for higher order goldstino terms and $L_{\alpha}^{\beta} =  \delta_{\alpha}^{\beta}\;  \tilde{\text{D}} - \frac{i}{2}\, (\sigma^m \bar \sigma^n)_{\alpha}^{\ \beta} v_{mn}$. Notice that the term without derivatives in the expression for the gaugino coincides with the one found in \eqref{eq:c2:solgaugino}, as already anticipated in there.

Using these constrained superfields, the following parameterisation for the vector superfield is proposed
\begin{equation}
\label{VX}
V= \tilde V + \sqrt 2 \frac{X\bar X}{D^2 X}+ \sqrt 2 \frac{X\bar X}{\bar D^2 \bar X} \, , 
\end{equation}
which indeed contains the complete amount of degrees of freedom of an unconstrained vector superfield, namely one real vector, one fermion and one real scalar. 
This parameterisation is again constructed with the same spirit of those proposed for the chiral superfield $\Phi$: the degrees of freedom associated to the breaking of supersymmetry are contained in $X$ and they are being separated from all the remaining fields, which are now embedded into $\tilde V$.
It is important to note that, if also $X$ is parametrised according to \eqref{XZA}, the contribution of $\mathcal{A}_2$ disappears from \eqref{VX}, because
\begin{equation}
\frac{X\bar X}{D^2 X} + \frac{X\bar X}{\bar D^2 \bar X} = 4 Z \bar Z \mathcal{A}_1.
\end{equation}
This was not obvious a priori but it is complementary to the result obtained in the previous subsection. 
In fact when parametrising (the auxiliary field part of) a vector superfield in terms of constrained chiral superfields, one clear obstacle arises because the auxiliary field of a vector superfield is real, while the auxiliary field of a chiral superfield is complex. 
The problem related to this mismatching of auxiliary degrees of freedom is immediately solved for all those systems in which the imaginary part of the auxiliary field $F$ of the superfield $X$ does not get an independent vacuum expectation value. 
As shown in the previous subsection, this occurs whenever the parameter $f$ giving the vacuum expectation value of $F$ is real, which is not a very restrictive condition.

At this point the parameterisation \eqref{VX} can be inserted into the Lagrangian \eqref{LWV} and, using the constraints \eqref{constrXW}, it is possible to obtain
\begin{equation}
\begin{aligned}
\label{LLXV2}
\mathcal{L}  &= \frac14 \left(\int d^2\theta\, \tilde W^\alpha \tilde W_\alpha +c.c.\right)+ \xi \int d^4 \theta\, \tilde V \, \\[3mm]
&+ \frac14 \int d^4 \theta \, X \bar X \left( 2 + \frac{D^2 X}{\bar D^2 \bar X} + \frac{\bar D^2 \bar X}{D^2 X} 
\right)  - \frac{\xi\sqrt 2}{4 }\left(\int d^2 \theta\, X+c.c.\right) . 
\end{aligned}
\end{equation}
As explained in the previous section, one can harmlessly introduce a new term proportional to (\ref{ZZA2}), which vanishes on-shell, so that the Lagrangian takes the more pleasant form
\begin{equation}
\label{LLXV}
\begin{aligned}
\mathcal{L}  &= \frac14 \left(\int d^2\theta\, \tilde W^\alpha \tilde W_\alpha +c.c.\right)+ \xi \int d^4 \theta\, \tilde V\\
& +  \int d^4 \theta \, X \bar X- \frac{\xi\sqrt 2}{4 }\left(\int d^2 \theta\, X+c.c.\right) . 
\end{aligned}
\end{equation}
The Lagrangian \eqref{LLXV} manifestly describes a theory of non-linearly realised supersymmetry, where the goldstino in $X$ interacts with the real vector field in $\tilde V$. 
This theory is equivalent to the original model \eqref{LWV}, where only the vector superfield $V$ is present and supersymmetry is spontaneously broken by the Fayet--Iliopoulos term. 
In other words it has been shown that, for pure D-term supersymmetry breaking, the theory can be parametrised in terms of a nilpotent chiral superfield $X$ accommodating the goldstino. Even in the case in which no chiral superfields are present from the beginning, therefore, a nilpotent $X$ can emerge from the model.

This analysis can be easily generalised in order to include matter couplings and possibly an abelian gauge symmetry
\begin{equation}
\label{GEN1}
\begin{aligned}
\mathcal{L} = &  
\frac{1}{4} \left(\int d^2\theta \, {\cal F}(\Phi) \,  W^\alpha W_\alpha +c.c.\right) 
+ \xi \int d^4 \theta\, V 
+ \int d^4 \theta \sum_i \, \bar \Phi^i  {e}^{q_i V} \Phi^i 
\\[2mm]
& 
- \int d^4 \theta \sum_{ij} \, \mu_{ij} \left( \bar \Phi^i  {e}^{q_i V} \Phi^i \right) \left( \bar \Phi^j  {e}^{q_j V} \Phi^j \right)
+ \left(\int d^2 \theta\, {\cal W}(\Phi^i) + c.c. \right) \, .  
\end{aligned}
\end{equation} 
The details of the parameterisation of the various fields depend on the supersymmetry breaking pattern and especially on the low energy spectrum. A systematic analysis would be then highly model dependent, but it is however possible to recognise some general features of these systems. Notice in fact that, when scalars are present in the infrared regime and some of the chiral superfields are splitted as
\begin{equation}
	\Phi^i = H^i + Y^i,
\end{equation}
an additional contribution to the superpotential appears of the form\footnote{The Wess--Zumino gauge $X \tilde V=0$ proposed in \cite{Komargodski:2009rz} is used.}
\begin{equation}
	 -\frac{\sqrt 2}{4} X \left[ \xi  + \sum_i q_i  |H^i|^2	+  \sum_{ij} \mu_{ij} (q_i + q_j)  |H^i|^2 |H^j|^2 \right] ,
\end{equation}
where the condition $X \bar{D}_{\dot \alpha} \bar{H}^I = 0$ has been used. 
It should be pointed out also that, for a generic choice of gauge kinetic function $\mathcal{F}$, the normalisation of the kinetic term of the nilpotent chiral superfield $X$ gets modified to
\begin{equation}
\label{GEN2}
\int d^4 \theta \left( \frac{{\cal F} + \bar {\cal F}}{2} \right) X \bar X .
\end{equation} 

The discussion of this subsection is concluded. The major results which has been obtained is that a model with pure D-term breaking and no chiral superfields can be recast into an equivalent one in which supersymmetry is spontaneously broken by the auxiliary field $F$ of a nilpotent chiral superfield $X$.

\subsection{Mixed F-term and D-term breaking} 
It is finally considered the case involving a mixing between F-term and D-term breaking contributions. In this class of models it is shown that, even if a nilpotent chiral $X$ is not present in the initial formulation of the theory, the degrees of freedom can again be organised in order to let $X$ emerge, independently from the decoupling of the massive modes.

A simple system to start with is described by the following Lagrangian 
\begin{equation}
\label{ggL}
\begin{aligned}
\mathcal{L} = &  
\frac{1}{4} \left(\int d^2\theta \, \left\{1 + \frac{\Phi}{M} \right\} \,  W^\alpha W_\alpha +c.c.\right) 
+ \xi \int d^4 \theta\, V 
\\ 
& + \int d^4 \theta \Phi \bar \Phi  + \left(\int d^2 \theta\, \left(f \Phi+\frac{m}{2}\Phi^2\right) + c.c. \right) \,,
\end{aligned}
\end{equation}
where the parameters $f$, $\xi$, $m$ and $M$ are chosen to be real, for simplicity.
At the component level, the F-term and D-term contributions to the scalar potential are respectively
\begin{equation}
{\cal V}_\text{F} = f^2 +2\,mfa +m^2(a^2+b^2)\ , \qquad {\cal V}_\text{D} = \frac{ \xi^2}{8\left(1+\frac{a}{M}\right) } \, , 
\end{equation}
where the scalar component $\phi=\Phi|$ has been decomposed into its real and imaginary parts  $\phi = a+ i b$. 
The purpose of this example is to get a low energy effective theory in which the goldstino is interacting with the vector, as in the case of pure D-term breaking. 
The difference with respect to that example, however, is that this time both the gaugino and the fermion in $\Phi$ are going to get a mass. 
The goldstino  is going then to be the massless combination of this two fermions, while the massive combination is decoupling in the low energy limit. 
The analysis is differentiated between two regimes of supersymmetry breaking: one where the F-term is dominating and the other where the D-term is dominating.
While some details of the two regimes change, both examples can be covered by the same parameterisation of the linear multiplets in terms of constrained superfields.

The regime where the F-term source of supersymmetry breaking is dominating is obtained when 
\begin{equation}
M  \gg \sqrt{f} \gg \sqrt \xi \gg E,
\end{equation}
where $E$ is expressing the energy range of validity of the effective theory, while the linear theory (\ref{ggL}) is valid for $ M \gg E > \sqrt{f}$.
For simplicity, given the aforementioned hierarchy and since the interest is in the qualitative behaviour of the model in the low energy, the parameters are tuned such that
\begin{equation}
{\xi}=4\sqrt{M^3 m}, \qquad f=M^2.
\end{equation}
With this particular choice the potential has a minimum at
\begin{equation}
\langle a\rangle = 0, \qquad \langle b \rangle =0.
\end{equation}
In this minimum the mass spectrum consists of one massless vector, two real scalars of mass $m_a^2=4m^2 + 4mM$ and $m_b^2 = 2m^2$ respectively, one goldstino and one massive fermion of mass $m_f^2 = \frac{(M-2m)^2}{16}$.
At low energies therefore only the vector field and the goldstino are expected to survive.
Since both $V$ and $\Phi$ contribute to the spontaneous breaking of supersymmetry, the goldstino and the massive fermion are given by combinations of $\chi^\Phi_\alpha$, the fermion in $\Phi$, and $\lambda_\alpha$, the gaugino. They are indeed:
\begin{equation}
\label{eigenvectors}
\begin{aligned}
&-\frac{i}{\sqrt 2}\sqrt\frac{M}{m}\chi^\Phi_\alpha +\lambda_\alpha \quad \sim \quad \text{goldstino},\\[3mm]
&-i\sqrt{2}\sqrt{\frac{m}{M}}\chi^\Phi_\alpha + \lambda_\alpha \quad \sim \quad \text{massive fermion}.
\end{aligned}
\end{equation}
For F-term to be dominating, $M \gg m$ is required and thus the goldstino mostly resides in $\Phi$.

On the other hand, the regime in which D-term dominates the spontaneous breaking of supersymmetry breaking is when
\begin{equation}
M \gg \sqrt{\xi} \gg \sqrt f \gg E.
\end{equation}
Once again, for simplicity, this situation is covered by setting
\begin{equation}
{\xi}=4\sqrt{m^3 M}, \qquad f=m^2.
\end{equation}
With this particular choice the potential has a minimum at
\begin{equation}
\langle a\rangle = 0, \qquad \langle b \rangle =0.
\end{equation}
In this minimum the mass spectrum is made up of one massless vector, two real scalars of mass $m_a^2=\frac{2m^2(2m+M)}{M}$ and $m_b^2 = 2m^2$, one goldstino and one massive fermion of mass $m_f^2 = \frac{m^2(m-2M)^2}{16 M^2}$.
Also in this case, in the low energy limit, only the vector field and the goldstino are expected to survive. 
Since both $V$ and $\Phi$ contribute to the spontaneous breaking of supersymmetry, the goldstino and the massive fermion are given by combinations of $\chi^\Phi_\alpha$, the fermion in $\Phi$, and $\lambda_\alpha$, the gaugino. They are indeed:
\begin{equation}
\label{eigenvectors2}
\begin{aligned}
&-\frac{i}{\sqrt 2}\sqrt\frac{m}{M}\chi^\Phi_\alpha +\lambda_\alpha \quad \sim \quad \text{goldstino},\\
&-i\sqrt{2}\sqrt{\frac{M}{m}}\chi^\Phi_\alpha + \lambda_\alpha \quad \sim \quad \text{massive fermion}.
\end{aligned}
\end{equation}
For D-term to be dominating, $M \gg m$ is required and thus the goldstino mostly resides in $V$.
 
The model in the two regimes is now analysed using the strategy of the previous subsections, in order to decouple degrees of freedom and to construct effective theories which are valid in the low energy.
In principle various different parameterisations of the unconstrained superfields in terms of the constrained ones can be adopted. 
To obtain the desired decoupling, in which the aforementioned massive modes are not present in the infrared, a convenient choice for the parameterisation is
\begin{equation}
\label{ggR}
\begin{aligned}
V & = \hat V + \sqrt 2 \frac{ Y \bar X }{\bar D^2 \bar X} + \sqrt 2 \frac{ \bar Y X }{D^2 X} \,  ,
\\
\Phi & = X + S \, ,
\end{aligned}
\end{equation} 
where $Y$ and $\hat V$ are respectively a chiral and a vector superfields satisfying 
\begin{equation}
X Y = 0  , \qquad X \bar X D^2 Y  = 0  
 , \qquad  X \hat W_\alpha = 0   ,  
\end{equation}
while the chiral superfields $X$ and $S$ satisfy  \eqref{XH1}. 
In particular, the only independent component of $Y$ is the fermion $\chi^Y_\alpha$, which is going to be aligned with the massive fermion.
By inserting the parameterisation \eqref{ggR} in the Lagrangian \eqref{ggL}, the model can be written entirely in terms of constrained superfields as
\begin{equation}
\label{ggL2}
\begin{aligned}
\mathcal{L} = &  
\frac{1}{4} \left(\int d^2\theta \, \left\{1 + \frac{X + S}{M} \right\} \,  
W^2 (\hat V, Y, X) +c.c.\right) 
+ \xi \int d^4 \theta\, \hat V \\
&- \frac{\sqrt 2 \xi}{4} \left( \int d^2 \theta \, Y + c.c. \right) 
 + \int d^4 \theta \left(X\bar X + S \bar S\right) \\
&
+  \left(\int d^2 \theta \, \left(f (X + S)+\frac{m}{2}(XS+S^2)\right)+c.c.\right)  , 
\end{aligned}
\end{equation} 
where the vector superfield $V$ in $W_\alpha$ is parametrized as in \eqref{ggR}.

In the zero-momentum limit, this Lagrangian reduces to
\begin{equation}
\begin{aligned}
\mathcal{L} &=-f^2 -\frac{\xi^2}{8} + |F+m\bar s+f|^2+\frac12\left(\text D +\frac{\xi}{2}\right)^2-m^2 s \bar s 
\\
& +\frac{1}{4M}\left[\text D^2- 4 m M\, f\right](s+\bar s)
+ \frac{1}{16 M}\left(F(\chi^Y)^2 + \bar F(\bar \chi^Y)^2\right)\\
&+\text{terms with derivatives}.
\end{aligned}
\end{equation}
Both in the D-term dominated scenario and in the F-term dominated one, the minimisation of the action at zero-momentum is obtained by setting 
\begin{equation}
F = - f, \qquad \text D = -\frac{\xi}{2} = -2\sqrt{m M \, f}, \qquad s=0,
\end{equation}
because of the large mass for $s$. In turns this also introduces an effective large mass for $\chi^Y_\alpha$, which is then stabilised at
\begin{equation}
\chi^Y_\alpha=0.
\end{equation}
In the effective theory it is therefore possible to set $S = Y = 0$, which are directly implied by $s = \chi^Y =0$.

While the decoupling of the scalar $s$ is straightforward, the decoupling of the massive fermion needs to be commented further. 
Because of the constraint $X\hat W_\alpha = 0$ that has been imposed, the fermion $\hat \lambda_\alpha$ is removed from the spectrum and expressed as
\begin{equation}
\hat \lambda_\alpha = i\, {\rm D}\frac{G_\alpha}{\sqrt{2}F}+\dots,
\end{equation}
where dots stand for terms with derivatives. 
The fermion $\lambda_\alpha$ in $V$, which is the gaugino of the linearly realised theory, is going to have then a contribution coming from $\hat \lambda_\alpha$, one coming from $G_\alpha$ and one coming from $\chi^Y_\alpha$, due to the fact that the parameterisation \eqref{ggR} has been used.
On the other hand, a massive fermion and a goldstino are present in the theory and they are given in \eqref{eigenvectors} and \eqref{eigenvectors2}. It turns out however that, in the zero-momentum limit, $\chi^Y_\alpha$ is aligned with the massive fermion and therefore when it is removed from the spectrum, what remains is automatically aligned with the goldstino.
The resulting low energy model is finally
\begin{equation}
\mathcal{L}=\frac14 \left(\int d^2 \theta\, \hat W^2 +c.c.\right) + \xi \int d^4\theta \,\hat V+\int d^4\theta\, X\bar X+f\left(\int d^2\theta\, X + c.c.\right)
\end{equation}
and describes a massless vector interacting with a goldstino. 
Notice that Lagrangian is similar to \eqref{LLXV}, even though it has been obtained from a different UV model.

The discussion on the emergence of a nilpotent chiral superfield $X$ in the low energy limit of generic models with spontaneous breaking of supersymmetry is concluded. Evidence has been given for the fact that the existence of such $X$ can be assumed without loss of generality. A systematic way for parametrising multiplets of linear supersymmetry in terms of constrained superfields has also been discussed. In the next chapter non-linear realisation of local supersymmetry are introduced.

\section{Discussion: End of the first part}

In the first part of this thesis four-dimensional models with spontaneously broken supersymmetry have been studied. It has been shown that the approach with constrained superfields captures all the properties of their low energy descriptions and, in particular, that the goldstino interactions can always be described by means of a nilpotent chiral superfield $X$.

The results can be summarised as follows. 
When supersymmetry is spontaneously broken and non-linearly realised, without loss of generality the goldstino sector can be described by means of a chiral superfield $X$ satisfying
\begin{equation}
\nonumber
X^2=0 \, . 
\end{equation} 
In the presence of matter couplings, any matter component field can be eliminated from a given matter superfield $Q_L$ by imposing
\begin{equation} 
\nonumber
X \bar X \, Q_L = 0 \, . 
\end{equation} 
These are the only two ingredients which are needed to construct any desired spectrum in the low energy regime.
Another interesting result which has been discussed is the fact that a model with pure D-term breaking and no chiral superfields can be recast into an equivalent one with F-term breaking, sourced by a nilpotent chiral superfield $X$.

\chapter{Supergravity}
\label{c4:sugra}

In this chapter the analysis of models with spontaneously broken and non-linearly realised local supersymmetry is initiated. Being a broad subject and also an active field of research, a thorough review would be out of the purposes of the present work. This chapter can therefore be thought of as an introduction to the topic, while the reader interested in additional details is addressed to the references cited hereinafter.  For simplicity, only the minimal case of $\mathcal{N}=1$ supergravity in four dimensions is considered. Generalisations to extended supersymmetry can be found for example in \cite{Kuzenko:2015rfx, Kuzenko:2017zla, Kuzenko:2017gsc,Antoniadis:2018blk}

Even though some results on non-linear realisations in supergravity are already contained in \cite{Lindstrom:1979kq,Samuel:1982uh}, the recent interest in the topic starts with \cite{Farakos:2013ih,Antoniadis:2014oya,Ferrara:2014kva}. One of the motivations behind these works is the construction of models with spontaneously broken local supersymmetry in which, as a consequence of the decoupling of the massive modes, the setup is considerably simplified with respect to the unbroken phase and applications to inflation, for example, can be plainly analysed (see \cite{Ferrara:2016ajl} for a recent review). 
In \cite{Farakos:2013ih}, the decoupling of sgoldstino fields is discussed and a supergravity Lagrangian for the goldstino, without scalars, is given. It is shown, in particular, how constraints that lead to non-linear realisations of supersymmetry emerge as consequence of the equations of motion of the goldstino superfield, when considering the aforementioned decoupling limit.
One of the first applications to inflation is contained in \cite{Antoniadis:2014oya}, in which a model with two chiral superfields, namely an axion superfield and a nilpotent goldstino superfield $X$, is mapped to the supergravity embedding of the Starobinski model \cite{Starobinsky:1980te, Farakos:2013cqa, Ferrara:2014ima}. 
In \cite{Ferrara:2014kva}, a more general analysis of inflationary models is presented in the framework of constrained superfields. It is shown, in particular, how to obtain an effective description of the setup proposed by Kachru, Kallosh, Linde and Trivedi (KKLT) in \cite{Kachru:2003aw, Kachru:2003sx} for constructing de Sitter vacua within string theory, by working entirely in the supergravity approximation and by using a nilpotent chiral superfield to describe the supersymmetry breaking sector.

A possible way to build theories with spontaneously broken and non-linearly local supersymmetry is to generalise to curved superspace the model \eqref{eq:c1:LXNL} in which the goldstino is described by a nilpotent $X$. This is the approach that is going to be adopted in the following section. In \cite{Bergshoeff:2015tra, Bergshoeff:2016psz}, superconformal methods have been employed in this spirit in order to obtain Lagrangians in which the spectrum is manifestly not supersymmetric, as it does not contain scalars, while in \cite{Hasegawa:2015bza} matter couplings have also been considered. In \cite{Bandos:2015xnf} the goldstino brane construction has been employed in order to obtain analogous models.

Even though the scalar potential admits in principle any value for the cosmological constant, these models are usually denoted ``de Sitter supergravity'', because the main motivation behind them is again related to inflation or to the studying of the current phase of the Universe. In the following, a different class of supergravity models with spontaneously broken supersymmetry is presented. They have been called ``minimal constrained supergravity'' since, beside for what concerns the supersymmetry breaking sector, non-linear realisations are present also in the gravity sector, where the auxiliary fields are eliminated by constraints and the field content is reduced to its minimum. 
A subclass of these models, moreover, admits by construction only positive values for the vacuum energy.

In supergravity, in contrast to the rigid supersymmetric case, the goldstino is a pure gauge degree of freedom and not a physical mode. It can therefore be eliminated from any given model with a supersymmetry transformation and, as a consequence, the gravitino acquires a mass in the vacuum. This is the so called superhiggs mechanism \cite{Cremmer:1982en} and it is going to be reviewed in the following. 
Its net effect is that, when local supersymmetry is spontaneously broken and in order to simplify the interactions, it is always possible to restrict the physical analysis to the unitary gauge, in which the goldstino field is set to zero.
For the same reasons, it should be possible to construct models with spontaneously broken and non-linearly local supersymmetry without any field or superfield dedicated to the description of the goldstino. This has been done in \cite{Dudas:2015eha, Antoniadis:2015ala}, in which supergravity models are obtained by using solely the superfields of the gravity sector. In particular, the non-linear realisation of local supersymmetry is implemented by imposing constraints on the curvature superfield $\mathcal{R}$. These models, as shown in the original works, can be mapped to the ones in which the goldstino is described in terms of a nilpotent chiral superfield and thus they are not discussed in the following.

\section{Coupling the goldstino to supergravity}
\label{c3:sec:VAsugra}
In this section the nilpotent chiral superfield $X$ describing the supersymmetry breaking sector is coupled to supergravity. The formalism adopted is that of curved superspace in the conventions of \cite{WessBagger} and with reduced Planck units, namely $8\pi G_N=1$. The minimal model with only $X$, besides the superfields of the gravity sector, is discussed first. The main result is a Lagrangian for the goldstino without scalars which, in general, admits any value for the cosmological constant, even if this kind of models are mainly motivated for the study of de Sitter solutions \cite{Kallosh:2015tea, Schillo:2015ssx}. The generalisation to the case of two chiral superfields $X$ and $Y$, subjected to the constraints $X^2=0$ and $XY=0$, is also presented in some detail. A different class of supergravity models \cite{Cribiori:2016qif} is then analysed in which non-linear realisations are also contained in the gravity sector of the theory, as a consequence of constraints which are imposed on the superfields describing the geometry of curved superspace.

\subsection{The minimal model}
In this subsection the goldstino Lagrangian \eqref{eq:c1:LXNL} is coupled to local supersymmetry. With respect to the rigid supersymmetric case, the complete calculation of a supergravity Lagrangian, from superspace to components, is more involved. For this reason and in order to keep the present discussion as plain as possible, the main steps are summarised in the appendix \ref{appB:sugraL}.
The results there contained are employed systematically through the present and the next chapters.

Given a set of chiral superfields 
\begin{equation}
\Phi^i = A^i + \sqrt 2 \Theta^\alpha \chi^{ i}_\alpha+\Theta^2 F^{ i}, 
\end{equation}
a Lagrangian describing their couplings which is invariant under local supersymmetry can be written in curved superspace as
\begin{equation}
\label{c3:eq:LSUGRAGEN}
\mathcal{L}=\int d^2 \Theta \, 2 \mathcal{E} \left[-\frac18 \left(\bar{\mathcal{D}}^2-8\mathcal{R}\right)\Omega(\Phi, \bar \Phi)+ W(\Phi)\right]+c.c.,
\end{equation}
where $\Omega(\Phi, \bar \Phi)$ is a real function of the chiral superfields, eventually related to the K\"ahler potential $K(\Phi, \bar \Phi)$ by
\begin{equation}
\Omega(\Phi, \bar \Phi) = -3\, e^{-K(\Phi, \bar \Phi)/3},
\end{equation}
while $W(\Phi)$ is the holomorphic superpotential.
The curvature superfield $\mathcal{R}$ is chiral and it contains information on the geometry of curved superspace. It is used in the Lagrangian to construct the chiral projector $-1/4\left(\bar{\mathcal{D}}^2-8\mathcal{R}\right)$ of supergravity, which generalises that of rigid supersymmetry, $-1/4 \bar{D}^2$. The component expansion of $\mathcal{R}$ is 
\begin{equation}
	\label{c3:eq:calR}
	\begin{aligned}
	 	{\cal R} &=  - \frac16 \bigg\{M + \Theta(2\sigma^{ab} \psi_{ab} - i \sigma^a \bar\psi_a \,M + i \psi_a b^a)  \\
		&+  \Theta^2\left(-\frac12 R + i \bar \psi^a\bar{\sigma}^b \psi_{ab} + \frac23 |M|^2 + \frac13 b^2 - i\, {\cal D}_a b^a \right. \\
		&+  \left.  \frac12 \bar{\psi}\bar{\psi}\,M-\frac12 \psi_a \sigma^a \bar\psi_c \,b^c + \frac18\, \epsilon^{abcd}(\bar\psi_a \bar \sigma_b \psi_{cd} + \psi_a \sigma_b \bar \psi_{cd})\right)\bigg\},
	\end{aligned}
\end{equation}
where $R$ is the Ricci scalar, while $M$ and $b_a$ are respectively the complex scalar and the real vector auxiliary field employed in the old-minimal off-shell formulation of supergravity. The chiral density $\mathcal{E}$, instead, contains information on the vielbein and its expansion is 
\begin{equation}
	\label{c3:eq:calE}
	2 {\cal E} = e \left\{1 + i\,\Theta \sigma^a \bar\psi_a - \Theta^2 (\bar M + \bar\psi_a \bar{\sigma}^{ab} \bar\psi_b)\right\}.
\end{equation}

At this point, the discussion is specialised to the case of interest, in which only one chiral superfield $X$ is present and the functions $\Omega$ and $W$ are given by 
\begin{equation}
\label{c3:eq:OmegaWXnilpotent}
\Omega(X, \bar X) =  X \bar X-3, \qquad W(X) = W_0 + f X,
\end{equation}
where $W_0$ and $f$ are complex parameters. This is the most generic model up to two derivatives involving only $X$, if the latter is subjected to the nilpotent constraint $X^2=0$. The solution to this constraint in supergravity is the same as in global supersymmetry and therefore the component expansion of $X$ is
\begin{equation}
X = \frac{G^2}{2F}+\sqrt 2\,  \Theta^\alpha G_\alpha + \Theta^2 F,
\end{equation}
or, equivalently, its projections are
\begin{equation}
\begin{aligned}
X| &= \frac{G^2}{2F},\\
\mathcal{D}_\alpha X| &= \sqrt 2\, G_\alpha,\\
\mathcal{D}^2 X| &= -4 F. 
\end{aligned}
\end{equation}

The procedure to calculate the supergravity Lagrangian \eqref{c3:eq:LSUGRAGEN} for the par\-ti\-cu\-lar case in which $\Omega$ and $W$ are given by \eqref{c3:eq:OmegaWXnilpotent} is now presented. It is discussed, in particular, how to incorporate the information about the constraint $X^2=0$, namely the fact that the scalar in $X$ is given by $X| = \frac{G^2}{2F}$.
Following the steps outlined in the appendix \ref{appB:sugraL} and using the fact that
\begin{equation}
\begin{aligned}
&\Omega_X = \bar X, \qquad \Omega_{X \bar X}= \delta_{X \bar X}, \qquad \Omega_{X X}=0,\\
&(\log \Omega)_{\bar X}=\frac{3}{X \bar X-3}, \qquad 	\,\ (\log \Omega)_{X \bar X} = - \frac{3}{(X\bar X-3)^2},
\end{aligned}
\end{equation}
the Lagrangian \eqref{c3:eq:LSUGRAGEN} has the form
\begin{equation}
\mathcal{L} = \mathcal{L}_\text{kin}+ \mathcal{L}_\text{aux} + \mathcal{L}_\text{2f}+\mathcal{L}_\text{4f},  
\end{equation}
where
\begin{align}
\nonumber
e^{-1}\mathcal{L}_\text{kin}&=\frac16 (X\bar X-3) R-\frac{1}{12}(X\bar X-3) \epsilon^{klmn}\left(\bar \psi_k \bar \sigma_l \psi_{mn}-\psi_k\sigma_l \bar\psi_{mn}\right)\\
\nonumber
&- \partial_m X \partial^m \bar X-\frac i2 \left(G \sigma^m \mathcal{D}_m \bar G+\bar G \bar \sigma^m \mathcal{D}_m G\right)\\
&+\frac14 \epsilon^{klmn}\left(\bar X\partial_k X-X \partial_k \bar X\right)\psi_l\sigma_m\bar\psi_n\\
\nonumber
&-\frac{\sqrt 2}{2} \left(\bar \psi_m \bar \sigma^n \sigma^m \bar G \partial_n X+\psi_m \sigma^n \bar \sigma^m G \partial_n \bar X\right)\\
\nonumber
&+\frac{\sqrt 2}{3}\left(\bar X G \sigma^{mn}\psi_{mn}+X \bar G\bar \sigma^{mn}\bar \psi_{mn}\right),
\end{align}

\begin{align}
\nonumber
e^{-1} \mathcal{L}_\text{aux}&=\frac 19 (X\bar X -3) \bigg|M - 3 \frac{X}{X\bar X-3} \bar F\bigg|^2-\frac{3}{X\bar X-3}F \bar F\phantom{AAAAAAA}\\
\nonumber
&- \frac19 (X\bar X-3) b_a b^a-\frac i3 \left(\bar X \partial_m X - X\partial_m \bar X\right)b^m\\
&-\frac 16 G \sigma^a \bar G b_a + \frac i6 \sqrt 2\left(\bar X \psi_a G - X\bar \psi_a \bar G\right)b^a\\
\nonumber
&- \bar M (W_0+fX) - M (\bar W_0 + \bar f \bar X) + f F + \bar f \bar F,
\end{align}

\begin{align}
e^{-1} \mathcal{L}_\text{2f} &= - (W_0 + f X) \bar \psi_a \bar \sigma^{ab} \bar \psi_b- (\bar W_0 + \bar f \bar X) \psi_a \sigma^{ab} \psi_b \phantom{A}\\
\nonumber
&-\frac{\sqrt 2}{2}i f G \sigma^a \bar \psi_a-\frac{\sqrt 2}{2}i \bar f \bar G \bar \sigma^a \psi_a,
\end{align}

\begin{align}
\nonumber
e^{-1} \mathcal{L}_\text{4f} &=\frac14 \left[(\psi_m \sigma_n \bar \psi^m)(G \sigma^n \bar G)+i \epsilon^{klmn}(\psi_k \sigma_l \bar \psi_m)(G \sigma_n \bar G)\right]\\
&+\frac{\sqrt 2}{8} \epsilon^{klmn}\psi_k\sigma_l\bar \psi_m\left(\bar X \psi_n G-X\bar\psi_n\bar G\right)\\
\nonumber
&-\frac{\sqrt 2}{4}i \left(\bar X \psi_m\sigma^{mn}G+X \bar \psi_m \bar \sigma^{mn}\bar G\right)(\psi_k\sigma^k \bar\psi_n-\psi_n\sigma^k\bar\psi_k).
\end{align}
In all these formulae, a projection to the lowest component on $X$ is always understood.

At this stage, the information about the constraint can be inserted. This amounts to substitute $X|= \frac{G^2}{2F}$ and to expand the resulting Lagrangian in powers of the fermion $G_\alpha$, which is going to be the goldstino. For simplicity, however, terms are kept only up to $\mathcal{O}(G^2)$.
Notice that it is crucial to perform such a step before taking the equations of motion for the auxiliary field $F$ since, when supersymmetry is non-linearly realised, the lowest component $X|$ depends on $F$.
Due to the fact that $X^2=0$ and since terms at most quadratic in $G_\alpha$ are considered, some simplifications occur, as for example
\begin{equation}
\begin{aligned}
&\log(\Omega)_{\bar X} = -\frac{X}{3}, \qquad \log(\Omega)_{X \bar X} = -\frac13 + \mathcal{O}(G^4),\\
&X \bar X = \mathcal{O}(G^4), \qquad \,\ -\frac{3}{X\bar X-3}= 1 + \mathcal{O}(G^4),
\end{aligned}
\end{equation}
and the Lagrangian becomes
\begin{align}
\label{c3:eq:LVAoffshell}
\nonumber
e^{-1}\mathcal{L}&=-\frac12 R+\frac14 \epsilon^{klmn}(\bar \psi_k \bar \sigma_l \psi_{mn}- \psi_k \sigma_l \bar \psi_{mn})\\
\nonumber
&-\frac i2 (G \sigma^m \mathcal{D}_m \bar G+ \bar G \bar \sigma^m \mathcal{D}_m G)\\
\nonumber
&-\frac{\sqrt 2}{2}i f G \sigma^a \bar \psi_a -\frac{\sqrt 2}{2}i \bar f \bar G \bar \sigma^ a\psi_a\\
\nonumber
&+\frac14 \left[(\psi_m \sigma_n \bar \psi^m)(G \sigma^n \bar G)+i \epsilon^{klmn}(\psi_k \sigma_l \bar \psi_m)(G \sigma_n \bar G)\right]\\
&-\frac13 M \bar M -\frac13 \left(\frac12 \bar M G^2 \frac{\bar F}{F}+ \frac12 M \bar G^2 \frac{F}{\bar F} \right)\\
\nonumber
&+\frac13 b_a b^a + |F+\bar f|^2- |f|^2\\
\nonumber
&-\left(W_0 + f \frac{G^2}{2F}\right)\left(\bar M + \bar \psi_a \bar \sigma^{ab} \bar \psi_b\right)\\
\nonumber
&-\left(\bar W_0 + \bar f \frac{\bar G^2}{2 \bar F}\right)\left(M + \psi_a \sigma^{ab} \psi_b\right)\\
\nonumber
&+ \mathcal{O}(G^3)
\end{align}
Up to this fermionic order, the kinetic terms of all the fields in the Lagrangian are already canonically normalised and therefore the Weyl rescaling mentioned in appendix \ref{appB:sugraL} is not needed. In other words, as a consequence of to the non-linear realisation of supersymmetry, such a rescaling is only affecting terms of order $\mathcal{O}(G^4)$. This is one instance of the simplifications that can occur when local supersymmetry is spontaneously broken and non-linearly realised.

It is now possible to take the equations of motion for the auxiliary fields in order to obtain the on-shell expression of the Lagrangian.
Notice that, while in the linearly realised regime the coupling among the auxiliary fields is due to the particular formulation that is adopted, for example usually it is present when using superspace but it is not when using superconformal methods, in the case under investigation such a coupling is always going to be present, since its very origin resides in the non-linear realisation of supersymmetry. Even if the auxiliary fields are coupled, their equations of motion can be solved with an iterative procedure, which gives
\begin{align}
M &= - 3 W_0 + \frac{f}{\bar f} G^2+ \mathcal{O}(G^3),\\
\label{c3:eq:eomba}
b_a &= \frac14 G \sigma_a \bar G + \mathcal{O}(G^3),\\
F &= - \bar f - \frac14 \frac{\bar f}{f^2} \bar G^2 \psi_a \sigma^{ab}\psi_b+W_0 \frac{\bar f}{f^2}\bar G^2+ \frac{\bar W_0}{2\bar f} G^2+ \mathcal{O}(G^3).
\end{align}
Notice that these expressions are different from the analogous in the case in which supersymmetry is linearly realised. Plugging them back, the following on-shell Lagrangian is produced
\begin{equation}
\label{c3:eq:LVASUGRAonshell}
\begin{aligned}
e^{-1}\mathcal{L} &=-\frac12 R+\frac14 \epsilon^{klmn}(\bar \psi_k \bar \sigma_l \psi_{mn}- \psi_k \sigma_l \bar \psi_{mn})\\
&-\frac i2 (G \sigma^m \mathcal{D}_m \bar G+ \bar G \bar \sigma^m \mathcal{D}_m G)-\frac{\sqrt 2}{2}i f G \sigma^a \bar \psi_a -\frac{\sqrt 2}{2}i \bar f \bar G \bar \sigma^ a\psi_a\\
&+\frac14 \left[(\psi_m \sigma_n \bar \psi^m)(G \sigma^n \bar G)+i \epsilon^{klmn}(\psi_k \sigma_l \bar \psi_m)(G \sigma_n \bar G)\right]\\
&-W_0 \bar \psi_a \bar \sigma^{ab} \bar \psi_b - \bar W_0 \psi_a \sigma^{ab}\psi_b +\frac{f}{\bar f}\frac{G^2}{2}\bar \psi_a \bar \sigma^{ab} \bar \psi_b + \frac{\bar f}{f} \frac{\bar G^2}{2} \psi_a \sigma^{ab} \psi_b\\
& - \bar W_0 G^2 \frac{f}{\bar f}- W_0 \bar G^2 \frac{\bar f}{f} -\left(|f|^2-3|W_0|^2\right)
+ \mathcal{O}(G^3),
\end{aligned}
\end{equation}
where in the last line the scalar potential $\mathcal{V}= \left(|f|^2-3|W_0|^2\right)$ appears.
This is a model with spontaneously broken and non-linearly realised local supersymmetry, in which the field $G_\alpha$ has the role of the goldstino, as it can be understood by the fact that its supergravity transformation is non-homogeneous
\begin{equation}
\delta_\epsilon G_\alpha =-\sqrt 2 \bar f \epsilon + \dots.
\end{equation}
Such a model reduces to \eqref{eq:c1:LXNL} in the flat space limit.

Before deepening further the role of the goldstino in supergravity, few comments are in order.
Notice first of all that, in the model under investigation, even though the scalar potential can assume any value, nevertheless the parameters can be tuned in order for the cosmological constant to be positive and the vacuum to be de Sitter. In this respect, once one or more matter superfields are coupled, the system can be employed for example to study inflation in a simplified setup, in which the sgoldstino has been already integrated out. By imposing constraints on matter superfields then, it is possible to eliminate also some of the matter component fields, using the same techniques that have been introduced in the previous chapters. In particular, the general constraint \eqref{eq:c2:genconstr} can be implemented directly into supergravity without any modification. Constraints can also be imposed at the Lagrangian level, by using a Lagrange multiplier \cite{Ferrara:2016een}, in order to work in a manifestly linear regime until the equations of motions are considered. 
Notice finally that a technical difference with respect to the linearly realised regime is that the knowledge of the K\"ahler potential $K$ and the superpotential $W$ is not sufficient in presence of the nilpotent Goldstino multiplet to produce the full supergravity action, since the equations of motion of the auxiliary fields are modified by the constraint.

\subsection{Superhiggs mechanism and unitary gauge}
In this subsection it shown that, by using supergravity transformations, which are symmetries of \eqref{c3:eq:LVASUGRAonshell}, it is possible to eliminate from the Lagrangian any coupling involving the goldstino $G_\alpha$. This is a consequence of the fact that, when supersymmetry is local, the goldstino is a pure gauge degrees of freedom and it contains no physical information. For this reason, in the following sections the condition $G_\alpha=0$ is going to be imposed in most of the situations as a convenient gauge choice. It is called unitary gauge and it allows a drastic simplification of the interactions, which makes models with spontaneously broken local supersymmetry far more accessible. 

The starting point is the Lagrangian \eqref{c3:eq:LVASUGRAonshell} which, by using integration by parts, can be recast into the form
\begin{equation}
\begin{aligned}
\label{c3:eq:LSH1}
\mathcal{L} &= -\frac12 e R-e \mathcal{V}+e \epsilon^{klmn} \bar \psi_k \bar \sigma_l \mathcal{D}_m\psi_n-\frac{2i}{|f|^2}e\zeta \sigma^m \mathcal{D}_m \bar \zeta\\
&+e\left[-i \bar \zeta \bar \sigma^a \psi_a - \bar W_0 \psi_a \sigma^{ab}\psi_b-2 \frac{\bar W_0}{|f|^2} \zeta^2+c.c.\right]\\
&+\text{four fermions interactions},  
\end{aligned}
\end{equation}
where the fermion
\begin{equation}
\zeta_\alpha = \frac{\sqrt 2}{2} f G_\alpha
\end{equation}
has been introduced for convenience. Interactions involving four fermions are neglected for simplicity. The idea is to perform, onto this Lagrangian, specific supersymmetry transformations along the fermion $\zeta_\alpha$, in order to eliminate completely the couplings involving it. The required transformations are
\begin{align}
\delta e^a_m &= i \alpha \left(\psi_m \sigma^a \bar \zeta-\zeta \sigma^a \bar \psi_m\right),\\
\delta \psi_m &=-2 \alpha \mathcal{D}_m \zeta -i \alpha W_0 \sigma^m \bar \zeta + \mathcal{O}(\zeta^3),
\end{align}
where $\alpha$ is a real parameter to be fixed. For reasons that are going to be clear in a while, also the second order transformation of the vielbein is needed \cite{Ferrara:2016ntj}, up to two fermions
\begin{equation}
\begin{aligned}
\delta^2 e^a_m &= i \alpha \left(\delta \psi_m \sigma^a \bar \zeta + \delta \bar \psi_m \bar \sigma^a \zeta\right)\\
&= \alpha^2\left(-2i\mathcal{D}_m \zeta \sigma^a \bar \zeta-2i\mathcal{D}_m \bar \zeta \bar \sigma^a \zeta+W_0 \bar \zeta^2 e^a_m + \bar W_0 \zeta^2 e^a_m\right)+ \mathcal{O}(\zeta^3).
\end{aligned}
\end{equation}
The application of these transformations onto the bosonic sector of \eqref{c3:eq:LSH1} gives
\begin{equation}
\label{c3:eq:SHbos}
\begin{aligned}
\mathcal{L}_\text{bos} &= -\frac12 e R- e\mathcal{V}\\
&=-\frac12 ER-E\mathcal{V}+E\left(R^m_a-\frac12 R e^m_a\right)\bigg\{-i\alpha\left(\Psi_m \sigma^a \bar \zeta-\zeta \sigma^a\bar\Psi_m\right)\\
&+\frac{\alpha^2}{2}\left(-2i \mathcal{D}_m \zeta \sigma^a \bar \zeta-2i \mathcal{D}_m \bar \zeta \bar \sigma^a \zeta+W_0 \bar \zeta^2 E^a_m +\bar W_0 \zeta^2E^a_m\right)\bigg\}\\
&-E\mathcal{V}\bigg\{-i\alpha\left(\Psi_m \sigma^m \bar \zeta-\zeta \sigma^m\bar\Psi_m\right)\\
&+\frac{\alpha^2}{2}\left(-2i \mathcal{D}_m \zeta \sigma^m \bar \zeta-2i \mathcal{D}_m \bar \zeta \bar \sigma^m \zeta+4 W_0 \bar \zeta^2 + 4 \bar W_0 \zeta^2\right)
\bigg\}
\end{aligned}
\end{equation}
where $E^a_m$ and $\Psi_m$ are the transformed fields, namely $\Psi_m= \psi_m + \delta \psi_m$, and the commutator algebra $\epsilon^{kmab}\bar \sigma_m \mathcal{D}_a\mathcal{D}_b \zeta=\frac i2 \left(R^k_a \bar \sigma^a -\frac12 R\bar \sigma^k\right)\zeta$ has been used. 
Along with the same logic, the gravitino mass term becomes
\begin{equation}
\label{c3:eq:SHmgrav}
\begin{aligned}
\psi_a \sigma^{ab}\psi_b &= \Psi_a \sigma^{ab} \Psi_b + \alpha\left(4  \Psi_a \sigma^{ab}\mathcal{D}_b \zeta-3i W_0 \Psi_a \sigma^a \zeta\right)\\
&+\alpha^2\left( 6i  W_0 \zeta \sigma^a \mathcal{D}_a \bar \zeta-4  \zeta \sigma^{ab} \mathcal{D}_a \mathcal{D}_b\zeta+ 6  W_0^2 \bar \zeta^2\right),
\end{aligned}
\end{equation}
while the gravitino-goldstino mixing term instead is
\begin{equation}
\label{c3:eq:SHgipsi}
\bar \zeta \bar \sigma^a\psi_a = \bar \zeta \bar \sigma^a \Psi_a + 2\alpha \bar \zeta \bar \sigma^a \mathcal{D}_a \zeta-4i \alpha W_0 \bar \zeta^2
\end{equation}
and finally the gravitino kinetic term gives
\begin{equation}
\label{c3:eq:SHgravkin}
\begin{aligned}
\epsilon^{klmn}\bar \psi_k \bar \sigma_l \mathcal{D}\psi_n &= \epsilon^{klmn}\bar \psi_k \bar \sigma_l \mathcal{D}_m \bar \psi_n\\
&+\left[i \alpha\left(R^k_m-\frac12 R e^k_m\right)\psi_k \bar \sigma^m \zeta+c.c. \right]\\
&+\left(4 \alpha \bar W_0 \bar \psi_a \sigma^{ab}\mathcal{D}_b \zeta+c.c.\right)\\
&+i \alpha^2 \left(R^k_m-\frac12 R e^k_m\right)\left(\mathcal{D}_k \zeta \sigma^m \bar \zeta + \mathcal{D}_k \bar \zeta \bar \sigma^m \zeta\right)\\
&+\left[\alpha^2 \left(R^k_m -\frac12 R e^k_m\right)\bar W_0 \zeta \sigma_k \bar \sigma^m \zeta+c.c.\right]\\
&+6i \alpha^2 |W_0|^2 \zeta \sigma^m \mathcal{D}_m \bar \zeta.
\end{aligned}
\end{equation}
Putting together \eqref{c3:eq:SHbos}, \eqref{c3:eq:SHmgrav}, \eqref{c3:eq:SHgipsi} and \eqref{c3:eq:SHgravkin}, the following Lagrangian is obtained
\begin{equation}
\begin{aligned}
\label{c3:eq:LSH2}
\mathcal{L}&=-\frac 12 e R- e\mathcal{V}\\
&+e\left(R^m_a -\frac12 (R+2\mathcal{V}) e^m_a\right)\bigg[-i\alpha \psi_m \sigma^a \bar \zeta\\
&+\frac{\alpha^2}{2}\left(-2i \mathcal{D}_m\zeta \sigma^a \bar \zeta+ \bar W_0 \zeta^2 e^a_m\right)+c.c.\bigg]\\
&+e \bigg\{\epsilon^{klmn}\bar\psi_k \bar \sigma_l \mathcal{D}_m \psi_n +\left[i \alpha\left(R^k_m-\frac12 R e^k_m\right)\psi_k \bar \sigma^m \zeta+c.c. \right]\\
&+\left(4 \alpha \bar W_0 \bar \psi_a \sigma^{ab}\mathcal{D}_b \zeta+c.c.\right)\\
&+i \alpha^2 \left(R^k_m-\frac12 R e^k_m\right)\left(\mathcal{D}_k \zeta \sigma^m \bar \zeta + \mathcal{D}_k \bar \zeta \bar \sigma^m \zeta\right)\\
&+\left[\alpha^2 \left(R^k_m -\frac12 R e^k_m\right)\bar W_0 \zeta \sigma_k \bar \sigma^m \zeta+c.c.\right]+6i \alpha^2 |W_0|^2 \zeta \sigma^m \mathcal{D}_m \bar \zeta \bigg\}\\
&-\frac{2i}{|f|^2}e \zeta \sigma^m \mathcal{D}_m \bar \zeta+e\bigg\{-i\bar \zeta \bar \sigma^a \psi_a - 2\alpha i\bar \zeta \bar \sigma^a \mathcal{D}_a \zeta-4 \alpha W_0 \bar \zeta^2\\
&-\bar W_0 \bigg[\psi_a \sigma^{ab} \psi_b + \alpha\left(4  \psi_a \sigma^{ab}\mathcal{D}_b \zeta-3i W_0 \psi_a \sigma^a \zeta\right)\\
&+\alpha^2\left( 6i  W_0 \zeta \sigma^a \mathcal{D}_a \bar \zeta-4  \zeta \sigma^{ab} \mathcal{D}_a \mathcal{D}_b\zeta+ 6  W_0^2 \bar \zeta^2\right)\bigg]\\
&-2 \frac{W_0}{|f|^2}\bar \sigma^2 + c.c.\bigg\}+\text{four fermions interactions},
\end{aligned}
\end{equation}
where capital letters for the transformed field are omitted. At this point, several cancellations occur and the crucial ones are commented below.
\begin{itemize}
\item[\tiny $\bullet$] The parameter $\alpha$ is fixed by requiring that the mixing between the gravitino and the goldstino vanishes. The condition is
\begin{equation}
i e \psi_a \sigma^a \bar \zeta \left(\mathcal{V}\alpha + 1+ 3\alpha |W_0|^2 \right)+c.c.=0,
\end{equation}
which gives
\begin{equation}
\alpha = -\frac{1}{\mathcal{V}+ 3 |W_0|^2}.
\end{equation}
For the particular model under consideration, therefore, $\alpha = -1/|f|^2$.
\item[\tiny $\bullet$] The mass term for the goldstino must cancel. This is motivated by the fact that the mass of the goldstino is protected by the Goldstone theorem. The terms to be considered are
\begin{equation}
2e \bar W_0 \zeta^2 \left(\alpha^2 \mathcal{V}-2\alpha-3 \alpha^2 |W_0|^2-\frac{1}{|f|^2}\right)+c.c.=0
\end{equation}
and they cancel as expected.
\item[\tiny $\bullet$] All the contributions to the kinetic term for the goldstino must cancel as well, since they cannot propagate any physical degrees of freedom in supergravity. Up to integration by parts, the terms involved in the cancellation are
\begin{equation}
2 e i \mathcal{D}_m \zeta \sigma^m \bar \zeta \left(\alpha^2 \mathcal{V}+ \alpha + 3 \alpha^2 |W_0|^2\right)=0
\end{equation}
and they vanish again as expected. Notice that integration by parts is simplified by the fact that the scalar potential $\mathcal{V}$ is constant, since the theory does not contain scalars. This situation is different from the analogous in the superhiggs mechanism in the linearly realised regime, in which the vacuum condition $\partial_m \mathcal{V}=0$ is usually used in order to simplify the calculation. 
\item[\tiny $\bullet$] Terms containing the geometric factor $\left(R^m_a -\frac12 R e^m_a\right)$ cancel among themselves and vanish trivially in flat space. Notice that one of these term arises from the contribution $\zeta \sigma^{ab}\mathcal{D}_a \mathcal{D}_b \zeta$ in \eqref{c3:eq:SHmgrav} and several of them come from \eqref{c3:eq:SHgravkin}. Their presence is the reason why a second order transformation on the vielbein has been performed in the bosonic sector.
\end{itemize}

After these cancellations are taken into account, all the goldstino couplings disappear and the Lagrangian \eqref{c3:eq:LSH2} drastically simplifies to
\begin{equation}
\begin{aligned}
\label{c3:eq:LVAUnitarygauge}
e^{-1}\mathcal{L}&= -\frac12  R +\frac14 \epsilon^{klmn}(\bar \psi_k \bar \sigma_l \psi_{mn}- \psi_k \sigma_l \bar \psi_{mn})\\
&- \bar W_0 \psi_a \sigma^{ab} \psi_b - W_0 \bar \psi_a \bar \sigma^{ab} \bar \psi_b- \mathcal{V}.
\end{aligned}
\end{equation}
It is possible to prove, in an analogous way, that also the goldstino couplings in the terms with four fermions cancel among themselves and in fact the result \eqref{c3:eq:LVAUnitarygauge} is exact up to all fermionic orders. It can be obtained directly from the original Lagrangian \eqref{c3:eq:LVASUGRAonshell} by adopting the unitary gauge choice
\begin{equation}
G_\alpha=0 \qquad \text{unitary gauge},
\end{equation}
which is consistent since the goldstino in supergravity is a pure gauge degrees of freedom. This gauge choice is going to be imposed in most of the calculations in the rest of the work. To conclude it is important to notice that, after the superhiggs mechanism has occurred and the goldstino has disappeared, the gravitino has become massive, with mass given by $W_0$. The interpretation of $W_0$ as a gravitino mass in \eqref{c3:eq:LSH1}, on the contrary, is not straightforward due to the presence of a mixing term between $\zeta_\alpha$ and $\psi_{\alpha\, a}$. Such term has been cancelled by the superhiggs mechanism.
Notice finally that, being massive, in principle also the gravitino could be decoupled in the low energy regime. A superfield constraint along this direction has been proposed in \cite{Farakos:2017mwd}.

\section{Two chiral superfields}
\label{sec:XYsugra}
In this section a generalisation of the previous model is considered, in which the nilpotent goldstino superfield $X$ is coupled to one chiral matter superfield $Y$. These superfields are constrained such that
\begin{equation}
\label{c3:eq:constrXYsugra}
X^2=0, \qquad XY=0
\end{equation}
and their component expansions are given by \eqref{eq:c2:solY2}, with the replacement $\theta_\alpha \to \Theta_\alpha$. In the unitary gauge they reduce to
\begin{equation}
X = \Theta^2 F, \qquad Y = \sqrt 2\, \Theta^\alpha \chi_\alpha+ \Theta^2 F^Y.
\end{equation}
The most generic model involving this two superfields is given by the K\"ahler potential and superpotential \cite{DallAgata:2015pdd, DallAgata:2015zxp}
\begin{equation}
\begin{aligned}
K &= X \bar X + Y \bar Y+ a\left(X \bar Y^2 + \bar X Y^2\right)\\
&+ b\left(Y \bar Y^2+\bar Y Y^2\right)+c \bar Y^2 \bar Y^2\\
W&= W_0 + fX + gY+ h Y^2,
\end{aligned}
\end{equation}
where $a,b,c,f,g$ and $h$ are parameters. The supergravity Lagrangian can be obtained by following the steps reported in the appendix \ref{appB:sugraL}.
The function $\Omega(X,Y,\bar X, \bar Y)$ is given by
\begin{equation}
\begin{aligned}
\Omega &= -3 \,e^{-\frac{K}{3}}\\
&= -3+K-\frac 16 K^2\\
&=-3+X \bar X + Y \bar Y+ a\left(X \bar Y^2 + \bar X Y^2\right) \\
&+b\left(Y \bar Y^2+\bar Y Y^2\right)+c \bar Y^2 \bar Y^2-\frac16 Y^2 \bar Y^2,
\end{aligned}
\end{equation}
where the constraints \eqref{c3:eq:constrXYsugra} have been used. In the unitary gauge it is possible to calculate
\begin{equation}
\begin{aligned}
{\Omega}_{X\bar X}|={\Omega}_{Y\bar Y}|&=1, \\
(\log \Omega)_{X \bar X}|=(\log \Omega)_{Y \bar Y}| = \Omega^{-1}|&=-3, \\
\Omega_{YY\bar X}| &= 2a, \\
\Omega_{YY\bar Y}| &= 2b.
\end{aligned} 
\end{equation}
The off-shell expression of the Lagrangian therefore is
\begin{equation}
\mathcal{L} = \mathcal{L}_\text{kin}+ \mathcal{L}_\text{aux} + \mathcal{L}_\text{2f}+\mathcal{L}_\text{4f},  
\end{equation}
where
\begin{align}
\nonumber
e^{-1}\mathcal{L}_\text{kin}&=-\frac12 R+\frac14 \epsilon^{klmn}\left(\bar \psi_k \bar \sigma_l \psi_{mn}-\psi_k\sigma_l \bar\psi_{mn}\right)\\
&-\frac i2 \left(\chi \sigma^m \mathcal{D}_m \bar \chi+\bar \chi \bar \sigma^m \mathcal{D}_m \chi\right),
\end{align}

\begin{align}
\nonumber
e^{-1} \mathcal{L}_\text{aux}&=-\frac13 M\bar M+ F \bar F + F^Y \bar F^Y\\
\nonumber
&+\frac13 b_a b^a-\frac 16 \chi \sigma^a \bar \chi b_a\\
&-\chi^2 (a \bar F+b\bar F^Y)- \bar \chi^2 (a F + bF^Y)\\
\nonumber
&- \bar M W_0 - M \bar W_0 + f F + \bar f \bar F+g F^Y + \bar g \bar F^Y,
\end{align}

\begin{align}
\nonumber
e^{-1} \mathcal{L}_\text{2f} &= -h \chi^2 -\bar h \bar \chi^2- W_0 \bar \psi_a \bar \sigma^{ab} \bar \psi_b- \bar W_0 + \bar f \bar X \psi_a \sigma^{ab} \psi_b\\
&-\frac{\sqrt 2}{2}i g \chi \sigma^a \bar \psi_a-\frac{\sqrt 2}{2}i \bar g \bar \chi \bar \sigma^a \psi_a,
\end{align}

\begin{align}
\nonumber
e^{-1} \mathcal{L}_\text{4f} &=\frac14 \left[(\psi_m \sigma_n \bar \psi^m)(\chi \sigma^n \bar \chi)+i \epsilon^{klmn}(\psi_k \sigma_l \bar \psi_m)(\chi \sigma_n \bar \chi)\right]\\
&+\frac14 \left(4c-\frac23\right)\chi^2 \bar \chi^2
\end{align}
Putting everything together it becomes
\begin{equation}
\label{c3:eq:LXYoffshell}
\begin{aligned}
e^{-1}\mathcal{L}&=-\frac12 R+\frac14 \epsilon^{klmn}\left(\bar \psi_k \bar \sigma_l\psi_{mn}-\psi_k\sigma_l\bar \psi_{mn} \right)\\
&-\frac i2 \left(\chi \sigma^m \mathcal{D}_m \bar \chi+ \bar \chi \bar \sigma^m \mathcal{D}_m \chi\right)-\frac{\sqrt 2}{2}i g\, \chi \sigma^a \bar \psi_a - \frac{\sqrt 2}{2}i \bar g \, \bar \chi \bar \sigma^a \psi_a\\
&-(h+a \bar F + b \bar F^Y)\chi^2 -(\bar h + a F + b F^Y) \bar \chi^2\\
&- W_0 \bar \psi_a \bar \sigma^{ab} \psi_b- \bar W_0 \psi_a \sigma^{ab} \psi_b\\
&+\frac14 \left[(\psi_m \sigma_n \bar \psi^m)(\chi \sigma^n \bar \chi)+i \epsilon^{klmn}(\psi_k \sigma_l \bar \psi_m)(\chi \sigma_n \bar \chi)\right]\\
&-\frac 18 \left(1-8c\right)\chi^2 \bar \chi^2\\
&-\frac13 |M+3 W_0|^2+\frac 13 \left(b_a-\frac14 \chi \sigma_a \chi\right)^2\\
&+|F+\bar f|^2 + |F^Y + \bar g|^2- (|f|^2+|g^2|-3 |W_0|^2)
\end{aligned}
\end{equation}
and, since all the kinetic terms are canonically normalised in this gauge, there is no need either for the Weyl rescaling or for the gravitino shift mentioned in the appendix \ref{appB:sugraL}. This simplification is again due to the fact the supersymmetry is spontaneously broken and non-linearly realised.

The equations of motion of the auxiliary fields are
\begin{align}
M &= -3 W_0,\\
b_a &= \frac14\chi \sigma_a \bar \chi,\\
F &= -\bar f+a \chi^2,\\
F^Y &=- \bar g+ b \chi^2
\end{align}
and, by substituting them back in \eqref{c3:eq:LXYoffshell}, the following on-shell Lagrangian is obtained
\begin{equation}
\label{c3:eq:LSUGRAXYonshell}
\begin{aligned}
e^{-1}\mathcal{L}&=-\frac12 R+\frac14 \epsilon^{klmn}\left(\bar \psi_k \bar \sigma_l\psi_{mn}-\psi_k\sigma_l\bar \psi_{mn} \right)\\
&-\frac i2 \left(\chi \sigma^m \mathcal{D}_m \bar \chi+ \bar \chi \bar \sigma^m \mathcal{D}_m \chi\right)-\frac{\sqrt 2}{2}i g\, \chi \sigma^a \bar \psi_a - \frac{\sqrt 2}{2}i \bar g \, \bar \chi \bar \sigma^a \psi_a\\
&-(h-af-bg)\chi^2-(\bar h -a \bar f - b \bar g)\bar \chi^2\\
&+\frac14 \left[(\psi_m \sigma_n \bar \psi^m)(\chi \sigma^n \bar \chi)+i \epsilon^{klmn}(\psi_k \sigma_l \bar \psi_m)(\chi \sigma_n \bar \chi)\right]\\
&-\frac 18 \left[1-8(c-a^2-b^2)\right]\chi^2 \bar \chi^2-\mathcal{V},
\end{aligned}
\end{equation}
where 
\begin{equation}
\mathcal{V} = |f|^2+|g^2|-3 |W_0|^2
\end{equation}
is the scalar potential. As in the previous case, in this supergravity theory without scalars, the scalar potential is constant.

Notice that the fermion mass matrix is different from the standard one of supergravity.
This is a general fact: the presence of the non-linear realisation is adding bilinear contributions in the fermions and, as a consequence, their spectrum is changed. The modified masses can be calculated directly from the Lagrangian or, alternatively, using the mass formula 
\begin{equation}
M_{ij} = e^{K/2}\left[D_iD_j W- \frac23 \frac{D_i W D_j W}{W}\right]_{s=0}+2\partial_k \mathcal{V}|_{s^k=0}\frac{\partial s^k}{\partial(\chi^i \chi^j)}
\end{equation}
given in \cite{DallAgata:2015pdd}, where $s^k=\{X, Y\}$ in the model under consideration. Mass formulae with non-linear supersymmetry are also analysed in \cite{Murli:2018gai}.

\section{Minimal Constrained Supergravity}
\label{c3:sec:MCS}

In the previous sections the strategy to construct models with spontaneously broken and non-linearly realised local supersymmetry has been introduced. The non-linearly realisation has been entirely encoded within the goldstino sector, described by a nilpotent chiral superfield $X$. In the minimal case, the model is called pure supergravity \cite{Antoniadis:2014oya,Bergshoeff:2015tra,Hasegawa:2015bza}, since it contains only the graviton and the gravitino as physical modes. It has been shown indeed that, when supersymmetry is local, the goldstino is a pure gauge degrees of freedom. 
While these models, which are mainly motivated from inflation \cite{Kallosh:2014via,DallAgata:2014qsj,Ferrara:2015cwa}, have been dubbed ``de Sitter'' supergravities, their cosmological constant is a function of two parameters, related to the supersymmetry breaking scale $f$ and to the gravitino mass $W_0$, in order that, depending on these values, it can be positive, negative or vanishing.

In this section supergravity models, whose physical spectra are also given by the graviton and the gravitino in the minimal case, but where supersymmetry is non-linearly realized even on the gravity multiplet itself are going to be constructed.
It is shown that the presence of the non-linear realisation within the gravity sector produces interactions and Lagrangians that differ from those in \cite{Antoniadis:2014oya,Bergshoeff:2015tra,Hasegawa:2015bza} and depend on three independent physical inputs: the supersymmetry breaking scale, the gravitino mass and the cosmological constant.
While this problem has been already tackled from a different perspective in \cite{Delacretaz:2016nhw}, with the purpose of constructing a supersymmetric effective field theory for inflation and using a supersymmetric generalization of the CCWZ construction \cite{Callan:1969sn}, the analysis presented in the following employs the language of superfields, allowing for more general and different constraints. Additional details on the relationship between this approach and that of \cite{Delacretaz:2016nhw} can be found in \cite{Cribiori:2016qif}.

\subsection{Constraining the auxiliary fields of gravity}
The old-minimal off-shell formulation of the supergravity multiplet can be given by means of three different superfields ${\cal E}$, $\mathcal{R}$ and ${B}_{a}$, two of which have already been introduced in \eqref{c3:eq:calE} and \eqref{c3:eq:calR}.\footnote{In \cite{WessBagger} the superfield $B_a$ is denoted with $G_a$ and it is defined $B_{\alpha \dot\alpha}=\sigma^m_{\alpha \dot \alpha}e_m^a b_a$. For completeness, it is worth mentioning that there exists also another superfield describing the gravity multiplet, $\bar W_{\dot \alpha \dot \beta \dot \gamma}$, which is chiral and symmetric in its indices, but it does not play any role in the present discussion.} 
The real superfield $B_{a}$ has the auxiliary vector field $b_a$ as lowest component, $B_a| = -\frac13 b_a $, and it is related to ${\cal R}$ by means of the superspace Bianchi identity
\begin{equation}
{\cal D}^{\alpha} B_{\alpha \dot \alpha} = \bar {\cal D}_{\dot \alpha} \bar{\cal R}.
\end{equation}
The complete component expression of $B_{a}$ can be found in \cite{Ferrara:1988qx,Ferrara:2017yhz}, where it is calculated in the superconformal setup.
To summarise, the objects $\mathcal{R}$ and $B_a$ are respectively a chiral and a real superfield which contain the two auxiliary fields of gravity in the lowest component:
\begin{equation}
\mathcal{R}|=-\frac16 M, \qquad B_a| = -\frac13 b_a.
\end{equation}
When local supersymmetry is spontaneously broken, it is then possible to apply the general constraint \eqref{eq:c2:genconstr} and eliminate the auxiliary fields of the gravity multiplet instead of integrating them out.\footnote{The general constraint \eqref{eq:c2:genconstr} can be applied in any model of spontaneously broken local supersymmetry in which the goldstino is described by a nilpotent $X$. In global supersymmetry it has been shown that this is always the case. The analogous proof in supergravity follows without any additional complication. Details can be found in \cite{Cribiori:2017ngp, Cribiori:2017laj}.} The technical aspects of this step are addressed below, while the physical consequences are going to be commented in the next subsection. 

Since $\mathcal{R}$ is chiral, in order to remove its lowest component it is sufficient to impose
\begin{equation}
X \mathcal{R}=0,
\end{equation}
which fixes $M$ as a function of the goldstino, the gravitino, the Riemann curvature and of the auxiliary field $b_a$. The solution in superspace is
\begin{equation}
\mathcal{R} = -2 \frac{\mathcal{D}^\alpha X \mathcal{D}_\alpha \mathcal{R}}{\mathcal{D}^2 \mathcal{R}}- X \frac{\mathcal{D}^2 \mathcal{R}}{\mathcal{D}^2 X}.
\end{equation}
Such constraint, as noted in \cite{DallAgata:2015zxp} for other constraints on auxiliary fields, implies that the final form of the potential is going to be different from the one of standard supergravity models. Both in global and in local supersymmetry, in fact, a scalar potential is generated by the integration of the auxiliary fields via their equations of motion. If a modification occurs in the auxiliary field sector, therefore, the scalar potential could be modified as well. The previous constraint can actually be replaced with the slightly more general form
\begin{equation}
\label{XRc}
X \left({\cal R} + \frac{c}{6}\right) = 0,
\end{equation}
where $c$ is a complex parameter on which the solution for $M$ is going to depend.
For a generic chiral superfield, this constraint simply adds a constant vacuum expectation value to the lowest component of the constrained multiplet, but for the case in which such a constrained multiplet is the supergravity curvature superfield $\mathcal{R}$, \eqref{XRc} has non-trivial implications on the effective Lagrangian.
As usual, (\ref{XRc}) can be solved by inspecting its top component.
Given the peculiar structure of the ${\cal R}$ superfield, which contains the auxiliary field $M$ in various places, its solution can be found only after an iterative procedure, which gives 
\begin{equation}
\label{c3:eq:solM}
\begin{aligned}
M &=  N\left(1-\frac{i}{\sqrt2 F} { G}\sigma^a \bar \psi_a\right)\\
&+ c\left(1-\frac{i}{\sqrt2 F} { G} \sigma^a \bar \psi_a-\frac{1}{2 F^2}({ G} \sigma^b \bar\psi_b)^2-\frac{1}{4F^2} \bar\psi^2\,{ G^2}\right)\\
&-\frac{c}{3 F^2} \,{G^2}\bigg[\bar N\left(1-\frac{i}{\sqrt2 \bar{F}}\bar{G}\,\bar\sigma^a\psi_a\right)\\
&+\bar{c}\left(1-\frac{i}{\sqrt2 \bar{F}} \bar{G}\,\bar\sigma^a\psi_a-\frac{1}{2 \bar F^2}(\bar{ G}\,\bar \sigma^b\psi_b)^2-\frac{1}{4 \bar F^2} \bar{ G}^2 \,\psi \psi\right)- \frac{|c|^2}{3 \bar F^2}\, \bar{G}^2\bigg],
\end{aligned}	
\end{equation}
where
\begin{equation}
\begin{aligned}
N &= \frac{\sqrt2}{F} {G} \sigma^{ab}\psi_{ab} + \frac{i}{\sqrt2 F} b^a \,{ G}\psi_a +\frac{1}{4 F^2} { G^2} { R} 	+ \frac{i}{2 F^2} G^2 {\cal D}_a b^a \\
& + \frac{1}{2 F^2} { G^2}\bigg[\frac12 \psi_m \sigma^m \bar\psi_n b^n-\frac13 b^2\\
&-i\, \bar\psi^m \bar\sigma^n \psi_{mn}-\frac18\epsilon^{klmn}\left(\bar\psi_k\bar\sigma_l\psi_{mn}+\psi_k\sigma_l\bar\psi_{mn}\right)\bigg]
\end{aligned}
\end{equation}
has been defined. Notice that, in the unitary gauge, the expression for $M$ drastically simplifies to
\begin{equation}
M=c.
\end{equation}

The other auxiliary field of supergravity, namely the real vector $b_a$, resides in the lowest component of $B_a$. Since this superfield is real, the desired constraint to be imposed on it is
\begin{equation}
\label{Bconst}
X \bar{X} B_{\alpha \dot \alpha} = 0.
\end{equation}
This is consistent because $B_{\alpha \dot \alpha}$ contains $b_a$ naked and not through its field-strength.
The superspace solution of this constraint is
\begin{equation}
\begin{aligned}
B_a &= -2 \frac{\bar{\cal D}_{\dot\alpha}\bar X}{\bar{\cal D}^2 \bar X}\bar{\cal D}^{\dot \alpha}B_a - \bar X \frac{\bar{\cal D}^2 B_a}{\bar{\cal D}^2\bar X}\\
&-2\frac{{\cal D^\alpha}X}{{\cal D}^2 X\,\bar{\cal D}^2\bar X} {\cal D}_\alpha \bar{\cal D}^2(\bar X B_a) -\frac{X}{{\cal D}^2 X \bar{\cal D}^2 \bar X}{\cal D}^2 \bar{\cal D}^2 (\bar X B_a).
\end{aligned}
\end{equation}
and it produces an expression for $b_a$ that is a function of the graviton, the gravitino and the goldstino
\begin{equation}
\label{c3:eq:solbaconstr}
\begin{aligned}
b_a&= \frac{\sqrt 2}{8}\frac{1}{\bar F}\left(\bar G \bar \sigma^a \sigma^b \bar \sigma^c \psi_{bc}\right)-\frac{\sqrt 2}{8}\frac{1}{F}\left(\bar \psi_{bc} \bar \sigma^b \sigma^c \bar \sigma^a G\right)\\
&-\frac{3 \sqrt 2}{4}\frac{1}{\bar F}\left(\bar G \bar \sigma^c \psi_{ca}\right)+\frac{3 \sqrt 2}{4}\frac{1}{ F}\left(\bar \psi_{ac} \bar \sigma^c G\right)\\
&-\frac{\sqrt 2i}{4}\frac{M}{\bar F}\left(\bar G \bar \psi_a\right)+\frac{\sqrt 2i}{4}\frac{\bar M}{ F}\left(G \psi_a\right) + \mathcal{O}(G^2).
\end{aligned}
\end{equation}
This expression vanishes trivially in the unitary gauge.

Before addressing the discussion of physical models, notice that the constraint \eqref{XRc} and \eqref{Bconst} are closely analogous to the second constraint in \eqref{Rocekconstr}. They all have the role, in fact, of eliminating the auxiliary fields from a given Lagrangian, substituting them with composite expressions of the goldstino and of the remaining degrees of freedom. Their solution, in particular, is equivalent to the standard procedure of solving the equations of motion.

\subsection{A new class of models}

In this subsection the consequences of the constraints \eqref{XRc} and \eqref{Bconst} are discussed at the level of supergravity Lagrangians. The model which is first going to be considered is again the most general with non-linear supersymmetry, up to two derivatives, involving only one chiral superfield $X$, subject to the nilpotent constraint. It is defined by the K\"ahler potential and superpotential:
\begin{equation}
K = X \bar{X}, \qquad W = W_0 + f \, X,
\end{equation}
where $W_0$ and $f$ are complex parameters. The Lagrangian in superspace is given by
\begin{equation}
\label{supL}
{\cal L} = \frac{3}{8} \int d^2 \Theta \,  2 {\cal E} \,  (\bar {\cal D}^2 - 8 {\cal R} )  
e^{- \frac{1}{3} K(X,\bar X)}+ \int d^2 \Theta \,  2 {\cal E} \, W + c.c.,
\end{equation}
which, when using the nilpotent constraint $X^2=0$, simplifies to
\begin{equation}
\begin{aligned}
\label{c3:eq:MCSL}
\mathcal{L}&=-3 \int d^2 \Theta 2\mathcal{E}\mathcal{R}-\frac18 \int d^2 \Theta 2 \mathcal{E} X \left(\bar {\cal D}^2-8 \mathcal{R}\right)\bar X\\
&+\int d^2 \Theta \,  2 {\cal E} \, W + c.c..
\end{aligned}
\end{equation}
The first term is the pure Einstein--Hilbert Lagrangian for supergravity, while the second and the third contain respectively the kinetic term and the superpotential contribution for the nilpotent $X$. The off-shell component expansion of this model is given in \eqref{c3:eq:LVAoffshell} and, in the unitary gauge, it reduces to
\begin{equation}
\begin{aligned}
e^{-1}\mathcal{L}&=-\frac12 R+\frac14 \epsilon^{klmn}\left(\bar \psi_k \bar \sigma_l \psi_{mn}-\psi_k\sigma_l \bar \psi_{mn}\right)-\\
&- \frac13 M \bar M + \frac13 b_a b^a + |F+\bar f|-|f|^2\\
&- W_0 \left(\bar M + \bar \psi_a \bar \sigma^{ab} \bar \psi_b\right)- \bar W_0 \left(M+ \psi_a \sigma^{ab}\psi_b\right).
\end{aligned}
\end{equation}

At this point several possibilities with different outcomes can occur. The case in which both $M$ and $b_a$
 are not constrained has been addressed already in section \ref{c3:sec:VAsugra} and it is not going to be commented further. A different strategy is to impose only \eqref{XRc}, while leaving $b_a$ unconstrained. To obtain the on-shell Lagrangian one has to substitute first \eqref{c3:eq:eomba} and \eqref{c3:eq:solM} into \eqref{c3:eq:LVAoffshell} and then to integrate out $F$ by solving its equations of motion, which are modified by the contributions coming from the solutions for $M$ and $b_a$. Another possibility is to constraint both $M$ and $b_a$ and to substitute \eqref{c3:eq:solM} and \eqref{c3:eq:solbaconstr} into \eqref{c3:eq:LVAoffshell}. The integration of $F$ can then proceed as usual but, depending on the situation, its expression is going to differ for couplings involving the goldstino. The important fact is that, up to field redefinitions, these last two possibilities are equivalent and in the unitary gauge they lead to the on-shell Lagrangian
\begin{equation}
\begin{aligned}
e^{-1}\mathcal{L}&=-\frac12 R+\frac14 \epsilon^{klmn}\left(\bar \psi_k \bar \sigma_l \psi_{mn}-\psi_k\sigma_l \bar \psi_{mn}\right)-\\
&- W_0 \bar \psi_a \bar \sigma^{ab} \bar \psi_b - \bar W_0 \psi_a \sigma^{ab}\psi_b-\mathcal{V},
\end{aligned}
\end{equation}
where the scalar potential is
\begin{equation}
\mathcal{V}= \Lambda_\text{susy}- 3 |W_0|^2
\end{equation}
and
\begin{equation}
\Lambda_\text{susy} = |f|^2 + \left|\frac{c}{\sqrt3} + \sqrt3\, W_0\right|^2.
\end{equation}
is the scale of supersymmetry breaking.

Notice that there are three independent parameters in the Lagrangian, corresponding to the gravitino mass $W_0$, the scale of supersymmetry breaking $\Lambda_\text{susy}$ and the cosmological constant $\Lambda$.
As anticipated, the cosmological constant does not follow the traditional form as in ordinary supergravity, where it is given by the difference of the $F$-terms squared minus three times the mass of the gravitino squared.
The constraint on the supergravity scalar auxiliary fields implies indeed a different expression, whose numerical value can be positive, negative or zero, depending on the choice of the parameters $f$, $W_0$ and $c$.
When $c=0$ a pure de Sitter supergravity, with a cosmological constant $\mathcal{V} = |f|^2$ is obtained.
When $c = -3 {W}_0$ on the other hand, a cosmological constant in the standard form $\mathcal{V} = |f|^2 - 3 |W_0|^2$ is reproduced and the supersymmetry breaking scale is directly related to $f$, namely $\Lambda_\text{susy} = |f|^2$. 
This is expected from models where supersymmetry is realised linearly on the gravity multiplet and the auxiliary field $M$ gets replaced by its equation of motion $M = -3\, W_0 + \dots$\,.

The simplest generalisation of the previous model is the case in which a matter chiral superfield $Y$ is coupled to $X$, such that
\begin{equation}
X^2=0, \qquad XY=0.
\end{equation}
In the unitary gauge, the off-shell Lagrangian is still given by \eqref{c3:eq:LXYoffshell} but differences in the couplings are going to appear on-shell. 
In the case in which only $M$ is constrained by \eqref{XRc}, the on-shell Lagrangian has again the form \eqref{c3:eq:LSUGRAXYonshell}, but the scalar potential is now 
\begin{equation}
\label{c3:eq:VmattMCS}
\mathcal{V} = \Lambda_\text{susy} - 3 |W_0|^2,
\end{equation}
where
\begin{equation}
\Lambda_\text{susy} = \bigg|\frac{c}{\sqrt 3}+ \sqrt 3 W_0\bigg|^2 + |f|^2+ |g|^2
\end{equation}
has been defined.
A more interesting situation is when, beside $M$, also $b_a$ is constrained by \eqref{Bconst}. In this case, the equations of motion of the auxiliary fields in the unitary gauge are
\begin{align}
F&=-\bar f + a \chi^2,\\
F^Y&=-\bar g+b \chi^2, 
\end{align}
while the constraints \eqref{XRc} and \eqref{Bconst} fix 
\begin{align}
M &= c,\\
b_a&=0.\\
\end{align}
The resulting on-shell Lagrangian is 
\begin{equation}
\begin{aligned}
e^{-1}\mathcal{L}&=-\frac12 R+\frac14 \epsilon^{klmn}\left(\bar \psi_k \bar \sigma_l\psi_{mn}-\psi_k\sigma_l\bar \psi_{mn} \right)\\
&-\frac i2 \left(\chi \sigma^m \mathcal{D}_m \bar \chi+ \bar \chi \bar \sigma^m \mathcal{D}_m \chi\right)-\frac{\sqrt 2}{2}i g\, \chi \sigma^a \bar \psi_a - \frac{\sqrt 2}{2}i \bar g \, \bar \chi \bar \sigma^a \psi_a\\
&-(h-af-bg)\chi^2-(\bar h -a \bar f - b \bar g)\bar \chi^2\\
&+\frac14 \left[(\psi_m \sigma_n \bar \psi^m)(\chi \sigma^n \bar \chi)+i \epsilon^{klmn}(\psi_k \sigma_l \bar \psi_m)(\chi \sigma_n \bar \chi)\right]\\
&-\frac 16 \left[1-6(c-a^2-b^2)\right]\chi^2 \bar \chi^2-\mathcal{V},
\end{aligned}
\end{equation}
where $\mathcal{V}$ is given by \eqref{c3:eq:VmattMCS}. A difference has appeared in the term with four fermions, with respect to the supergravity Lagrangian \eqref{c3:eq:LSUGRAXYonshell} in which supersymmetry is non-linearly realised only in the goldstino sector.

In this section supergravity models where supersymmetry is non-linearly realized by constraining the goldstino superfield but also the auxiliary fields of the supergravity multiplet have been considered. The resulting theories depend on three parameters related to the cosmological constant, the mass of the gravitino and the supersymmetry breaking scale.
These models are different from those appearing in the literature and the difference is  manifest when matter couplings are considered.
Notice finally that, while the CCWZ procedure adopted in \cite{Delacretaz:2016nhw} is so general that it is not guaranteed a priori that the resulting effective does not break unitary, in the new class of models constructed in this section with constrained superfields, the unitarity bound for the gravitino \cite{Deser:2001dt} is automatically satisfied for any value of the parameter $c$.

\subsection{Comment on K\"ahler invariance and non-linear local supersymmetry}

In the class of models discussed in the previous section, constraints have been imposed on the auxiliary fields $M$ and $b_a$ of the gravity multiplet. As a consequence of the constraint \eqref{XRc} on $M$, the scalar potential cannot be put in the standard form
\begin{equation}
\mathcal{V} = e^K \left(g^{i \bar \jmath} D_i W \bar D_{\bar \jmath} \bar W- 3 W\bar W\right),
\end{equation}
since the negative contribution, proportional to the superpotential, comes precisely from the integration of $M$. The constraint \eqref{Bconst} on $b_a$, instead, has consequences on the nature of the scalar manifold, when matter couplings are considered. Since in $\mathcal{N}=1$ supergravity in four dimensions both the fermions and the superpotential transform under K\"ahler transformations, the scalar manifold is not simply K\"ahler but it is restricted to be K\"ahler--Hodge. This fact is commented further in the next chapter but, already at this stage, it is possible to understand how the constraint \eqref{Bconst} influences it. From the appendix \ref{appB:sugraL} it is possible to notice that, in standard supergravity, the on-shell value of $b_a$ provides the correct connection associated to a K\"ahler--Hodge manifold. When constraints are imposed on the gravity sector, instead, the auxiliary field $b_a$ is substituted with the expression \eqref{c3:eq:solbaconstr}, which differs from \eqref{appB:eq:eomb} and which is vanishing in the unitary gauge. In this class of models, therefore, the fermions and the superpotential becomes trivial sections of the K\"ahler bundle and the scalar manifold is expected to be an arbitrary K\"ahler manifold rather than restricted to be K\"ahler--Hodge. This observation leads to the construction of the supergravity models presented in section \ref{c4:sec:kahlinvsugra}, where K\"ahler invariance is restored as in rigid supersymmetry. It is important to notice however that, even if in the worst scenario K\"ahler invariance is lost when supersymmetry is non-linearly realised, the metric of the scalar manifold still remains positive definite, as shown in \cite{Ferrara:2017bnq}.

\subsection{On the validity regime of the effective theory}

The models presented in this chapter are four-dimensional low energy effective theories with spontaneously broken local supersymmetry. In general, when considering effective theories, interactions have to be small with respect to some cut-off parameter and the appearence of a strong regime can be problematic. It is important then to understand which is such a cut-off parameter at which the effective description breaks down. For simplicity the analysis is performed on a Minkowski background and only the pure theory without matter couplings is considered. In this subsection, the Plank mass $M_P$ is restored in the relevant expressions and the gravitino mass $m_{3/2}= \exp(K/2M_P^2) \bar W$ is defined on the vacuum. 

One possible signal of the breakdown of the effective theory is the breaking of the unitarity condition on tree-level scattering amplitudes. In supergravity the worst energy behaviour is expected for the scattering amplitude of two gravitini into two gravitini. To calculate this it is possible to use a theorem concerning the equivalence between gravitino and goldstino physical amplitudes \cite{Casalbuoni:1988kv}, which states that the S matrix elements for helicities $\pm 1/2$ gravitini and other physical particles are asymptotically equal, up to corrections of order $\mathcal{O}(m_{3/2}/\sqrt s)$, where $\sqrt s$ is the energy at which the scattering occurs, to the corresponding S matrix elements where each $\pm 1/2$ gravitino is replaced by the corresponding goldstino. As a consequence, one can just consider the following effective Lagrangian describing non-linear goldstino couplings \eqref{c3:eq:LVASUGRAonshell}
\begin{equation}
\mathcal{L}= - f^2 + i \partial_m \bar G \bar \sigma^m G- m_{3/2} G^2 - \bar m_{3/2} \bar G^2+ \left(\frac{\partial^2 G^2}{4f^2}+ \mathcal{O}\left(\frac{G^2}{M_P^2}\right)\right)\bar G^2+ \dots,
\end{equation}
where dots stand for terms which are further suppressed with $f$ or $M_P$. It is important to notice that the goldstino and the gravitino mass coincide in this model, otherwise the aforementioned theorem on the equivalence between physical amplitudes could not be used. By looking at terms with four goldstini, which are expected to have the worst energy behaviour, it is possible to notice that, since non-derivative interactions are suppressed with $M_P$, the relevant coupling in order to study the validity regime of the effective theory is
\begin{equation}
\sim \frac{1}{f^2} \bar G^2 \partial^2 G^2.
\end{equation}
This gives a contribution to the amplitude of the type \cite{Casalbuoni:1988sx}
\begin{equation}
\mathcal{A} \sim \frac{s^2}{f^2},
\end{equation}
which, in turn, implies the unitarity bound
\begin{equation}
\sqrt s \lesssim \sqrt f \sim \sqrt{m_{3/2}M_P}.
\end{equation}
This is the same bound found in \cite{Casalbuoni:1988sx} but its origin is different. In the linearly realised theory, in fact, the bound stems from the contribution of non-derivative four goldstino terms with sgoldstino exchange. In the present case, instead, the sgoldstino is removed from the spectrum and the bound originates from four goldstino derivatives terms, coming from the non-linear realisation of supersymmetry. In other words, the non-linearly realised terms play the same physical role of their linear counterparts. When the unitarity bound is saturated, strong interactions start to occur as, for example, the formation of bound states. One candidate for this new states is the bilinear $G^2$, which corresponds indeed to the removed sgoldstino in the non-linear regime.

\chapter{Alternative K\"ahler invariance in supergravity}
\label{c4:altsugra}

In this chapter non-linear realisations of supersymmetry are employed in order to construct matter-coupled effective theories, with spontaneously broken local supersymmetry, where the manifold of the matter fields is K\"ahler but, in general, it is not K\"ahler--Hodge. In contrast to the standard formulation of supergravity where, when considering matter couplings within the superspace approach, K\"ahler invariance is not present from the beginning but it is implemented with a Weyl rescaling, in the construction proposed in this chapter such a rescaling is not needed and K\"ahler invariance follows directly as for the global supersymmetric case. The resulting Lagrangian, therefore, mimics a global supersymmetric model with matter, which has been coupled to gravity. This framework can be adopted for constructing models for inflation inspired by rigid supersymmetry, in which all the obstructions due to supergravity are overcome.

The material presented in this chapter has never been published before, but it enjoys connections with previously analysed constructions. In the discussion of minimal constrained supergravity models, in fact, it has been stressed how the elimination of the auxiliary fields of gravity could spoil the K\"ahler invariance of the theory. By following the procedure given within this chapter, it is then possible to restore such an invariance, even though in a form which is different from the standard one of supergravity. The net effect of this construction is to produce theories in which the scalar potential does not assume the usual form. In this respect, these models are similar in spirit to those discussed in \cite{Farakos:2018sgq, Aldabergenov:2018nzd}.

\section{Linear realisations and K\"ahler invariance}
In this section the properties of K\"ahler invariance within linearly realised supersymmetric theories coupled to matter are reviewed, both in the global and in the local case. The discussion starts from the simple situation in which matter is given by chiral superfields and then the presence of a real linear superfield is considered, which is going to be of utmost importance for the construction proposed in the next section. Due to the properties of this superfield, in fact, it is going to be possible to implement K\"ahler invariance in supergravity in an alternative way, which is much similar to the form it assumes in global supersymmetry.

\subsection{K\"ahler invariance in global and local supersymmetry} 

The matter sector of rigid supersymmetric theories is commonly described by a set of chiral superfields $\Phi^i$:
\begin{equation}
\Phi^i = A^i + \sqrt 2 \theta \chi^i + \theta^2 F^i . 
\end{equation}
The most generic supersymmetric Lagrangian, up to two derivatives and involving only chiral superfields, is of the form
\begin{equation}
\label{LKW}
\begin{aligned}
{\cal L} &= \int d^4 \theta \, K(\Phi,\bar \Phi) + \left( \int d^2 \theta \, W(\Phi) + c.c. \right)\\
&=-g_{i \jmath} \partial_m A^i \partial^m \bar A^{\jmath}- i g_{i \jmath} \bar \chi^{\jmath} \bar \sigma^m D_m \chi^i\\
&-\frac12 D_i W_j \chi^i \chi^j-\frac12 \bar D_{\imath}  \bar W_{\bar \jmath} \bar \chi^{\imath} \bar \chi^{\jmath}+\frac14 R_{i \imath j \jmath} \chi^i \chi^j \bar \chi^{\imath}  \bar \chi^{\jmath}\\
&- g^{i \jmath}  W_i  \bar W_{\jmath},
\end{aligned}
\end{equation}
where the  K\"ahler potential $K$ and the superpotential $W$ are respectively a real and a holomorphic function of the chiral superfields. The quantity $g_{i \jmath}=\frac{\partial^2 }{\partial A^i \partial \bar A^{\jmath}}K|$ is the metric of the scalar manifold.

It is known that, in order for supersymmetry to be preserved, the manifold of the scalar fields is restricted to be K\"ahler. As a consequence, the Lagrangian \eqref{LKW} has to be invariant under K\"ahler transformations
\begin{equation} 
\label{1deltaK}
K(\Phi,\bar \Phi) \rightarrow K(\Phi,\bar \Phi) + \mathcal{F}(\Phi) + \bar{\mathcal{F}}(\bar \Phi)  \, ,  
\end{equation} 
where $\mathcal{F}$ is a holomorphic function of the chiral superfields. This property can be checked to work at the component level, or directly in superspace, where it is just a consequence of the fact that the integration of a holomorphic quantity in full rigid superspace vanishes, namely
\begin{equation}
\int d^4 \theta \left( \mathcal{F} + \bar{\mathcal{F}} \right) = 0 . 
\end{equation}

Implementing K\"ahler invariance in supergravity is less straightforward. For example the Lagrangian
\begin{equation}
\int d^4 \theta \, E \, K(\Phi,\bar \Phi), 
\end{equation}
which would generalise naturally the one of global supersymmetry, is not invariant under \eqref{1deltaK} since, even ignoring total derivative terms, this time the superspace integration of a holomorphic function is not vanishing,
\begin{equation}
\int d^4 \theta \, E \, \left( \mathcal{F} + \bar{\mathcal{F}} \right) \ne 0,
\end{equation}
due to the presence of the real superspace density $E$.
In this sense, K\"ahler invariance in supergravity does not follow directly from the superspace construction, but it has to be restored afterwards with a specific procedure, whose main steps are given in appendix \ref{appB:sugraL}. As a consequence, the generic supergravity Lagrangian with chiral matter couplings in superspace takes the form 
\begin{equation}
\label{standardsugra}
{\cal L} =  \int d^2 \Theta \, 2 {\cal E} \, \left[ \frac38 
\left( \bar{\cal D}^2 - 8 {\cal R} \right) 
e^{ - \frac K3}  + W \right] + c.c. \, . 
\end{equation} 
Notice first of all that, with respect to the global supersymmetric case, matter couplings are implemented through an exponential of the K\"ahler potential, instead of with a linear coupling. 
This is a consequence of the fact that the Lagrangian is invariant under K\"ahler transformations only if they are accompanied by a super-Weyl transformation of the vielbein $E_M^{\ \ A}$ of the form \cite{WessBagger} 
\begin{equation}
\begin{aligned}
\delta E_M^{\ \ a} = & \frac16 (\mathcal{F}  + \bar{\mathcal{F} } )  E_M^a \, , 
\\
\delta E_M^{\ \ \alpha} = &  \frac16 (2 \bar{\mathcal{F} } - \mathcal{F}  ) E_M^{\ \ \alpha} 
+ \frac{i}{12} E_M^{\ \ b} (\epsilon \sigma_b)^{\alpha}_{\ \dot \alpha}  \bar{\cal D}^{\dot \alpha} \bar{\mathcal{F} }.
\end{aligned}
\end{equation}
The super-Weyl transformation of the gravity sector induces a super-Weyl transformation on the matter sector. At the component level indeed the chiral rotation of the gravitino induces an analogous local rotation on the other fermions in the theory, namely on the fermionic component of the chiral superfields. In this sense in supergravity fermions are non-trivial sections of a $U(1)$-bundle over the scalar manifold.

The K\"ahler potential and the fermionic fields are defined locally on the scalar manifold. Different local patches support different definitions, which are eventually related by K\"ahler transformations and chiral rotations respectively. 
Inconsistencies might arise when considering triple intersections of three local patches. The requirement that the three different definitions of the fermions coincide in the common region translates into a global condition for the scalar manifold to be K\"ahler of restricted type, also called K\"ahler--Hodge \cite{Witten:1982hu}.

In addition, in order for the complete theory to be K\"ahler--Weyl invariant, it is necessary to require the superpotential to transform as 
\begin{equation}
\label{c4:eq:eW}
W \rightarrow  We^{-\mathcal{F}} \, . 
\end{equation} 
This property of the superpotential is new with respect to the global supersymmetric case and it manifests itself already in the bosonic sector, in particular in the scalar potential 
\begin{equation}
\mathcal{V} = e^K \left[ g^{i\jmath } D_i W  D_{\jmath} \bar W  - 3 W \bar W \right] \, , 
\end{equation}
where the connection $K_i$ appears explicitly in the covariant derivative
\begin{equation}
\label{c4:eq:DW}
D_i W = W_i + K_i W \, .
\end{equation}
The transformation \eqref{c4:eq:eW} and the K\"ahler transformation of $K_i$, in fact, compensate each others. Following the reverse logic, one can think of the rigid supersymmetric case as the trivial situation in which the superpotential has vanishing charge with respect to the connection $K_i$.

To summarise, K\"ahler invariance of the supergravity Lagrangian \eqref{standardsugra} is a consequence of the combined transformations
\begin{align}
K\, &\to \, K + \mathcal{F}+ \bar{\mathcal{F}}, & W \,&\to\, W e^{-\mathcal{F}},\\
\nonumber
d^4 \theta E \, &\to\, d^4 \theta E e^{\frac{\mathcal{F}+\bar{\mathcal{F}}}{3}}, & d^2\Theta \,2 \mathcal{E}\, &\to \, d^2 \,\Theta 2 \mathcal{E}\, e^{\mathcal{F}}
\end{align}

In the following, non-linear realisations are going to be used in order to construct supergravity models where K\"ahler invariance is realised as in global supersymmetry. In particular, the fermions are going to be trivial sections over the $U(1)$ bundle, with a vanishing charge, the Weyl rescaling is not going to be needed and the superpotential is going to be invariant under K\"ahler transformations. As in the rigid supersymmetric case, the scalar potential is then going to be constructed solely out of the partial derivatives $W_i$. For these reasons, the scalar manifold can effectively be thought of as simply  K\"ahler and not restricted to be K\"ahler--Hodge.

\subsection{Real linear superfields and K\"ahler invariance}
In the previous discussion the crucial differences concerning K\"ahler invariance between global and local supersymmetry have been reviewed. 
In this subsection it is shown how it is possible to implement K\"ahler invariance in supergravity, for a specific class of models, in a form which is closer to the global supersymmetric case. Since the key ingredient for the construction proposed below is a real linear superfield, its main properties are reviewed first \cite{Adamietz:1992dk}.

A real linear superfield $L$ is a real scalar superfield,
\begin{equation}
\label{c4:eq:Lreal}
L = L^*,
\end{equation}
subject to the constraints
\begin{equation}
\label{c4:eq:D2L}
\begin{aligned}
(\bar {\cal D}^2 - 8 {\cal R} ) \, L &= 0, \\
({\cal D}^2 - 8 \bar{\cal R} ) \, L \, &=0.
\end{aligned}
\end{equation}
The independent component fields it contains are a real scalar $\varphi$, a Weyl fermion $\omega_\alpha$ and a real antisymmetric tensor $B_{mn}$. They are defined by the following projections
\begin{equation}
\begin{aligned}
\varphi\, &= \, L|, \\
\omega_\alpha \, &=\, \frac{1}{\sqrt{2}}\mathcal{D}_\alpha L|,\\
\sigma^m_{\alpha\dot\alpha}H_{m} + \frac23 \varphi \, b_{\alpha \dot \alpha} \, &=\, -\frac12 [ {\cal D}_\alpha \, , \bar {\cal D}_{\dot \alpha} ] L |,
\end{aligned}
\end{equation}
where the antisymmetric tensor $B_{mn}$ is encoded inside
\begin{equation}
H_m = \frac{1}{3!} \epsilon_{mnkl} \partial^n B^{kl}.
\end{equation}

Given a set of chiral superfields $\Phi^i$ with K\"ahler potential $K(\Phi, \bar \Phi)$, it is possible to construct a coupling between $K$ and $L$ in which K\"ahler invariance is manifest. The discussion is performed at the superspace level, where K\"ahler transformations work in a more straightforward manner, but the component forms of the main results are going to be given as well. 
Under K\"ahler transformations the K\"ahler potential changes according to
\begin{equation}
\label{c4:eq:KLR}
K \rightarrow K + \mathcal{F} + \bar{\mathcal{F}},
\end{equation}
where $\mathcal{F}$ is an holomorphic function of the chiral superfields.
Consider therefore the following superspace coupling between $K$ and $L$: 
\begin{equation}
\label{c4:eq:KL}
{\cal L}_{LK} = - \frac{1}{8} \int d^2 \Theta \,  2 {\cal E} \,  (\bar {\cal D}^2 - 8 {\cal R} )  
\, L \, K  +c.c. \, . 
\end{equation}
By using the superspace identity \cite{Hindawi:1995qa}
\begin{equation}
\int d^2\Theta\, 2 {\cal E}\, (\bar {\cal D}^2 - 8 {\cal R} ) \, (U - U^*) + c.c. \,=\, \text{total derivative},
\end{equation}
which is valid for any superfield $U$, it is possible show that
\begin{equation}
\label{c4:eq:identityL}
\int d^2 \Theta \,  2 {\cal E} \,  (\bar {\cal D}^2 - 8 {\cal R} ) \, L \left( \mathcal{F} + \bar{\mathcal{F}} \right) + c.c. = \text{total derivative} ,  
\end{equation}
which implies that \eqref{c4:eq:KL} is invariant under K\"ahler transformations \eqref{c4:eq:KLR}, up to boundary terms. This is precisely the requirement which is missing in order for the standard supergravity Lagrangian to be invariant under K\"ahler transformations.
For a generic K\"ahler potential $K( \Phi , \bar \Phi)$ the bosonic sector of \eqref{c4:eq:KL3} is
\begin{equation}
\label{c4:eq:KL5}
e^{-1} {\cal L}_{LK} = \varphi \, g_{i \jmath} \, F^i \bar F^{\jmath}  
- \varphi \, g_{i \jmath} \, \partial_m A^i \partial^m \bar A^{\jmath} 
- \frac{i}{2} H^m \left( K_{\imath} \, \partial_m \bar A^{\imath} 
- K_i \, \partial_m A^i \right)  
\end{equation} 
and it is indeed K\"ahler invariant up to a total derivative. Notice the presence of a non-trivial coupling between $H_m$ and the composite $U(1)$ K\"ahler connection 
\begin{equation}
{\cal A}_m = \frac{i}{2} \left( K_{\imath} \, \partial_m \bar A^{\imath} - K_i \, \partial_m A^i \right) . 
\end{equation} 
This coupling is invariant under \eqref{c4:eq:KLR}:
\begin{equation}
\delta \left(H^m \mathcal{A}_m \right) = -\frac i2 \partial_m \mathcal{F} H^m + c.c. = \frac i2 \mathcal{F} \partial_m H^m+c.c. = 0
\end{equation}
up to total derivatives, since $\partial_m  H^m =0$ is the Bianchi identity for $H_m$.

A supergravity Lagrangian which incorporates \eqref{c4:eq:KL} is of the type
\begin{equation}
\label{c4:eq:Lgeneric}
\mathcal{L} = - \int d^4 \theta E f(L) + \int d^4\theta E LK+ \left(\int d^2\Theta 2{\cal E} W + c.c\right), 
\end{equation}
where $f(L)$ is a generic scalar function of the real linear superfield $L$. The first superspace integral contains the degrees of freedom of the gravity sector, coupled to $L$, while the second and the third are respectively the K\"ahler potential and the superpotential contribution of the chiral model,  coupled to the real linear superfield. The bosonic sector of \eqref{c4:eq:Lgeneric} is
\begin{equation}
\begin{aligned}
e^{-1}\mathcal{L} &=
-\frac{1}{6} R \left(f(\varphi)-f_\varphi\,\varphi\right) 
-\frac{1}{9} M \bar{M} \left( f(\varphi)+f_{\varphi\varphi}\,\varphi^2-f_\varphi\,\varphi\right)\\
&+ \frac19 b^a\,b_a\left(f(\varphi)+f_{\varphi\varphi}\,\varphi^2-f_\varphi\, \varphi\right)
+\frac14 f_{\varphi\varphi}\, H^m H_m 
+\frac13 f_{\varphi \varphi}\,\varphi \,H^m b_m\\
&- \frac14 f_{\varphi\varphi}\, \partial^m \varphi \, \partial_m \varphi
+\varphi \, g_{i \jmath} \, F^I \bar F^{\jmath}  
- \varphi \, g_{i \jmath} \, \partial_m A^i \partial^m \bar A^{\jmath} \\
&- \frac{i}{2} H^m \left( K_{\imath} \, \partial_m \bar A^{\imath} 
- K_i \, \partial_m A^i \right)
- W \bar M - \bar W M + W_i F^i + \bar W_{\imath} \bar F^{\imath},
\end{aligned}
\end{equation}
where the shorthand notation $f_\varphi = \frac{\partial }{\partial \varphi}f(L)\big|$,  $f_{\varphi\varphi} = \frac{\partial}{\partial \varphi}\frac{\partial}{\partial \varphi}f(L)\big|$  has been used.

To better understand the origin of the proposed coupling \eqref{c4:eq:KL3} and the reason why it is invariant under K\"ahler trasformations, it can be instructive to look at the dual description. It is known, in fact, that a real linear superfield can be dualised into a chiral superfield. 
To perform the duality, the linearity constraints on $L$, namely \eqref{c4:eq:D2L}, have to be relaxed and imposed eventually at the Lagrangian level using a Lagrange multiplier. Consider therefore the following modified Lagrangian
\begin{equation}
\label{LgenericSS}
\mathcal{L} = - \int d^4 \theta E f(L) + \int d^4\theta E L \left( K + S + \bar S \right) + \left(\int d^2\Theta \,2{\cal E}\, W+ c.c\right) , 
\end{equation}
where $L$ is real, but otherwise unconstrained, while $S$ is a chiral superfield with the role of Lagrange multiplier. By integrating out $S$ the constraints \eqref{c4:eq:D2L} are obtained and the model reduces to \eqref{c4:eq:Lgeneric}. On the other hand, by taking the superspace equations of motion for the unconstrained $L$ gives
\begin{equation}
\label{LKS}
- f_L  + K + S + \bar S = 0 ,
\end{equation}
which can be solved to express $L$ as a function of $K + S + \bar S$. 
Inserting back the solution, the resulting Lagrangian has the form 
\begin{equation}
\label{c4:eq:LgenericSS}
\mathcal{L} =  \int d^4\theta E \, {\mathbb{ F}} \left[ K( \Phi, \bar \Phi  ) + S + \bar S \right] 
+ \left(\int d^2\Theta \,2{\cal E}\, W(\Phi) + c.c\right)  ,
\end{equation}
where $\mathbb{F}$ indicates the Legendre transform. The origin of the K\"ahler invariance of the coupling \eqref{c4:eq:KL3} is now manifest. The chiral superfield $S$, which is dual to the linear superfield $L$, has the role of compensator in the dual picture: it absorbs the variation of $K$ under K\"ahler transformations and makes the theory K\"ahler invariant.
For completeness, the bosonic sector of this model is given below, even though it is not going to be used in the following 
\begin{equation}
\begin{aligned}
e^{-1} {\cal L} = & - \frac12 R - \mathbb{F}' g_{i \jmath} \partial_m A^i \partial^m \bar A^{\jmath}\\
&- {\mathbb{F}}'' \left( [\partial_m S + K_i \partial_m A^i ] [ \partial^m \bar S + K_{\jmath} \partial^m \bar A^{\jmath} ] \right) 
\\
& - e^\mathbb{F} \left( \frac{1}{\mathbb{F}'} g^{i \jmath } \partial_{\jmath} \bar W \partial_i W - 3 W \bar W 
+ \frac{\mathbb{F}' \mathbb{F}' }{\mathbb{F}'' } W \bar W \right) .
\end{aligned}
\end{equation}

The model \eqref{c4:eq:Lgeneric} can be generalised to include also an abelian vector superfield $V$. The resulting Lagrangian is
\begin{equation}
\label{c4:eq:Lgeneric+V}
\begin{aligned}
\mathcal{L} &= - \int d^4 \theta E f(L) + \int d^4\theta E LK+ \left(\int d^2\Theta 2{\cal E} W + c.c\right)\\
&+\frac14\left(\int d^2\Theta 2 \mathcal{E} \mathcal{W}^\alpha \mathcal{W}_\alpha+c.c.\right)+2 \xi \int d^4 \theta E VL, 
\end{aligned}
\end{equation}
where $\mathcal{W}_\alpha= -\frac14 (\bar{\mathcal{D}}^2-8 R)\mathcal{D}_\alpha V$ is the superfield strength of $V$ and $\xi$ is the Fayet--Iliopoulos parameter. This model is both K\"ahler and gauge invariant. Under gauge transformations, the vector superfield changes according to 
\begin{equation}
V \to V + \Phi + \bar \Phi,
\end{equation}
where $\Phi$ is a chiral superfield. The superfield strength $\mathcal{W}_\alpha$ is gauge invariant by construction, while the coupling proportional to $\xi$ is gauge invariant due to the presence of $L$ and as a consequence of the identity \eqref{c4:eq:identityL}.
The bosonic sector of \eqref{c4:eq:Lgeneric+V} is 
\begin{equation}
\begin{aligned}
e^{-1}\mathcal{L} &=
-\frac{1}{6} R \left(f(\varphi)-f_\varphi\,\varphi\right) 
-\frac{1}{9} M \bar{M} \left( f(\varphi)+f_{\varphi\varphi}\,\varphi^2-f_\varphi\,\varphi\right)\\
&+ \frac19 b^a\,b_a\left(f(\varphi)+f_{\varphi\varphi}\,\varphi^2-f_\varphi\, \varphi\right)
+\frac14 f_{\varphi\varphi}\, H^m H_m 
+\frac13 f_{\varphi \varphi}\,\varphi \,H^m b_m\\
&- \frac14 f_{\varphi\varphi}\, \partial^m \varphi \, \partial_m \varphi
+\varphi \, g_{i \jmath} \, F^I \bar F^{\jmath}  
- \varphi \, g_{i \jmath} \, \partial_m A^i \partial^m \bar A^{\jmath} \\
&- \frac{i}{2} H^m \left( K_{\imath} \, \partial_m \bar A^{\imath} 
- K_i \, \partial_m A^i \right)
- W \bar M - \bar W M + W_i F^i + \bar W_{\imath} \bar F^{\imath}\\
&+\frac12 {\rm D}^2-\frac14 v_{mn} v^{mn}+\xi \varphi {\rm D}-\xi v^n H_n.
\end{aligned}
\end{equation}
Notice finally that the coupling proportional to $\xi$ is very similar in form to the Fayet--Iliopoulos D-term of global supersymmetry \cite{Fayet:1974jb}. This fact is going to be commented further in subsection \ref{c4:sec:newDterm} and in the following chapter \ref{c5:dSvacuafromsugra}.

\section{Non-linear realizations and K\"ahler invariance}
\label{c4:sec:kahlinvsugra}
In the model discussed above K\"ahler invariance in supergravity is implemented as in rigid supersymmetry, at the cost of having introduced an additional linear superfield $L$ which contains physical and propagating degrees of freedom. As it has been shown in the previous chapters, however, if the spontaneous breaking of supersymmetry is assumed, by using non-linear realisations it is possible to construct composite expressions out of the goldstino. Since in supergravity the latter is not a physical mode, such expressions would not contain new propagating degrees of freedom. In the present section, the nilpotent goldstino superfield $X$ is therefore used to construct a composite object having the properties of a real linear superfield. This composite real linear superfield is then used to reproduce the results of the previous section: a theory of supergravity is constructed, where K\"ahler invariance is realised as in global supersymmetry but where, this time, the degrees of freedom are solely those of supergravity coupled to matter.

\subsection{Chiral models} 
Given the chiral goldstino superfield $X$, such that $X^2=0$, it is possible to construct a composite real linear superfield $L$ with the specific property
\begin{equation}
\label{c4:eq:LLG1}
L | = 1 + {\cal O}(G, \bar G) \, . 
\end{equation}
The expression
\begin{equation}
\label{LL3}
L = {\cal D}^\alpha \left( \bar{\cal D }^2 - 8 {\cal R} \right) {\cal D}_\alpha \left[ \frac{X \bar X}{{\cal D}^2 X \bar{\cal D}^2 \bar X} \right]\, ,  
\end{equation} 
in fact satisfies 
the conditions \eqref{c4:eq:Lreal}, \eqref{c4:eq:D2L} and \eqref{c4:eq:LLG1}.
The lowest component of this composite $L$ has the form 
\begin{equation}
\begin{aligned}
L| = & \, 1
- \frac{i}{2f} \bar \gamma \bar \sigma^a \psi_a
-\frac{i}{2f}\gamma \sigma^a \bar \psi_a
+ \frac{1}{3f^2}\gamma^2 \bar M
+ \frac{1}{3f^2} \bar \gamma^2 M
\\
& + \frac{1}{4 f^2} \gamma^2 \bar \psi_b \bar \sigma^e
\sigma^b \bar \psi_e
+ \frac{1}{4f^2} \bar \gamma^2 \psi_e \sigma^b \bar
\sigma^e \psi_b
- \frac{1}{2f^2} \gamma \sigma^a \bar \gamma b_a
\\
& + \frac{i}{f^2} \bar \gamma \bar \sigma^a e_{a}^{\ m} D_m \gamma
+ \frac{i}{f^2} \gamma \sigma^a e_{a}^{\ m} D_m \bar \gamma
\\
& -  \frac{1}{2f^2} (\gamma \sigma^a \bar \psi_e) (\psi_a
\sigma^e \bar \gamma)
+ \frac{1}{2f^2} (\psi_b \sigma^b \bar \gamma) (\gamma \sigma^a
\bar \psi_a)
+ {\cal O}(\gamma^3) \, , 
\end{aligned}
\end{equation}
where the fermion
\begin{equation} 
\gamma_\alpha = f\frac{G_\alpha}{\sqrt{2}F} 
\end{equation} 
has been defined. In the unitary gauge, the independent components of the composite real linear superfield are
\begin{align}
L|&=1,\\
\mathcal{D}_\alpha L| &= -\frac i2 \sigma^a_{\alpha \dot \alpha} \bar \psi_a^{\dot \alpha},\\
 -\frac12 \left[\mathcal{D}_\alpha, \bar{\mathcal{D}}_{\dot \alpha}\right] L|&= \frac 23 b_{\alpha \dot \alpha}-\frac14 \psi_{a\, \alpha} \left(\bar \psi_b \bar \sigma^b \sigma^a\right)_{\dot \alpha}-\frac14 \left(\sigma^a \bar \sigma^b \psi_b\right)_\alpha \bar \psi_{a\, \dot \alpha}\\
\nonumber
&-\frac12 \sigma^a_{\alpha \dot \beta} \bar \psi_b^{\dot \beta} \psi_a^\beta\sigma^b_{\beta \dot \alpha}
\end{align}

By using this ingredient, along with the same logic of the previous section, it is possible to construct a supergravity model where supersymmetry is spontaneously broken and non-linearly realised and where K\"ahler invariance is implemented as in global supersymmetric theories. Such a model is not going to contain any additional physical degrees of freedom associated to the real linear superfield $L$, since the latter is a composite object built out of the goldstino. 
Consider first the superspace term \eqref{c4:eq:KL3}
\begin{equation}
\label{c4:eq:KL3}
{\cal L}_{LK} = - \frac{1}{8} \int d^2 \Theta \,  2 {\cal E} \,  (\bar{\cal D}^2 - 8 {\cal R} )  
L K  +c.c. \,. 
\end{equation} 
Inserting the explicit expression \eqref{LL3} for the composite real linear superfield $L$, the resulting component form up to two fermions and in the unitary gauge is
\begin{equation}
\begin{aligned}
e^{-1}\mathcal{L}_{LK} &= -g_{i\bar\jmath}\, \partial_m A \partial^m \bar A^{\bar \jmath} + g_{i\jmath} F^i \bar F^{\bar \jmath}\\
&-\frac i2 g_{i\bar\jmath} \left(\chi^i \sigma^m \mathcal{D}_m \bar \chi^{\bar \jmath}+ \bar \chi^{\bar \jmath} \bar \sigma^m \mathcal{D}_m \chi^i\right)\\
&+\frac i2 (\chi^i \sigma^m \bar \chi^{\bar \imath})(K_{ij \bar \imath}\partial_m A^j- K_{i \bar \imath \bar \jmath}\partial_m \bar A^{\bar \jmath})\\
&-\frac{\sqrt 2}{4}i g_{i\bar\jmath} F^i (\psi_c \sigma^c \bar \chi^{\bar \jmath})-\frac{\sqrt 2}{4}ig_{i\jmath} \bar F^{\bar \jmath}(\bar \psi_c \bar \sigma^c \chi^i)\\
&+\frac{\sqrt 2}{2}g_{i\jmath}\left[e^a_m \partial^m \bar A^{\bar \jmath} (\psi_a \chi^i)+e^a_m \partial^m A^i (\bar \psi_a \bar \chi^{\bar \jmath})\right]\\
&-\frac{\sqrt 2}{4}g_{i\bar\jmath}\left[e_a^m \partial_m \bar A^{\bar \jmath} (\psi_c \sigma^a \bar \sigma^c \chi^i)+e_a^m \partial_m A^i (\bar \psi_c \bar \sigma^a \sigma^c \bar \chi^{\bar \jmath})\right]\\
&-\frac12 K_{ij\bar \imath} \bar F^{\bar \imath}(\chi^i \chi^j)-\frac12 K_{i \bar \imath \bar \jmath}F^i (\bar \chi^{\bar \imath} \bar \chi^{\bar \jmath})\\
&+\text{terms with four fermions}.
\end{aligned}
\end{equation}
This is manifestly K\"ahler invariant and it can be coupled directly to gravity. A complete supergravity Lagrangian in which matter couplings are embedded with this term is 
\begin{equation}
\label{c4:eq:nL1}
{\cal L} =  - 3  \int d^2 \Theta \,  2 {\cal E} \, {\cal R} 
- \frac{1}{8} \int d^2 \Theta \,  2 {\cal E} \,  (\bar{\cal D}^2 - 8 {\cal R} )  
  LK  
+ \int d^2 \Theta \,  2 {\cal E} \, W +c.c. 
\end{equation} 
and its invariance under K\"ahler transformations
\begin{equation}
\label{c4:eq:KSS}
K \rightarrow K + \mathcal{F} + \bar{\mathcal{F}} \, , 
\end{equation}
has been discussed in the previous section at the superspace level. 
In particular the superpotential $W$ has not to transform as \eqref{c4:eq:eW}.\footnote{The only requirement which is needed for the superpotential $W$ is that it should always have a linear term in the superfield $X$, such that the supersymmetry breaking condition $\langle F \rangle \neq 0$ is satisfied.} 
Up to two fermions, the component expansion of the supergravity Lagrangian \eqref{c4:eq:nL1} is 
\begin{equation}
\begin{aligned}
\label{c4:eq:altsugra}
e^{-1}\mathcal{L} = & -\frac12 R+\frac14 \epsilon^{klmn} \left(\bar \psi_k \bar \sigma_l\psi_{mn}-\psi_k \sigma_l \bar \psi_{mn}\right)\\
&-g_{i\bar\jmath} \,\partial_m A \partial^m \bar A^{\bar \jmath} + g_{i\jmath} F^i \bar F^{\bar \jmath}\\
&-\frac13 M \bar M+ \frac13 b_a b^a- \bar M W - M \bar W + W_i F^i + \bar W_{\bar \imath} \bar F^{\bar \imath}\\
&-\frac i2 g_{i\bar\jmath} \left(\chi^i \sigma^m \mathcal{D}_m \bar \chi^{\bar \jmath}+ \bar \chi^{\bar \jmath} \bar \sigma^m \mathcal{D}_m \chi^i\right)\\
&+\frac i2 (\chi^i \sigma^m \bar \chi^{\bar \imath})(K_{ij \bar \imath}\partial_m A^j- K_{i \bar \imath \bar \jmath}\partial_m \bar A^{\bar \jmath})\\
&-\frac{\sqrt 2}{4}i g_{i\bar\jmath} F^i (\psi_c \sigma^c \bar \chi^{\bar \jmath})-\frac{\sqrt 2}{4}ig_{i\jmath} \bar F^{\bar \jmath}(\bar \psi_c \bar \sigma^c \chi^i)\\
&+\frac{\sqrt 2}{2}g_{i\jmath}\left[e^a_m \partial^m \bar A^{\bar \jmath} (\psi_a \chi^i)+e^a_m \partial^m A^i (\bar \psi_a \bar \chi^{\bar \jmath})\right]\\
&-\frac{\sqrt 2}{4}g_{i\bar\jmath}\left[e_a^m \partial_m \bar A^{\bar \jmath} (\psi_c \sigma^a \bar \sigma^c \chi^i)+e_a^m \partial_m A^i (\bar \psi_c \bar \sigma^a \sigma^c \bar \chi^{\bar \jmath})\right]\\
&-\frac12 K_{ij\bar \imath} \bar F^{\bar \imath}(\chi^i \chi^j)-\frac12 K_{i \bar \imath \bar \jmath}F^i (\bar \chi^{\bar \imath} \bar \chi^{\bar \jmath})\\
&-\frac12 W_{ij} \chi^i \chi^j-\frac12 \bar W_{\bar \imath \bar \jmath} \bar \chi^{\bar \imath} \bar \chi^{\bar \jmath}\\
&-\bar W \psi_a \sigma^{ab} \psi_b-W \bar \psi_a \bar \sigma^{ab} \bar \psi_b\\
&-\frac{\sqrt 2}{2}i W_i \chi^i \sigma^a \bar \psi_a- \frac{\sqrt 2}{2}i \bar W_{\bar \imath} \bar \chi^{\bar \imath} \bar \sigma^a \psi_a\\
&+\text{terms with four fermions}
\end{aligned}
\end{equation}
and, once again, this is manifestly K\"ahler invariant.
Notice that, in contrast to the standard formulation of supergravity, this Lagrangian appears to be directly in the Einstein frame and the Weyl rescaling reported in the appendix \ref{appB:sugraL} is not needed. 
Notice also the important fact that the integration of the auxiliary field $b_a$ is gaussian and it sets $b_a=0$. In the standard formulation of supergravity, on the contrary, the equations of motion \eqref{appB:eq:eomb} of $b_a$ give the composite connection associated to the K\"ahler--Hodge nature of the scalar manifold. As expected, this connection is not appearing in the present model, since the K\"ahler--Hodge condition has not to be imposed. In addition, by comparing \eqref{c4:eq:altsugra} and \eqref{appB:eq:gensugraL}, it is possible to observe that terms in $M$, $b_a$ and $F$ which are present in the latter, are not present in the former. These terms in fact would spoil K\"ahler invariance and then, in order to recover it, the Weyl rescaling would be needed.

At this point two different directions can be followed in order to eliminate the auxiliary fields and to produce the scalar potential. In both approaches the auxiliary fields $F^i$ of the matter sector are going to be integrated out through their equations of motion
\begin{equation}
g_{i \bar \jmath}F^i + \bar W_{\bar \jmath}-\frac{\sqrt 2}{4}i g_{i \bar \jmath}(\bar \psi_a \bar \sigma^a \chi^i)-\frac12 K_{ij \bar \jmath}(\chi^i \chi^j)=0.
\end{equation}

If also the auxiliary fields $M, b_a$ of the gravity sector are integrated out
\begin{align}
M&=-3 W,\\
b_a&=0,
\end{align}
the following scalar potential is produced 
\begin{equation}
\mathcal{V} = \left(g^{i \bar \jmath} W_i \bar W_{\bar \jmath}- 3 W \bar W\right).
\end{equation}
This is similar to the scalar potential of standard supergravity \eqref{appB:eq:V} with the important exceptions that the exponential factor $e^K$ is not present and the derivatives on the superpotential are not covariant. These are consequences of the fact that, in the present model, the superpotential is not transforming under K\"ahler transformations and, therefore, the expression for $\mathcal{V}$ is directly invariant under \eqref{c4:eq:KSS}.

A different theory can be obtained by eliminating the auxiliary fields of gravity using non-linear realisations of supersymmetry. In the spirit of the models presented in section \ref{c3:sec:MCS}, constraints can be imposed on the gravity sector in order to remove its auxiliary fields. Imposing the constraint \eqref{Bconst} on $b_a$ would not produce novelties in the unitary gauge, since its solution coincides with the equations of motion. On the contrary, imposing the constraint \eqref{XRc} gives $M=c$ and the scalar potential becomes
\begin{equation}
\mathcal{V} = \left(\Lambda_{\text{susy}}- 3 W \bar W\right),
\end{equation}
where 
\begin{equation}
\Lambda_{\text{susy}}= g^{i \bar \jmath}W_i \bar W_{\bar \jmath}+\left|\frac{c}{\sqrt 3}+ \sqrt 3 W\right|^2
\end{equation}
contains information on the supersymmetry breaking scale. Notice that, in the particular case in which $c=0$, the scalar potential reduces to
\begin{equation}
\mathcal{V} = g^{i \bar \jmath}W_i \bar W_{\bar \jmath}.
\end{equation}
This is precisely the same form it assumes in global supersymmetry, with the only difference that, in the present case, it can never be vanishing, since supersymmetry has to be always broken in the vacuum.

\subsection{Gauged chiral models} 
\label{c4:sec:gaugings}

In the model presented in the previous section, analytic isometries of the scalar manifold can be gauged. If supersymmetry is linearly realised on the matter sector, the gauging procedure is closely similar to the global case. Such a procedure is explained in detail in \cite{WessBagger} but, for completeness, the main steps are summarised below.

To each analytic isometry of the scalar manifold it is possible to associate a holomorphic Killing vector
\begin{equation}
\begin{aligned}
{\cal X}^{(a)} \,&=\, {\cal X}^{i (a)}(A^j) \frac{\partial}{\partial A^i},\\
\bar{\cal X}^{(a)} \,&=\, \bar{\cal X}^{\bar \imath (a)}(\bar A^{\bar \jmath}) \frac{\partial}{\partial \bar A^{\bar \imath}}
\end{aligned}
\end{equation}
and a real prepotential function $\mathcal{P}^{(a)}(A,\bar A)$ obeying
\begin{equation}
\begin{aligned}
g_{i \bar \jmath}\bar {\cal X}^{ \bar \jmath (a)}  &= i \frac{\partial \mathcal{P}^{(a)}}{\partial A^i},\\
g_{i\bar \jmath} {\cal X}^{i (a) }  &= -i   \frac{\partial \mathcal{P}^{(a)}}{\partial \bar A^{\bar \jmath}}.
\end{aligned}
\end{equation}
Being defined through differential equations, the $\mathcal{P}^{(a)}$ are determined up to real constants.
The Killing vectors close the algebra of the isometry group $\mathcal{G}$
\begin{equation}
\begin{aligned}
[{\cal X}^{(a)}, {\cal X}^{(b)}] \, &= \, - f^{abc} {\cal X}^{(c)},\\
[\bar{\mathcal{X}}^{(a)}, \bar{\cal X}^{(b)}] \,&=\, - f^{abc} \bar{\cal X}^{(c)},\\
[\mathcal{X}^{(a)},\bar{\mathcal{X}}^{(b)}]\, &=\, 0,
\end{aligned}
\end{equation}
where Latin indices between parenthesis run over the dimensions of $\mathcal{G}$.
The variation of the K\"ahler potential under an analytic isometry is given by
\begin{equation}
\label{deltaK}
\delta K = [\epsilon^{(a)} {\cal X}^{(a)} + \bar\epsilon^{(a)}\bar{\cal X}^{(a)}]K = \epsilon^{(a)} \mathcal{F}^{(a)} + \bar\epsilon^{(a)}\bar{\mathcal{F}}^{(a)} - i (\epsilon^{(a)}-\bar\epsilon^{(a)}) \mathcal{P}^{(a)},
\end{equation}
where the holomorphic functions $\mathcal{F}^{(a)} = \mathcal{X}^{(a)}K + i \mathcal{P}^{(a)}$ have been introduced. Notice that if the parameter $\epsilon$ is chosen to be real, then \eqref{deltaK} reduces to a K\"ahler transformation. Promoting this relation to superspace gives
\begin{equation}
\delta K = \Lambda^{(a)} \mathcal{F}^{(a)}+\bar\Lambda^{(a)} \bar{\mathcal{F}}^{(a)}-i (\Lambda^{(a)}-\bar\Lambda^{(a)})\mathcal{P}^{(a)},
\end{equation}
where $\Lambda^{(a)}$ are chiral superfields. A counterterm function $\Gamma$ is now introduced in superspace such that its gauge variation is by construction
\begin{equation}
\delta \Gamma = i (\Lambda^{(a)}-\bar\Lambda^{(a)})\mathcal{P}^{(a)}.
\end{equation}
Replacing then in the theory
\begin{equation}
K \rightarrow K + \Gamma,
\end{equation}
the gauge variation of the resulting Lagrangian is going to have the form of a K\"ahler transformation, since
\begin{equation}
\delta(K+ \Gamma) = \Lambda^{(a)} \mathcal{F}^{(a)}+\bar\Lambda^{(a)} \bar{\mathcal{F}}^{(a)}.
\end{equation}
If the theory is K\"ahler invariant then, as a consequence of this procedure it is going also to be gauge invariant.
The explicit form of $\Gamma$ for a general gauging can be found in \cite{WessBagger}. It turns out that it is a function of the chiral matter superfields $\Phi^i$, $\bar \Phi^{\bar \jmath}$ and also of the vector superfield $V = V^{(a)} T^{(a)}$, where $T^{(a)}$ are the hermitian generators of the isometry group ${\cal G}$. In the Wess--Zumino gauge this function reduces to
\begin{equation}
\Gamma = V^{(a)} \mathcal{P}^{(a)}+\frac12 g_{i \bar \jmath} \, {\cal X}^{i(a)} \bar{\cal X}^{\bar \jmath (b)} V^{(a)} V^{(b)}.
\end{equation}

Following this procedure it is possible to construct the gauged version of the supergravity Lagrangian \eqref{c4:eq:nL1}. It is sufficient to replace
\begin{equation}
K \rightarrow K + V^{(a)} \mathcal{P}^{(a)}+\frac12 g_{i \bar \jmath} \, {\cal X}^{i(a)} \bar{\cal X}^{\bar \jmath (b)} V^{(a)} V^{(b)}
\end{equation}
and to add the superspace coupling which produces the kinetic terms for the vector $V$.  The resulting supergravity Lagrangian is
\begin{equation}
\begin{aligned}
\mathcal{L} &= - 3  \int d^2 \Theta \,  2 {\cal E} \, {\cal R} +\frac14 \left(\int d^2\Theta 2\mathcal{E}\, \mathcal{W}^\alpha \mathcal{W}_\alpha+c.c.\right)\\
&- \frac{1}{8} \int d^2 \Theta \,  2 {\cal E} \,  (\bar {\cal D}^2 - 8 {\cal R} )  
\, L \, (K + \Gamma)  
+ \int d^2 \Theta \,  2 {\cal E} \, W +c.c. , 
\end{aligned}
\end{equation}
where the first line contains the kinetic terms of the gravity multiplet and of the vector multiplet, while the second line contains the matter couplings. In particular the superfield strength $\mathcal{W}_\alpha=-\frac14 (\bar{\mathcal{D}}^2-8R)\mathcal{D}_\alpha V$ has been defined  and the vector superfield $V$ inside $\Gamma$ is rescaled by a factor two, $V\to 2V$, for convenience. The complete Lagrangian is K\"ahler and gauge invariant by construction.
The bosonic sector is
\begin{equation}
\begin{aligned}
e^{-1}\mathcal{L}&=-\frac12 R-\frac 14 v_{mn}^{(a)}v^{mn(a)}-g_{i \bar \jmath}\, \mathcal{D}_m A^i \bar{\mathcal{D}}^m \bar A^{\bar \jmath}\\
&-\frac13 M \bar M +\frac 13 b_a b^a - \bar M W- M \bar W \\
&+\frac12 D^{(a)}D^{(a)}+g_{i \bar \jmath}\, F^i \bar F^{\bar \jmath}+D^{(a)}\mathcal{P}^{(a)}+W_i F^i +\bar W_{\bar \jmath}\bar F^{\bar \jmath},
\end{aligned}
\end{equation}
where the gauge covariant derivative $\mathcal{D}_m A^i = \partial_m A^i - v_m^{(a)} \mathcal{X}^{i(a)}$ has been defined. This Lagrangian appears directly in the Einstein frame and, as for its ungauged version, the Weyl rescaling is not needed in the unitary gauge.

By integrating out the auxiliary fields the following on-shell bosonic Lagrangian is produced
\begin{equation}
\label{c4:eq:gaugedLonshell}
e^{-1}\mathcal{L}=-\frac12 R-\frac 14 v_{mn}^{(a)}v^{mn(a)}-g_{i \bar \jmath}\, \mathcal{D}_m A^i \bar{\mathcal{D}}^m \bar A^{\bar \jmath}-\mathcal{V}
\end{equation}
where the scalar potential is 
\begin{equation}
\mathcal{V} = \frac12 \mathcal{P}^{(a)} \mathcal{P}^{(a)}+\left(g^{i\bar \jmath} W_i \bar W_{\bar \jmath}- 3 W\bar W\right).
\end{equation}
Notice again its similarity with the scalar potential of rigid supersymmetry. It is possible, alternatively, to constrain the auxiliary fields of the gravity multiplet, by imposing \eqref{XRc} and \eqref{Bconst}. The resulting on-shell theory has the form of \eqref{c4:eq:gaugedLonshell} but this time the scalar potential is going to be
\begin{equation}
\mathcal{V} = \frac12 \mathcal{P}^{(a)} \mathcal{P}^{(a)}+\left(\Lambda_{\text{susy}}- 3 W\bar W\right),
\end{equation}
where
\begin{equation}
\Lambda_{\text{susy}}=g^{i\bar \jmath} W_i \bar W_{\bar \jmath}+\left| \frac{c}{\sqrt 3}+\sqrt 3 W\right|^2
\end{equation}
gives the supersymmetry breaking scale. Notice that for $c=0$ this reduces to 
\begin{equation}
\mathcal{V} = \frac12 \mathcal{P}^{(a)} \mathcal{P}^{(a)}+g^{i\bar \jmath} W_i \bar W_{\bar \jmath},
\end{equation}
which is precisely the form of the scalar potential of a gauged chiral model in global supersymmetry, with the only exception that, in the present case, it can never be vanishing.

The supergravity models constructed in this section can be employed to the studying of inflationary scenarios. Due to the particular form of their scalar potential, in fact, in which no exponential factor $e^K$ appears, they can accommodate the slow-roll conditions more directly then standard supergravity.

\subsection{An alternative D-term in supergravity}
\label{c4:sec:newDterm}
To implement the Fayet--Iliopoulos D-term into supergravity is not straightforward, as it is going to be discussed in the next chapter. A necessary requirement, in fact, is the gauging of the R-symmetry of the theory, which leads to the appearance of the vector multiplet $V$ through an exponential factor $e^{-\frac23 \xi V}$, in order to preserve gauge invariance. When supersymmetry is spontaneously broken and non-linearly realised, however, more possibilities can occur. In order to understand the point and to set the ground for the forthcoming discussion of the following chapter, consider again as an example the model \eqref{c4:eq:Lgeneric+V} in the particular case in which $f(L)=3$ and $K=0$ for simplicity:
\begin{equation}
\label{c4:eq:Lgeneric+V2}
\begin{aligned}
\mathcal{L} &= - 3\int d^4 \theta E + \frac14\left(\int d^2\Theta 2 \, \mathcal{E} \, \mathcal{W}^\alpha \mathcal{W}_\alpha+c.c.\right)+2 \xi \int d^4 \theta E VL.
\end{aligned}
\end{equation}
This model contains a Fayet--Iliopoulos term linear in $V$ which is gauge invariant even in supergravity, due to the presence of the real linear superfield $L$. In the case in which this real superfield is composite and it is constructed out of the goldstino, as in \eqref{LL3}, then only the degrees of freedom of gravity and of the vector multiplet are present.
The bosonic sector is indeed going to be
\begin{equation}
\begin{aligned}
e^{-1}\mathcal{L} &= -\frac12 R -g_{i \bar\jmath}\partial_m A^i \partial^m \bar A^{\bar \jmath}-\frac14 v_{mn}v^{mn}\\
&-\frac13 M \bar M+\frac13 b_a b^a+g_{i \bar \jmath} F^i \bar F^{\bar \jmath}+\frac12 D^2 + \xi D
\end{aligned}
\end{equation}
and it contains a term linear in $D$. After the elimination of the auxiliary fields, a constant scalar potential is produced of the type $\mathcal{V}= \frac{\xi^2}{2}$, which is precisely the form it assumes in the case of a Fayet--Iliopoulos D-term breaking.

In the next chapter this model is going to be derived in a different way, generalised to the case of matter couplings and inserted eventually in the more general context of the construction of de Sitter vacua from supergravity and string theory.

\chapter{de Sitter vacua from supergra\-vity}
\label{c5:dSvacuafromsugra}

The interest in de Sitter solutions is supported by observations, which indicate that, in the part of the Universe we are living in, the vacuum energy has an extremely small but positive value, of order $\Lambda \sim 10^{-122}$ in Planck units. 
In the case in which quintessence model were disfavoured by data \cite{Cicoli:2018kdo,Akrami:2018ylq}, the decision of studying stable or metastable de Sitter vacua would be strongly motivated. In particular, it would be essential to understand whether they are ultimately realisable withing string theory or not \cite{Danielsson:2018ztv, Obied:2018sgi, Denef:2018etk, Roupec:2018mbn, Kallosh:2018nrk, Bena:2018fqc}.

In this chapter the construction of de Sitter vacua in four-dimensional $\mathcal{N}=1$ supergravity and the relationship with non-linearly realised supersymmetry are investigated. A complete review of the subject would be out of the purposes of the present thesis and the attention is then focused only on one of the most studied setups, which has been proposed by Kachru, Kallosh, Linde and Trivedi \cite{Kachru:2003aw,Kachru:2003sx} for constructing de Sitter vacua from string theory compactifications. 
It is discussed, in particular, how $\mathcal{N}=1$ supergravity in four dimensions can be used to capture the physical properties of KKLT and which are the necessary ingredients in order to obtain a consistent low energy effective theory. 

It is known that linearly realised supersymmetry cannot be preserved on a de Sitter background \cite{Pilch:1984aw}. When considering non-linear realisations, however, a Lagrangian with spontaneously broken local supersymmetry which allows for de Sitter solutions can nevertheless be constructed, as shown in chapter \ref{c4:sugra}. In this class of models, the goldstino plays a crucial role in uplifting the vacuum energy with a positive contribution.  For this reason, in the present chapter it is first reviewed how an effective description for KKLT can be obtained in four dimensions with the use of non-linear supersymmetry in the language of constrained superfields, in the case in which the goldstino is described by a nilpotent chiral superfield $X$ \cite{Ferrara:2014kva}.
An alternative low energy effective theory for KKLT is then proposed in which the goldstino is embedded into a vector multiplet. This description has off-shell linearly realised supersymmetry and no constraints are indeed imposed on the superfields in the model. The construction is based on a novel embedding of the Fayet--Iliopouolos D-term into supergravity which, on the contrary to the standard situation, does not require the gauging of the R-symmetry. The relationship between the two effective descriptions for KKLT is then investigated and it is shown how it is possible to map one into the other.

\section{KKLT setup and non-linear realisations of supersymmetry}
In this section the basic ingredients involved in the KKLT setup are reviewed and it is shown how non-linear supersymmetry implemented with a nilpotent goldstino superfield can capture its low energy effective description. In the original KKLT proposal, type IIB string theory compactified on a Calabi--Yau threefold is considered. In this framework,  when looking for four-dimensional $\mathcal{N}=1$ descriptions, the K\"ahler potential has a no-scale structure for the moduli, while the superpotential receives contributions both from fluxes and from non-perturbative effects. 
Once the complex structure moduli and the axion-dilaton have been stabilised using fluxes and consequently integrated out, what remains is an effective theory for the K\"ahler modulus $T$, which governs the size of the internal manifold. This field can be described in terms of a chiral superfield with K\"ahler potential and superpotential with a no-scale structure:
\begin{equation}
K = -3 \log(T+\bar T), \qquad W=W_0.
\end{equation}
Even though the string theory origin is known and explained in the original works, for all the purposes of the effective description, $W_0$ can be interpreted as a parameter.  This model, due to its no-scale structure, has a vanishing scalar potential at any point of the moduli space. In order to stabilise also the scalar in $T$, non-perturbative corrections can be considered. As explained in \cite{Kachru:2003aw}, they lead to a modification of the type
\begin{equation}
\label{c5:eq:LKKLT}
K=-3 \log(T+\bar T),\qquad W = W_0 + A e^{- a T},
\end{equation}
where again $a$ and $A$ are parameters with a specific string theory origin. This step is very delicate, as discussed for example \cite{Kachru:2018aqn} and in the reference therein, since the insertion of non-perturbative effects into a tree level effective Lagrangian could spoil the consistency of the resulting description.
For the rest of the discussion, however, it is going to be assumed that such a procedure is consistent and the model \eqref{c5:eq:LKKLT} is therefore going to be studied by means of $\mathcal{N}=1$ supergravity in four dimensions.

From a direct inspection of the scalar potential
\begin{equation}
\begin{aligned}
\mathcal{V}_{KKLT} &= e^K\left( g^{T\bar T} D_T W D_{\bar T} \bar W - 3 W \bar W\right)\\
&=\frac{a A e^{-a (T+\bar T)}}{3(T+\bar T)^2} \left[ A (6 + a(T+\bar T))+3 (e^{a T}+e^{a \bar T})W_0\right]
\end{aligned}
\end{equation}
it is possible to realise that it admits stable and supersymmetric anti de Sitter vacua. Considering in fact the case in which the axion inside $T$ is set to zero for simplicity, $\text{Im}\, T=0$, the scalar potential reduces to
\begin{equation}
\mathcal{V}_{KKLT} = \frac{a A e^{- a \text{Re}\, T}}{2(\text{Re}\, T)^2}\left[A e^{- a\text{Re}\, T}+ W_0 + \frac13 a A \text{Re}\, T e^{- a \text{Re}\, T}\right].
\end{equation}
The supersymmetric vacuum condition $D_T W=0$ is then solved by
\begin{equation}
W_0=- A e^{- a\text{Re}\, T}\left(1+\frac23 a \text{Re}\, T\right)
\end{equation}
and the vacuum energy becomes
\begin{equation}
\langle \mathcal{V} \rangle = \langle e^K (- 3 W^2) \rangle = -\frac{a^2 A^2 e^{-2 a \text{Re}\, T_{crit}}}{6 \text{Re}\, T_{crit}} < 0,
\end{equation}
which is negative in the region $\text{Re}\, T >1$ in which the effective description is valid. To obtain a de Sitter vacuum a new ingredient has to be introduced. The authors of \cite{Kachru:2003aw}, without jeopardising the stabilisation of the moduli, added to this setup the contribution of one anti D3 brane. In particular, they predict a modification of the scalar potential due to this new object of the type
\begin{equation}
\label{c5:eq:VdS}
\mathcal{V} = \mathcal{V}_{KKLT} + \frac{\mu^4}{(T+\bar T)^2},
\end{equation}
where $\mu$ is a parameter in the effective theory which can be interpreted from string theory as well. By inspection of this modified scalar potential it is possible to realise that it admits de Sitter vacua. 

The problem of understanding which is the modification, if any, of the four-dimensional effective theory \eqref{c5:eq:LKKLT} producing the scalar potential \eqref{c5:eq:VdS}, remained open until the work of \cite{Ferrara:2014kva}, in which such an effective theory has been constructed explicitly.
The crucial observation is that, as argued in the original paper \cite{Kachru:2003aw} and confirmed by other works \cite{Kallosh:2014wsa, Bergshoeff:2015jxa, Vercnocke:2016fbt, Kallosh:2016aep, Aalsma:2017ulu, Aalsma:2018pll}, the anti D3 brane is breaking supersymmetry spontaneously. As a consequence, a goldstino has to be present in the spectrum of the low energy effective theory and, after embedding it into a nilpotent chiral superfield $X$, it is sufficient to consider the slightly modified model
\begin{equation}
\label{c5:eq:KKLTnilpotent}
K = -3 \log(T+\bar T- X\bar X),\qquad W = W_0 +Ae^{- a T}-\mu^2 X.
\end{equation}
At $X^2=0$ this model reproduces exactly the scalar potential \eqref{c5:eq:VdS}. This fact triggered the interest in constrained superfields and non-linear realisations of supersymmetry as an efficient tool to study physical properties of inflation or of the current phase of the Universe \cite{Kallosh:2018wme}. 

In the effective theory \eqref{c5:eq:KKLTnilpotent} supersymmetry is spontaneously broken by F-term breaking and it is non-linearly realised. It can be questioned however if there exist alternative descriptions for the same setup, but with different features. In the following section an example in this direction is going to be presented, namely an effective description for the KKLT setup in which supersymmetry is broken by D-term, instead of F-term, but it remains linearly realised off-shell. This D-term represents a novel embedding of the Fayet--Iliopoulos D-term into supergravity and it does not require the gauging of the R-symmetry.

\section{New D-term and de Sitter vacua}
\label{c5:sec:newDterm}
In this section the standard embedding of the Fayet--Iliopoulos D-term into supergravity is reviewed first and and then a novel embedding is presented. It does not require the gauging of the R-symmetry and, for this reason, it can avoid typical restrictions \cite{Villadoro:2005yq}. It is shown eventually how this new D-term can be related to the KKLT setup discussed before.

\subsection{Standard Fayet--Iliopoulos D-term in supergravity}
D-term breaking is the spontaneous breaking of supersymmetry in which the real auxiliary field $D$ of a vector multiplet $V$ acquires a non-vanishing vacuum expectation value. The first example is due to Fayet and Iliopoulos \cite{Fayet:1974jb} in global supersymmetry, while its extension to the local case has been constructed by Freedman \cite{Freedman:1976uk}, who noticed that, in order to promote the model of Fayet and Iliopoulos to supergravity, a necessary requirement is the existence of a $U(1)$ R-symmetry gauged by the vector $v_m$ of a vector multiplet. In this subsection the model of Freedman is reviewed in the language of superspace, in order to compare it with the novel embedding of the D-term presented in the following.

In section \ref{c4:sec:gaugings} it has been noticed that the prepotentials $\mathcal{P}^{(a)}$ associated to gauged analytic isometries are defined up to real constants. This constants correspond precisely to the Fayet--Iliopoulos parameters. Consider indeed the minimal situation in which there is only one constant prepotential $\mathcal{P}=\xi$ and the associated killing vector is vanishing $\mathcal{X}=0$. The gauged supergravity Lagrangian in superspace, \eqref{eq:appB:LSUGRAG}, when only one vector multiplet is involved and no matter superfields are present, reduces to\footnote{With respect to \eqref{eq:appB:LSUGRAG}, the vector $V$ is rescaled by a factor of two, $V \to 2 V$, for convenience.}
\begin{equation}
\mathcal{L} = -3 \int d^4\theta E e^{-\frac23 \xi V}+\frac14 \left(\int d^2\Theta \, 2\mathcal{E}\, \mathcal{W}^\alpha \mathcal{W}_\alpha +c.c.\right).
\end{equation}
As it is going to be justified in a while, the gaugino $\lambda_\alpha$ can be set to zero for simplicity, so that the component form of this Lagrangian reduces to
\begin{equation}
\begin{aligned}
\label{c5:eq:FreedOff}
e^{-1}\mathcal{L} &= -\frac12 R -\frac14 v_{mn} v^{mn}+\frac12 \epsilon^{klmn}\left(\bar \psi_k \bar \sigma_l \hat{\mathcal{D}}_m \psi_n-\psi_k \sigma_l \hat{\mathcal{D}}_m \bar \psi_n\right)\\
&-\frac13 M \bar M +\frac13 b_a b^a +\frac12 D^2+\xi D+\frac23 \xi v_a b^a +\frac13 \xi^2 v_a v^a,
\end{aligned}
\end{equation}
where a modified covariant derivative on the gravitino has been defined
\begin{equation}
\label{c5:eq:hatDpsi}
\hat{\mathcal{D}}_m \psi_n = \mathcal{D}_m \psi_n +\frac i2 \xi v_m \psi_n.
\end{equation}
By taking the equations of motion of the auxiliary fields
\begin{align}
M&=0,\\
b_a &= -\xi v_a,\\
D&=-\xi,
\end{align}
it is possible to realise that supersymmetry is spontaneously broken in the vacuum by the auxiliary field $D$ of the vector superfield. Since the gaugino $\lambda_\alpha$ is the only fermion in this minimal model, it is the goldstino of the broken supersymmetry and therefore the condition $\lambda_\alpha=0$ imposed before corresponds to the unitary gauge choice, which can always be adopted consistently.
The on-shell expression of the Lagrangian in this gauge is
\begin{equation}
e^{-1}\mathcal{L} = -\frac12 R -\frac14 v_{mn} v^{mn}+\frac12 \epsilon^{klmn}\left(\bar \psi_k \bar \sigma_l \hat{\mathcal{D}}_m \psi_n-\psi_k \sigma_l \hat{\mathcal{D}}_m \bar \psi_n\right)-\frac12 \xi^2,
\end{equation}
where a constant positive definite scalar potential has been generated as a consequence of the supersymmetry breaking. Notice that, in the modified covariant derivative \eqref{c5:eq:hatDpsi}, the vector $v_m$ has the role of a connection for a $U(1)$ symmetry which is rotating the gravitino. This is a confirmation for the fact that the implementation of the Fayet--Iliopoulos D-term into supergravity requires the gauging of the R-symmetry.

The previous model can be coupled to chiral matter. The generalised superspace Lagrangian is
\begin{equation}
\mathcal{L} = -3 \int d^4\theta E e^{-\frac13 (K+2\xi V)}+\frac14 \left(\int d^2\Theta\, 2\mathcal{E} \mathcal{W}^\alpha \mathcal{W}_\alpha + c.c.\right).
\end{equation}
Considering only the bosonic sector, for simplicity, its component form can be calculated using the ingredients in appendix \ref{appB:sugraL} and it is
\begin{equation}
\begin{aligned}
e^{-1}\mathcal{L}&=\frac16 \Omega R - \Omega_{i\bar\jmath}\partial_m A^i \partial^m \bar A^{\bar\jmath}-\frac14 v_{mn}v^{mn}\\
&+\frac19 \Omega |M-3(\log \Omega)_{\bar \imath} \bar F^{\bar \imath}|^2 + \Omega \log \Omega_{i \bar\jmath} F^i \bar F^{\bar \jmath}-\frac19 \Omega b_a b^a\\
&-\bar M W- M \bar W+W_i F^i +\bar W_{\bar \imath}\bar F^{\bar\imath}+\frac12 D^2 - \frac{\xi}{3} \Omega D\\
&-\frac{i}{3}\left(\Omega_i \partial_m A^i-\Omega_{\bar \imath}\partial_m \bar A^{\bar \imath}\right)(b^m+\xi v^m)-\frac29 \Omega v_a b^a-\frac19 \xi^2 \Omega v_a v^a,
\end{aligned}
\end{equation}
where $\Omega = -3 e^{-K/3}$. The equations of motions of the auxiliary fields give
\begin{align}
M -3 (\log \Omega)_{\bar \imath}\bar F^{\bar \imath}&=9 W \Omega^{-1},\\
b_a +\xi v_a&= -\frac32 i \left(\Omega_i \partial_m A^i-\Omega_{\bar \imath}\partial_m \bar A^{\bar \imath}\right)\Omega^{-1},\\
F^i&=-e^{\frac{K}{3}}g^{i\bar\jmath}\bar D_{\bar \jmath}\bar W,\\
D &=\frac{\xi}{3}\Omega = -\xi e^{-\frac{K}{3}}
\end{align}
and plugging them back into the Lagrangian results in 
\begin{equation}
\label{c5:eq:LFI+matt}
\begin{aligned}
e^{-1}\mathcal{L} &= -\frac12 e^{-\frac{K}{3}}R +\frac12 e^{\frac{K}{3}}\partial_m\Omega \partial^m \Omega\\
&-\frac14 v_{mn} v^{mn}-g_{i\bar \jmath}e^{-\frac{K}{3}}\partial_m A^i \partial^m \bar A^{\bar\jmath}\\
&-e^{\frac{K}{3}}\left(g^{i\bar\jmath}D_i W \bar D_{\bar\jmath}\bar W-3W\bar W \right)-\frac{\xi^2}{2}e^{-\frac{2}{3}K}.
\end{aligned}
\end{equation}
In order to bring this Lagrangian to the Einstein frame, a Weyl rescaling is needed
\begin{equation}
\label{c5:eq:Weylrescaling}
e^a_m\to e^{\frac K6} e^a_m 
\end{equation}
and the final result is 
\begin{equation}
e^{-1}\mathcal{L} = -\frac12 R-\frac14 v_{mn} v^{mn} - g_{i\bar\jmath}\partial_m A^i\partial^m\bar A^{\bar\jmath}-\mathcal{V},
\end{equation}
where
\begin{equation}
\mathcal{V} = e^K\left(g^{i\bar\jmath} D_i W\bar D_{\bar\jmath}\bar W-3 W \bar W\right)+\frac{\xi^2}{2}.
\end{equation}
This is the bosonic sector of the matter-coupled supergravity Lagrangian with a Fayet--Iliopoulos D-term. It can be compared to the Lagrangian \eqref{c4:eq:gaugedLonshell} constructed in the previous chapter, in which K\"ahler invariance has been implemented in an alternative way and, for this reason, the form of the F-term contribution to the scalar potential is different, while the D-term part does not change.

Before concluding this subsection some comments on the consequences and the restrictions of the implementation of the Fayet--Iliopoulos D-term into supergravity are in order. First of all notice that, since there has to be a gauged R-symmetry in the theory, no constant term in the superpotential is allowed, as it would break such R-symmetry explicitly. Due to this, a gravitino mass term of the type $W_0 \psi_a \sigma^{ab}\psi_b$ cannot be produced. Recall also the fact that, in supergravity, F-term breaking and D-term breaking are not independent. To understand the point, define the K\"ahler invariant combination
\begin{equation}
\mathcal{G} = K + \log W + \log \bar W.
\end{equation}
Since any supergravity Lagrangian can always be expressed in terms of $\mathcal{G}$, instead of $K$ and $W$ \cite{WessBagger}, in order for the model to be gauge invariant it is sufficient to require that $\mathcal{G}$ is gauge invariant, namely
\begin{equation}
\delta_{(a)} \mathcal{G} = \delta_{(a)} K +\frac{\delta_{(a)} W}{W} + \frac{\delta_{(a)} \bar W}{\bar W}\equiv 0.
\end{equation}
Even in those situations in which $K$ is already gauge invariant, it is still possible to transform the superpotential with a constant phase
\begin{equation}
\delta_{(a)} W = i \xi_{(a)} W
\end{equation}
without spoiling the condition $\delta_{(a)} \mathcal{G}=0$. These $\xi_{(a)}$ are precisely the Fayet--Iliopoulos parameters, which are therefore related to the F-term contribution, $W_i$, by 
\begin{equation}
\frac{\mathcal{X}_{(a)}^i W_i}{W} = i \xi_{(a)}.
\end{equation}
As a consequence of the gauged R-symmetry, it is also believed that the Fayet--Iliopoulos D-term cannot arise in any consistent quantum theory of gravity, since it would require the existence of a global symmetry in the ultraviolet regime \cite{Komargodski:2009pc}.

\subsection{New D-term without gauged R-symmetry}
In this subsection a different embedding of the Fayet--Iliopoulos D-term into supergravity is presented, which avoids the aforementioned restrictions. In particular, it does not require the gauging of the R-symmetry and it is manifestly gauge invariant already at the Lagrangian level.
The idea is to construct a supersymmetric model with a vector multiplet such that, at the component level, a linear term in the auxiliary field $D$ is present. When supersymmetry is linearly realised, the proposal of Freedman seems to be the only solution compatible with gauge invariance, but if non-linear realisations of supersymmetry are considered, more possibilities can occur. An example in this direction has already been constructed in section \ref{c4:sec:newDterm}. An alternative proposal along with the same spirit and which does not involve any additional composite superfield is
\begin{equation}
\begin{aligned}
\label{c5:eq:NewDtermsuperspace}
\mathcal{L} &= -3\int d^4\theta E +\frac14\left(\int d^2\Theta\, 2\mathcal{E}\, \mathcal{W}^\alpha \mathcal{W}_\alpha + c.c.\right)\\
&-8\xi \int d^4\theta E \frac{\mathcal{W}^2\bar{\mathcal{W}}^2}{\mathcal{D}^2\mathcal{W}^2 \bar{\mathcal{D}}^2\bar{\mathcal{W}}^2} \mathcal{D}^\alpha \mathcal{W}_\alpha,
\end{aligned} 
\end{equation}
which is manifestly gauge invariant. The first line contains the kinetic terms for gravity and for the vector multiplet, while in the second line a new coupling involving the vector multiplet is present which enjoys the desired properties: it is not gauging the R-symmetry and its component expansion contains a term $\xi D$. Couplings similar to this has been considered in \cite{Cecotti:1986gb}. Setting the gaugino $\lambda_\alpha$ to zero for simplicity, the component expansion of this Lagrangian is
\begin{equation}
\begin{aligned}
e^{-1}\mathcal{L} &= -\frac12 R -\frac14 v_{mn} v^{mn}+\frac12 \epsilon^{klmn}\left(\bar \psi_k \bar \sigma_l {\mathcal{D}}_m \psi_n-\psi_k \sigma_l {\mathcal{D}}_m \bar \psi_n\right)\\
&-\frac13 M \bar M +\frac13 b_a b^a +\frac12 D^2+\xi D.
\end{aligned}
\end{equation}
By comparing it with \eqref{c5:eq:FreedOff} it is possible to observe that, in the present case, the vector $v_m$ is not appearing as a gauge connection in the covariant derivative of the gravitino, since the R-symmetry is not gauged. As a consequence, it would be possible to add consistently a constant parameter in the superpotential, which would produce then a mass term for the gravitino and a negative contribution to the scalar potential. The equations of motion of the auxiliary fields are
\begin{align}
M &= 0,\\
b_a&=0,\\
D&=-\xi
\end{align}
and supersymmetry is again spontaneously broken by the auxiliary field $D$. The choice $\lambda_\alpha=0$ corresponds then to the unitary gauge. The on-shell Lagrangian is
\begin{equation}
e^{-1}\mathcal{L} = -\frac12 R -\frac14 v_{mn} v^{mn}+\frac12 \epsilon^{klmn}\left(\bar \psi_k \bar \sigma_l {\mathcal{D}}_m \psi_n-\psi_k \sigma_l {\mathcal{D}}_m \bar \psi_n\right)-\frac{\xi^2}{2}.
\end{equation}
Being no gauged R-symmetry, this model evades the restrictions mentioned at the end of the previous subsection.

It is possible to couple the new D-term to matter in the form of chiral superfields. The Lagrangian in superspace is
\begin{equation}
\label{c5:eq:NewDterm+matter}
\begin{aligned}
\mathcal{L} &= - 3 \int d^4\theta E  e^{-\frac K3} +\left(\int d^2 \Theta \, 2 \mathcal{E} \, W + c.c.\right)\\
&+\frac14\left(\int d^2\Theta \, 2\mathcal{E}\, \mathcal{W}^\alpha \mathcal{W}_\alpha+c.c.\right) -8\xi \int d^4\theta E \frac{\mathcal{W}^2\bar{\mathcal{W}}^2}{\mathcal{D}^2\mathcal{W}^2 \bar{\mathcal{D}}\bar{\mathcal{W}}^2} \mathcal{D}^\alpha \mathcal{W}_\alpha.
\end{aligned}
\end{equation}
Considering for simplicity only the bosonic sector, the component expansion is
\begin{equation}
\begin{aligned}
e^{-1}\mathcal{L}&=\frac16 \Omega R - \Omega_{i\bar\jmath}\partial_m A^i \partial^m \bar A^{\bar\jmath}-\frac14 v_{mn}v^{mn}\\
&+\frac19 \Omega |M-3(\log \Omega)_{\bar \imath} \bar F^{\bar \imath}|^2 + \Omega \log \Omega_{i \bar\jmath} F^i \bar F^{\bar \jmath}-\frac19 \Omega b_a b^a\\
&-\bar M W- M \bar W+W_i F^i +\bar W_{\bar \imath}\bar F^{\bar\imath}+\frac12 D^2 +\xi D\\
&-\frac{i}{3}\left(\Omega_i \partial_m A^i-\Omega_{\bar \imath}\partial_m \bar A^{\bar \imath}\right)b^m,
\end{aligned}
\end{equation}
and the equations of motion of the auxiliary fields give
\begin{align}
M -3 (\log \Omega)_{\bar \imath}\bar F^{\bar \imath}&=9 W \Omega^{-1},\\
b_a &= -\frac32 i \left(\Omega_i \partial_m A^i-\Omega_{\bar \imath}\partial_m \bar A^{\bar \imath}\right)\Omega^{-1},\\
F^i&=-e^{\frac{K}{3}}g^{i\bar\jmath}\bar D_{\bar \jmath}\bar W,\\
D &=-\xi.
\end{align}
The on-shell Lagrangian is therefore
\begin{equation}
\begin{aligned}
e^{-1}\mathcal{L} &= -\frac12 e^{-\frac{K}{3}}R +\frac12 e^{\frac{K}{3}}\partial_m\Omega \partial^m \Omega\\
&-\frac14 v_{mn} v^{mn}-g_{i\bar \jmath}e^{-\frac{K}{3}}\partial_m A^i \partial^m \bar A^{\bar\jmath}\\
&-e^{\frac{K}{3}}\left(g^{i\bar\jmath}D_i W \bar D_{\bar\jmath}\bar W-3W\bar W \right)-\frac{\xi^2}{2}
\end{aligned}
\end{equation}
and notice that, with respect to the analogous formula \eqref{c5:eq:LFI+matt}, there is a difference in the term proportional to $\xi$. As a consequence of such a difference, after performing the Weyl \eqref{c5:eq:Weylrescaling} rescaling in order to go to the Einstein frame, the result is
\begin{equation}
\label{c5:eq:LnewDterm+mattonshell}
e^{-1}\mathcal{L} = -\frac12 R-\frac14 v_{mn} v^{mn} - g_{i\bar\jmath}\partial_m A^i\partial^m\bar A^{\bar\jmath}-\mathcal{V},
\end{equation}
where the scalar potential is
\begin{equation}
\mathcal{V} = e^K\left(g^{i\bar\jmath} D_i W\bar D_{\bar\jmath}\bar W-3 W \bar W\right)+\frac{\xi^2}{2}e^{\frac 23 K}.
\end{equation}
Notice that an exponential factor containing the K\"ahler potential has appeared in the D-term contribution to scalar potential of the theory. This factor is going to play a major role in the next subsection. Notice also that, in this new embedding of the Fayet--Iliopoulos model, albeit spontaneously broken, supersymmetry is linearly realised off-shell and no constraints are imposed on the superfields. For the new D-term to be well defined, finally, $\xi\neq0$ is a necessary requirement since supersymmetry has always to be broken in the vacuum. Similar constructions have been proposed in \cite{Kuzenko:2018jlz}, while applications to inflation and cosmology have been discussed in \cite{Aldabergenov:2017hvp, Antoniadis:2018cpq, Antoniadis:2018oeh, Abe:2018plc}.

\subsection{An effective description for KKLT}
In the present subsection it is shown how the new D-term can reproduce the effective theory \eqref{c5:eq:VdS}, which captures the low energy physics of the KKLT construction. In order to obtain this, it is sufficient to consider the case in which, besides the vector multiplet, only one matter chiral superfield $T$ is present, with K\"ahler potential and superpotential  given by \eqref{c5:eq:LKKLT}:
\begin{equation}
K = - 3 \log (T+\bar T),\qquad W = W_0 + A e^{-a T},
\end{equation}
which are the result of the moduli stabilisation in string theory and which lead to a model admitting stable and supersymmetric anti de Sitter vacua.
Inserting these into the Lagrangian \eqref{c5:eq:LnewDterm+mattonshell} for the matter-coupled D-term gives idirectly the scalar potential
\begin{equation}
\mathcal{V} = \mathcal{V}_{KKLT}+\frac{\xi^2}{2}\frac{1}{(T+\bar T)^2},
\end{equation}
which has precisely the form \eqref{c5:eq:VdS}. Having reproduced the same scalar potential, it is possible to argue that the new D-term is capturing the low energy effective description of the KKLT setup to construct de Sitter vacua in string theory, in the strong warping region \cite{Bandos:2016xyu}.
 With respect to the description \eqref{c5:eq:KKLTnilpotent}, supersymmetry is linearly realised off-shell, in the sense that the non-linear realisation enters only through the equations of motions of the auxiliary fields, and it is spontaneously broken by the auxiliary field $D$ of a vector multiplet. In particular no constraints are imposed on the superfields in the model.

Since two different models have been presented which describe the same physical system, it should be possibile to related one to the other. This is done in the next subsection at the superspace level.

\section{Relationship with non-linear supersymmetry}
In this section it is shown that the new D-term presented in the previous section is equivalent to the model \eqref{c5:eq:KKLTnilpotent} in which supersymmetry is broken by a chiral nilpotent goldstino superfield $X$. This is another instance of the fact that, as discussed in  \ref{c2:eq:DtermisFterm}, when supersymmetry is non-linearly realised it is possible to map a model with D-term breaking into an equivalent one with F-term breaking.
It is discussed first how the new D-term can be recast into an F-term in the pure case, when matter superfields are not present. The same procedure is then performed in the case of matter couplings and the model \eqref{c5:eq:KKLTnilpotent} is obtained.

Following the discussion presented in subsection \ref{c2:eq:DtermisFterm}, the first step is to introduce a nilpotent chiral superfield $X$ in which the goldstino is embedded and in terms of which the vector superfield can be parametrised:
\begin{equation}
V = \tilde V + \sqrt 2 \frac{X\bar X}{\mathcal{D}^2 X}+\sqrt 2 \frac{X\bar X}{\bar{\mathcal{D}}^2 \bar X},
\end{equation}
where the constraints
\begin{equation}
X^2 = 0,\qquad X\tilde{\mathcal{W}}_\alpha=0,\qquad X\bar X \mathcal{D}^\alpha \tilde{\mathcal{W}}_\alpha=0
\end{equation}
have been imposed. As a consequence, $X$ contains the goldstino and the auxiliary field which acquires a non-vanishing vacuum expectation value, while $\tilde V$ contains only the vector field as independent degree of freedom. 
By inserting this parameterisation into the second term in \eqref{c5:eq:NewDtermsuperspace}, the kinetic terms for $\tilde V$ and for $X$ are generated:
\begin{equation}
\begin{aligned}
\frac14 \left(\int d^2\Theta \, 2\mathcal{E}\, \mathcal{W}^\alpha \mathcal{W}_\alpha + c.c.\right) &= \frac14 \int d^4\theta E X\bar X\left(2+\frac{\mathcal{D}^2 X}{\bar{\mathcal{D}}^2 \bar X}+\frac{\bar{\mathcal{D}}^2 \bar X}{\mathcal{D}^2 X}\right)\\
&+\frac14 \left(\int d^2\Theta \,  2\mathcal{E}\, \tilde{\mathcal{W}}^\alpha \tilde{\mathcal{W}}_\alpha +c.c.\right),
\end{aligned}
\end{equation}
where $\tilde{\mathcal{W}}_\alpha = -1/4 (\bar{\mathcal{D}}^2-8\mathcal{R}) \mathcal{D}_\alpha \tilde{V}$.
As for the rigid supersymmetric case analysed in section  \ref{c2:eq:DtermisFterm}, this kinetic term for $X$ contains non-standard higher-derivatives interactions, which however are going to be eliminated in a while. By performing the same procedure on the new D-term coupling, a superpotential linear in $X$ is produced
\begin{equation}
-8\xi \int d^4\theta E \frac{\mathcal{W}^2\bar{\mathcal{W}}^2}{\mathcal{D}^2\mathcal{W}^2 \bar{\mathcal{D}}^2\bar{\mathcal{W}}^2} \mathcal{D}^\alpha \mathcal{W}_\alpha = -\frac{\sqrt 2}{2}\xi \left(\int d^2\Theta \, 2\mathcal{E}\, X+c.c.\right),
\end{equation}
which breaks supersymmetry spontaneously. To perform the calculation, the superspace identity
\begin{equation}
\frac{\mathcal{W}^2\bar{\mathcal{W}}^2}{\mathcal{D}^2\mathcal{W}^2 \bar{\mathcal{D}}^2\bar{\mathcal{W}}^2} = \frac{X\bar X}{\mathcal{D}^2 X \bar{\mathcal{D}}^2 \bar X}
\end{equation}
has been used. Putting all these ingredients together, the supergravity Lagrangian \eqref{c5:eq:NewDtermsuperspace} in terms of $X$ and $\tilde V$ becomes
\begin{equation}
\begin{aligned}
\mathcal{L} &= -3 \int d^4\theta E +\frac14 \left(\int d^2\Theta \,  2\mathcal{E}\, \tilde{\mathcal{W}}^\alpha \tilde{\mathcal{W}}_\alpha +c.c.\right)\\
&+\frac14 \int d^4\theta E X\bar X\left(2+\frac{\mathcal{D}^2 X}{\bar{\mathcal{D}^2}\bar X}+\frac{\bar{\mathcal{D}^2}\bar X}{\mathcal{D}^2 X}\right)\\
&-\frac{\sqrt 2}{2}\xi \left(\int d^2\Theta \, 2\mathcal{E}\, X+c.c.\right).
\end{aligned}
\end{equation}
Since in this Lagrangian there is a term linear in $X$ in the superpotential and since $\xi$ is real, it is possible to apply the discussion of subsection \eqref{c2:sec:XZA} and to restore a contribution of the type  \eqref{ZZA2}, which cancels the undesired higher derivatives. As a consequence, the model becomes finally
\begin{equation}
\begin{aligned}
\mathcal{L} &= -3 \int d^4\theta E +\frac14 \left(\int d^2\Theta \,  2\mathcal{E}\, \tilde{\mathcal{W}}^\alpha \tilde{\mathcal{W}}_\alpha +c.c.\right)\\
&+\int d^4\theta E X\bar X-\frac{\sqrt 2}{2}\xi \left(\int d^2\Theta \, 2\mathcal{E}\, X+c.c.\right)\\
&=-3 \int d^4\theta E e^{-\frac{X\bar X}{3}}+\frac14 \left(\int d^2\Theta \,  2\mathcal{E}\, \tilde{\mathcal{W}}^\alpha \tilde{\mathcal{W}}_\alpha +c.c.\right)\\
&-\frac{\sqrt 2}{2}\xi \left(\int d^2\Theta \, 2\mathcal{E}\, X+c.c.\right).
\end{aligned}
\end{equation}
This is a Lagrangian for $X$ and $\tilde V$, in which supersymmetry is broken by F-term and which is on-shell equivalent to the original system \eqref{c5:eq:NewDtermsuperspace}. The same procedure can be performed also when matter couplings to a chiral model $\hat{K}(\Phi, \bar \Phi)$, $\hat{W}(\Phi)$ are introduced. In this case, following the same steps as before, the Lagrangian \eqref{c5:eq:NewDterm+matter} becomes
\begin{equation}
\begin{aligned}
\mathcal{L} &= -3 \int d^4\theta E e^{-\frac{\hat{K}}{3}} +\left(\int d^2\Theta \, 2\mathcal{E}\, \hat{W}\right)\\
 &+\frac14 \left(\int d^2\Theta \,  2\mathcal{E}\, \tilde{\mathcal{W}}^\alpha \tilde{\mathcal{W}}_\alpha +c.c.\right)-\frac{\sqrt 2}{2}\xi \left(\int d^2\Theta\, 2\mathcal{E}\, X + c.c.\right).
\end{aligned}
\end{equation}
By redefining the K\"ahler potential and the superpotential according to
\begin{equation}
K = -3\log\left(e^{-\frac{\hat K}{3}}-\frac13 X\bar X\right),\qquad W = -\frac{\sqrt 2}{2}\xi X +\hat{W},
\end{equation}
the model assumes the standard form 
\begin{equation}
\begin{aligned}
\mathcal{L} &= -3 \int d^4\theta E e^{-\frac{K}{3}}+\left(\int d^2\Theta \, 2 \mathcal{E}\, W+c.c.\right)\\
&+\frac14 \left(\int d^2\Theta \,  2\mathcal{E}\, \tilde{\mathcal{W}}^\alpha \tilde{\mathcal{W}}_\alpha +c.c.\right).
\end{aligned}
\end{equation}
This Lagrangian is on-shell equivalent to \eqref{c5:eq:NewDterm+matter}. Considering then the simple case in which only one chiral matter superfield $T$ is present with
\begin{equation}
\hat{K} = - 3 \log(T+\bar T),\qquad \hat W =W_0 + A e^{-a T},
\end{equation}
the KKLT effective description \eqref{c5:eq:KKLTnilpotent} is finally recovered
\begin{equation}
K = -3\log\left(T+\bar T - \frac13 X\bar X\right),\qquad W=W_0 + A e^{-a T}-\frac{\sqrt 2}{2}\xi X.
\end{equation}
This concludes the proof that the new D-term is capturing the low energy effective theory of the KKLT setup and that it is equivalent to the description of the same setup in terms of constrained superfields.

\section{Discussion: End of the second part}
In the second part of this thesis four-dimensional models with spontaneously broken local supersymmetry have been studied. The construction of supergravity Lagrangians with a nilpotent goldstino multiplet has been reviewed and matter coup
lings have been considered. Non-linear realisations of supersymmetry have also been used to construct new models which evade some of the restrictions which are present in the linear regime. The class of supergravity Lagrangian with alternative K\"ahler invariance and the new embedding of the Fayet--Iliopoulos D-term without the gauging of the R-symmetry are examples in this direction.

It is possible to conclude that, while on one hand non-linear realisations of supersymmetry represent a powerful tool in the construction of effective theories, which can describe several physical phenomena even at energies accessible with the present experiments, on the other hand they can give precious hints on how a certain theory in the ultraviolet regime is constraining the physics in the low energy.

\begin{appendices}

\chapter{The Samuel--Wess formalism}
\label{appASW}

In this appendix the goldstino superfields $\Lambda_\alpha$ and $\Gamma_\alpha$, introduced in \cite{Ivanov:1978mx, Samuel:1982uh} and employed in chapter \ref{c1:susybreakingsec}, are reviewed together with their properties. The discussion starts from rigid supersymmetry and then some of the results are generalised to supergravity.

\section{Supersymmetry}
Consider the Volkov--Akulov goldstino $\lambda_\alpha$, with non-linear and non-homogeneous supersymmetry transformation
\begin{equation}
\label{appA:eq:deltalambda}
\delta_\epsilon \lambda_\alpha = f \epsilon_\alpha - \frac if (\lambda \sigma^m \bar \epsilon- \epsilon \sigma^m \bar \lambda)\partial_m \lambda_\alpha.
\end{equation}
This transformation mixes $\lambda_\alpha$ with its complex conjugate. It is possible however to introduce chiral coordinates
\begin{equation}
y^m = x^m - \frac{i}{f^2}\lambda (y) \sigma^m \bar \lambda(y)
\end{equation}
in terms of which a new goldstino field $\gamma_\alpha$ can be defined
\begin{equation}
\lambda_\alpha(y) = \gamma_\alpha(x).
\end{equation}
The explicit form of such a relation can be determined by Taylor expanding and by using the anticommuting nature of the fields which are involved. The result is
\begin{equation}
\begin{aligned}
\gamma_\alpha &= \lambda_\alpha-\frac{i}{f^2}v^m \partial_m \lambda_\alpha-\frac{1}{2f^4} v^m v^n \partial_m \partial_n \lambda_\alpha-\frac{1}{f^4} v^m (\partial_m v^n)\partial_n \lambda_\alpha\\
& + \frac{i}{f^6} v^l (\partial_l v^m )(\partial_m v^n)\partial_n \lambda_\alpha + \frac{i}{2f^6} v^l v^m (\partial_l \partial_m v^n)\partial_n \lambda_\alpha,
\end{aligned}
\end{equation}
where $v^m =\frac 1f (\lambda \sigma^m \bar \epsilon- \epsilon \sigma^m \bar \lambda)$.
In contrast to that of $\lambda_\alpha$, the supersymmetry transformation of $\gamma_\alpha$ is chiral
\begin{equation}
\label{appA:eq:deltagamma}
\delta_\epsilon \gamma_\alpha =f \epsilon_\alpha -\frac{2i}{f} \gamma \sigma^m \bar \epsilon \partial_m \gamma_\alpha.
\end{equation}

These two goldstino fields can be embedded into superspace as lowest components of two spinor superfields $\Lambda_\alpha$ and $\Gamma_\alpha$
\begin{equation}
\label{appA:eq:lowcomp}
\Lambda_\alpha|= \lambda_\alpha, \qquad \Gamma_\alpha|=\gamma_\alpha,
\end{equation}
which satisfy the defining properties
\begin{equation}
\begin{aligned}
\label{appA:eq:DLambda}
D_\alpha \Lambda_\beta & = f\,\epsilon_{\beta\alpha}+\frac{i}{f}\,\sigma^m_{\alpha\dot\beta}\bar\Lambda^{\dot\beta}\partial_m\Lambda_\beta,\\
\bar D_{\dot \alpha} \Lambda_\beta & = -\frac{i}{f}\,\Lambda^\rho \sigma^m_{\rho\dot\alpha}\partial_m\Lambda_\beta,
\end{aligned}
\end{equation}
and 
\begin{equation}
\begin{aligned}
\label{appA:eq:DGamma}
D_\alpha \Gamma_\beta & = f\,\epsilon_{\beta\alpha},\\
\bar D_{\dot \alpha} \Gamma_\beta & = -\frac{2i}{f}\,\Gamma^\rho \sigma^m_{\rho\dot\alpha}\partial_m\Gamma_\beta.
\end{aligned}
\end{equation}
These are the superspace generalisation of the supersymmetry transformations \eqref{appA:eq:deltalambda} and \eqref{appA:eq:deltagamma} and indeed they close the supersymmetry algebra
\begin{equation}
\{D_\alpha, D_\beta\}=0, \qquad \{D_\alpha, \bar D_{\dot \beta} \}=-2i \sigma^m_{\alpha \dot \beta} \partial_m.
\end{equation}
The relations \eqref{appA:eq:lowcomp}, \eqref{appA:eq:DLambda} and \eqref{appA:eq:DGamma} contain all the necessary information to calculate all the higher components of $\Lambda_\alpha$ and $\Gamma_\alpha$. To pass from one superfield to the other it is possible to use the superspace redefinition
\begin{equation}
\Gamma_\alpha = -2f \frac{D_\alpha \bar D^2 (\Lambda^2 \bar \Lambda^2)}{D^2 \bar D^2 \Lambda^2 \bar\Lambda^2}
\end{equation}
and a supersymmetric Lagrangian for them is of the type
\begin{equation}
\mathcal{L} = -\frac{1}{f^2} \int d^4 \theta \Lambda^2 \bar \Lambda^2 =- \frac{1}{f^2}\int d^4 \theta \Gamma^2 \bar \Gamma^2.
\end{equation}
Some useful superspace relations are
\begin{align}
\Lambda^2 \bar \Lambda^2 &= \Gamma^2 \bar \Gamma^2,\\
D^2 \Lambda^2 &= -4 f^2 + 4i \partial_m \Lambda \sigma^m \bar \Lambda +\frac{2}{f^2}\bar \Lambda^2 \partial^m \Lambda \partial_m \Lambda\\
\nonumber
&-\frac{4}{f^2} \Lambda \partial_m \Lambda \bar \Lambda \partial^m \bar \Lambda -\frac{2}{f^2} \bar \Lambda^2 \Lambda \partial^2 \Lambda,\\
\bar D^2 (\Lambda^2 \bar \Lambda^2) &= - 4 \Lambda^2 f^2 + 4i \Lambda^2 \partial_m \Lambda \sigma^m \bar \Lambda + \frac{4}{f^2}\bar \Lambda^2 \Lambda^2 \partial_m \Lambda \sigma^{mn}\partial_n \Lambda.
\end{align}

\section{Supergravity}
In curved superspace the superfield $\Gamma_\alpha$, such that $\Gamma_\alpha|=\gamma_\alpha$, is defined by
\begin{equation}
\begin{aligned}
D_\beta \Gamma_\alpha &= \epsilon_{\alpha \beta}\left(f-\frac 2f \bar{\mathcal{R}} \,\Gamma^2\right),\\
\bar D^{\dot \beta} \Gamma^\alpha &= \frac{2i}{f} (\bar \sigma^b \Gamma)^{\dot \beta}\mathcal{D}_b \Gamma^\alpha+\frac{1}{2f} \Gamma^2 B^{\dot \beta \alpha},
\end{aligned}
\end{equation}
where $\mathcal{D}_a \Gamma_\alpha = e^m_b \mathcal{D}_m \Gamma_\alpha-\frac12 \psi_a^\beta \mathcal{D}_\beta \Gamma_\alpha -\frac12 \bar \psi_{b \dot \beta}\bar{\mathcal{D}}^{\dot \beta}\Gamma_\alpha$ is the supercovariant derivative in superspace. The superfield $\Gamma_\alpha$ satisfies the supersymmetry algebra
\begin{equation}
\{ \mathcal{D}_A, \mathcal{D}_B \} \Gamma^\alpha = \Gamma^\beta {R_{AB\beta}}^\alpha - T_{AB}^C \mathcal{D}_C \Gamma^\alpha.
\end{equation}
Some useful superspace relations are
\begin{align}
\mathcal{D}_\alpha \mathcal{D}_\beta \Gamma_\gamma &= -4 \bar{\mathcal{R}} \epsilon_{\gamma \beta}\Gamma_\alpha,\\
\bar{\mathcal{D}}_{\dot \alpha}\mathcal{D}_\beta\Gamma_\gamma&=-\Lambda^2 \bar{\mathcal{R}} \epsilon_{\gamma \beta}\left(\frac{4i}{f^2}(\mathcal{D}_a \Lambda \sigma^a)_{\dot \alpha}+\frac 2f \mathcal{D}^\delta B_{\delta \dot \alpha}\right),\\
\mathcal{D}_\beta\bar{\mathcal{D}}_{\dot \alpha}\Gamma_\gamma &=-\bar{\mathcal{D}}_{\dot \alpha}\mathcal{D}_\beta\Gamma_\gamma-2i \sigma^b_{\beta \dot \alpha}\mathcal{D}_b \Gamma_\gamma+\epsilon_{\gamma \beta}\Gamma^\delta B_{\delta \dot \alpha}- \Gamma_\beta B_{\delta \dot \alpha},\\
\bar{\mathcal{D}}_{\dot\beta}\bar{\mathcal{D}}_{\dot\alpha}\Gamma_\gamma &= \frac12 \epsilon_{\dot \alpha \dot \beta}\bar{\mathcal{D}}^2 \Gamma_\gamma,\\
\nonumber
\bar{\mathcal{D}}^2 \Gamma_\gamma &= \frac{8}{f^2}(\Gamma \sigma^a \bar \sigma^b \mathcal{D}_a \Gamma)\mathcal{D}_b \Gamma_\gamma- 8\mathcal{R}\, \Gamma_\gamma\\
&+\frac{2}{f^2}\Gamma^2\bigg(-2 \mathcal{D}^a \mathcal{D}_a \Gamma_\gamma-6i G^b \mathcal{D}_b \Gamma_\gamma\\
\nonumber
&+ f \bar{\mathcal{D}}_{\dot \delta}B^{\dot \delta \delta}\epsilon_{\delta \gamma}+ i (\mathcal{D}_b \Gamma \sigma^b B)_\gamma\bigg),\\
\mathcal{D}_\alpha \Gamma^2 &= 2 f \Gamma_\alpha,\\
\bar{\mathcal{D}}_{\dot\alpha}\Gamma^2 &= \frac{2i}{f}\Gamma^2 \mathcal{D}_b \Gamma^\alpha \sigma^b_{\alpha \dot \alpha},\\
\mathcal{D}^2 \Gamma^2 &= -4f^2+  8 \bar R \Gamma^2,\\
\bar{\mathcal{D}}^2 \Gamma^2&=16 \Gamma^2 \left(\frac{1}{f^2}\mathcal{D}_a \Gamma \sigma^{ab}\mathcal{D}_b\Gamma-\mathcal{R}\right).
\end{align}

\chapter{Supergravity Lagrangian}
\label{appB:sugraL}
In this appendix the main steps which are needed in the construction of supergravity Lagrangians from superspace to components are summarised.
The case of standard linearly realised supergravity is presented first and then some ingredients are given, which have been used to build the alternative models of chapter \ref{c4:altsugra}.

\section{Standard supergravity}
Given a generic real function $\Omega(\Phi,\bar \Phi)$ of a set of chiral superfields $\Phi^i$, the superfield
\begin{equation}
\Xi=\left(\bar{\mathcal{D}}^2-8 \mathcal{R}\right) \Omega
\end{equation}
is chiral and it can be expanded in curved superspace as
\begin{equation}
\begin{aligned}
\Xi &= \left(\bar{\mathcal{D}}^2-8\mathcal{R}\right)\Omega\big|
+\Theta^\alpha \mathcal{D}_\alpha\left(\bar{\mathcal{D}}^2-8\mathcal{R}\right)\Omega\big|-\frac14 \Theta^2 \mathcal{D}^2 \left(\bar{\mathcal{D}}^2-8\mathcal{R}\right)\Omega\big|.
\end{aligned}
\end{equation}
Its projections are
\begin{align}
\left(\bar{\mathcal{D}}^2-8\mathcal{R}\right)\Omega\big|  &= \frac43 M \Omega  -4 \Omega_{\bar \imath }\bar F^{\bar \imath}+ 2\, \Omega_{\bar \imath \bar \jmath}\, \bar \chi^{\bar \imath} \bar \chi^{\bar \jmath},
\end{align}
\begin{align}
\nonumber
\mathcal{D}_\alpha\left(\bar{\mathcal{D}}^2-8\mathcal{R}\right)\Omega \big| &= 
\frac 43 i \, \Omega \psi_{\alpha a} b^a
-\frac43 i \,\Omega\, \sigma^a_{\alpha\dot\alpha}\bar\psi_a^{\dot\alpha}M
+\frac83 \,\Omega\, ({\sigma^{ab}})_\alpha^\beta \psi_{ab\,\beta} \\
\nonumber
&+\frac43 \sqrt 2 \Omega_i M \chi^i_\alpha -4\sqrt 2 i \,\Omega_{\bar \imath}\, \sigma^c_{\alpha\dot\alpha} e^m_c \mathcal{D}_m\bar \chi^{\dot\alpha\bar \imath}\\
& + 4i \, \Omega_{\bar \imath} \, \sigma^a_{\alpha\dot\alpha}\bar \psi_a^{\dot\alpha}\bar F^{\bar \imath}
-4 \Omega_{\bar \imath}\, \sigma^b_{\alpha\dot\alpha}\bar \sigma^{c\, \dot\alpha\beta}\psi_{b\,\beta}\,e^m_c \mathcal{D}_m \bar A^{\bar \imath}\\
\nonumber
&+2\sqrt 2 \, \Omega_{\bar \imath}\, \sigma^b_{\alpha\dot\alpha}\bar \sigma^{c\, \dot\alpha\beta}\psi_{b\,\beta}(\bar \psi_c \bar\chi^{\bar \imath})
-\frac23 \sqrt 2 \,\Omega_{\bar \imath }\sigma^a_{\alpha\dot\alpha}\bar\chi^{\dot\alpha\,\bar \imath}\, b_a\\
\nonumber
&+4i \, \Omega_{\bar \imath \bar \jmath}\, \sigma^b_{\alpha \dot \alpha}\bar \chi^{\dot\alpha\, \bar \imath}(\bar \psi_b \bar\chi^{\bar \jmath})
-4 \sqrt 2 \,\Omega_{i\bar \jmath}\chi^{i}_\alpha \bar F^{\bar \jmath}\\
\nonumber
&-4\sqrt 2 i\, \Omega_{\bar \imath \bar \jmath} \sigma^a_{\alpha\dot\alpha}\bar \chi^{\dot\alpha\, \bar \imath}\, e^m_a \mathcal{D}_m \bar A^{\bar \jmath}+2\sqrt 2 \Omega_{i \bar \imath \bar \jmath}\,\chi_\alpha^i (\bar \chi^{\bar \imath}\bar \chi^{\bar \jmath}),
\end{align}
\begin{align}
\nonumber
\mathcal{D}^2&\left(\bar{\mathcal{D}}^2-8\mathcal{R}\right)\Omega\big|  = \frac83 \Omega R-\frac{16}{3}i\, \Omega\, \bar\psi^m \bar \sigma^n\psi_{mn}-\frac23 \Omega\, \epsilon^{klmn}\left(\bar \psi_k \bar \sigma_l\psi_{mn}+\psi_k\sigma_l\bar\psi_{mn}\right)\\
\nonumber
&+\frac{16}{3}i \Omega \,e^m_a \mathcal{D}_m b^a-\frac{32}{9}\Omega\, M \bar M-\frac{16}{9}\Omega\, b_a b^a-\frac83 \Omega\, (\bar\psi^a\bar\psi_a)M\\
\nonumber
&+\frac83 \, \Omega \, \psi_m \sigma^m \bar \psi_n b^n+\frac{16}{3} \sqrt 2 \Omega_{i}\, \chi^i \sigma^{ab}\psi_{ab}+\frac83 \sqrt 2i \Omega_i \chi^i\psi_a \, b^a\\
\nonumber
&-\frac83 \sqrt 2i \Omega_i \, \chi^i \sigma^a \bar \psi_a M +\frac83 M\, \Omega_{ij}\, \chi^i \chi^j-\frac{16}{3}M \Omega_i F^i\\
\nonumber
&-\frac43 \sqrt 2 i\, \Omega_{\bar \imath}\bar \psi_a \bar \sigma^a \sigma^c \bar\chi^{\bar \imath} b_c+\frac83 \sqrt 2 i \, \Omega_{\bar \imath}\bar \psi_a \bar\chi^i b^a+\frac83 \sqrt 2 \Omega_{\bar \imath}\, \bar \psi_{mn}\bar \sigma^{mn}\bar\chi^{\bar \imath}\\
\nonumber
&+\frac{32}{3}\bar M \Omega_{\bar \imath }\bar F^{\bar \imath}+8 \,\Omega_{\bar \imath} \bar F^{\bar \imath}(\bar \psi_a \bar \psi^a)-4 \sqrt 2 i\, \Omega_{\bar \imath}\, \bar \psi^c \bar \sigma^a \psi_c (\bar \psi_a\bar \chi^{\bar \imath})\\
&+8i \,\Omega_{\bar \imath}\,\bar \psi_c \bar\sigma^a \psi^c \, e^m_a \mathcal{D}_m \bar A^{\bar \imath}-8\sqrt 2\,\Omega_{\bar \imath}\, \bar\psi^a e^m_a \mathcal{D}_m \bar \chi^{\bar \imath}-\frac{16}{3}\sqrt 2 i\, \Omega_{\bar \imath} b^a (\bar \psi_a \bar \chi^{\bar \imath})\\
\nonumber
&+\frac{32}{3}i \Omega_{\bar \imath}\, b^a e^m_a \mathcal{D}_m \bar A^{\bar \imath}-8 \sqrt 2 \Omega_{\bar \imath}\, e^m_a \mathcal{D}_m (\bar\psi^a \bar\chi^{\bar \imath}) + 16 \Omega_{\bar \imath}\, e^m_a \mathcal{D}_m\left(e_n^a \mathcal{D}^n \bar A^{\bar \imath}\right)\\
\nonumber
&-\frac83 \Omega_{i\bar \jmath}\, \chi^i \sigma^a \bar \chi^{\bar \jmath}b_a+8 \sqrt 2 i \Omega_{i\bar \jmath}\chi^i \sigma^c \bar \psi_c \bar F^{\bar \jmath}+8 \Omega_{i \bar \jmath}\, \chi^i \sigma^c \bar \sigma^ a\psi_c (\bar \psi_a \chi^{\bar \jmath})\\
\nonumber
&-8 \sqrt 2 \Omega_{i \bar \jmath}\, \chi^i \sigma^c \bar \sigma^a \psi_c e^m_a \mathcal{D}_m \bar A^{\bar \jmath}-16 i \Omega_{i \bar \jmath}\, \chi^i \sigma^c e^m_c \mathcal{D}_m \bar \chi^{\bar \jmath}+ 16 \Omega_{i \bar \jmath } \,F^i \bar F^{\bar \jmath}\\
\nonumber
&-8 \Omega_{i j \bar \imath}\,(\chi^i \chi^j)\bar F^{\bar \imath}+ 8 \Omega_{\bar \imath \bar \jmath}\, (\bar \psi_a \chi^{\bar \imath})(\bar \psi^a \chi^{\bar \jmath})-16 \sqrt 2 \Omega_{\bar \imath \bar \jmath}\, e^m_a \mathcal{D}_m \bar A^{\bar \imath}(\bar \psi^a \chi^{\bar \jmath})\\
\nonumber
&+16 \Omega_{\bar \imath \bar \jmath} \,g^{mn} \mathcal{D}_m \bar A^{\bar \imath}\mathcal{D}_n \bar A^{\bar \jmath}-\frac{16}{3}\Omega_{\bar \imath \bar \jmath} \, \bar M (\bar \chi^{\bar \imath}\bar \chi^{\bar \jmath}) + 8 \sqrt 2 i \, \Omega_{i \bar \imath \bar \jmath}\, \chi^i \sigma^a \bar \chi^{\bar \imath}(\bar \psi_a \chi^{\bar \jmath})\\
\nonumber
&-16 i \Omega_{i \bar \imath \bar \jmath}\, \chi^i \sigma^a \bar \chi^{\bar \jmath}\, e^m_a \mathcal{D}_m \bar A^{\bar \jmath}
-8 \Omega_{i \bar \imath \bar \jmath} F^i (\bar \chi^{\bar \imath}\bar\chi^{\bar \jmath})+4 \Omega_{ij\bar \imath\bar \jmath}(\chi^i \chi^j)(\bar \chi^{\bar \imath}\bar \chi^{\bar \jmath}).
\end{align}

A generic supergravity Lagrangian with chiral matter couplings can be written in curved superspace as 
\begin{equation}
\label{eq:appB:GENSUGRAL}
\mathcal{L}=\int d^2 \Theta \, 2 \mathcal{E} \left[-\frac18 \left(\bar{\mathcal{D}}^2-8\mathcal{R}\right)\Omega(\Phi, \bar \Phi)+ W(\Phi)\right]+c.c.,
\end{equation}
where $\Omega$ is going to be related to the K\"ahler potential of the chiral model, while $W$ is the superpotential. The component form of this Lagrangian is 
\begin{equation}
\label{appB:eq:gensugraL}
\mathcal{L}= \mathcal{L}_\text{kin}+\mathcal{L}_\text{aux}+\mathcal{L}_\text{2f}+\mathcal{L}_\text{4f},
\end{equation}
where
\begin{align}
\nonumber
e^{-1}\mathcal{L}_\text{kin}&=\frac16 \Omega R-\frac{1}{12}\Omega \epsilon^{klmn}\left(\bar \psi_k \bar \sigma_l \psi_{mn}-\psi_k\sigma_l \bar\psi_{mn}\right)\\
\nonumber
&- \Omega_{i \bar \jmath}\partial_m A^i \partial^m \bar A^{\bar \jmath}-\frac i2 \Omega_{i\bar \jmath}\left(\chi^i \sigma^m \mathcal{D}_m \bar \chi^{\bar \jmath}+\bar \chi^{\bar \jmath} \bar \sigma^m \mathcal{D}_m \chi^i\right)\\
\nonumber
&+\frac14 \epsilon^{klmn}\left(\Omega_i\partial_k A^i-\Omega_{\bar \imath} \partial_k \bar A^{\bar \imath}\right)\psi_l\sigma_m\bar\psi_n\\
&+\frac i2 \chi^i \sigma^m \bar \chi^{\bar \imath}\left(\Omega_{i j \bar \imath }\partial_m A^{j}-\Omega_{i \bar \imath \bar \jmath} \partial_m \bar A^{\bar \jmath}\right)\\
\nonumber
&-\frac{\sqrt 2}{2}\Omega_{i \bar \jmath}\left(\bar \psi_m \bar \sigma^n \sigma^m \bar \chi^{\bar \jmath}\partial_n A^i+\psi_m \sigma^n \bar \sigma^m \chi^i\partial_n \bar A^{\bar \jmath}\right)\\
\nonumber
&+\frac{\sqrt 2}{3}\left(\Omega_i \chi^i \sigma^{mn}\psi_{mn}+\Omega_{\bar \imath}\bar \chi^{\bar \imath}\bar \sigma^{mn}\bar \psi_{mn}\right),
\end{align}

\begin{align}
\nonumber
e^{-1} \mathcal{L}_\text{aux}&=\frac 19 \Omega |M - 3 (\log \Omega)_{\bar \imath} \bar F^{\bar \imath}|^2+\Omega (\log \Omega)_{i \bar \jmath} F^i \bar F^{\bar \jmath}\\
\nonumber
&- \frac19 \Omega b_a b^a-\frac i3 \left(\Omega_i \partial_m A^i - \Omega_{\bar \jmath}\partial_m \bar A^{\bar \jmath}\right)b^m\\
\nonumber
&-\frac 16 \Omega_{i \bar \jmath}\chi^i \sigma^a \bar \chi^{\bar \jmath}b_a + \frac i6 \sqrt 2\left(\Omega_i \psi_a \chi^i - \Omega_{\bar \imath}\bar \psi_a \bar \chi^{\bar \imath}\right)b^a\\
&+\frac 16 M \Omega_{ij}\chi^i \chi^j+\frac 16 \bar M \Omega_{\bar \imath \bar \jmath}\bar \chi^{\bar \imath} \bar \chi^{\bar \jmath}\\
\nonumber
&-\frac12 \Omega_{ij \bar \imath}\chi^i \chi^j\,\bar F^{\bar \imath}-\frac12 \Omega_{i \bar \imath \bar \jmath}\bar \chi^{\bar \imath}\bar \chi^{\bar \jmath}\,F^i\\
\nonumber
&- \bar M W - M \bar W + W_i F^i + \bar W_{\bar \imath} \bar F^{\bar \imath},
\end{align}

\begin{align}
\nonumber
e^{-1} \mathcal{L}_\text{2f} &= -\frac12 W_{ij} \chi^i \chi^j-\frac12 \bar W_{\bar \imath \bar \jmath}\bar \chi^{\bar \imath} \bar \chi^{\bar \jmath}\\
&- W \bar \psi_a \bar \sigma^{ab} \bar \psi_b- \bar W \psi_a \sigma^{ab} \psi_b\\
\nonumber
&-\frac{\sqrt 2}{2}i W_i \chi^i \sigma^a \bar \psi_a-\frac{\sqrt 2}{2}i \bar W_{\bar \imath}\bar \chi^{\bar \imath}\bar \sigma^a \psi_a,
\end{align}

\begin{align}
\nonumber
e^{-1} \mathcal{L}_{4f} &= \frac 14 \Omega_{ij \bar \imath \bar \jmath}(\chi^i \chi^j)(\bar \chi^{\bar \imath}\bar \chi^{\bar \jmath})\\
\nonumber
&-\frac14 \Omega_{ij}(\psi_a\sigma^{ab}\psi_b)(\chi^i \chi^j)-\frac14 \Omega_{\bar \imath \bar \jmath}(\bar \psi_a \bar \sigma^{ab} \bar \psi_b)(\bar \chi^{\bar \imath}\bar \chi^{\bar \jmath})\\
&+\frac14 \Omega_{i \bar \jmath}\left[(\psi_m \sigma_n \bar \psi^m)(\chi^i \sigma^n \bar \chi^{\bar \jmath})+i \epsilon^{klmn}(\psi_k \sigma_l \bar \psi_m)(\chi^i \sigma_n \bar \chi^{\bar \jmath})\right]\\
\nonumber
&+\frac{\sqrt 2}{8} \epsilon^{klmn}\psi_k\sigma_l\bar \psi_m\left(\Omega_i \psi_n \chi^i-\Omega_{\bar \imath}\bar\psi_n\bar\chi^{\bar \imath}\right)\\
\nonumber
&-\frac{\sqrt 2}{4}i \left(\Omega_i \psi_m\sigma^{mn}\chi^i+\Omega_{\bar \imath}\bar \psi_m \bar \sigma^{mn}\bar\chi^{\bar \imath}\right)(\psi_k\sigma^k \bar\psi_n-\psi_n\sigma^k\bar\psi_k)
\end{align}
In these expressions, $\Omega$ and its derivatives have to be interpreted as projections to their lowest component.

In order to obtain the on-shell form of the previous Lagrangian, the auxiliary fields have to be integrated out. Their equations of motion can be calculated at the component level and they are 
\begin{align}
M -3 (\log \Omega)_{\bar \imath} \bar F^{\bar \imath}&= 9 W \Omega^{-1}+\frac32 \Omega^{-1}\Omega_{\bar \imath \bar \jmath}\bar \chi^{\bar \imath} \bar \chi^{\bar \jmath},\\
\label{appB:eq:eomb}
b_a &= -\frac32 i \left(\Omega_i \partial_a A^i - \Omega_{\bar \jmath}\partial_a \bar A^{\bar \jmath}\right) \Omega^{-1}-\frac 34 \Omega_{i \bar \jmath}\chi^i \sigma_a \bar \chi^{\bar \jmath}\Omega^{-1}\\
\nonumber
&+\frac 34 i \sqrt 2 \left(\Omega_i \psi_a \chi^i - \Omega_{\bar \imath}\bar \psi_a \bar \chi^{\bar \imath}\right) \Omega^{-1},\\
\Omega \left(\log \Omega\right)_{i \bar \jmath} F^i &= 3 (\log \Omega)_{\bar \jmath}\bar W- \bar W_{\bar \jmath}\\
\nonumber
&-\frac12 (\log \Omega)_{\bar \jmath}\Omega_{ij} \chi^i \chi^j+\frac12 \Omega_{ij \bar \jmath}\chi^i \chi^j.
\end{align}
It is alternatively possible to calculate them at the superspace level, obtaining
\cite{Farakos:2018sgq}
\begin{align}
W -\frac{1}{12}\left(\bar{\mathcal{D}}^2-8 \mathcal{R}\right)\Omega&=0\\
G_{\alpha \dot \alpha}-\frac14 \Omega^{-1}\left([\mathcal{D}_\alpha, \bar{\mathcal{D}}_{\dot \alpha}]\Omega-3 \Omega_{i \bar \jmath}\mathcal{D}_\alpha \Phi^i \bar{\mathcal{D}}_{\dot \alpha}\bar \Phi^{\bar \jmath}\right)&=0\\
W_i -\frac14 \left(\bar{\mathcal{D}}-8\mathcal{R}\right)\Omega_i&=0
\end{align}
Notice indeed that, by projecting to the lowest component, the latter reduce to the former. After that these expressions for the auxiliary fields are substituted back into the Lagrangian, new terms with four fermions are generated. The next step is to restore the canonical normalisation for the physical fields, namely to go to the so called Einstein frame. The Ricci scalar can be canonically normalised with a Weyl rescaling of the vielbein of the type
\begin{equation}
e^a_m \rightarrow e^a_m e^{\frac{K}{6}},
\end{equation} 
where $K(\Phi, \bar \Phi)$ is related to $\Omega$ by
\begin{equation}
\Omega = - 3 \,e ^{-\frac{K}{3}}
\end{equation}
and it is going to be the K\"ahler potential, which encodes the geometry of the scalar manifold. To restore the correct matter field normalisations then, a field-dependent redefinition of the spin-$\frac12$ fermions is needed
\begin{equation}
\chi_i \rightarrow e^{-\frac{K}{12}}\chi_i, \qquad \psi_m \rightarrow e^{\frac{K}{12}}\psi_m
\end{equation} 
and eventually a shift of the gravitino
\begin{equation}
\psi_m \rightarrow \psi_m + \frac{i}{3 \sqrt 2} \sigma_m \bar \chi^{\bar \imath}K_{\bar \imath} .
\end{equation}
After all these transformations are performed, the following on-shell supergravity Lagrangian is obtained
\begin{equation}
\begin{aligned}
e^{-1}\mathcal{L} &= -\frac12 {R}- g_{i\bar \jmath}\partial_m A^i \partial^m \bar A^{\bar \jmath}\\
&-i g_{i \bar \jmath}\bar\chi^{\bar \jmath}\bar\sigma^m \mathcal{D}_m \chi^i + \epsilon^{klmn}\bar\psi_k \bar\sigma_l \mathcal{D}_m\bar\psi_n\\
&-\frac{\sqrt 2}{2}g_{i \bar \jmath}\partial_n \bar A^{\bar \jmath} \chi^i \sigma^m \bar \sigma^n \psi_m - \frac{\sqrt 2}{2}g_{i \bar \jmath}\partial_n A^{i} \bar\chi^{\bar \jmath} \bar\sigma^m \sigma^n \bar\psi_m\\
&+\frac14 g_{i \bar \jmath}\left(i \epsilon^{klmn}\psi_k\sigma_l\bar\psi_m +\psi_m \sigma^n \bar\psi^m\right)\chi^i \sigma_n \bar\chi^{\bar \jmath}\\
&-\frac18 \left(g_{i \bar \imath}g_{j \bar \jmath}-2 R_{i \bar \imath j \bar \jmath}\right)\chi^i \chi^j \bar \chi^{\bar \imath} \bar \chi^{\bar \jmath}\\
&- e^{\frac{K}{2}}\bigg(\bar W \psi_a \sigma^{ab}\psi_b+ W \bar \psi_a \bar \sigma^{ab}\bar\psi_b\\
&+\frac i2 \sqrt 2 D_i W \chi^i \sigma^a \bar \psi_a + \frac i2 \sqrt 2 D_{\bar \imath} \bar W \bar \chi^{\bar \imath} \bar \sigma^a \psi_a\\
&+\frac 12 \mathcal{D}_i D_j W \chi^i \chi^j+ \frac12 \mathcal{D}_{\bar \imath}D_{\bar \jmath}\bar W \bar \chi^{\bar \imath}\bar \chi^{\bar \jmath}
\bigg) -\mathcal{V},
\end{aligned}
\end{equation}
where the scalar potential is
\begin{equation}
\label{appB:eq:V}
\mathcal{V} = e^{K}\left(g^{i \bar \jmath} D_i W D_{\bar \jmath}\bar W- 3 W \bar W\right)
\end{equation}
and the covariant derivatives are defined as follows
\begin{align}
\mathcal{D}_m \chi^i &= \partial_m \chi^i + \chi^i \omega_m + \Gamma^i_{jk}\partial_m A^j \chi^k -\frac14 (K_j \partial_m A^j-K_{\bar \jmath}\partial_m \bar A^{\bar \jmath})\chi^i,\\
\mathcal{D}_m \psi_n &= \partial_m \psi_n + \psi_n \omega_m +\frac14 (K_j \partial_m A^j -  K_{\bar \jmath}\partial_m \bar A^{\bar \jmath})\psi_n,\\
D_i W &= W_i + K_i W,\\
\mathcal{D}_i D_j W &= W_{ij} + K_{ij}W+K_i D_j W+K_j D_i W- K_i K_jW-\Gamma^k_{ij}D_k W.
\end{align}

This Lagrangian is invariant under K\"ahler transformations
\begin{align}
K &\rightarrow K + \mathcal{F} + \bar{\mathcal{F}},\\
\chi^i &\rightarrow e^{\frac i2 \text{Im} \mathcal{F}}\chi^i,\\
\psi_n &\rightarrow e^{-\frac i2 \text{Im} \mathcal{F}}\psi_n,\\
W &\rightarrow e^{-\mathcal{F}}W,\\
D_i W&\rightarrow e^{-\mathcal{F}}D_iW,
\end{align}
where $\mathcal{F}(\Phi)$ is a generic holomorphic function. Notice that in supergravity, on the contrary to the case of global supersymmetry, the fermions and the superpotential have to transform under K\"ahler transformations, in order for the complete Lagrangian to be K\"ahler invariant.  In particular, they are sections of a line bundle over the K\"ahler manifold which, in turns, is restricted to be K\"ahler--Hodge.

Analytic isometries of the scalar manifold can be gauged. To each gauged analytic isometry it is associated a holomorphic Killing vector
\begin{equation}
\begin{aligned}
{\cal X}^{(a)} \,&=\, {\cal X}^{i (a)}(A^j) \frac{\partial}{\partial A^i},\\
\bar{\cal X}^{(a)} \,&=\, \bar{\cal X}^{\bar \imath (a)}(\bar A^{\bar \jmath}) \frac{\partial}{\partial \bar A^{\bar \imath}}
\end{aligned}
\end{equation}
and a real prepotential function $\mathcal{P}^{(a)}(A,\bar A)$, satisfying the differential equations
\begin{equation}
\begin{aligned}
g_{i \bar \jmath}\bar {\cal X}^{ \bar \jmath (a)}  &= i \frac{\partial \mathcal{P}^{(a)}}{\partial A^i},\\
g_{i\bar \jmath} {\cal X}^{i (a) }  &= -i   \frac{\partial \mathcal{P}^{(a)}}{\partial \bar A^{\bar \jmath}}.
\end{aligned}
\end{equation}
The Killing vectors close the algebra of the isometry group $\mathcal{G}$
\begin{equation}
\begin{aligned}
[{\cal X}^{(a)}, {\cal X}^{(b)}] \, &= \, - f^{abc} {\cal X}^{(c)},\\
[\bar{\mathcal{X}}^{(a)}, \bar{\cal X}^{(b)}] \,&=\, - f^{abc} \bar{\cal X}^{(c)},\\
[\mathcal{X}^{(a)},\bar{\mathcal{X}}^{(b)}]\, &=\, 0,
\end{aligned}
\end{equation}
where Latin indices between parenthesis run over the dimension of $\mathcal{G}$.
In order to implement gauge invariance in the model, a counterterm function $\Gamma$ is introduced such that the gauge transformation of the combination $K + \Gamma$ has the form of a K\"ahler transformation. The explicit expression for $\Gamma$ can be found in \cite{WessBagger} but, in the Wess--Zumino gauge, it reduces to
\begin{equation}
\Gamma = V^{(a)}\mathcal{P}^{(a)} + \frac12 g_{i\jmath}\mathcal{X}^{i(a)}\mathcal{X}^{i(b)}V^{(a)}V^{(b)}.
\end{equation}
Its components in the Wess--Zumino gauge are
\begin{align}
\Gamma| &= 0,\\
\mathcal{D}_\alpha \Gamma| &= \bar{\mathcal{D}}_{\dot \beta} \Gamma| = \mathcal{D}_\alpha \mathcal{D}_\beta \Gamma| = \bar{\mathcal{D}}_{\dot \alpha} \bar{\mathcal{D}}_{\dot \beta} \Gamma|=0,\\
\mathcal{D}_\alpha \bar{\mathcal{D}}_{\dot \alpha}\Gamma| &= - \sigma^m_{\alpha \dot \alpha}v_m^{(a)}\mathcal{P}^{(a)},\\
\mathcal{D}_\alpha \bar{\mathcal{D}}^2 \Gamma| &=4i \lambda_\alpha^{(a)}\mathcal{P}^{(a)}+2i (\psi_c \sigma^m \bar \sigma^c)^\beta \epsilon_{\alpha \beta}\,v_m^{(a)}\mathcal{P}^{(a)}\\
\nonumber&+2i\sqrt 2 g_{i \bar \jmath}\,\mathcal{X}^{i\, (a)}v_m^{(a)}(\bar \chi^{\bar \jmath}\bar \sigma^m)^\beta \epsilon_{\alpha \beta},\\
\mathcal{D}^2 \bar{\mathcal{D}}_{\dot \alpha}\Gamma| &= -4i \bar \lambda_{\dot \alpha}^{(a)}\mathcal{P}^{(a)}+2\sqrt 2 i g_{i \bar \jmath}\,\bar{\mathcal{X}}^{\bar \jmath \, (a)}v_m^{(a)}\chi^{i\, \beta}\sigma^m_{\beta \dot \alpha},\\
\nonumber
\mathcal{D}^2\bar{\mathcal{D}}^2 \Gamma| &= 8 D^{(a)}\mathcal{P}^{(a)}-8 i e^m_c \mathcal{D}_m v^{c\,(a)}\mathcal{P}^{(a)}+\frac{16}{3}v_c^{(a)}\mathcal{P}^{(a)}b^c\\
\nonumber
&+16 g_{i \bar \jmath}\,\mathcal{D}_m \bar A^{\bar \jmath}\mathcal{X}^{i\, (a)}v^{m\,(a)}-4g_{i \bar \jmath}\, \mathcal{X}^{i\, (a)}\bar{\mathcal{X}}^{\bar \jmath\, (b)}v_m^{(a)}v^{m\, (b)}\\
&-4 \bar \psi_c \bar \sigma^c \lambda^{(a)}\mathcal{P}^{(a)}+4 \bar \lambda^{(a)}\bar \sigma^c \psi_c \mathcal{P}^{(a)}+4 \psi^b \sigma^c \bar \psi_b v_c^{(a)}\mathcal{P}^{(a)}\\
\nonumber
&+ 8 \sqrt 2 g_{i \bar \jmath}\, \bar{\mathcal{X}}^{\bar \jmath\,(a)}(\lambda^{(a)}\chi^i)+8 \sqrt 2 g_{i \bar \jmath}\, {\mathcal{X}}^{i\,(a)}(\bar\lambda^{(a)}\bar\chi^{\bar \jmath})\\
\nonumber
&+4 \sqrt 2 g_{i \bar \jmath}\,\bar{\mathcal{X}}^{\bar \jmath\, (a)}v_m^{(a)}(\chi^i \sigma^c \bar \sigma^m \psi_c)-8 \sqrt 2 g_{i \bar \jmath} \mathcal{X}^{i\, (a)} v^{m\, (a)}e^c_m (\bar \psi_c \bar \chi^{\bar \jmath})\\
\nonumber
&+8 \mathcal{P}^{(a)}_{i \bar \jmath}v^{(a)}_m (\bar \chi^{\bar \jmath}\bar\sigma^m \chi^i)
\end{align}
The gauge variations of the quantities $K$, $\Gamma$ and $W$ are
\begin{align}
\delta K &= \Lambda^{(a)}\mathcal{F}^{(a)}+\bar{\Lambda}^{(a)}\bar{\mathcal{F}}^{(a)}-i\left(\Lambda^{(a)}-\bar{\Lambda^{(a)}}\right)\mathcal{P}^{(a)},\\
\delta \Gamma &= i\left(\Lambda^{(a)}-\bar{\Lambda^{(a)}}\right)\mathcal{P}^{(a)},\\
\delta W &= \Lambda^{(a)}\mathcal{X}^{(a)}W = -\Lambda^{(a)}\mathcal{F}^{(a)}W,
\end{align}
where the holomorphic function
\begin{equation}
\mathcal{F}^{(a)} = \mathcal{X}^{(a)}K+i \mathcal{P}^{(a)}
\end{equation}
has been defined. The combination $K+\Gamma$ therefore transforms as
\begin{equation}
\delta(K+\Gamma) = \Lambda^{(a)}\mathcal{F}^{(a)}+\bar{\Lambda}^{(a)}\bar{\mathcal{F}}^{(a)}
\end{equation}
and, after the replacement
\begin{equation}
K \to K + \Gamma,
\end{equation}
the Lagrangian \eqref{eq:appB:GENSUGRAL}, with the inclusion of the vector multiplet kinetic term, becomes
\begin{equation}
\begin{aligned}
\label{eq:appB:LSUGRAG}
\mathcal{L} &= \int d^2 \Theta\, 2 \mathcal{E} \left[\frac38 (\bar{\mathcal{D}}-8\mathcal{R})e^{-\frac13(K+\Gamma)}\right]\\
&+\int d^2\Theta\,2 \mathcal{E} \left(\frac{1}{16}\mathcal{W}^\alpha \mathcal{W}_\alpha + W\right)+c.c.
\end{aligned}
\end{equation}
Its total gauge variation is
\begin{equation}
\begin{aligned}
\delta \mathcal{L} &= \int d^2 \Theta \, 2 \mathcal{E}\, \left[-\frac18 (\bar{\mathcal{D}}^2-8\mathcal{R})\bigg[\Lambda^{(a)}\mathcal{F}^{(a)}+\bar{\Lambda}^{(a)}\bar{\mathcal{F}}^{(a)}\right]e^{-\frac13(K+\Gamma)}\\
&+\Lambda^{(a)}\mathcal{X}^{(a)}W\bigg]+c.c.
\end{aligned}
\end{equation}
and this can be absorbed with a super-Weyl transformation, which is a symmetry of the superspace Lagrangian. The component form of the gauge invariant Lagrangian \eqref{eq:appB:LSUGRAG} and additional details can be found in \cite{WessBagger}.

\section{Alternative K\"ahler-invariant supergravity}
In this section the necessary ingredients to construct the models contained in chapter \ref{c4:altsugra} are given.
Considering a real linear superfield $L$, such that
\begin{equation}
\begin{aligned}
L&=L^*,\\
(\overline {\cal D}^2 - 8 {\cal R} ) \, L &= 0
\end{aligned}
\end{equation}
 and a generic real function $K(\Phi,\bar \Phi)$ of a set of chiral superfields $\Phi^i$, it is possible to construct the chiral superfield
\begin{equation}
\label{appBSigma}
\begin{aligned}
\Sigma &= \left(\bar{\mathcal{D}}^2-8\mathcal{R}\right) LK\\
& =\left(\bar{\mathcal{D}}^2-8\mathcal{R}\right) LK|+\Theta^\alpha \mathcal{D}_\alpha \left(\bar{\mathcal{D}}^2-8\mathcal{R}\right) LK|\\
&-\frac14 \Theta^2 \mathcal{D}^2\left(\bar{\mathcal{D}}^2-8\mathcal{R}\right) LK|
\end{aligned}
\end{equation}
In the case in which $L$ is the composite object
\begin{equation}
L = \mathcal{D}^\alpha \left(\bar{\mathcal{D}}^2-8\mathcal{R}\right)\mathcal{D}_\alpha \left[\frac{X\bar X}{\mathcal{D}^2 X \bar{\mathcal{D}}^2 \bar X}\right],
\end{equation}
where $X^2=0$ is assumed, its components in the unitary gauge are
\begin{align}
L|&=1,\\
\mathcal{D}_\alpha L| &= -\frac i2 \sigma^a_{\alpha \dot \alpha} \bar \psi_a^{\dot \alpha},\\
\bar{\mathcal{D}}_{\dot \alpha}L|&=\frac i2 \psi_a^\alpha \sigma^a_{\alpha \dot \alpha},\\
\mathcal{D}_\alpha \bar{\mathcal{D}}_{\dot \alpha}L| &= -\frac23 b_{\alpha \dot \alpha}+\frac12 \psi_{a\, \alpha} \left(\bar \psi_b \bar \sigma^b \sigma^a\right)_{\dot \alpha}+\frac12 \sigma^a_{\alpha \dot \beta} \bar \psi_b^{\dot \beta} \psi_a^\beta\sigma^b_{\beta \dot \alpha},\\
 -\frac12 \left[\mathcal{D}_\alpha, \bar{\mathcal{D}}_{\dot \alpha}\right] L|&= \frac 23 b_{\alpha \dot \alpha}-\frac14 \psi_{a\, \alpha} \left(\bar \psi_b \bar \sigma^b \sigma^a\right)_{\dot \alpha}-\frac14 \left(\sigma^a \bar \sigma^b \psi_b\right)_\alpha \bar \psi_{a\, \dot \alpha}\\
\nonumber
&-\frac12 \sigma^a_{\alpha \dot \beta} \bar \psi_b^{\dot \beta} \psi_a^\beta\sigma^b_{\beta \dot \alpha}\\
\mathcal{D}^2 \bar{\mathcal{D}}_{\dot\alpha}L|&=- \frac i3 \bar{\psi}^a_{\dot \alpha} b_a-\frac 43 i \bar M \psi_a^\alpha \sigma^a_{\alpha \dot\alpha}-2i (\bar \psi_b \bar \sigma^{ba})_{\dot \alpha} b_a\\
\nonumber
&+\frac 83 \bar \psi_{ab\, \dot \beta} {\bar \sigma^{ab\, \dot \beta}}_{\phantom{aaa}\dot \alpha}+2 e^m_a \left(\mathcal{D}_m \bar \psi_b \bar \sigma^b \sigma^a\right)_{\dot \alpha}\\
\nonumber
&+i(\bar \psi_c \bar \psi_d)(\psi_b \sigma^b \bar \sigma^d \sigma^c)_{\dot \alpha}-i (\psi_d \sigma^b \bar \psi_c)(\bar \psi_b \bar \sigma^d \sigma^c)_{\dot \alpha}.
\end{align}
Those of $\Sigma$ are then
\begin{equation}
 \left(\bar{\mathcal{D}}^2-8\mathcal{R}\right) LK| = \sqrt 2 i K_{\bar \imath} \psi_a \sigma^a \bar \chi^{\bar \imath}- 4 K_{\bar \imath} \bar F^{\bar \imath}+ 2 K_{\bar \imath \bar \jmath} \bar \chi^{\bar \imath} \bar \chi^{\bar \jmath},
\end{equation}
\begin{equation}
\begin{aligned}
\mathcal{D}_\alpha  &\left(\bar{\mathcal{D}}^2-8\mathcal{R}\right) LK|=-2\sqrt 2  b_{\alpha \dot \alpha} K_{\bar \imath} \bar \chi^{\bar \imath\, \dot \alpha}+6i \sigma^a_{\alpha \dot \beta} \bar\psi_a^{\dot \beta} K_{\bar \imath} \bar F^{\bar \imath}\\
&-2 K_{\bar \imath} (\psi_b \sigma^b \bar \sigma^a)^\beta \epsilon_{\alpha \beta} e^m_a \mathcal{D}_m \bar A^{\bar \imath}-4 K_{\bar \imath} (\psi_b \sigma^a \bar \sigma^b)^\beta \epsilon_{\alpha \beta} e^m_a \mathcal{D}_m \bar A^{\bar \imath}\\
&+4 \sqrt 2 i \epsilon_{\alpha \beta}K_{\bar \imath} e^m_c (\mathcal{D}_m \bar \chi^{\bar \imath} \bar \sigma^c)^\beta+\sqrt 2 K_{\bar \imath} \psi_{a\,\alpha}(\bar \psi_b \bar \sigma^b \sigma^a \bar \chi^{\bar \imath})\\
&+ \sqrt 2 K_{\bar \imath} (\psi_b \sigma^b \bar \sigma^a)^\beta \epsilon_{\alpha \beta}(\bar \psi_a \bar \chi^{\bar \imath})+2\sqrt 2 K_{\bar \imath} (\psi_b \sigma^a \bar \sigma^b)^\beta \epsilon_{\alpha \beta}(\bar \psi_a \bar \chi^{\bar \imath})\\
&+ \sqrt 2 K_{\bar \imath} \sigma^a_{\alpha \dot \alpha} \bar \psi_b^{\dot \alpha}(\psi_a \sigma^b \bar \chi^{\bar \imath})+2i K_{i \bar \jmath} \chi^i_\alpha (\psi_a \sigma^a \bar \chi^{\bar \jmath})-4\sqrt 2 K_{i \bar \jmath}\chi^i_\alpha \bar F^{\bar \jmath}\\
&-i K_{\bar \imath \bar \jmath} \sigma^a_{\alpha \dot \beta} \bar \psi_a^{\dot \beta}(\bar \chi^{\bar \imath} \bar \chi^{\bar \jmath})+4 \sqrt 2 i K_{\bar \imath \bar \jmath}(\bar \chi^{\bar \imath}\bar \sigma^a)^\beta \epsilon_{\alpha \beta}e^m_a \mathcal{D}_m \bar A^{\bar \jmath}\\
&-4 i K_{\bar \imath \bar \jmath}(\bar \chi^{\bar \imath}\bar \sigma^a)^\beta \epsilon_{\alpha \beta}(\bar \psi_a \bar \chi^{\bar \jmath})+2\sqrt 2 K_{i \bar \imath \bar \jmath}\chi^i_\alpha (\bar \chi^{\bar \imath} \bar \chi^{\bar \jmath}),
\end{aligned}
\end{equation}
\begin{align}
\nonumber
\mathcal{D}^2 &\left(\bar{\mathcal{D}}^2-8\mathcal{R}\right) LK| = 16 K_{\bar \imath \bar \jmath} \mathcal{D}_m \bar A^{\bar \imath} \mathcal{D}^m \bar A^{\bar \jmath}+ 16 K_{i \bar \jmath} F^i \bar F^{\bar \jmath}+\frac{48}{3}K_{\bar \imath} \bar F^{\bar \imath } \bar M\\
\nonumber
&+ 4 \sqrt 2 K_{\bar \imath} e^m_a \mathcal{D}_m \bar \psi_b \bar \sigma^b \sigma^a \bar \chi^{\bar \imath}+ 8 \sqrt 2 K_{\bar \imath} \bar \psi_{ab} \bar \sigma^{ab} \bar \chi^{\bar \imath}- 4 \sqrt 2 i K_{\bar \imath}(\bar \psi_b \bar \sigma^b \sigma^a \bar \chi^{\bar \imath})b_a \\
\nonumber
&-4 \sqrt 2 i \bar M K_{\bar \imath}(\psi_a \sigma^a \bar \chi^{\bar \imath})-8 K_{i \bar \jmath}(\chi^i \sigma^a \bar \chi^{\bar \jmath})b_a+ 4i K_{\bar \imath} \psi_b \sigma^c \bar \sigma^a \sigma^b \bar \psi_c e^m_a \mathcal{D}_m \bar A^{\bar \imath}\\
\nonumber
&-4 \sqrt 2 i K_{i \bar \jmath} F^i (\psi_a \sigma^a \bar \chi^{\bar \jmath})- 4 \sqrt 2 K_{i \bar \jmath}(\psi_a \sigma^a \bar \sigma^b \chi^i)e^m_b \mathcal{D}_m \bar A^{\bar \jmath}- 8 \bar M K_{\bar \imath \bar \jmath} (\bar \chi^{\bar \imath }\bar \chi^{\bar \jmath})\\
\nonumber
&+4\sqrt 2 K_{\bar \imath \bar \jmath} (\bar \chi^{\bar \imath} \bar \sigma^a \sigma^b \bar \psi_b)e^m_a \mathcal{D}_m \bar A^{\bar \jmath}-8 \sqrt 2 K_{\bar \imath}(\bar \psi^a e^m_a \mathcal{D}_m \bar \chi^{\bar \imath})\\
\nonumber
&+12 \sqrt 2 i K_{i \bar \jmath} \bar F^{\bar \jmath}(\chi^i \sigma^a \bar \psi_a)+ 4 \sqrt 2 K_{\bar \imath}e^m_a (\mathcal{D}_m \bar \chi^{\bar \imath} \bar \sigma^a \sigma^b \bar \psi_b)\\
\nonumber
&+16 K_{\bar \imath}e^a_m \mathcal{D}^m(e^n_a \mathcal{D}_n \bar A^{\bar \imath})-4 K_{\bar \imath} \bar F^{\bar \imath}(\bar \psi_a \bar \sigma^a \sigma^b \bar \psi_b)\\
\nonumber
&-8 \sqrt 2 K_{\bar \imath}e^m_a \mathcal{D}_m(\bar \psi^a \bar \chi^{\bar \imath})-8i K_{\bar \imath}(\psi_b \sigma^a \bar \psi^b)e^m_a \mathcal{D}_m \bar A^{\bar \imath}+ 8 K_{\bar \imath} \bar F^{\bar \imath}(\bar \psi_a \bar \psi^a)\\
&-16 \sqrt 2 K_{\bar \imath \bar \jmath}e^m_a \mathcal{D}_m \bar{A}^{\bar \imath} (\bar \psi^a \bar \chi^{\bar \jmath})+16 i K_{i \bar \jmath}e^m_a \mathcal{D}_m \bar \chi^{\bar \imath} \bar \sigma^a \chi^j\\
\nonumber
&-8\sqrt 2 K_{i \bar \jmath}(\psi_c \sigma^a \bar \sigma^c \chi^i) e^m_a \mathcal{D}_m \bar A^{\bar \jmath}-8 K_{\bar \imath ij}\bar F^{\bar \imath}(\chi^i \chi^j)\\
\nonumber
&-8 K_{i \bar \imath\bar \jmath}F^i (\bar \chi^{\bar \imath} \bar \chi^{\bar \jmath})+16 i K_{i \bar \imath \bar \jmath}(\bar \chi^i \bar \sigma^a \chi^i)e^m_a \mathcal{D}_m \bar A^{\bar \jmath}\\
\nonumber
&+2 \sqrt 2 i K_{\bar \imath}(\bar \psi_c \bar \psi_d)(\psi_b \sigma^b \bar \sigma^d \sigma^c \bar \chi^{\bar \imath})-2 \sqrt 2 i K_{\bar \imath}(\psi_d \sigma^b \bar \psi_c)(\bar \psi_b \bar \sigma^d \sigma^c \bar \chi^{\bar \imath})\\
\nonumber
&-2 \sqrt i K_{\bar \imath}(\psi_b \sigma^c \bar \sigma^a \sigma^b \bar \psi_c)(\bar \psi_a \bar \chi^{\bar \imath})+4 \sqrt 2 i K_{\bar \imath}(\psi_b \sigma^c \bar \psi^b)(\bar \psi_c \bar \chi^{\bar \imath})\\
\nonumber
&-4 K_{i \bar \jmath}(\psi_c \chi^i)(\bar \psi_b \bar \sigma^b \sigma^c \bar \chi^{\bar \jmath})+4 K_{i \bar \jmath}(\chi^i \sigma^b \bar \psi_c)(\psi_b \sigma^c \bar \chi^{\bar \jmath})\\
\nonumber
&+4 K_{i \bar \jmath}(\psi_c \sigma^c \bar \sigma^a \chi^i)(\bar \psi_a \bar \chi^{\bar \jmath})-4 K_{\bar \imath \bar \jmath}(\bar \chi^{\bar \imath}\bar \sigma^a \sigma^c \bar \psi_c)(\bar \psi_a \bar \chi^{\bar \jmath})\\
\nonumber
&+ 8K_{\bar \imath \bar \jmath}(\bar \psi_a \bar \chi^{\bar \imath})(\bar \psi^a \bar \chi^{\bar \jmath})+ 8 K_{i \bar \jmath}(\psi_c \sigma^a \bar \sigma^c \bar \chi^i)(\bar \psi_a \bar \chi^{\bar \jmath})\\
\nonumber
&+2\sqrt 2 i K_{ij \bar \imath}(\chi^i \chi^j)(\psi_c \sigma^c \bar \chi^{\bar \imath})-2\sqrt 2 i K_{\bar \imath ij}(\chi^i \sigma^a \bar \psi_a)(\bar \chi^{\bar \imath}\bar \chi^{\bar \jmath})\\
\nonumber
&-8 \sqrt 2 i K_{\bar \imath \bar \jmath i}(\bar \chi^{\bar \imath} \bar \sigma^a \chi^i)(\bar \psi_a \bar \chi^{\bar \jmath})+4 K_{\bar \imath \bar \jmath i j}(\chi^i \chi^j)(\bar \chi^{\bar \imath}\bar \chi^{\bar \jmath}).
\end{align}

In order to calculate the Lagrangian in which analytic isometries of the scalar manifold are gauged, the components of the chiral superfield 
\begin{equation}
\begin{aligned}
\Delta &= (\bar{\mathcal{D}}^2-8R)L\Gamma\\
&=(\bar{\mathcal{D}}^2-8R)L\Gamma|+ \Theta^\alpha \mathcal{D}_\alpha (\bar{\mathcal{D}}^2-8R)L\Gamma|-\frac14 \Theta^2 \mathcal{D}^2 (\bar{\mathcal{D}}^2-8R)L\Gamma|
\end{aligned}
\end{equation}
are needed. In the Wess--Zumino gauge they are
\begin{align}
(\bar{\mathcal{D}}^2-8R)L\Gamma| &= 0,\\
\mathcal{D}_\alpha (\bar{\mathcal{D}}^2-8R)L\Gamma| &= 4 i \lambda_\alpha^{(a)}\mathcal{P}^{(a)}-2 \sqrt 2 g_{i \bar \jmath}\, \mathcal{X}^{i {(a)}}v_m^{(a)} \sigma_{\alpha \dot \beta}^m \bar \chi^{\bar \jmath \dot\beta}\\
\nonumber
&-2i \sigma_{\alpha \dot \alpha}^m \bar \sigma^{c\, \dot\alpha \beta}\psi_{c\,\beta} v_m^{(a)}\mathcal{P}^{(a)}+2i \sigma_{\alpha \dot \alpha}^c \bar \sigma^{m\, \dot\alpha \beta}\psi_{c\,\beta} v_m^{(a)}\mathcal{P}^{(a)}\\
\nonumber
\mathcal{D}^2(\bar{\mathcal{D}}^2-8R)L\Gamma| &= 8 D^{(a)}\mathcal{P}^{(a)} + 16 g_{i \bar \jmath}\, v^{(a)m}\mathcal{D}_m \bar A^{\bar \jmath}\mathcal{X}^{i (a)}\\
\nonumber
&-4 g_{i \bar \jmath}\, \mathcal{X}^{i (a)} \bar{\mathcal{X}}^{\bar \jmath (b)} v_m^{(a)}v^{m (b)}
-8i e^m_c \mathcal{D}_m v^{c(a)}\mathcal{P}^{(a)}\\
\nonumber
&+ 8 (\lambda^{(a)}\sigma^c \bar \psi_c)\mathcal{P}^{(a)}+4 (\psi^b \sigma^c \bar \psi_b)v_c^{(a)}\mathcal{P}^{(a)}\\
\nonumber
&+2 (\psi_c \sigma^b \bar \sigma^m \sigma^c \bar \psi_b) v_m^{(a)}\mathcal{P}^{(a)}\\
&+2 \sqrt 2 g_{i \bar \jmath}\,\bar{\mathcal{X}}^{\bar \jmath (a)}v_m^{(a)}(\psi_c \sigma^c \bar \sigma^m \chi^i)\\
\nonumber
&+2 \sqrt 2 g_{i \bar \jmath}\,\mathcal{X}^{i(a)} v_m^{(a)}(\bar \chi^{\bar \jmath}\bar \sigma^m  \sigma^b \bar \psi_b)\\
\nonumber
&+4 \sqrt 2 g_{i \bar \jmath} \,\bar{\mathcal{X}}^{\bar \jmath (a)}v_m^{(a)}(\chi^i \sigma^c \bar \sigma^m \psi_c)\\
\nonumber
&-8 \sqrt 2 g_{i \bar \jmath}\,\mathcal{X}^{i(a)} v^{(a)m}e^c_m (\bar \psi_c \bar \chi^{\bar \jmath})+ 8 \sqrt 2 g_{i \bar \jmath}\, \bar{\mathcal{X}}^{\bar \jmath (a)} (\lambda^{(a)}\chi^i)\\
\nonumber
&+ 8 \sqrt 2 g_{i \bar \jmath}\, \mathcal{X}^{i(a)} (\bar \lambda^{(a)}\bar \chi^{\bar \jmath})+ 8 \mathcal{P}^{(a)}_{i \bar \jmath}v_m^{(a)}(\bar \chi^{\bar \jmath}\bar \sigma^m \chi^i)
\end{align}
The gauge invariant form of the matter couplings in superspace is then 
\begin{equation}
\begin{aligned}
\mathcal{L} = -\frac 18 \int d^2 \Theta 2 \mathcal{E} (\bar{\mathcal{D}}^2-8\mathcal{R})[L(K+\Gamma)]+c.c.
\end{aligned}
\end{equation}
where the vector $V$ in $\Gamma$ is rescaled of a factor two, $V\to 2 V$, for convenience. Its bosonic sector is
\begin{equation}
e^{-1}\mathcal{L}_{bos}=-g_{i \bar \jmath}\, \mathcal{D}_m A^i \bar{\mathcal{D}}^m \bar A^{\bar \jmath}+g_{i \bar \jmath}\, F^i \bar F^{\bar \jmath}+D^{(a)}\mathcal{P}^{(a)},
\end{equation}
where $\mathcal{D}_m A^i$ is the gauge covariant derivative
\begin{equation}
\mathcal{D}_m A^i = \partial_m A^i - v_m^{(a)} \mathcal{X}^{i(a)}. 
\end{equation}
\end{appendices}

\newpage\thispagestyle{empty}$ $ 
\newpage
\pagestyle{acknowledgement}

\section*{\center Acknowledgements}

\vspace{1cm}
I would like to thank my supervisor, Gianguido Dall'Agata, for his essential assistance and support during the three years of my Ph.D..\\

\noindent
I would like to thank, on a personal level, Fotis Farakos for his precious help. \\

\noindent
I would like to thank the group at the KU Leuven, especially Antoine van Proeyen, for the kind hospitality when I was visiting in fall 2017. \\

\noindent
I would like to thank all the friends, collaborators and colleagues I have met in these years. \\

\noindent
I would like to thank Sofia and my family for their constant presence.
\clearpage\null\thispagestyle{empty}

\pagestyle{biblio}
\bibliographystyle{jhep}
\phantomsection
\providecommand{\href}[2]{#2}\begingroup\raggedright\endgroup

\end{document}